\documentclass[12pt,a4paper,dvips]{article}
\usepackage{graphicx,epsfig}
\usepackage{hhline}

\usepackage{amsmath,amssymb}
\usepackage{times}
\usepackage[varg]{txfonts}
\DeclareMathAlphabet{\mathbold}{OML}{txr}{b}{it}

\usepackage{array,multirow,dcolumn}
\usepackage[mathlines,displaymath]{lineno}
\usepackage{rotating}

\usepackage[numbers,square,comma,sort&compress]{natbib}
\usepackage{hypernat}
\usepackage{textcomp}
\newcommand{\tablecaption}{%
\caption}

\newcolumntype{.}{D{.}{.}{-1}}
\newcolumntype{-}{D{-}{-}{-1}}

\usepackage[dvipsnames]{color}
\definecolor{rltred}{rgb}{0.75,0,0}
\definecolor{rltgreen}{rgb}{0,0.5,0}
\definecolor{rltblue}{rgb}{0,0,0.5}

\newcounter{pdfadd}    
\usepackage[hyperindex,bookmarks,bookmarksnumbered,breaklinks,a4paper,unicode]{hyperref}
\hypersetup{%
  pdftitle        = {Measurement of the inclusive ep Scattering Cross Section
 at low Q2 and x at HERA},
  urlcolor        = rltblue,       
  urlbordercolor  = 0 0 0.5,
  filecolor       = rltblue,       
  filebordercolor = 0 0 0.5,
  linkcolor       = rltred,        
  linkbordercolor = 0.75 0 0,
  citecolor       = rltgreen,      
  citebordercolor = 0 0.5 0,
  pagecolor       = rltgreen,      
  pagebordercolor = 0 0.5 0,
  menucolor       = rltgreen,      
  menubordercolor = 0 0.5 0,
  colorlinks    = true,
  pdfauthor     = {H1 Collaboration},
  pdfsubject    = { },
  pdfkeywords   = {High-Energy Physics, Particle Physics, Proton Structure, DIS}
}

\newlength{\dinwidth}
\newlength{\dinmargin}
\newcommand{\empz}{\mbox{$E$$-$$P_z$}}
\newcommand{\qqe}{\mbox{$Q^2_e$}}
\newcommand{\qqs}{\mbox{$Q^2_\Sigma$}}
\newcommand{\ys}{\mbox{$y_\Sigma$}}
\newcommand{\xs}{\mbox{$x_\Sigma$}}
\newcommand{\ee}{\mbox{$E_e^{\prime}$}}
\newcommand{\thetae}{\mbox{$\theta_e$}}
\newcommand{\thetah}{\mbox{$\theta_h$}}
\newcommand{\gp}{\mbox{$\gamma p$}}
\newcommand{\pt}{\mbox{$P_\perp$}}
\newcommand{\piz}{\mbox{$\pi^0$}}
\newcommand{\Pth}{\mbox{$P_{\perp}^h$}}

\newcommand{\electron}{\mbox{positron}}

\newcommand{\Fig}{\mbox{figure}}
\newcommand{\Tab}{\mbox{table}}
\newcommand{\Eq}{\mbox{equation}}
\newcommand{\Sec}{\mbox{section}}

\newcommand{\FFig}{\mbox{Figure}}

\newcommand{\EEq}{\mbox{Equation}}
\newcommand{\SSec}{\mbox{Section}}

\newcommand{\Figs}{\mbox{figures}}
\newcommand{\Tabs}{\mbox{tables}}
\newcommand{\Eqs}{\mbox{equations}}
\newcommand{\Secs}{\mbox{sections}}

\newcommand{\FFigs}{\mbox{Figures}}

\renewcommand{\perp}{{\rm T}}

\newcommand{\MB}{\mbox{NVX}}
\newcommand{\SVX}{\mbox{SVX}}

\newcommand{\thetamaxsvx}{178^{\circ}}
\newcommand{\thetamaxmb}{176.5^{\circ}}

\newcommand{\permil}{\textperthousand}

\def\dof{\mathop{n_{\rm dof}}\nolimits}

\def\DeltaM{\mathop{ w_{i,e}}\nolimits}
\def\DeltaMK{\mathop{ w_{k,e}}\nolimits}

\newcommand\T{\rule{0pt}{2.3ex}}
\newcommand\B{\rule[-1.ex]{0pt}{0pt}}

\begin{document}

\makeatletter \def\NAT@space{} \makeatother

\begin{titlepage}
 
\noindent
DESY 08-171 \hfill ISSN 0418-9833 \\
August 2009

\vspace*{3.5cm}

\begin{center}
\begin{Large}

{\bfseries Measurement of the Inclusive $\mathbold{ep}$ Scattering
 Cross Section \\
  at Low $\mathbold{Q^2}$ and $\mathbold{x}$ at HERA}

\vspace*{2cm}

H1 Collaboration

\end{Large}
\end{center}

\vspace*{2cm}

\begin{abstract} \noindent
A  measurement of the inclusive $ep$ scattering cross section is presented 
in the region of low momentum transfers,
$0.2$\,GeV$^2$ $\leq Q^2 \leq 12$\,GeV$^2$, and low Bjorken $x$,
$5 \cdot 10^{-6} \lesssim x \lesssim 0.02$.
The result is  based on two data sets collected in dedicated runs by the H1
Collaboration at HERA at beam energies of $27.6$\,GeV and $920$\,GeV for positrons
and protons, respectively. A combination with data previously published by H1
leads to a  cross section measurement of  a few percent accuracy.
A kinematic reconstruction method exploiting radiative $ep$ 
events extends the measurement to lower $Q^2$ and larger $x$.
The data are compared with theoretical models which apply to
the transition region from
photoproduction to deep
inelastic scattering.
\end{abstract}

\vspace*{1.5cm}

\begin{center}
{\slshape Accepted by EPJ C}
\end{center}

\end{titlepage}

\begin{flushleft}

F.D.~Aaron$^{5,49}$,           
C.~Alexa$^{5}$,                
V.~Andreev$^{25}$,             
B.~Antunovic$^{11}$,           
S.~Aplin$^{11}$,               
A.~Asmone$^{33}$,              
A.~Astvatsatourov$^{4}$,       
S.~Backovic$^{30}$,            
A.~Baghdasaryan$^{38}$,        
E.~Barrelet$^{29}$
% ...

\bigskip{\it
% ...
$ ^{2}$ Vinca  Institute of Nuclear Sciences, Belgrade, Serbia \\
% ...
$ ^{4}$ Inter-University Institute for High Energies ULB-VUB, Brussels;
          Universiteit Antwerpen, Antwerpen; Belgium$^{ c}$ \\
 $ ^{5}$ National Institute for Physics and Nuclear Engineering (NIPNE) ,
          Bucharest, Romania \\
% ...
$ ^{11}$ DESY, Hamburg, Germany \\
% ...
$ ^{25}$ Lebedev Physical Institute, Moscow, Russia$^{ e}$ \\
% ...
$ ^{29}$ LPNHE, Universit\'{e}s Paris VI and VII, IN2P3-CNRS,
          Paris, France \\
 $ ^{30}$ Faculty of Science, University of Montenegro,
          Podgorica, Montenegro$^{ e}$ \\
% ...
$ ^{33}$ Dipartimento di Fisica Universit\`a di Roma Tre
          and INFN Roma~3, Roma, Italy \\
% ...
$ ^{38}$ Yerevan Physics Institute, Yerevan, Armenia \\
% ...
$ ^{49}$ Also at Faculty of Physics, University of Bucharest,
          Bucharest, Romania \\
% ...
$ ^c$ Supported by FNRS-FWO-Vlaanderen, IISN-IIKW and IWT
      and  by Interuniversity
Attraction Poles Programme,
      Belgian Science Policy \\
% ...
$ ^e$ Supported by the Deutsche Forschungsgemeinschaft \\
% ...
}
\end{flushleft}
 

\newpage

\def\frchi{155.3}
\def\frnpt{149}
\def\frnpa{  5}
\def\frdzv{ 0.75}
\def\frdze{ 0.03}
\def\frdov{ 0.052}
\def\frdoe{ 0.002}
\def\frdtv{ 1.08}
\def\frdte{ 0.00}
\def\frdev{-1.16}
\def\frdee{ 0.03}
\def\frqzv{ 0.093}
\def\frqze{ 0.010}
\def\frrrv{ 0.56}
\def\frrre{ 0.07}
\def\frdiagchi{128.0}
\def\frdiagnpt{149}
\def\frdiagnpa{  5}
\def\frdiagdzv{ 0.73}
\def\frdiagdze{ 0.04}
\def\frdiagdov{ 0.053}
\def\frdiagdoe{ 0.003}
\def\frdiagdtv{ 1.08}
\def\frdiagdte{ 0.00}
\def\frdiagdev{-1.17}
\def\frdiagdee{ 0.04}
\def\frdiagqzv{ 0.097}
\def\frdiagqze{ 0.016}
\def\frdiagrrv{ 0.64}
\def\frdiagrre{ 0.05}
\def\frsnchi{145.7}
\def\frsnnpt{142}
\def\frsnnpa{  5}
\def\frsndzv{ 0.77}
\def\frsndze{ 0.03}
\def\frsndov{ 0.052}
\def\frsndoe{ 0.002}
\def\frsndtv{ 1.08}
\def\frsndte{ 0.00}
\def\frsndev{-1.16}
\def\frsndee{ 0.02}
\def\frsnqzv{ 0.091}
\def\frsnqze{ 0.010}
\def\frsnrrv{ 0.65}
\def\frsnrre{ 0.09}
\section{Introduction\label{sec:introduction}}
Deep inelastic lepton-hadron scattering (DIS) is pivotal for the understanding
of the structure of the nucleon and of the dynamics of parton interactions.
Since the discovery of Bjorken scaling~\cite{Bloom:1969kc}
and its violation~\cite{Fox:1974ry} at fixed target experiments,
DIS measurements have made essential contributions to in the development
of the theory of strong interactions, Quantum Chromodynamics (QCD).
Major progress in the exploration of strong interactions
has been achieved at the electron\footnote{Unless
explicitly stated, the generic name ``electron'' is used throughout this paper
to denote both electron and positron.}-proton collider HERA, operating at the 
energy frontier.  Measurements performed at HERA
are essential for predictions of the physics at the forthcoming
proton-proton collider, the Large Hadron Collider (LHC).

The high centre-of-mass energy of the $ep$ scattering 
at HERA leads to  a wide kinematic range
extending to large values of the modulus of the four-momentum transfer squared,
denoted  $Q^2$, and to very small values of the Bjorken $x$ variable.
At the HERA beam energies of $E_e=27.6$\,GeV for the electron 
and $E_p=920$\,GeV for the proton, Bjorken $x$ values as small
as $10^{-4}$ ($10^{-6}$) are accessible for $Q^2$ of $10$\,GeV$^2$
($0.1$\,GeV$^2$). 

A salient feature of the 
structure function $F_2(x,Q^2)$, discovered  by the
H1~\cite{h1firstf2} and  ZEUS~\cite{zeusfirstf2} collaborations
with the very first  HERA data, is its strong rise for 
$x \rightarrow 0$.  
In terms of parton distribution functions, this can be directly 
interpreted as a strong rise of the sea quark density towards small $x$.
Similarly the increase of $F_2(x,Q^2)$ with $Q^2$ at fixed small $x$ reveals
a strongly rising behaviour of the gluon density towards low $x$. 
This is obtained in perturbative QCD (pQCD)
analyses of DIS 
data~\cite{Adloff:2003uh,Chekanov:2002pv,Pumplin:2002vw,Alekhin:2002fv}
using the derivative $\partial F_2 / \partial \ln Q^2$,
which is related to the gluon and quark densities as prescribed by the DGLAP evolution 
equations~\cite{Gribov:1972ri,Gribov:1972rt,Lipatov:1974qm,Dokshitzer:1977sg,Altarelli:1977zs}.

The DGLAP 
approach, in which only $\alpha_S  \ln Q^2$ terms are summed,
may not apply at  lowest $x$ values 
 as terms 
involving powers of $\alpha_s  \ln(1/x)$ become large.
The parton dynamics at low $x$ may  be better approximated by different evolution
equations, such as BFKL~\cite{Kuraev:1976ge,Kuraev:1977fs,Balitsky:1978ic},
CCFM~\cite{Ciafaloni:1987ur,Catani:1989yc,Catani:1989sg,Marchesini:1994wr}
or non-linear equations~\cite{Gribov:1984tu,Mueller:1985wy,Balitsky:1995ub,Kovchegov:1999yj,Braun:2000wr,Iancu:2000hn,Bartels:2004ef,Bartels:2007dm}.
The non-linear effects, arising due to the large gluon density and
corresponding for example to gluon-gluon recombination, could tame 
the rise of $F_2$  at low $x$. Further clarification of  low $x$ parton 
dynamics requires data of the highest precision, in a wide range of $x$ and $Q^2$.

For $Q^2 \lesssim 2$\,GeV$^2$, as the strong coupling constant
$\alpha_s(Q^2)$ increases, the higher order corrections to the 
perturbative expansion become large and lead to the 
breakdown of the pQCD calculations. 
Measurements at low $Q^2$ and low $x$ thus probe this transition 
in which quarks and gluons cease to be relevant degrees of 
freedom. This onset of 
soft  hadron physics is described by phenomenological, 
often QCD-inspired models  (see \cite{Donnachie:2002en} for a review).

An attractive view of virtual photon-proton scattering has been developed 
with the colour dipole model\,\cite{Nikolaev:1990ja}. It originated from 
the observation that in the proton rest frame, at low $x$ the photon may 
fluctuate into a quark-antiquark pair
with a lifetime $\propto 1/x$,
long before the interaction with the proton~\cite{Ioffe:1969kf,Gribov:1968gs}. 
Therefore the cross sections
can be expressed as a product of the square of the wavefunction
of the $q\overline{q}$ pair with a universal dipol-proton cross section.
Another phenomenological model, used here, describes $F_2(x,Q^2)$  
based on the idea of self-similarity
of the proton substructure at small $x$~\cite{fractal}.

Access to the smallest $x$ implies an extension of the
measurements to high values of the inelasticity $y$ where
the cross section becomes sensitive to the longitudinal
structure function $F_L(x,Q^2)$. This function completes the description 
of inclusive virtual photon-proton scattering, which involves transverse and
longitudinal photon polarisation states.
In the naive quark-parton model (QPM), $F_L$ is zero, while in  QCD it 
provides independent information~\cite{altmar} on the gluon
distribution and may become correspondingly large at low $x$.

This paper presents new measurements of the inclusive $ep$ cross section 
in the range
 $0.2 \leq Q^2 \leq 12$\,GeV$^2$ 
and 
$5 \cdot 10^{-6} \leq x \leq 0.02$.
The data were collected with the H1 detector in two $e^+p$ running
periods with dedicated settings of the inclusive electron triggers.
One data set (termed nominal vertex, ``\MB'') was collected in 
the year $1999$ and
corresponds to an integrated luminosity of $2.1\,$pb$^{-1}$. The other was
collected in the year $2000$, with the interaction region shifted along the proton beam
direction by $70$\,cm (termed shifted vertex, ``\SVX''), and corresponds
to $505\,$nb$^{-1}$.

Shifting the interaction region  allows detection of the scattered
electron at larger polar angles\footnote{In the H1 coordinate system 
the $z$ axis points along the outgoing proton beam direction termed
forward direction.
Therefore large electron polar angles $\theta_e$ close to $180^\circ$
correspond to very small angles with respect to the incoming 
electron direction. The coordinate system is right-handed.
The $x$ ($y$) axis is directed horizontally (vertically).}
which otherwise cannot be accessed in the main H1 apparatus
and thus provides acceptance in the region $Q^2 \lesssim 2$\,GeV$^2$.
In comparison to the previous H1 measurement with a shifted
vertex~\cite{Adloff:1997mf}, an increased precision is reached using the higher
luminosity of the new data and employing, in addition to the previous 
backward instrumentation of the H1 detector, 
an upgraded Backward Silicon Tracker (BST). 
The vertex reconstruction using the electron track in the BST allows the kinematic 
range to be extended at low $Q^2$ and low $y$.

The measurement region is further extended towards lower $Q^2$ and
higher $x$ values by
exploiting events with hard photons emitted collinearly to the electron beam
(Initial State Radiation or ISR).
Such events are treated as $ep$ events at an effectively reduced centre-of-mass
energy.
Unlike in the previous H1 ISR analysis~\cite{Ahmed:1995cf},
the emitted photons are not explicitly detected, but the missing momentum is
determined using momentum conservation.
For this method the BST charged particle validation of the scattered electron
is important to reduce the physics
background from photoproduction events, in which the scattered electron escapes 
undetected in the electron beam direction.

The measurement presented here is combined with previously published
data~\cite{h1alphas,Adloff:1997mf} taken at $E_p=820$\,GeV in the region
$Q^2 \ge 1.5$\,GeV$^2$  (\MB97) and  
in the region $Q^2\ge 0.35$\,GeV$^2$ employing 
a shifted vertex technique (SVX95). The data sets are combined 
taking into account their systematic error correlations.
The resulting accuracy reaches two percent precision
in the bulk region of the measurement providing the most precise
measurement in this kinematic domain. 

Data on $F_2$ extending to low $Q^2$ were published by the
ZEUS Collaboration using a detector mounted near the beam 
pipe~\cite{Breitweg:2000yn}. For $Q^2 \gtrsim 2$~GeV$^2$,
ZEUS data~\cite{Chekanov:2001qu} from the $820$\,GeV operation of HERA are 
also available. 

The paper is organised as follows: In \Sec\,\ref{sec:theory} basic  
definitions are given. In \Sec\,\ref{sec:models} models are introduced
which are subsequently compared to the data.
In \Sec\,\ref{sec:xsection} the methods to determine the DIS event
kinematics and the principle of the cross section measurement are presented.
In \Sec\,\ref{sec:det} the H1 apparatus is described with emphasis on the
components of key importance for the present measurement.
\SSec\,\ref{sec:eventrecon} presents the event selection and reconstruction,
followed by \Sec\,\ref{sec:mc} on the Monte Carlo (MC) simulation  of events.
In \Sec\,\ref{sec:dataanal} a detailed account of the analysis techniques
and uncertainties of the measurement is 
given, and the cross sections obtained from the
    1999 and 2000 data are presented. 
In \Sec\,\ref{sec:averaging} the data averaging method
 and the combination of the new data 
with the previous H1 data taken at $E_p = 820$\,GeV are presented.
\SSec\,\ref{sec:pheno} is devoted to a phenomenological analysis
of the $x$ dependence of $F_2$ and to extractions of the longitudinal
structure function $F_L$ and in \Sec\,\ref{sec:phenomod} the data are compared
to phenomenological models. A summary is given in \Sec\,\ref{sec:summary}.

\section{Definitions\label{sec:theory}}
In the low $Q^2$
 kinematic range of the present measurement, contributions
from $Z$ boson exchange to neutral current deep inelastic scattering
can be neglected. In the one-photon exchange approximation,
the double differential cross section for neutral current DIS
is given, in its reduced form $\sigma_r$, by
\begin{equation} \label{sigred}
\sigma_r = \frac { Q^4 x} { 2\pi \alpha^2 [1+(1-y)^2]} \cdot
\frac{{\rm d}^2\sigma }{{\rm d}x\,{\rm d}Q^2} = 
F_2(x,Q^2) -  f(y) \cdot  F_L(x,Q^2)
\end{equation}
%
with the fine structure constant denoted $\alpha$ and $f(y)=y^2/[1+(1-y)^2]$.
The inelasticity $y$ is related to  $Q^2$,  $x$ and the centre-of-mass
energy squared, $s=4 E_e E_p$, by $y=Q^2/sx$.
In the quark-parton model (QPM), $x$ denotes the fraction of the proton
momentum carried by the parton coupling to the exchanged boson.

The DIS cross section, \Eq~\ref{sigred}, is determined by two structure
functions, $F_2$ and $F_L$. 
These are related to the cross sections for the scattering 
of longitudinally and transversely
polarised photons off protons, $\sigma_L$ and $\sigma_T$. 
At low $x$, the relationships
\begin{equation} \label{sigL}
F_L =  \frac  {Q^2}  { 4 \pi^2 \alpha}  (1-x)\cdot \sigma_L ~,
\end{equation}
\begin{equation} \label{sigLT}
F_2 = \frac {Q^2 }  { 4 \pi^2 \alpha} (1-x) \cdot ( \sigma_L + \sigma_T) ~,
\end{equation}
%
hold to very good approximation.
Positivity of the longitudinal and transverse scattering
cross sections imposes the restriction $0 \leq F_L \leq F_2$.
Using the ratio $R(x,Q^2)$
\begin{equation} \label{AR}
R = \frac { \sigma_L} { \sigma_T}  = \frac{F_L}{F_2-F_L}~,
\end{equation}
the reduced cross section in \Eq~\ref{sigred} can be written as
\begin{equation} \label{sigrR}
 \sigma_r = F_2(x,Q^2) \cdot \left[1 -  f(y) \cdot \frac{R}{1+R}\right] ~.
\end{equation}
For  most of the kinematic domain, the reduced DIS neutral current scattering
cross section is well approximated by the $F_2$ structure function, since
$F_L$ leads to a sizeable effect only for large inelasticity values $y$. 

The reduced cross section $\sigma_r$ can be re-expressed as
\begin{equation}
 \sigma_r = \frac{Q^2 (1 - x)}{4 \pi^2 \alpha} \sigma_{\gamma^* p}^{\rm eff}~,
\end{equation}
with the effective virtual photon-proton cross section
\begin{equation} 
 \sigma_{\gamma^* p}^{\rm eff} = \sigma_T + [1 - f(y)]\,\sigma_L ~.
\end{equation}
The sum $\sigma_L + \sigma_T $ is referred to as the
total virtual photon-proton cross section, $\sigma_{\gamma^* p}^{\rm tot}$,
which is often expressed as a function of $Q^2$ and of the invariant mass
of the virtual photon-proton system, $W$.
For small $x$, $W$ can be calculated as $W = \sqrt{Q^2  (1-x) / x}$,
such that $W^2 \simeq sy$.
The total and the effective virtual photon-proton cross sections differ significantly
 only in the region of high $y$.

\section{Models \label{sec:models}}
The low $x$ data presented here extend to low values of $Q^2$ 
for which perturbative QCD is not applicable. 
The phenomenological models formulated for this transition region
reproduce the $W$ dependence of the $\gamma^* p$ cross section,
which is weak in the photoproduction
region \cite{Aid:1995bz}. A steep increase towards large values
of $W$ develops in the perturbative region, which is equivalent
to the rise of the proton structure function $F_2$ towards low $x$
at fixed $Q^2$.

In the context of the present measurement
colour dipole models ({\itshape e.g.}
 \cite{Gotsman:1997zw,Golec-Biernat:1998js,Forshaw:1999uf,Bartels:2002cj,Iancu:2003ge,Kowalski:2006hc})
 are particularly interesting 
because 
$F_L$ and $F_T=F_2 - F_L$ are both described by a single
characteristic dipole scattering cross section 
$\hat{\sigma}$ combined with either 
the longitudinal  or the transverse 
photon wavefunction.
The squares of the wavefunctions
of the $q \overline{q}$ fluctuations 
of longitudinally and transversally polarised photons 
are\,\cite{Nikolaev:1990ja}
\begin{equation}
\begin{split}
W_L(z,r,Q^2) &= \frac{\textstyle 6 \alpha}{\textstyle \pi^2}
 \sum_{i=1}^{ n_f}{e_i^2 Q^2 z^2(1-z)^2 K_0(\epsilon r)^2} ~,\\
\textstyle
W_T(z,r,Q^2) &= \frac{\textstyle 3 \alpha}{\textstyle 2\pi^2}
 \sum_{i=1}^{n_f}{e_i^2 [ (1 -2z(1-z)) \epsilon^2 K_1(\epsilon r)
 + m_i^2  K_0(\epsilon r)]} ~,
\end{split}
\label{waves}
\end{equation}
where 
$\epsilon^2 = m_i^2 + z(1-z) Q^2$, $m_i$ ($e_i$) is the mass (charge) of
quark $i$, $K_{0}(u)$ and $K_1(u)=-\partial_uK_0$ 
are modified Bessel functions, $r$ is the transverse separation of the $q \overline{q}$ pair
and $z$ denotes the fractional energy sharing between $q$ and $\overline{q}$. 
 In this approach the cross sections $\sigma_{T,L}$  
are obtained from integrals over the impact parameter space as
\begin{equation} \label{cdmx}
\sigma_{L,T}(x,Q^2) = \int {\rm d}^2r \int_0^1 {\rm d}z W_{L,T}(z,r,Q^2) \hat{\sigma}(x,r^2).
\end{equation} 
Colour dipole models differ
by the chosen
expressions for the  cross section $\hat{\sigma}$.
With the measurement
extending into the region of high $y$
one can confront the predictions of such models for the
two structure functions with the data. As an illustration,
the data are compared in this paper to two versions of the colour dipole
model, the original version 
by Golec-Biernat and W\"usthoff (GBW)\,\cite{Golec-Biernat:1998js}
and a more recent model, based on the Colour Glass Condensate
approach to the high parton density regime, by  Iancu, Itakura and Munier
(IIM)\,\cite{Iancu:2003ge}.

Two further models are used in this paper in order to
parameterise $F_2(x,Q^2)$. The fractal model is based on the observation
 that the proton structure
 at low $x$ exhibits self-similar properties for different $x$ and $Q^2$
 values. Two continuous, variable and correlated fractal dimensions are
 chosen to describe the self-similarity in $x$ and $Q^2$~\cite{fractal}.
In a more phenomenological approach $F_2$ is parameterised as
$x^{-\lambda(Q^2)}$. These two models are also compared  with 
the reduced cross section, $\sigma_r$, after making assumptions on $R$.

\section{Measurement of the DIS Cross Section\label{sec:xsection}}
\subsection{Reconstruction of Event Kinematics\label{ssec:kinematics}}
In the colliding beam experiments at HERA, the DIS event kinematics can be
reconstructed using the measurements of the 
scattered lepton, the hadronic final state, or a combination of the two.
This complementarity enlarges the kinematic coverage and provides 
an additional control of  the systematic uncertainties.

The energy of the scattered \electron\ \ee\ and its polar
angle \thetae\
are used in the ``electron method'' to determine the kinematics via
\begin{align}
 y_e &= \frac{2 E_e - \ee \left(1 - \cos \thetae \right)} {2 E_e}
 \equiv \frac{2E_e - \Sigma_e}{2 E_e} ~, \label{eq:ey} \\
\qqe &= \frac{{\ee }^2 \sin^2\thetae } {1 - y_e} ~, \qquad
 x_e = \frac{\qqe}{4\, E_p\, E_e\, y_e} ~. \label{eq:ex}
\end{align}

Using energy-momentum conservation, the event kinematics can also be
determined from the hadronic final state.
An important quantity is the difference between
the total energy and the total longitudinal momentum
\begin{equation} \label{eq:sigma}
 \empz \equiv E'_e \left(1-\cos \thetae\right) + 
 \sum_i \left(E_i - P_{z,i}\right) \equiv  \Sigma_e + \Sigma_h ~,
\end{equation}
where $E_i$ ($P_{z,i}$) is the reconstructed energy (longitudinal component
of the momentum) of a particle $i$ from the hadronic final state.
In the reconstruction masses are neglected for both the \electron\ and the
hadronic final state particles.
The measured $\empz$ is insensitive to 
losses in the  proton beam direction and is thus only weakly affected
by the incomplete reconstruction of the proton remnant.
For non-radiative events, the relation $\empz \simeq 2 E_e$ holds. 
This allows   $2E_e - \Sigma_e$ in \Eq~\ref{eq:ey} to be replaced 
by $\Sigma_h$ and leads to the introduction of the $y_h$ variable~\cite{yjb}
\begin{equation}\label{eq:yh}
 y_h  = \frac{\Sigma_h} {2 E_e} ~.
\end{equation}

For events in which a photon is emitted collinearly to the incoming
\electron, the radiated photon is not reconstructed in
the sub-detectors used to calculate $\empz$. In this case 
$(\empz)/2$ is equal to an ``effective''
incident \electron\ beam energy, reduced relatively
to the nominal beam energy by the momentum carried
by the radiated photon. This is employed  in the 
$\Sigma$ method, for which $2 E_e$ in \Eq~\ref{eq:yh}
is substituted by the measured \empz~\cite{ysigma}
\begin{equation}\label{eq:ys}
\ys = \frac{\Sigma_h}{\empz}~.
\end{equation}
For this method, $Q^2$ is calculated by replacing $y_e$ in
\Eq~\ref{eq:ex} by $\ys$, and Bjorken $x$ is calculated by substituting $y_e$,
$Q^2_e$ and $2E_e$ by $\ys$, $\qqs$ and $\empz$, respectively\footnote{%
Note that in previous H1 publications the nominal \electron\ beam energy was 
used instead of $(\empz)/2$ in the calculation of $\xs$.
The method of $x$ calculation used here is called the I$\Sigma$
method in \cite{ysigma}.}
\begin{equation}
\qqs = \frac{{\ee}^2 \sin^2 \thetae }{1 - \ys } ~, \qquad
\xs = \frac{\qqs }{2\, E_p\, \ys } \cdot \frac{1}{\empz} ~.
\end{equation}
By using a consistent set of the variables $x_\Sigma$, $y_\Sigma$ and $Q^2_\Sigma$,
the measurement also correctly reconstructs the kinematics for
events with initial state QED radiation.
Therefore, the method covers lower $Q^2$ and higher $x$ values, which become
 accessible due to the reduced centre-of-mass energy for these events.

The total transverse momentum of the hadronic final state  is
\begin{equation}
 \Pth = \left| \sum_i \mathbold{P}_{\,\perp,i} \right|
\end{equation}
where 
 $\mathbold{P}_{\,\perp,i}$ is the 
transverse momentum vector of the particle
$i$ and the sum runs over all particles. $\Pth$ is rather insensitive to particle losses  collinear to the
beam for a wide range of $y$. The combination of \Pth\ and $\Sigma_h$ defines
the hadronic scattering angle
\begin{equation}\label{eq:thh}
\tan \frac{\thetah}{2} = \frac{\Sigma_h}
{\Pth}~,
\end{equation}
which, within the QPM, follows  the direction of the struck quark.

In this analysis, both the electron and the $\Sigma$ methods are used 
for the cross section measurement. The electron method provides the better  
resolution in $x$ for large inelasticities $y > 0.1$, but the resolution 
degrades as $1/y$. Use of the $\Sigma$ method extends 
the measurement down to $y \sim 0.002$. Below this $y$ value, losses along
the proton beam direction become important and are difficult to estimate. 
The $\Sigma$ method as is used here
noticeably increases the kinematic coverage towards low $Q^2$ 
and high Bjorken $x$ due to initial state QED radiation.

\subsection{Determination of the DIS Cross Section} \label{ssec:xsection}

\begin{figure}[b]
\centerline{%
\epsfig{file=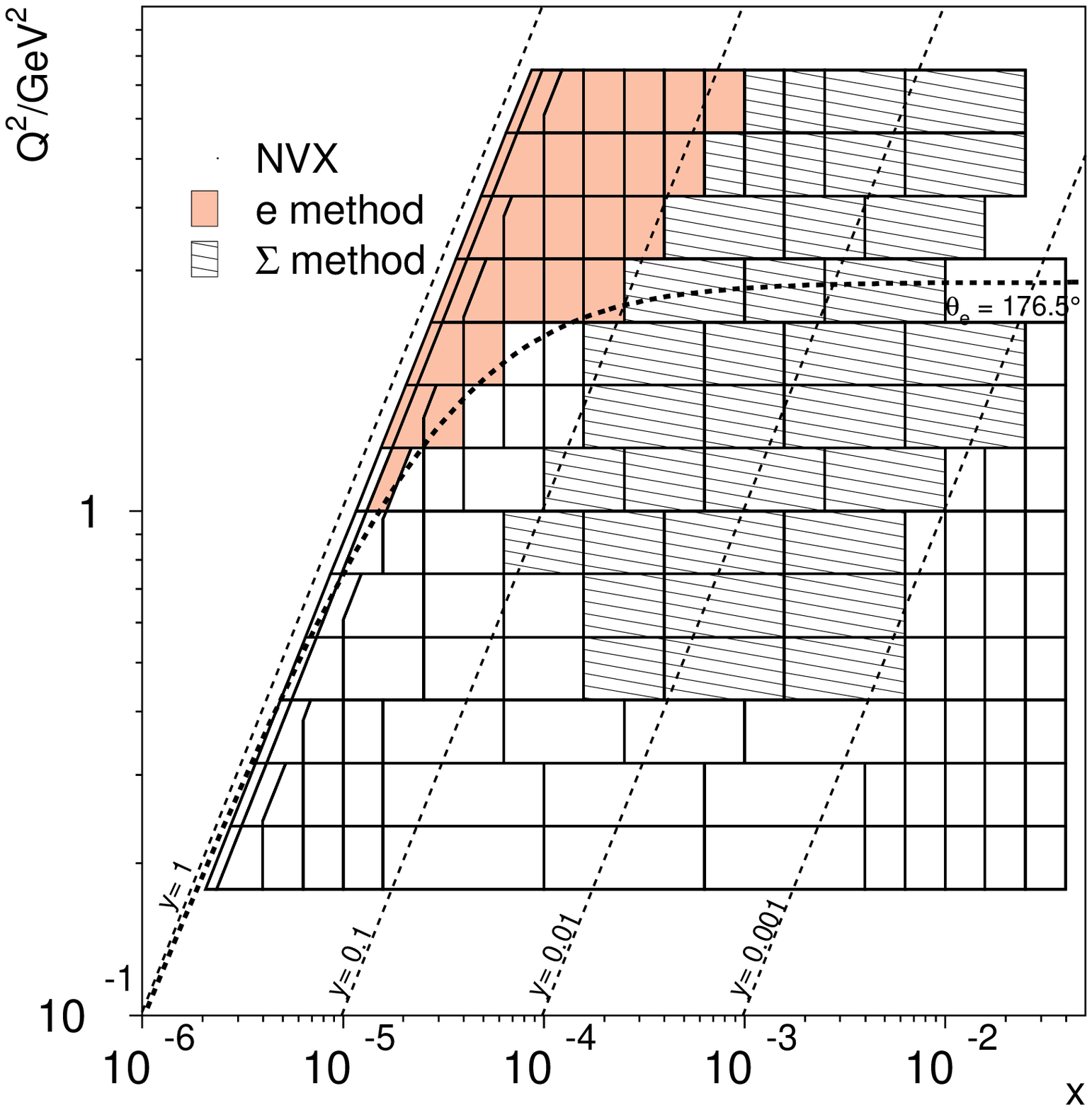,width=0.5\linewidth}%
\epsfig{file=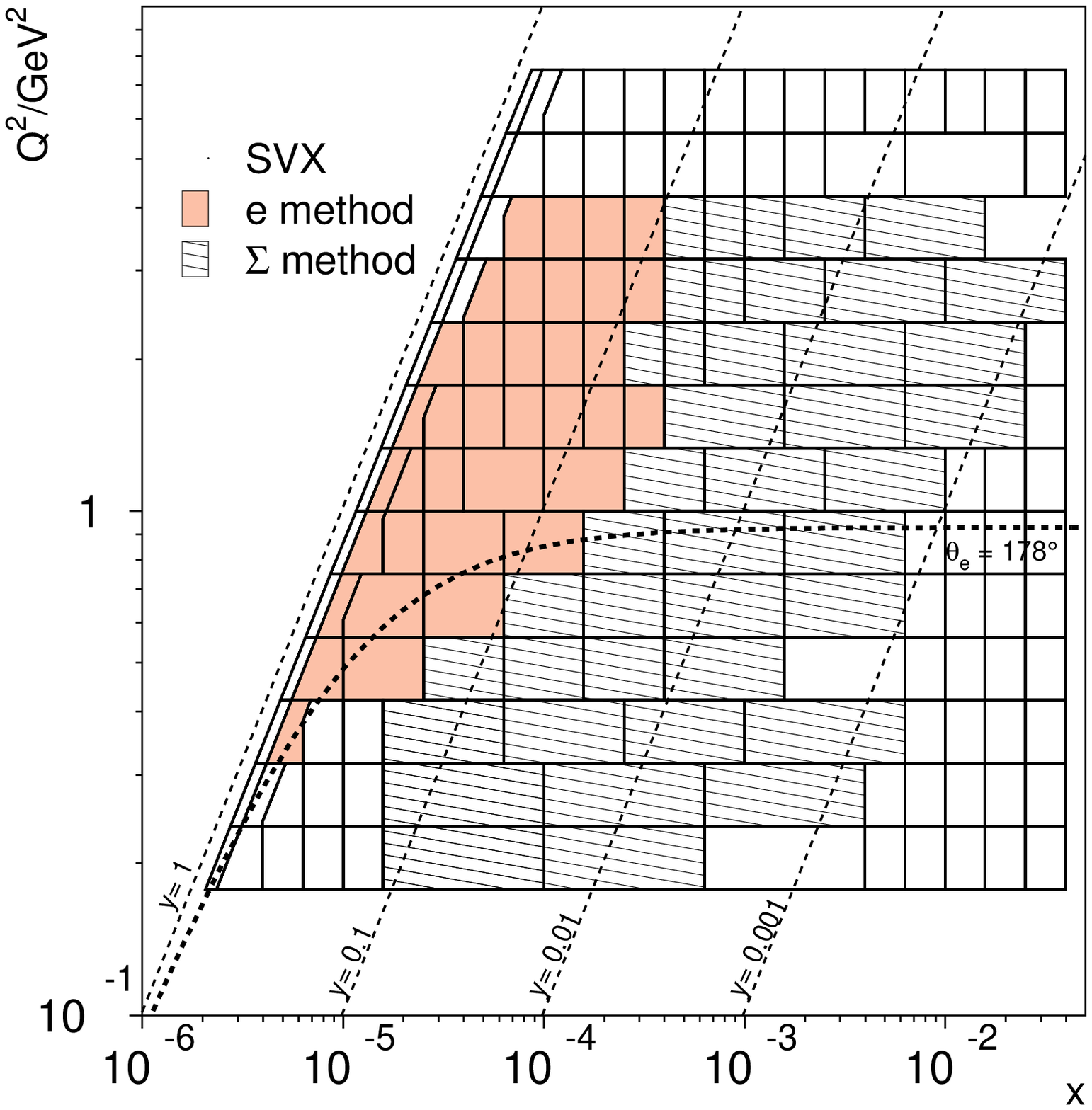,width=0.5\linewidth}
}
\caption{\label{fig:bins}
Illustration of the bins used for the cross section measurement in the
 {\MB} (left) and {\SVX} (right) analyses. 
Dashed lines of constant \thetae~ indicate approximate angular acceptance 
for both measurements. The dark (light) shaded area corresponds to the
bins where the electron ($\Sigma$) method is used for the measurement
of the cross section. The measurement in the bins
outside the angular acceptance range employ  the
$\Sigma$ method for ISR events.}
\end{figure}

The measurement of the double differential cross section is performed
in bins of $x$ and $Q^{2}$, or $y$ and $Q^{2}$, depending on the region
in the kinematic phase space, as shown in \Fig~\ref{fig:bins}.
The bin sizes and shapes as well as methods used for the 
kinematic reconstruction are chosen based on the following prescription:

\begin{itemize}
\item
In $Q^2$, a binning equidistant in $\log_{10} Q^2$ is chosen with eight bins
per decade, as in previous H1 publications~\cite{h1alphas}. This binning 
reflects the good $Q^2$ resolution of the H1 detector.
\item
The $x$ and $Q^2$ values at which the measured double differential 
cross section is quoted, also referred to as
bin centres,  are placed at an approximately logarithmic average value
within the bin boundaries for the $x$ and $Q^2$ binning, and at the linear
average for the $y$ binning.
\item
For high $y>0.6$, the electron method has an excellent kinematic resolution. 
In this region, the measured cross section is sensitive to the
longitudinal structure function $F_L$, which leads to a rapidly changing shape
of the reduced cross section as a function of $y$. Therefore a fine binning,
linear in $y$,  is chosen for $y>0.6$: two $y$ bins are used for each $Q^2$
interval with boundaries at $y=0.85$, $0.75$ and $0.6$.
\item
For $y<0.6$ the binning is defined in $x$.
The default $x$ binning is equidistant in $\log_{10}x$
with five bins per decade, as chosen previously~\cite{h1alphas}. 
The transition between the $x$ and $y$ binning is defined by the 
$y$ value of the nominal bin centre, $y_c$, for the transition bins:
for $y_c>0.6$, the bin is combined with the nearest $y$ bin and for $y_c\le 0.6$ 
it is combined with the nearest $x$ bin.  
\end{itemize}
The resolution in each bin is checked using a Monte Carlo simulation.
Two variables are calculated for this purpose, the purity 
$P=N_{\rm rec,gen}/N_{\rm rec}$ and the stability
 $S=N_{\rm rec,gen}/N_{\rm gen}$,
where $N_{\rm rec}$ ($N_{\rm gen}$) is the total number of reconstructed
(generated) Monte Carlo events in the bin and $N_{\rm rec,gen}$ is the number
of events which are both generated and reconstructed in the same bin.
The purity and stability are calculated for both the electron and the $\Sigma$
methods.
For the cross section measurement the method with the higher purity is used.
The choices are illustrated in \Fig~\ref{fig:bins}. 
The purity and stability typically exceed
$50\%$. If either the purity or the stability is below $25\%$
in a bin for the chosen reconstruction method, 
the bin is combined with an adjacent bin.
Bins with larger sizes can thus be created 
at the acceptance edges as shown in~\Fig~\ref{fig:bins}.

The $\theta_e = \thetamaxmb$ and $\theta_e = \thetamaxsvx$ lines in
\Fig~\ref{fig:bins} indicate the approximate angular acceptance limits
of the H1 detector 
for the nominal and the shifted vertex positions, respectively. 
In each plot measurement bins below 
$\theta_e$ lines are visible. The measurement in these bins becomes possible
using the $\Sigma$ method for events with initial state photon radiation
which effectively reduces the centre-of-mass energy. These bins are
further referred to as ISR bins. 
The $\Sigma$ method cannot be used at high $y$, where its resolution 
is poor, leading to large migrations of nominal energy events
into the ISR bins and thus to purities below the accepted value. 
This causes the gap between the ISR and electron method bins at high $y$.

The calculation of the reduced double differential $ep$ cross section is
performed by correcting the data
using the MC simulations. The following formula is applied to each analysis bin
\begin{equation}
\sigma_{r}\left(x_c,Q^2_c\right) = 
 \frac{N_{\rm data} - N_{\rm bg} }{A \epsilon\, \mathcal{L}_{\rm data}} \,
\frac{c_{\rm bc}}{1 + \delta_{\rm rc}} ~.
\label{eq:acc}
\end{equation}
Here, $(x_c,Q^2_c)$ is the bin centre,
$N_{\rm data}$ is the number of data events, 
$N_{\rm bg}$ is the number of  background events, 
estimated using MC simulations,
$A$ and $\epsilon$ are the detector acceptance and efficiency,
$\mathcal{L}_{\rm data}$ is the integrated luminosity,
$\delta_{\rm rc}$ are QED radiative corrections,
and $c_{\rm bc}$ are the corrections for finite bin size effects. 
The radiative and bin centre corrections can be determined using 
the Monte Carlo simulation. In this case, \Eq~\ref{eq:acc} becomes
\begin{equation}
\sigma_{r}\left(x_c,Q^2_c\right) =
 \frac{N_{\rm data} - N_{\rm bg}}{N_{\rm MC}} \,
 \frac{\mathcal{L}_{\rm MC}}{\mathcal{L}_{\rm data}} \,
\sigma^{\rm MC}_{r}\left(x_c,Q^2_c\right) ~,
\label{e:sigmamcm}
\end{equation}
where  $N_{\rm MC}$ is the number of signal MC events and
$\mathcal{L}_{\rm MC} = N_{\rm gen}/\sigma_{\rm gen}$ is the
Monte Carlo luminosity.
Here $N_{\rm gen}$
denotes the total number of generated MC events and $\sigma_{\rm gen}$
is the total integrated cross section for the MC generation.
The quantity $\sigma^{\rm MC}_{r}\left(x_c,Q^2_c\right)$ is
the reduced double differential cross section at the bin centre
calculated at the Born level with the same structure functions
as are used in the MC generation.

The correction for the detector acceptance using Monte Carlo modelling
requires the cross section model used in the simulation
to be sufficiently close to the data, such that migrations between the bins
are well reproduced. The cross section model should also describe the
kinematic region outside the measurement range, in particular at
low $y$ and low $Q^2$, to account for radiative corrections and long range
migrations.    
In practice, this is achieved using an iterative MC event reweighting
procedure which converges after one iteration for the measurement region.
First, the double differential cross section is measured following
\Eq~\ref{e:sigmamcm} using an initial approximation for the MC input cross
section. Next, the measured double differential cross section is fitted with
a new parameterisation using the fractal model 
and the analysis of the Monte Carlo events is repeated
with an additional weight factor, equal to the ratio of the new to the initial
double differential cross sections in each simulated event. For
the reweighting, the event kinematics are calculated using the generated
$x$ and $Q^2$ variables at the hadronic vertex, such that corrections due
to radiation from the lepton line are properly accounted for.
This reweighting procedure is used for the measurement region.
For the high $x>0.02$ domain, which lies outside the measurement region,
the ALLM parameterisation~\cite{Abramowicz:1997ms}
is used.

\section{H1 Detector\label{sec:det}}
\subsection{Overview}
\begin{figure}[b]
\centerline{\psfig{file=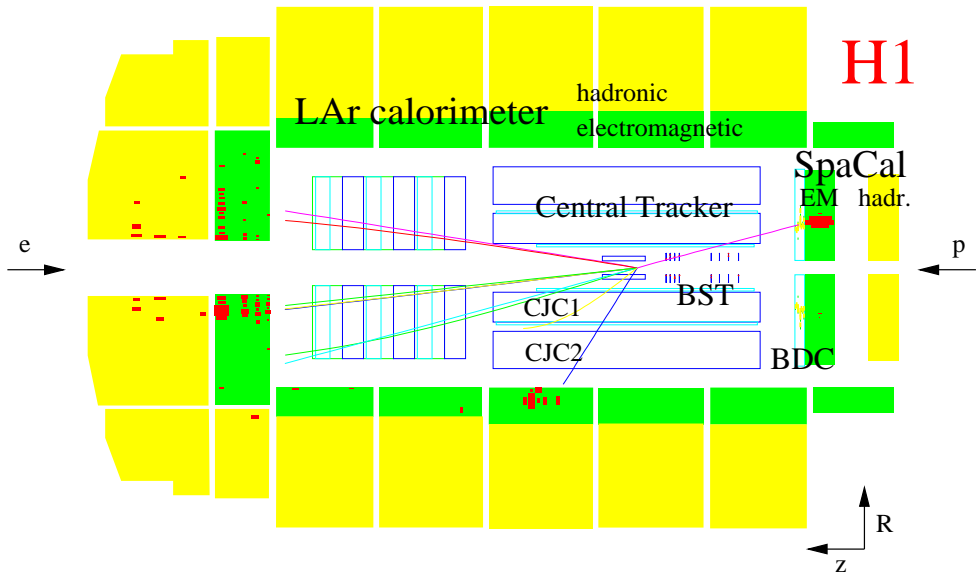,width=0.8\linewidth}}
\caption{\label{fig:h1event}A  low $Q^2$ event as reconstructed 
in the H1 detector.
The electron is scattered into the backward region. The electron trajectory is
reconstructed in the Backward Silicon Tracker (BST) and in the Backward Drift
Chamber (BDC). The electron energy is determined using the SpaCal calorimeter.
The hadronic final state is detected in the central and
forward tracking detectors, and in the LAr calorimeter.
}
\end{figure}

A complete description of the H1 detector is given in~\cite{h1det,h1det2}.
Here the  components used for the present measurement are discussed.
In \Sec~\ref{sec:backdet} the  detectors for the scattered electron
measurement are described in detail.
A schematic view of the H1 detector is given in \Fig~\ref{fig:h1event},
in which a typical low $Q^2$ event is shown.

Around the interaction region a set of tracking chambers, surrounded by 
electromagnetic and hadronic calorimeters, operates in a solenoidal magnetic 
field of $1.16$\,T. The tracking system is subdivided into forward, central
and backward tracking devices.
The nominal interaction point of the electron and proton beams lies about 
in the middle of the Central Tracker, at the origin of the coordinate
system. The interaction vertex positions have an approximately Gaussian
distribution in $z$ with $\sigma_{z} \approx 10$\,cm.
The calorimetry system consists of the Liquid Argon calorimeter (LAr) covering
the central and forward directions and the lead-scintillator spaghetti
calorimeter (SpaCal)~\cite{spacalha,spacala,spacalb,spacalc} measuring particles
scattered backwards.

The Central Tracker consists of four drift chambers,
two multi-wire proportional chambers (MWPCs) and a silicon tracking device.
The largest tracker components  are the two concentric drift chambers, 
CJC1 and CJC2, which have sense wires strung parallel to the beam axis
with the drift cells inclined at about $30^\circ$ with respect to the radial
direction, such that the drift direction of ionisation electrons is
approximately perpendicular to the wire plane.
The charge deposits are read out 
from both ends of each wire, providing particle identification via 
ionisation energy loss
and an approximate determination of the $z$ coordinate via the
charge asymmetry between the two wire end signals (``charge division'').

Tracks found in the CJC are linked to the hits found in two chambers equipped with
wires strung around the beam axis, following polygonal
support structures, dedicated to the precise 
measurement of $z$ coordinates. 
The inner $z$ chamber (CIZ) is located inside CJC1 and the outer
$z$ chamber (COZ) lies between CJC1 and CJC2. To reduce the number of
acceptable combinations with the CJC, the $z$ chambers also determine
 a $\phi$ coordinate using the charge division measurement.
The tracks are further constrained by linking to hits  in the central
silicon tracker (CST)~\cite{cst}. The CST consists of two layers of
double-sided silicon strip detectors surrounding the beam pipe, 
covering an angular range of $30^\circ < \theta < 150^\circ$ for tracks
passing through both layers. 

The two cylindrical proportional chambers, the CIP mounted inside CIZ, 
and the COP located between the COZ and CJC2,
are used together to identify tracks pointing to the
interaction vertex and thus to reduce background at the trigger level.
A combined CIP-COP signal is used 
in coincidence with the SpaCal to trigger events with low $E'_e$
(see \Sec~\ref{sec:online}).

The LAr calorimeter~\cite{larnim}, mounted in a large  cryostat, is 
 used in this analysis for the 
measurement of the hadronic energy. The angular coverage
of the calorimeter is $4^\circ < \theta < 154^\circ$
for an interaction vertex at $z=0$. The calorimeter consists
of an electromagnetic section with lead absorbers ($20-30$ radiation lengths)
and a hadronic section with steel absorbers.
The total depth is between $4.5$ and $8$ hadronic interaction lengths. 
 The LAr calorimeter is divided
along the $z$ direction into  wheels. The electromagnetic
section has eight wheels while the hadronic section has seven.
The calorimeter has a high degree
of spatial segmentation with a total of about $45000$ cells. 
Its hadronic energy resolution, as determined in test beam measurements~%
\cite{larnim2}, is 
$\sigma_E / E \approx 50\% / \sqrt{E/{\rm GeV}} \oplus 2\%$.

Two electromagnetic crystal calorimeters, a photon tagger (PT) and an electron tagger (ET),
located at $z = -103.1$\,m and $z = -33$\,m, respectively, are used to monitor
the luminosity via the measurement of the Bethe-Heitler process $ep\to \gamma ep$. The luminosity corresponding to the main interaction region 
can be separated from the additional (``satellite'') interaction regions 
using information from 
the scintillator hodoscopes of the time-of-flight system (TOF) and from the HERA proton pick-up (PPU) monitor,
a $34$~cm long stripline device located at $-3$~m from the interaction point.
The ET can be used to measure  the scattered electron in 
photoproduction processes, with $Q^2 \le 10^{-2}$\,GeV$^2$ and $0.2 < y < 0.7$.
The PT detects photons radiated collinearly to the
incoming electron direction. 

\subsection{Backward Detectors\label{sec:backdet}}
The measurement of the  inelastic $ep$ scattering cross section
at low $Q^2$ relies on the identification of the scattered
electron in the backward part of the H1 apparatus. 
The energy of the scattered electron is measured in the SpaCal calorimeter.
For the low $Q^2$ region under study, $\theta_e$ lies outside the angular 
acceptance of the Central Tracker.
The polar angle of the scattered electron can, however, be measured either 
by the Backward Silicon Tracker (BST), based solely on the electron track, 
or by a combination of the less precise Backward Drift Chamber (BDC) signal with 
the hadronic final state vertex, as reconstructed using the
Central Tracker.
The redundancy of the angular measurements provides additional
cross checks over a large angular range, whilst the BDC extends
the polar angle coverage to larger $\thetae$.

\subsubsection{Backward Silicon Tracker}
\begin{figure}
\centerline{\psfig{file=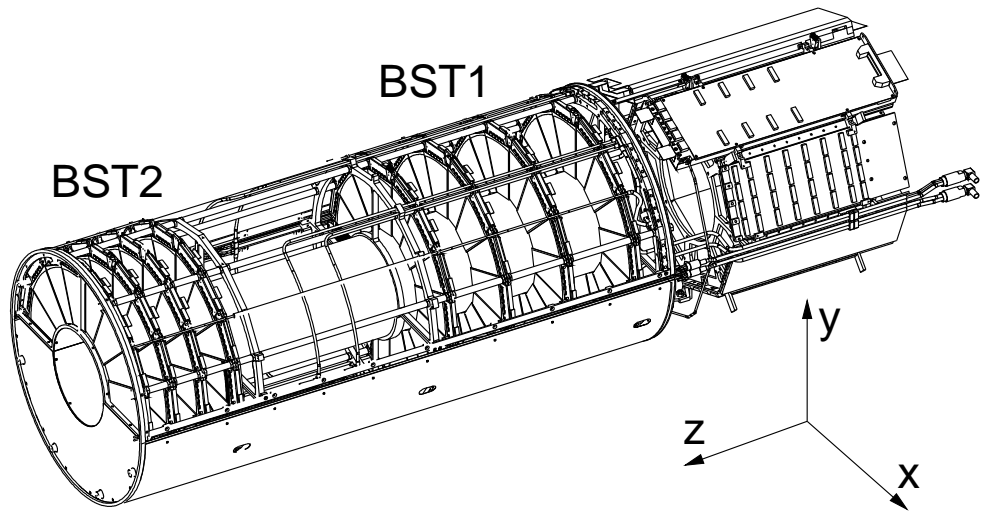,width=0.8\linewidth}}
\caption{\label{fig:bstscheme}
Schematic layout of the H1 Backward Silicon Tracker (BST).
The active area is composed of
eight wheels subdivided into two modules,
BST1 and BST2, of four wheels each. One wheel is made of
$16$ $r$ sensors and one $u$ sensor 
(mounted on the back side, not shown here).
Eight consecutive sensors in 1/16 of azimuth build a
BST sector. In $z$ the  module BST1 extends from
$- 73.2$ to $-95.7$\,cm, BST2 from $-35.9$ to $- 46.9$\,cm. 
Readout boards are placed in the rear section.
Also indicated are the electric
shielding and the water cooling pipes.}
\end{figure}

The BST in the configuration installed in $1999$~\cite{Henschel:2000tm} 
is schematically shown in \Fig~\ref{fig:bstscheme}. It 
consists of eight planes (disks) and $16$ azimuthal  sectors.
The planes are mounted perpendicularly to the beam axis 
and are arranged in two modules, BST1 and BST2, of four planes each. 
A first version of the BST with four planes is described in\,\cite{Eick:1996gv}.

\begin{figure}
\centerline{\psfig{file=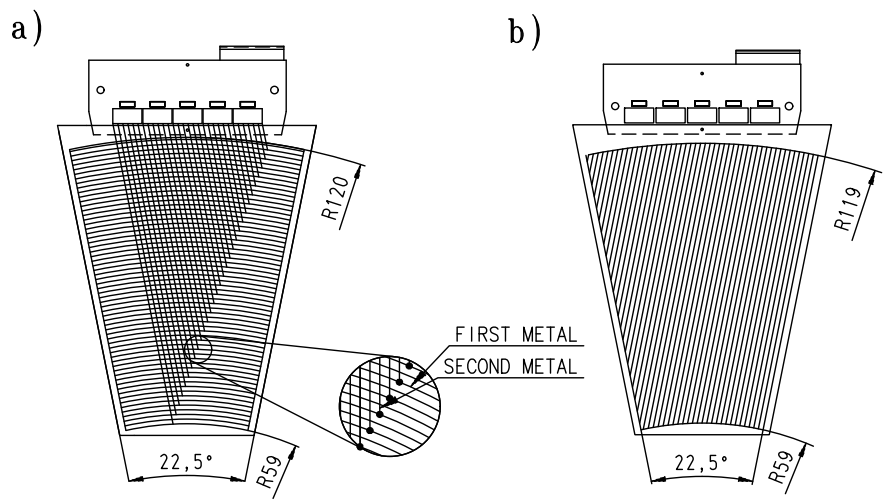,width=0.8\linewidth}}
\caption{\label{fig:bstsensors} The two types of silicon sensors used in the BST:
a) $r$ sensor, b) $u$ sensor, each with $640$ 
readout strips. The $r$ sensor has a double metal structure for the readout lines
to reach the top (outer radius) part where the five amplifiers are mounted
on the hybrid, as sketched.}
\end{figure}

Each BST plane is equipped with $16$ wedge shaped, single sided,
double metal, silicon strip  sensors of $250$ $\rm \mu m$ thickness. 
Each sensor contains $640$ sensitive $p$ strips which are concentric around the
beam axis with a pitch of $96$\,$\mu$m. The signals are 
amplified and temporarily stored by five  on-detector front-end
chips, called Analogue Pipeline Chips~\cite{Horisberger:1993ec} (APCs), until
a readout instruction is received~\cite{Haynes:1998xq}.
Using these ``$r$ sensors'' (\Fig~\ref{fig:bstsensors}a) the
track polar angle can be determined.
The acceptance range of the BST for the nominal vertex position is 
$164^\circ < \theta_e < 176^\circ$. 

In addition to the $r$ sensors, each plane contains one single sided, single
metal, silicon strip sensor, in the azimuthal
sector $45^\circ < \phi < 67.5^\circ$ mounted behind the 
$r$ sensor. This ``$u$ sensor''  has
$640$ sensitive strips parallel to the reference edge of the sensor
with a pitch of $75$\,$\mu$m (\Fig~\ref{fig:bstsensors}b).
It thus measures  hits
in $u$ coordinate space defined by $u = r\sin\phi_u$, where $\phi_u$ is the
azimuthal angle with respect to the reference edge of the sensor.
Combining the information from $r$ and $u$ sensors, it is
possible to measure the transverse momentum and determine
the  charge of a track in the BST. This feature is used in this analysis
to cross check the simulation of photoproduction background.

During data taking  an online
hit finding is performed. This takes into account individual
pedestals of each channel, which are dynamically updated. 
Coherent shifts in the amplitude of groups of strips, so called ``common
mode'', are also corrected for. For reconstructed tracks, the most probable
signal-to-noise values for the hits is about
$15$ for the $r$ sensors and $30$ for the $u$ sensors.
The single hit resolution is $20\,(15)$\,$\mu$m for the 
$r$ ($u$) coordinate.

\subsubsection{SpaCal and BDC} \label{sec:spabdc}
The SpaCal calorimeter covers the polar angle
range of $153^{\circ} < \theta < 177^{\circ}$ as measured from the nominal $z$ vertex
position. It consists of an
electromagnetic section~\cite{spacala,spacalb,spacalc}  with $1192$ cells of size $4.05\times 4.05\times 25$\,cm$^3$ 
in front of a hadronic section with 136 cells of size $11.9\times 11.9\times 25$\,cm$^3$.
The total amount of passive material traversed by particles from the
interaction vertex up to the SpaCal is of the order of one radiation
length.
The electromagnetic section comprises $27.5$ radiation lengths
and provides an electromagnetic energy resolution of
$\sigma_E / E = 7\% / \sqrt{E/{\rm GeV}} \oplus 1\%$.
The hadronic section~\cite{spacalha} is used for a coarse hadronic energy measurement and to
distinguish hadronic from electromagnetic showers.
The whole calorimeter comprises $2$ hadronic interaction lengths.
The energy resolution for hadrons amounts to $\sigma_E / E \sim 60\% 
/ \sqrt{E/{\rm GeV}}$.

The SpaCal cells consist of lead sheets with embedded scintillating fibres.
The fibres from each cell are bundled together and attached via light mixers
to photomultiplier tubes (PMTs). The stability of the PMT gain can be checked
using a dedicated LED system.

\begin{figure}
\centerline{
\epsfig{file=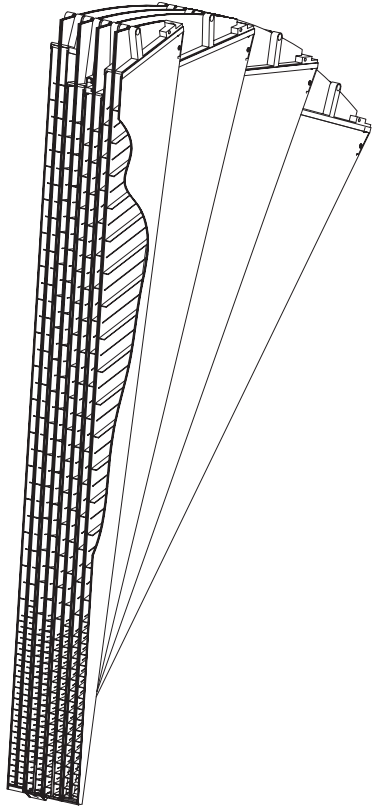,width=0.3\linewidth}
}
\caption{\label{fig:bdc3d}
Three-dimensional view of a section of the BDC illustrating the layer structure and 
the drift cell geometry. The chamber has a radial coverage from $6\,$cm to $71\,$cm. 
At a radius of about $22\,$cm  the segmentation is changed and a transition drift cell is 
introduced.}
\end{figure}

The backward drift chamber~\cite{SchwabTh}
is mounted in front of the SpaCal and has the
same angular acceptance. It consists of four double layers, each of them
divided azimuthally into eight sectors. 
A three dimensional view of a section of the BDC is given in
\Fig~\ref{fig:bdc3d}.
The sense wires are strung perpendicularly to the beam
axis and are fixed at the sector edges leading to an octagonal geometry with
almost radial drift directions. The drift cells are $1$\,cm wide in the inner
region and $3$\,cm wide in the outer. At the transition
from the inner to the outer region a special cell is introduced
 with $0.5$\,cm drift
distance at the inner side and $1.5$\,cm drift distance at the outer.
The cells within one double layer are shifted by half a drift cell in
the radial direction to solve the inner-outer hit assignment ambiguity.
The double layers are rotated by $11.25^{\circ}$ with respect to each other
to allow for reconstruction of
the azimuthal coordinate. In addition, this reduces the efficiency losses
at the sector edges.
The radial resolution for minimum ionising particles is $400$\,$\rm\mu m$.
The resolution in the azimuthal direction is about $2$\,mm.

\section{Data Collection and Reconstruction}\label{sec:eventrecon}
The H1 detector uses a multi-level trigger system for data collection in which
two hardware trigger levels are followed by a software filter farm. 
After improvements of the detector calibration and
the reconstruction code, the data are reprocessed offline.
This section describes the first analysis stages, including the online
data selection and the reconstruction algorithms. 

\begin{table}
\centerline{%
\begin{tabular}{cccc}
\hline \hline
Trigger & Energy      & Effective             & Tracking       \\
  Name  & Threshold   & $r_{\rm sp}$ Cut  & Condition      \\
\hline
 S3     & $12.0$\,GeV  &  $10$\,cm              &       ---      \\
 S0     & $6.5$\,GeV   &  $12$\,cm              &       ---      \\ 
 S9     & $2.0$\,GeV   &  $15$\,cm              &  CIP-COP track \\
\hline \hline
\end{tabular}
}
\tablecaption{\label{tab:trig}Overview of the main trigger conditions used 
for the {\MB} and {\SVX}~analyses. $r_{\rm sp} $ is the radial coordinate
of the SpaCal cluster.}
\end{table}

\subsection{Online Event Selection} \label{sec:online}

The online trigger conditions used in this analysis (\Tab\,\ref{tab:trig}) 
are based primarily on a localised energy deposition in
the electromagnetic section of the SpaCal (inclusive
electron trigger). Three different energy thresholds
are used. The trigger condition based on the lowest energy 
threshold (S9) collects events at the highest $y$.
Since a significant background contamination is present at low energies 
and radii,
the inner SpaCal region is excluded from S9.
To maintain an acceptable trigger rate, S9 contains an extra  
condition requiring the pattern of hit
pads in the CIP-COP proportional chambers to be consistent with 
at least one vertex pointing track.
This condition is similar to the requirement of a reconstructed
track from the hadronic final state in the Central Tracker.

\subsection{Track and Vertex Reconstruction in the Central Tracker}
The track reconstruction in the Central Tracker is initiated from the CJC 
hit measurements. Initially, candidate trajectories are found in the $xy$ plane 
using a fast circle fit algorithm~\cite{Karimaki:1991xv}. The $z$ coordinate is added
to the tracks based on charge division information. 
A linear fit in $S-z$ space
is performed where the $S$ coordinate measures an arc length
of the track in the $xy$ projection. Next, the tracks are fitted
to a common vertex in the $xy$ plane. At this stage detailed
corrections are applied for multiple scattering in the detector material and for
magnetic field non-uniformity. For the tracks which are determined
by the fit to originate from a common $xy$ vertex,
a combined $S-z$ fit is performed to determine an initial approximation
of the vertex $z$ position  and of the track polar angles.

The vertex-fitted CJC tracks are then combined with hits found
in the $z$ chambers, employing a robust global minimisation technique~\cite{glazovth}.
This link improves the $z$ vertex resolution from about $1$~cm to $1$~mm. The track $\theta$ resolution
is correspondingly improved from approximately $10$\,mrad to 
$1$\,mrad.
For the NVX sample, where the $z$ coordinate of the interaction
vertex is around zero,
the CJC tracks are also combined with hits found in the CST resulting in 
a vertex resolution of about $0.1$\,mm. CST hits are not used for the SVX
sample since the interaction vertex position is outside the CST acceptance.

\subsection{Reconstruction of the Electron Kinematics}
\subsubsection{Energy Reconstruction in the SpaCal Calorimeter \label{sec:RecSpacal}}

The reconstruction of the scattered electron kinematics is based on the measurement of  
a deposition of energy, termed a cluster, found in the electromagnetic SpaCal. 
The clustering algorithm searches for a cell with a local 
maximum in energy. The cluster is then built around this seed cell
by adding neighbouring cells with energies above the noise threshold.
The centre-of-gravity of the cluster is determined based
on all associated cells using a logarithmic energy weighting.  
To suppress background from hadrons and from decays of $\pi^0\to \gamma\gamma$
with the photons reconstructed in a single
cluster, a cluster radius estimator, $R_{\rm log}$, is used based
on logarithmic energy weighting. The background from
hadronic particles is further suppressed using a cut on the energy
deposit, $E_{had}$, in the hadronic section of the SpaCal behind the 
electromagnetic cluster.

The electron candidate cluster is required to be associated
to a track in one of the
backward trackers, BDC or BST, in order to reduce
background from backward photons and to  measure the polar 
angle \thetae\,accurately. For the determination of
\thetae, the trajectory of the scattered electron  is assumed to be 
a straight line in  $r-z$ coordinate space.

\subsubsection{Track Reconstruction in the BDC}
The BDC reconstruction of the electron scattering angle
$\theta_e$ starts from the line connecting
the SpaCal cluster and the Central Tracker vertex as an initial approximation.
The scattered electron azimuthal angle $\phi_{\rm e}$ 
is taken from the SpaCal cluster centre-of-gravity.
Only the BDC hits in the octant containing $\phi_{\rm e}$ 
are used for the $\theta_e$ reconstruction.

The $\thetae$ determination follows from a minimisation procedure.
A least squares track fit combines the Central Tracker 
vertex, the SpaCal cluster centre-of-gravity
energy, and all BDC measurements  in
a corridor of variable size $\Delta r$ around the current best estimate of the track 
direction. 
Initially, the corridor has a size of $5$\,cm. It is gradually reduced with improved
track parameters to about five times the BDC resolution. 
The SpaCal cluster is considered to be linked to the BDC track segment if there
are at least four hits from the eight layers 
remaining at the final iteration and if the radial distance between the track 
projected to the SpaCal $z$ coordinate and the SpaCal cluster is less than $2.5$\,cm.

\subsubsection{Track Reconstruction in the BST}
The reconstruction of the electron track in the BST uses the azimuthal location of the SpaCal cluster. 
The three adjacent  sectors which in azimuth are closest to $\phi_{\rm e}$ are selected. 
The $r$ coordinates of all BST hits in the selected sectors are projected
along the line defined by the hit and the SpaCal cluster 
to the BST plane closest to the SpaCal.
A clustering of the projected hits in this plane is then 
performed using a histogram technique. The line connecting
the position corresponding to the
peak in the histogram and the SpaCal cluster is
used as an initial approximation for the track.

The track finding then proceeds using an iterative 
minimisation technique with robust rejection of outliers, 
similar to the BDC reconstruction.  All hits in the selected  sectors
are included into a least squares minimisation. The contribution of
each hit is weighted with an
exponential suppression factor, which depends on the distance from the
hit to the track, and on an additional parameter, which
defines the width of an effective corridor around the track. For the first iteration, 
the width of the corridor is equal to the SpaCal spatial resolution.
For further iterations the width is gradually
reduced until it reaches five times the BST spatial resolution.
The event vertex $z$ coordinate is given by the distance of closest approach
of the BST track to the beam line.

For the sector equipped with the $u$ strip detectors, the reconstruction
of the azimuthal coordinate is also performed. At least three  
$u$ hits associated to linked $r$ hits are required. 
If multiple $u$ hits per plane are found, all possible track combinations 
are formed and the one best matching the SpaCal cluster 
is selected. To determine the space points, the $u$ hits are combined with
the $r$ hits extrapolated along the $r$ track to the $z$
position of the $u$ sensor. Then the transformation
$(r,u) \Rightarrow (x,y)$ is performed. A circle fit including the
position of the interaction vertex in $(x,y)$ determined by the beam spot size of $150$\,$\mu$m in $x$ and
$60$\,$\mu$m in $y$, yields the curvature
and therefore 
charge and the transverse momentum of the particle.

\subsection{Reconstruction of the Hadronic Final State\label{sec:hfs}}
The reconstruction of the hadronic final state uses information from the 
central tracker and the LAr and SpaCal calorimeters, excluding a
cone in the SpaCal calorimeter around the electron candidate cluster.
The cone axis is defined by the vertex position and the centre-of-gravity 
of the SpaCal cluster. The cone radius is $20$~cm at the
surface of the SpaCal electromagnetic section. The energy of the
cells  inside the cone is excluded from the hadronic final 
state calculation for both the electromagnetic and hadronic 
sections of the SpaCal.

Tracks pointing to the backward part of the H1 detector are excluded
 from the hadronic final state. Instead, the reconstructed SpaCal
 clusters outside the electron isolation cone are used.
In the central region, the Central Tracker and 
LAr signals are linked for each particle by matching the measurements in each detector.
For energies below $2$\,GeV, the tracker information is used
while for higher energies the calorimeter
information is used, as it provides
the better energy resolution.

The determination of $\Sigma_h$ is affected by the presence of extra 
activity in the calorimeters. The bias is particularly strong for small
$\Sigma_h$ and thus small $y_h$. For the SpaCal, this extra activity can be
induced by the scattered electron, with some energy leaking outside the
isolation cone or by a radiated photon emitted at a large angle.
The contribution of these sources of extra activity to $\Sigma_h$
is proportional to $\Sigma_e$ to good approximation. 
To reduce the influence of these effects, $10\%$ of $\Sigma_e$ is subtracted
from the total SpaCal $\Sigma_h$. If the result is negative, the SpaCal contribution
is set to zero. This procedure reduces the contribution of SpaCal
to $\Sigma_h$  to a negligible level 
for low $y$ events, as is expected from the event kinematics.

Channels affected by  electronic noise
 in the LAr are identified event by event 
using a dedicated topological algorithm.
LAr cells with an energy below $0.4$\,GeV ($0.8$\,GeV), which are separated from other
cells by more than $40$\,cm ($20$\,cm) in the central (forward) region of the
calorimeter are classified as noise and excluded from the 
$\Sigma_h$ and $\bf \Pth $ calculations.

\section{Monte Carlo  Simulations\label{sec:mc}}
In the simulation, 
DIS events are generated using the DJANGOH~1.4~\cite{django} event generator
which includes leading order QED radiative effects as implemented in
HERACLES~\cite{heracles}.
For the event generation, leading order parton distribution functions
define $F_2$ while $F_L$ is set to zero. The structure functions are subsequently
reweighted to the fractal model parameterisation of  $F_2$ 
and to $F_L$  following 
the procedure described in \Sec~\ref{ssec:xsection}.
The final state parton showers are simulated using the Colour Dipole
Model~\cite{CDMA,CDMB} as implemented in ARIADNE~4.1~\cite{ari}.
Events with a very low mass of the hadronic final state ($W < 5$\,GeV) are
simulated using SOPHIA~\cite{sophia},
which includes a detailed description of low mass final states,
including the resonance region. 
The fragmentation into hadrons is performed with JETSET~7.4~\cite{Sjostrand:1993yb}.
Photoproduction background is generated with the PHOJET~1.6~\cite{phojet0,phojet} 
program, which uses a two-component dual parton model~\cite{dpm}  including diffractive processes
and vector meson production.

The simulation of QED radiative corrections includes photon emission
from the lepton. Radiation from quarks,
which is estimated to be small for low $x$, is not simulated.
The simulation of QED radiative corrections is checked
using the analytical calculation package HECTOR~\cite{hector}. 
An agreement to better than $0.5\%$ is found in the kinematic range of this
measurement.

The generated events are passed through a simulation of the H1 detector
response based on the GEANT3~\cite{geant} package. Tracing of the particles in the 
trackers up to the  calorimeters 
is based on a detailed description of the detector material.
The response of the calorimeters to electromagnetic particles is simulated
using a fast shower parameterisation technique~\cite{gflash}, 
while the hadronic response is simulated using GHEISHA~\cite{gheisha}.

The level of noise and beam related background in the 
calorimeters is determined using 
events from dedicated runs with random triggers 
which are  overlaid on the simulated events.
Spurious hits in the BST are added
to the simulation based on randomly triggered events.

The MC events are subjected to the same reconstruction and analysis
procedure as the  data. Also, for consistency of the analysis,
the calibrations of the SpaCal and the LAr, as well as the
BST and BDC alignments, are performed for the reconstructed MC 
events in the same way as for the data.

\section{Data Analysis\label{sec:dataanal}}
At low $Q^2$ the DIS cross section is
large, and for the available integrated luminosity for this analysis the
statistical uncertainty of the measurement becomes smaller than the
systematic uncertainty. For low inelasticities, corresponding to a
large fraction of the measured phase space, the scattered electron
energy is large 
 and background contributions are negligible compared to
the genuine DIS signal. In this region a
set of selection criteria is imposed which is sufficient to
reconstruct the event kinematics in the least biased way. 
Whenever possible the electron trajectory is reconstructed
using the BST alone and only the SpaCal is used for triggering.

Events with the scattered electron outside the BST acceptance are reconstructed 
using the BDC and the Central Tracker vertex. The analysis is also extended
to the highest accessible $y$ values for which the  precision 
is limited by the uncertainty of 
photoproduction background. In this region several
additional electron identification criteria are imposed in order to
minimise the systematic uncertainty.

For the two data samples, NVX and SVX, a total of four separate analyses is
performed as summarised in 
\Tab~\ref{tab:ds}. The analyses differ in the triggers
and in the method employed for reconstructing \thetae.  The main
kinematic region of the \MB-BST data set, with \thetae\ 
measured in the BST, is analysed based on the trigger
S0. An extension to $0.75
<y <0.85$ is achieved using the trigger S9 (\MB-S9) and
requiring signals in both tracking detectors, BDC and  BST. For the 
{\SVX} sample, the main region of the phase space is covered by a BST-based
analysis with the trigger S0 (\SVX-BST). An extension to $\thetae =
\thetamaxsvx$ is achieved by adding data collected with the trigger S3
and including events with \thetae\ measured by a combination of the
Central Tracker vertex and BDC  information (\SVX-BDC).

The measurement is verified by performing a number of cross check
analyses exploiting the redundancy in the kinematic reconstruction and
the large overlap of the kinematic regions of different data sets.  The data
reconstructed with the BST are compared with those reconstructed
with the BDC. The results of the electron method are cross checked
with those of the $\Sigma$ method. Moreover, the measurement based on the
shifted vertex sample is compared to that based on the nominal vertex
sample.

In the following a detailed description of the different analyses is given. 
Further information can be found 
in~\cite{ecksteinth, lastovickath, vargasth, behrendtth}.

\begin{table}
\centerline{%
\begin{tabular}{c|c|l}
\hline \hline
{\bfseries Analysis} & {\bfseries Trigger} & \multicolumn{1}{c}{\bfseries Description}\\
\hline
\multicolumn{3}{c}{\T \B NVX ($z_{\rm vtx} \sim 0$\,cm)}
\\
\hline
 \MB-BST & S0      & Main analysis \\
 \MB-S9  & S9      & Extension to lower $\ee$ \\
\hline
\multicolumn{3}{c}{ \T \B  SVX ($z_{\rm vtx} \sim 70$\,cm)}\\
\hline
 \SVX-BST   & S0      & Main analysis \\
 \SVX-BDC   & S0, S3  & Extension to larger \thetae\ \\
\hline \hline
\end{tabular}
}
\tablecaption{\label{tab:ds}Inclusive analyses of DIS data.
The trigger conditions used to collect the data sets are described in
\Sec~\ref{sec:online}, \Tab~\ref{tab:trig}.
}
\end{table}

\subsection{Event Selection}

\subsubsection{Criteria}

An overview of the selection criteria used in the different analyses
is given in \Tab~\ref{tab:cuts}.
\begin{table}[tb]
\centerline{%
\begin{tabular}{ll}
\hline \hline
\multicolumn{1}{c}{\bfseries Description} & \multicolumn{1}{c}{\bfseries Cut}\\
  \hline
\multicolumn{2}{c}{ \T \B Common cuts}\\
\hline
Scattered electron energy & $\ee > 7$\,GeV; $4$\,GeV (\MB-S9)\\
Vertex $z$ position        & $|z_{\rm vtx}-z_{\rm nom}| < 35$\,cm \\
SpaCal cluster radius     & $R_{\rm log}<4$\,cm \\
Hadronic energy fraction  & $E_{\rm had}/\ee < 0.15$  \\
$P_\perp$ balance         & $P_{\perp}^h/P_{\perp}^e > 0.3$\\
  \hline
\multicolumn{2}{c}{ \T \B Electron method cuts}\\
\hline
\empz\ balance    & $ \empz > 35$\,GeV \\
\hline
\multicolumn{2}{c}{ \T \B BST analysis cuts for NVX-S9, NVX-BST and SVX-BST}\\
\hline
BST validation    & $N_{\rm link~BST} \ge 2$ (\MB); $3$ (\SVX)\\
BST--SpaCal radial match & $|\Delta r_{\rm BST-SpaCal}| < 1.5$\,cm\\
BST noise         & $N_{\rm hit~total} < 120$ (\MB); $200$ (\SVX)\\
\hline
\multicolumn{2}{c}{ \T \B BDC analysis cuts for SVX-BDC}\\
\hline
BDC validation         & $N_{\rm link~BDC} \ge 4$\\
BDC-SpaCal radial match & $|\Delta r_{\rm BDC-SpaCal}| < 2.5$\,cm\\
Central Tracker vertex & $N_{\rm track} \geq 1$\\
                       & $y_{\Sigma} \geq 0.03$ \\
\hline
\multicolumn{2}{c}{ \T \B Additional NVX-S9 analysis cuts}\\
\hline
BST--BDC radial match         & $|\Delta r_{\rm BST-BDC}|<0.75$\,cm \\
BST--CT $z_{\rm vtx}$ match          & $|z_{\rm VTX,BST} - z_{\rm VTX,CT}|/\sigma < 5.0$\\
Central Tracker vertex & $N_{\rm track} \geq 2$\\
\hline \hline
\end{tabular}
}
\tablecaption{\label{tab:cuts}Selection criteria used in the analysis.}
\end{table}
The background from
non-$ep$ interactions is suppressed by requiring the event vertex
($z_{\rm vtx}$) to be reconstructed within a distance of $\pm 35$\,cm from
the average $z$ position ($z_{\rm nom}$).
In order to be identified with the scattered electron, the highest energy 
cluster in the
electromagnetic SpaCal section\footnote{For the S9 analysis 
(\Tab~\ref{tab:ds})
the cluster with the maximum
transverse momentum $P_\perp^e$ is chosen instead
of the highest energy cluster.} has to satisfy the 
following criteria: (i) the cluster
centre-of-gravity lies in the region of high efficiency of the
corresponding trigger; (ii) the transverse cluster radius is consistent
with an electromagnetic particle, $R_{\rm log} <
4$\,cm; (iii) the energy deposition in the hadronic SpaCal section
behind the cluster is small, $E_{\rm had}/\ee < 0.15$; (iv) depending
on the analysis, the cluster is validated by a BST or a BDC track
segment. If the highest energy cluster does not satisfy one of these
 cuts, the next  highest energy cluster is
used. This procedure is repeated for up to three clusters with
energies above $7$\,GeV, or $4$\,GeV (\MB-S9).

The further event selection is based on a global balance between the
hadronic final state and the electron.  Events  for
which the hadronic final state is poorly  reconstructed are
rejected by demanding that the total hadronic transverse momentum
$P^h_\perp$ be at least $30\%$ of the electron transverse momentum
$P^e_\perp$. This efficiently removes migrations from very low $y$, which
lie outside the measurement region.
Events with large initial state radiation are excluded
from the electron method measurement by requiring $\empz > 35$\,GeV.
This condition is not used for the $\Sigma$ method, which takes
QED radiation properly into account.

The BST analyses include requirements on the minimum number of BST
hits linked to the electron track ($N_{\rm link~BST}$) and on the
matching of the BST track extrapolated to the \mbox{$z$ position} of
the SpaCal cluster, $\Delta r_{\rm BST-SpaCal} $.  Similarly, for the
BDC based analyses, a minimum number of linked BDC hits ($N_{\rm
  link~BDC} $) and radial BDC--SpaCal matching ($ \Delta r_{\rm
  BDC-SpaCal}$) are required.  In addition, the BST analyses require a
low  level of noise   by cutting on the variable $N_{\rm hit~total}$, the
total number of BST hits.  The BDC analyses demand the presence of
at least one central track ($N_{\rm track}$).

The S9 analysis extends the measurement to low $E_e'$, corresponding
to high $y$, where the largest
uncertainty stems from the large photoproduction background. To
suppress this background, both the BST and BDC track segments are
required to pass the same criteria as in the other analyses.  In
addition, a tight matching condition is applied for the two trackers using
$\Delta r_{\rm BST-BDC}$, the radial distance between the BDC and BST tracks
calculated at the BDC plane, as well as 
$|z_{\rm VTX,BST}-z_{\rm VTX,CT}|/\sigma$, the
distance in $z$ between the BST vertex and the Central Tracker vertex position 
divided by the uncertainty of this difference. Finally, to ensure a
high trigger efficiency for the analysed sample, at least two central
tracks must be reconstructed.

\subsubsection{Efficiency Determination}
\begin{figure}
\epsfig{file=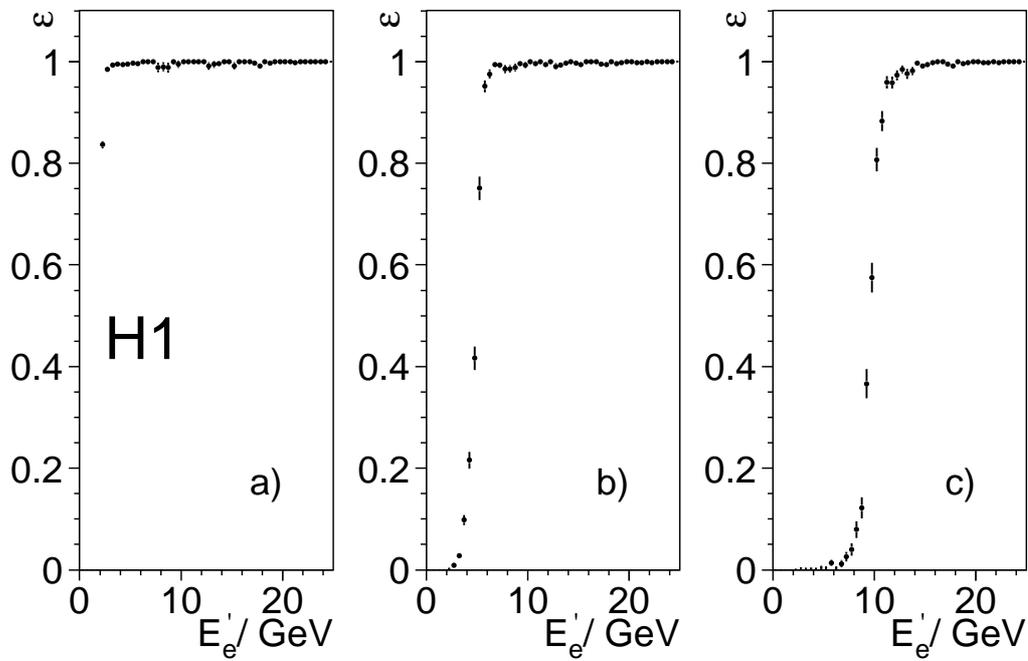,width=0.9\linewidth}
\caption{\label{fig:trigthresh}Efficiency of the
SpaCal electron triggers, 
S9 a), S0 b), and S3 c), used in this analysis, as a function of \ee.}
\end{figure}

The efficiencies of the triggers are determined using independently
triggered data samples.  For the SpaCal trigger conditions, 
most of the cells show a high ($>99.5\%$) efficiency
above the rather sharply defined threshold, see \Fig~\ref{fig:trigthresh}.
A few cells are identified which show  high thresholds.
They are excluded from the analysis by applying
geometrical cuts on the electron impact point reconstructed at the calorimeter
surface, which is calculated using $\theta_e$ and $\phi_e$.  
The efficiencies of the CIP-COP
conditions employed in the S9 trigger (see \Sec~\ref{sec:online}) are
studied as functions of \ee\ and the track multiplicity. 
Since the average reconstructed track multiplicity 
increases with $Q^2$, the inefficiency
diminishes from $3\%$ at $Q^2 = 1$\,GeV$^2$ to $2\%$ at
$Q^2=10$\,GeV$^2$.  The data are corrected for this inefficiency.  The
systematic uncertainty on the trigger efficiency is estimated to be
$1\%$ for S9 and $0.5\%$ for the other triggers.

The inefficiency of the software filter farm component of the trigger
is determined using a sample of the rejected events, recorded for cross checks.
The primary reason for the rejection is the online reconstruction of the
event vertex which occasionally wrongly classifies $ep$ events as non-$ep$
background.
The loss corresponds to $0.7\%$ for the {\MB} analysis and $0.8\%$
($0.5\%$) for the \mbox{\SVX-BST} (\mbox{\SVX-BDC}) analysis. This loss is
 consistent with  being uniform across the phase space and is applied as a global
correction with a systematic uncertainty of  size equal to the correction.

The efficiencies of the electron identification requirements (cluster
shape, hadronic fraction, BDC or BST validation) for high energies of
the scattered electron are evaluated using events passing all other
selection cuts but the one to be investigated.
This direct approach is applicable for $\ee > 20$\,GeV due to the negligible background.

For low $\ee$, the electron identification efficiency is studied after
the background subtraction. In addition, special background-free 
samples are used.  One  such sample comprises
initial state radiation events with the radiated photon detected in
the photon tagger. Background levels below $1\%$ are achieved in this
case, even for $\ee\sim 3$\,GeV, by requiring the sum of the 
photon tagger and  scattered electron energies to be close to the
electron beam energy. The efficiency of the electron identification
cuts does not vary much as a function of \ee. It is always above
$90\%$ and is well reproduced by the simulation.

\begin{figure}
\centerline{%
\epsfig{file=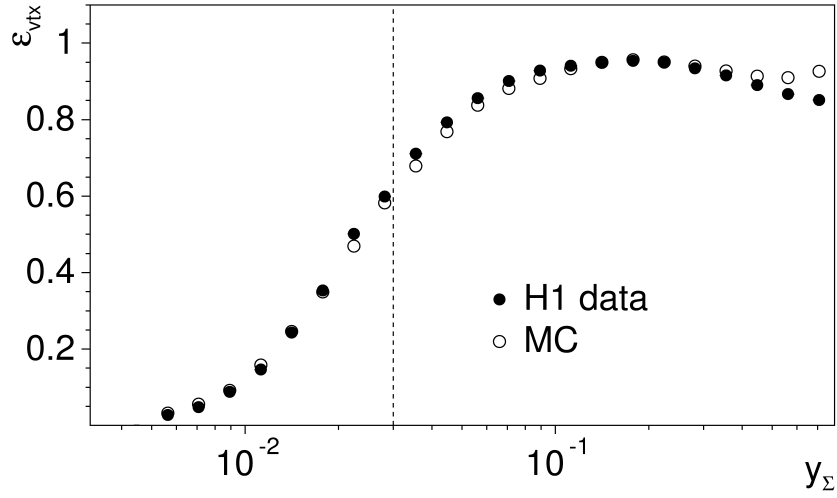,width=0.7\linewidth}
}
\caption{\label{fig:effzvtxct}Central Tracker vertex reconstruction efficiency as a function of $y_{\Sigma}$. 
The dashed line corresponds to the applied
selection criterion, $y_{\Sigma}>0.03$. In the analysis the cross section at high $y$ is
measured with $y_e$ instead of $y_{\Sigma}$.}
\end{figure}

The efficiency to find a central tracker vertex for the BDC analysis
is determined using events with a reconstructed BST vertex. As shown in
\Fig~\ref{fig:effzvtxct}, it is larger than $50\%$ for $y_{\Sigma}>0.03$
and $93\%$ for $y_{\Sigma}>0.1$. The BDC analysis is  
restricted to $y_{\Sigma}>0.03$. At
larger values of $y_{\Sigma}$ the efficiency decreases again, the
effect being more pronounced in the data than in the
simulation. The reason for this difference is a deficit of events with a
large rapidity gap in the DJANGOH model, as already observed in
\cite{Adloff:1997mf}. This is accounted for by applying a correction to
the MC simulation. The systematic error of the cross section resulting 
from this correction is found to decrease with increasing $Q^{2}$, from 
$10\%$ to $2\%$.

A special procedure is developed to determine the BST inefficiency.  Two
main sources of inefficiency are distinguished which are both closely
related to the readout procedure.  The first is a hit finding
inefficiency, which mostly depends on the performance of the front-end amplifier
readout chip (APC). This efficiency is determined for each APC
using BST tracks,
requiring hits reconstructed in all but the BST plane under investigation.
For most of the APCs the efficiency is high ($>95\%$), but about $5\%$
of APCs have efficiencies below $80\%$. A few
BST regions, with an APC efficiency below $40\%$, are excluded from the
analysis. 

Correlated readout losses constitute the second source of
inefficiency. In this case, signals are lost coherently in either
BST1, BST2 or in both modules. The main source of coherent losses
comes from timing desynchronisation at a level of
 about $5\%$ with some dependence on the $\phi$ sector.
The coherent losses are measured separately and combined for BST1 and
BST2 for each $\phi$ sector. They are measured using
a background-free DIS sample ($15$~GeV$<E'_e<$~$32$~GeV) with a well
reconstructed CT vertex and BDC track. All sources of
BST losses are incorporated into the simulation.

\begin{figure}
\centerline{%
\psfig{file=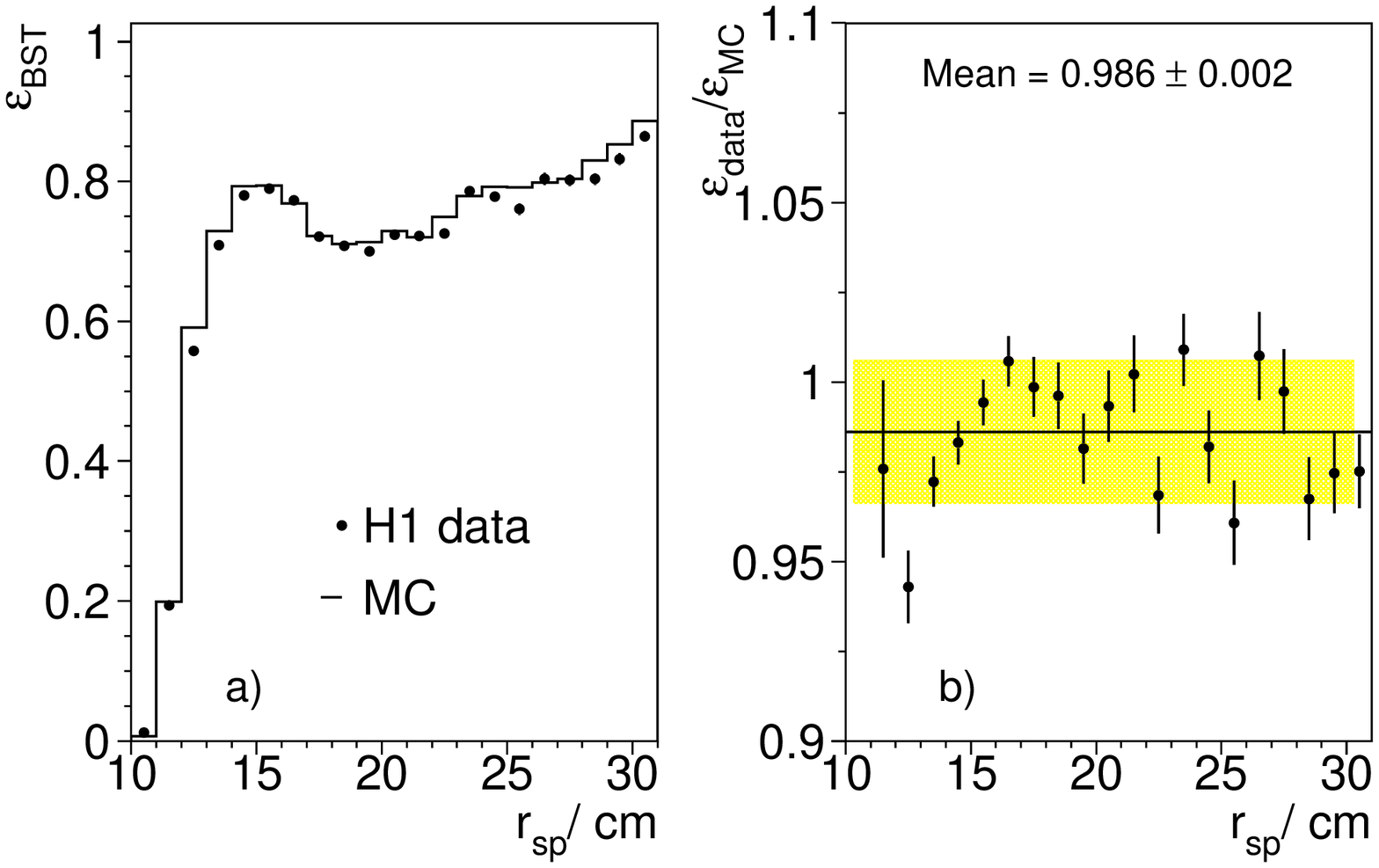,width=0.9\linewidth}
}
\caption{\label{fig:effbstmb}a) BST track segment finding efficiency as a function of the radial
position of the electron candidate in the  SpaCal, 
for the {\MB} data sample, b) ratio of data to MC efficiencies.}
\end{figure}

\begin{figure}
\centerline{%
\psfig{file=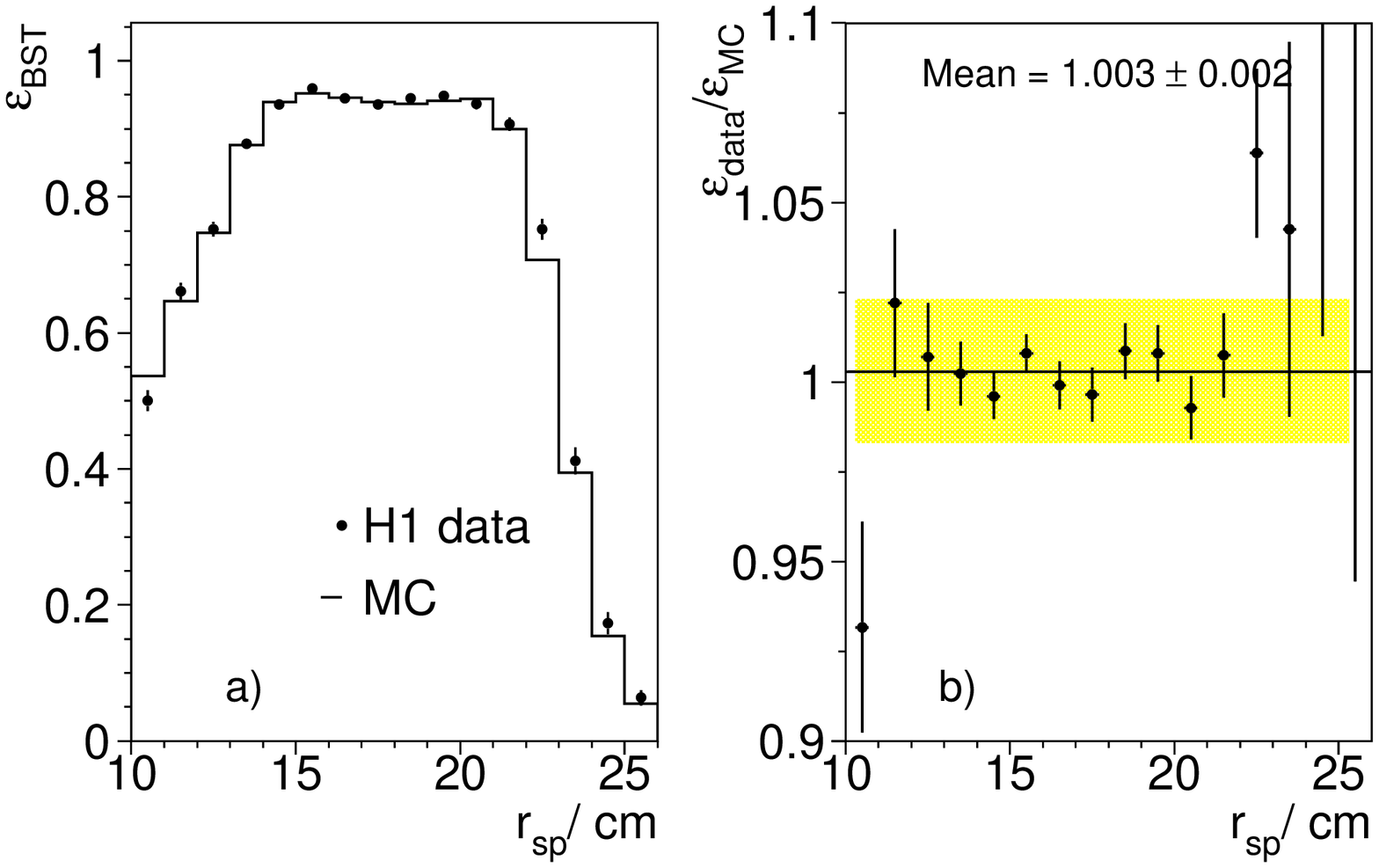,width=0.9\linewidth}
}
\caption{\label{fig:effbstsvx}a) BST track segment finding efficiency as a function of the radial
position of the electron candidate in the SpaCal, for the {\SVX} sample,
b) ratio of data to MC efficiencies.}
\end{figure}

The efficiency of the BST track segment finder, used to reconstruct
the event vertex within the nominal $z$ range and to validate
the SpaCal electron candidate, is checked globally for data and
for the DJANGOH simulation using events with
a well reconstructed central vertex and a BDC track segment. The
photoproduction background is suppressed by demanding 
$E'_e > 15\, \mathrm{GeV}$.  
In this procedure not only is the BST efficiency
examined, but the description of the BST acceptance and the
imperfections of the tracking algorithm are also
checked. \FFig~\ref{fig:effbstmb} and \Fig~\ref{fig:effbstsvx} show the
global BST efficiency as a function of the electron candidate's radial
position in the SpaCal for the {\MB} and {\SVX} samples, respectively.
Based on this comparison, the systematic uncertainty 
attributed to the description of the BST efficiency
is taken to be $2\%$ in both the {\MB} and the {\SVX} data
analyses. This value also includes uncertainties arising from
inefficiencies of the other electron identification 
criteria described above.

The efficiency of the selection criteria based on the
BDC, $N_{\rm link~BDC} \ge 4$ and
$|\Delta r_{\rm BDC-SpaCal}| < 1.5$\,cm, is determined
for events with \mbox{$E'_e > 20\, \mathrm{GeV}$} for data and for the
DJANGOH simulation. A correction to the simulated events is applied to
account for discrepancies which are largest in the narrow transition
region from small to large cells in the 
BDC.  
Events in this region are rejected from the
\MB-S9 analysis. The systematic uncertainty attributed to the BDC
efficiency amounts to $1.5\%$, also covering differences between data
and the MC simulation for other electron selection criteria.

\subsection{Subdetector Alignment and Calibration}

\subsubsection{Alignment} \label{sec:alignment}
\paragraph{Alignment of the Central Tracker, BDC and SpaCal.}

The relative alignment of the H1 sub-detectors and the alignment of the
detector with respect to the beam direction is performed in several
steps. The first step is the internal alignment of the Central
Tracker. The $x$ and $y$ coordinates are defined by the precisely known
positions of the CJC wires while the $z$ coordinate is defined by the
 COZ. Using cosmic muon tracks, the relative
positions of the inner and the outer CJC parts, the location of the
CIZ and  the parameters for the $z$ coordinate measurement
in the CJC are determined.  The
beam axis is reconstructed by measuring the $x$ and $y$ coordinates
of the interaction vertex as functions of its $z$ coordinate.

The alignment of the SpaCal and of the  BDC  is performed 
using high energy electron candidates, with $\ee > 20$\,GeV, linked
to a central track. The central track is required to have
at least two reconstructed CIZ hits and the $\theta$ uncertainty must
be smaller than $2$\,mrad.
The $x$ and $y$ offsets of the BDC and the SpaCal are measured by studying
the difference in the polar angle measurements for these electron 
candidates between the Central Tracker and the BDC,
$\Delta\theta = \theta_{\rm CT} -\theta_{\rm BDC}$,
and between
 the azimuthal angle measurements from the Central Tracker and the SpaCal,
$\Delta\phi = \phi_{\rm CT} - \phi_{\rm SpaCal}$,
as functions of the azimuthal angle $\phi_{\rm SpaCal}$.
The two methods find a consistent alignment  in the $x$
direction.
For the $y$ direction, the alignment is found to be different by $2$\,mm
between the $\Delta\theta$ and $\Delta\phi$ methods. The average of the two
values is used to correct for the misalignment.

\begin{figure}
\centerline{%
\psfig{file=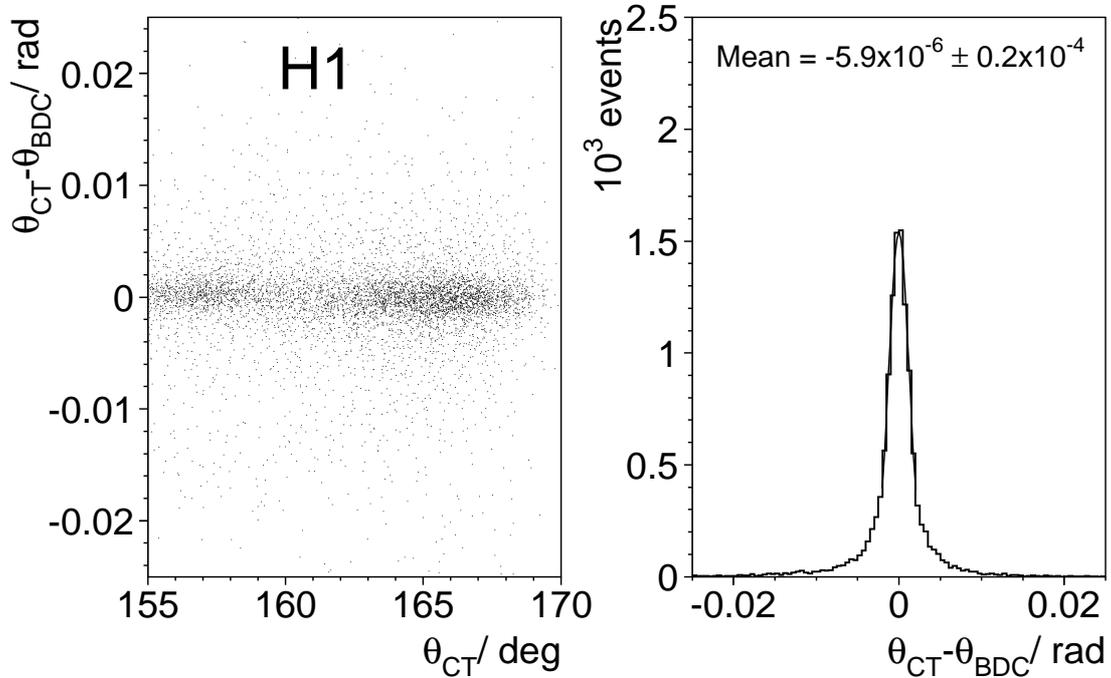,width=0.9\linewidth}
}
\caption{\label{fig:ctbdcalig}Relative alignment of the Central Tracker
(CT) and BDC.
Left: $\theta_{\rm CT}-\theta_{\rm BDC}$  versus $\theta_{\rm CT}$ 
after alignment.
Right: $\theta_{\rm CT}-\theta_{\rm BDC}$ after alignment.}
\end{figure}

The $z$ offset of the BDC is measured by studying $\Delta\theta$ versus
$\theta_{\rm CT}$.
The $z$ offset of the SpaCal is checked by comparing the $\theta$ measurements
in the BDC and in the SpaCal.
The tilts of the backward detectors are studied using $\Delta\theta$
versus $\theta_{\rm CT}$ for positive and negative $x$ and $y$ separately;
they are found to be negligible. 
\FFig~\ref{fig:ctbdcalig} shows the comparison of the $\thetae$
measurement in the Central Tracker and the BDC after alignment.

The SpaCal alignment with respect to the beam direction is
cross-checked using quasi-elastic QED Compton (QEDC) events. These are
$ep$ scattering events of the type $ep \rightarrow ep \gamma$ with a
hard photon radiated from the lepton line, the proton being scattered
quasi-elastically at low momentum transfer such that the outgoing
electron and photon are detected in the main detector, nearly
back-to-back in azimuth.
The QEDC process is selected by requiring two energy deposits in
the electromagnetic SpaCal section with energies above $4$\,GeV.  The
sum of both cluster energies is required to exceed $25$\,GeV.  The
back-to-back requirement is enforced by demanding $\cos \Delta\phi_{e\gamma} <
-0.9$ with $\Delta\phi_{e\gamma}$ being the azimuthal angle between the electron and the photon.
Elastic events are selected by demanding no tracks reconstructed
in the CJC and low activity in the calorimeters apart from the
selected electron and photon. This alignment agrees within $1$~mm
with the alignment obtained using central tracks.

The dominant uncertainty of the alignment stems from the difference in
the $y$ direction 
between the $\Delta\theta$ and $\Delta\phi$ methods.
Since the H1 detector is nearly $\phi$ symmetric, biases in $y$ reconstruction
do not lead to large shifts in the measured cross section.
To cover a potential global bias of the $\theta_e$ measurement, 
a systematic uncertainty of $0.5$\,mrad is assigned for the
polar angle reconstruction with the BDC and the Central Tracker.

\paragraph{BST alignment.}
In the global BST alignment, the position of the BST is determined
with respect to the H1 coordinate system. In the internal BST
alignment, radial offsets and rotations around the $z$ axis of the
individual wafers are determined.  The global and internal alignments
use the electron track reconstructed from
  the Central Tracker vertex and the BDC track
segment  as a reference and compare
it to the track segment found in the BST.

\begin{figure}
\centerline{%
\psfig{file=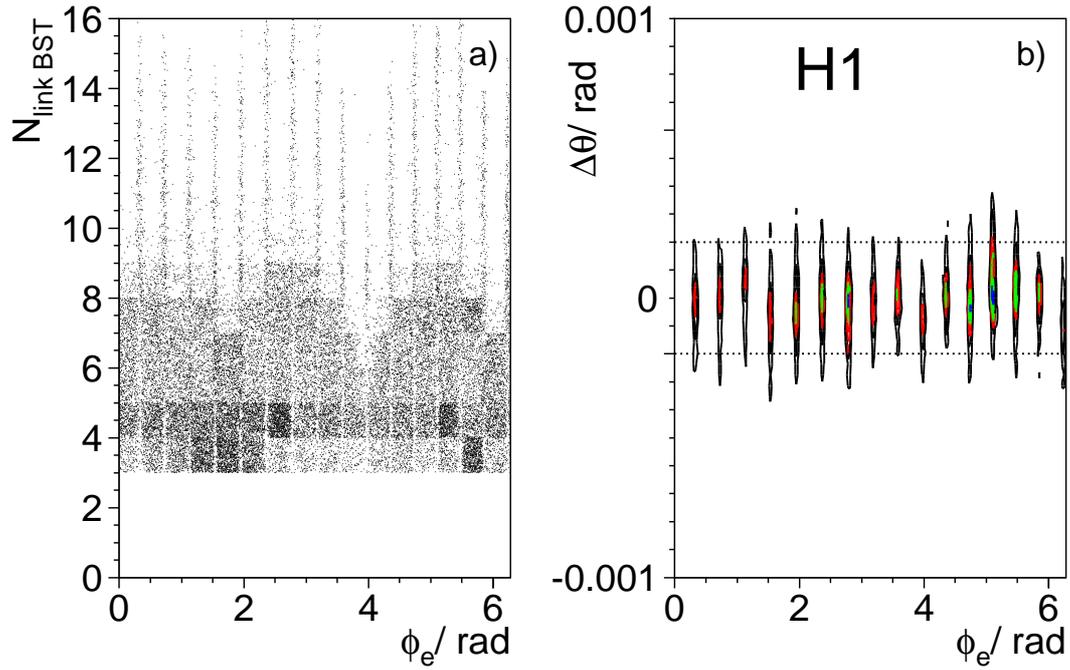,width=0.9\linewidth}
}
\caption{\label{fig:bstalig} {a)} Scatter plot of the number of BST
hits linked to the electron candidate BST track as a function of the azimuthal
angle $\phi_{\rm e}$ determined by the associated SpaCal cluster. 
At least three linked hits are required to define a track.
A number of linked hits exceeding eight corresponds  to a
track passing the azimuthal  BST 
wafer overlap region; {b)} Contours of equal density for the distribution of
$\Delta \theta = \theta_1 - \theta_2$, where $\theta_{1,2}$ are the polar
angles measured in the two  overlapping BST sectors, as a function
of $\phi_{\rm e}$. The horizontal dotted lines indicate  $\pm 0.2$\,mrad
as is used for the systematic uncertainty of the alignment.}
\end{figure}

During the detector assembly each sensor is fixed to its nominal
position with a mechanical precision of about $100\,\mu$m.
Remaining degrees of freedom are $128$ radial shifts and $128$ rotations of the
wafers.  For the \mbox{$r$ strip} sensors, these parameters
are determined for all detectors simultaneously using the
global minimisation package Millepede~\cite{Blobel:2002ax}.  The
degeneracy between shifts and rotations is resolved utilising the
wafer overlap regions. Typical shifts are less than $200\;\mu$m and
most rotations are less than $1$ mrad. \FFig~\ref{fig:bstalig}a) shows
the distribution of the number of BST linked hits as a function of
$\phi_{e}$. \FFig~\ref{fig:bstalig}b) shows the difference in the
$\thetae$ measurement between the two BST overlapping sectors in these cases,
after the BST alignment. An agreement to better than $0.2$\,mrad
is observed. Based on this study, the uncertainty on the scattered
angle reconstruction by the BST is taken to be $0.2$\,mrad.

The alignment of the $u$ strip detector is done in an analogous way. Here,
shifts perpendicular to the $u$ coordinate for the $8$ wafers are
determined simultaneously using the interaction vertex and the
BDC measurement as additional external constraints. The shifts of up
to about $100\,\mu$m are included in the external alignment.

\subsubsection{Electromagnetic Energy Calibration} \label{sec:elmcalib}
The largest uncertainty in the electromagnetic energy calibration
stems from fluctuations of the gain factors of the individual SpaCal
photomultiplier tubes.  During the data taking, an initial cross
calibration of the SpaCal cells was performed using cosmic muons. The
stability of the gains was controlled by means of a dedicated LED
system. First corrections to the gain factors were applied
using DIS events based on the position of the ``kinematic peak'' - an
enhancement in the \ee\ distribution close to the electron beam energy
which is characteristic of DIS at low $Q^2$ at HERA.

At the analysis stage, a cell-by-cell gain determination is performed
using the double angle (DA) calibration. The DA method is also used to
perform additional non-uniformity corrections taking into account
variations of the energy scale on the sub-cell size level. The SpaCal
energy non-linearity, caused particularly by the energy losses in dead
material in front of the calorimeter, is modelled
in detail~\cite{LendermannTh}
using the H1 detector simulation based on the GEANT program~\cite{geant}. 
The simulation is checked and corrected
using $\pi^0\to\gamma\gamma$ decays. Finally,  the energy scale is
checked using $J/\psi\to ee$ decays and QED Compton events.  All
calibration steps are described in the following.

\paragraph{Double Angle Calibration.}
The double angle calibration procedure makes use of 
kinematic peak events.
Large statistics are available in this kinematic domain with
negligible background contamination.  For $y < 0.1$
the hadronic methods of $y$ reconstruction (\Eqs~\ref{eq:yh} and \ref{eq:ys})
have superior resolution. In particular, the scattered electron energy
can be re-expressed in terms of the hadronic (\Eq~\ref{eq:thh})
and  electron scattering angles as
\begin{equation}
E_{\rm DA} = \frac{E_e\left(1-y_{\rm DA}\right)}
{\sin^2 \frac{\thetae}{2} }~, \qquad
y_{\rm DA} = \frac{\tan \frac{\thetah}{2}}
{\tan \frac{\thetah}{2}+\tan \frac{\thetae}{2}}~.
\end{equation}
In this method the scattered electron energy is calibrated
to the electron beam energy\footnote{The influence of  initial
state radiation, which effectively reduces the electron beam energy,
is small for this kinematic selection and is included in the simulation.}. 
The calibration corrects for
genuine miscalibration and also energy loss
in the dead material between the interaction point and
the calorimeter. The same calibration procedure is applied separately to the
data and the simulated events.

For the calibration, events with $\ee > 20$\,GeV are selected.  The event
vertex position and the scattered electron angle are measured using
the BST. A good $E_{\rm DA}$ resolution is achieved by requiring
$15^\circ < \thetah < 80^\circ$.
The calibration is performed by adjusting the gain factors of the
individual SpaCal cells, such that the energy of the cluster agrees
with the reference given by $E_{\rm DA}$. This is achieved in an
iterative procedure: for each selected event, a ratio $c_{\rm
  ev}=E_{\rm DA}/\ee$ is calculated.  The cluster energy is usually
shared among several cells; the contribution of each cell with an
energy $E_i$ is given by $W_i = E_i/\ee$.  A $W_i$ weighted average of
$c_{\rm ev}$ for each calorimeter cell is then calculated based on all
calibration events. This average is used to modify the gain factor at
the next iteration. The calibration procedure 
converges after three iterations.

The cell-by-cell calibration is followed by calibrations as a function
of (i) the distance between the
centre-of-gravity of the cluster and
the centre of the cell with highest energy to correct for biases
of the clustering algorithm,
(ii) $R_{\rm box} = \max(|x_{\rm sp}|,|y_{\rm
  sp}|)$, where $x_{\rm sp}$, $y_{\rm sp}$ are the $x$ and $y$
 cluster coordinates, in order to correct for energy losses in 
between SpaCal cells, and (iii) $r_{\rm sp} =
\sqrt{x_{\rm sp}^2+y_{\rm sp}^2}$, to correct for losses in the
dead material in front of the calorimeter.  These additional corrections
are applied sequentially.

\begin{figure}
\centerline{%
\epsfig{file=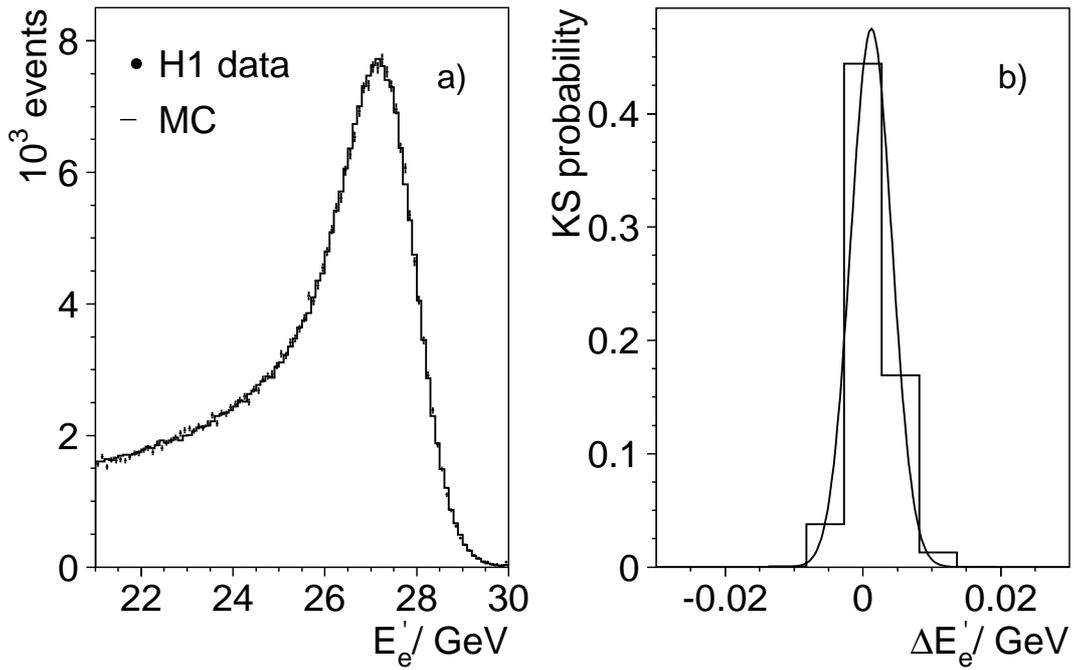,width=0.9\linewidth}
}
\caption{\label{fig:figpeak}{a)} Distribution of the scattered
electron energy $\ee$ for the {\MB} data sample; 
b)~Kolmogorov-Smirnov test probability distribution
as a function of the relative shift in the measured and simulated
energy distributions.}
\end{figure}

The results of the double angle calibration are checked by comparing
the distribution of the electron energy $\ee$ in the data and
the simulation for the standard selection in the kinematic peak region.
By comparing the widths of these distributions an additional 
Gaussian smearing of $1.1\%$ ($0.2\%$) is applied to the
electron energies in the simulated events for the {\MB} (\SVX) data 
set. The need for this smearing in the
MC may be due to short time scale drifts of the photomultiplier gain
factors which are not simulated, to imperfections in the shower
shape simulation or to  a deficiency in the passive material
 simulation.
For the {\MB} sample, the kinematic
peak comparison is presented in \FFig\,\ref{fig:figpeak}a). 
\FFig\,\ref{fig:figpeak}b) shows the Kolmogorov-Smirnov
(KS) test probability distribution as a function of the relative shift
in the energy distribution between the data and the simulation. Shifts
above $20$\,MeV are excluded, which corresponds to 
a relative energy scale agreement better than $0.1\%$.  
The systematic uncertainty
on the relative energy scale at the kinematic peak is taken to be
$0.2\%$ to account for the uncertainties of the HERA beam energy,
for uncertainties in the resolution adjustment in the simulation,
deficiencies of the double angle method and a 
residual variation of the level of agreement in the 
kinematic peak between data and MC  for different $Q^2$ bins.

\paragraph{Calibration using $\piz \to \gamma\gamma$.}
The double angle energy calibration determines the energy scale of the
individual SpaCal cells and radially dependent corrections of the energy
loss for scattered electron energies close to the electron beam
energy.  The deviations from the linearity of the SpaCal response are
measured using $\pi^0\,\to\,\gamma\gamma$ decays which probe 
much lower energies.

Events with two clusters in the electromagnetic SpaCal
section are selected. The larger of the two cluster energies is required to be
above $2.2$\,GeV, exceeding the trigger energy threshold, the
smaller cluster energy is required to be above $0.7$\,GeV. The event
vertex is determined from tracks reconstructed in the Central
Tracker. The two clusters are assumed to be produced by two photons.
The invariant mass, $M_{\gamma\gamma}$, is calculated
using the reconstructed cluster energies and positions.

\begin{figure}
\centerline{%
\psfig{file=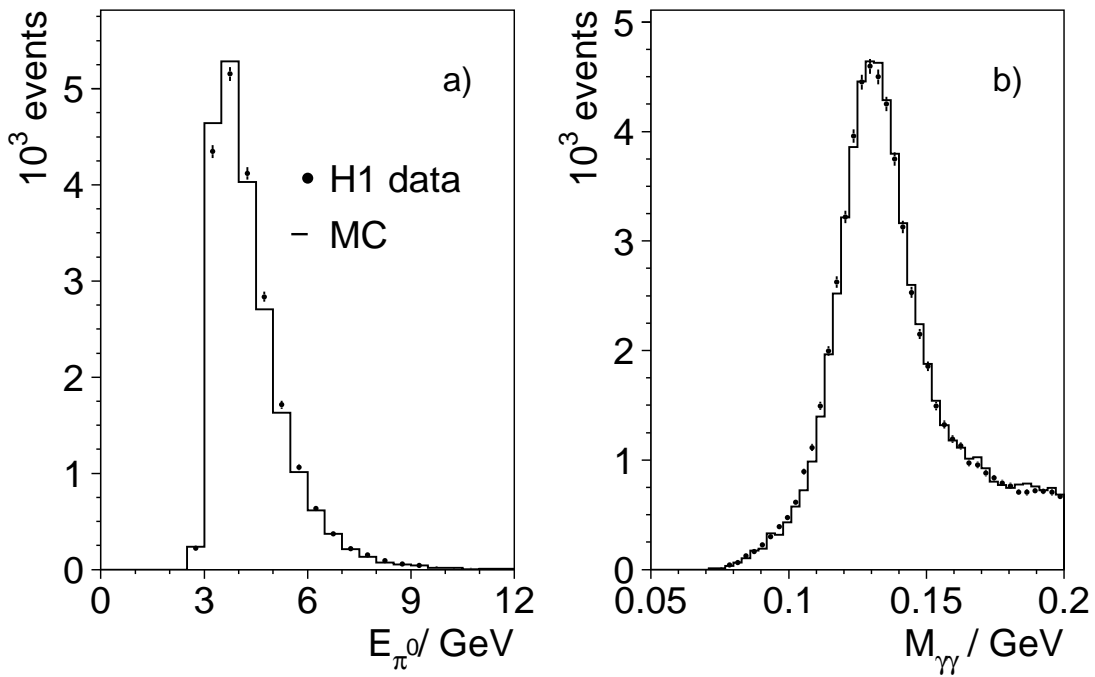,width=0.9\linewidth}
}
\caption{\label{fig:pzm}a) Energy distribution 
for $\pi^0$ candidates based on the {\MB} sample as triggered by
the low energy trigger, S9;
b) di-photon invariant mass distribution for $\pi^0$
candidates. The double angle calibration constants are applied to the
data and MC simulation.}
\end{figure}

The simulation of $\pi^0 \to \gamma\gamma$ decays is checked using the
PHOJET MC sample. A reasonable agreement of the simulation with
the data is observed for the total energy of the two clusters 
as shown in \Fig~\ref{fig:pzm}a).   
The simulated $\pi^0$ energy spectrum is
reweighted to that observed in the data, in order to
reproduce the opening angle and individual photon energy
distributions.

The two-photon mass distribution is shown in \Fig~\ref{fig:pzm}b).  A
prominent peak above the background is observed close
 to the nominal $\pi^0$ mass.
The peak is shifted to lower values, around $130$\,MeV
instead of $135$\,MeV. This difference is not reproduced by the MC
simulation after the double angle calibration. The figure shows the
distribution of simulated events after applying an additional
correction of $-3\%$ to the energy scale for them. 
The data and the simulation are then in
a good agreement. The shift of the peak is possibly caused by not
fully simulated energy losses in the dead material in front of the calorimeter.

The low $\gamma \gamma$ invariant mass and the relatively high photon
energy cuts in the study of $\pi^0 \to \gamma\gamma$ decays lead to a
rather small separation between the photon clusters in SpaCal, with an average
separation of only $13$\,cm. An overlap of the
adjacent clusters could lead to an energy scale shift.  Additional
studies are performed to estimate this effect.  The data sample is
split into sub-samples with approximately equal statistics based on
the larger or smaller cluster energy or on the cluster separation.  In
addition, the $M_{\gamma\gamma}$ distribution is studied as a function
of the projected $\pi^0$ location in the calorimeter, the latter being
calculated as an energy weighted sum of the two cluster positions.  In
all these studies the relative shifts of the energy scale in the data
versus the simulation are consistent within $1\%$ which is taken as a
systematic uncertainty of the energy determination at $\ee=2$~GeV.

A check of the relative energy scale using $\pi^0$ decays is also
performed for the {\SVX} sample. The larger distance from the decay
vertex to the calorimeter leads to  larger separations between photon
clusters, on average $18$\,cm.
The relative shift of the
$M_{\gamma\gamma}$ distribution between the data and the simulation
after the double angle calibration is $-2.7\%$ in this case,
consistent with the shift observed for the {\MB} sample.

The relative bias of the energy scale is corrected
in the data assuming a linear dependence on $E'_e$.
No correction is applied at $\ee=27.6$\,GeV and a correction of $+3\%$
is applied  at $\ee = 2$\,GeV.  The systematic uncertainty
of the energy scale determination is also assumed to follow a linear
dependence rising from $0.2\%$ at $\ee=27.6$\,GeV to $1\%$ at $2$\,GeV.

\paragraph{Tests of the SpaCal Energy Calibration.}
The  SpaCal energy response 
is checked using
$J/\psi \to ee$ decays and QED Compton $ep \to ep\gamma$ events.  The
$J/\psi$ candidates are selected by requiring exactly two electromagnetic
clusters with a total energy of less than $22$\,GeV. At least one of
the two clusters has to be linked to either a BST or a CJC track and
both clusters must be associated with a BDC track segment. 
Events with additional CJC tracks not associated to the electrons are
rejected, thus selecting events from elastic $J/\psi$ production. 
The event vertex is defined by the CJC or the BST tracks.

In this study, the SpaCal energy measurement is explicitly corrected
to the absolute scale
obtained from the mean ratio of the reconstructed to the generated electron
energy from the DJANGOH simulation. Both the double angle and $\pi^0$
calibration corrections are applied, so that the peak in the
di-electron invariant mass $M_{ee}$
distribution can be directly compared to the nominal $J/\psi$ mass,
$M_{J/\psi} = 3.096$\,GeV.

\begin{figure}
\centerline{%
\psfig{file=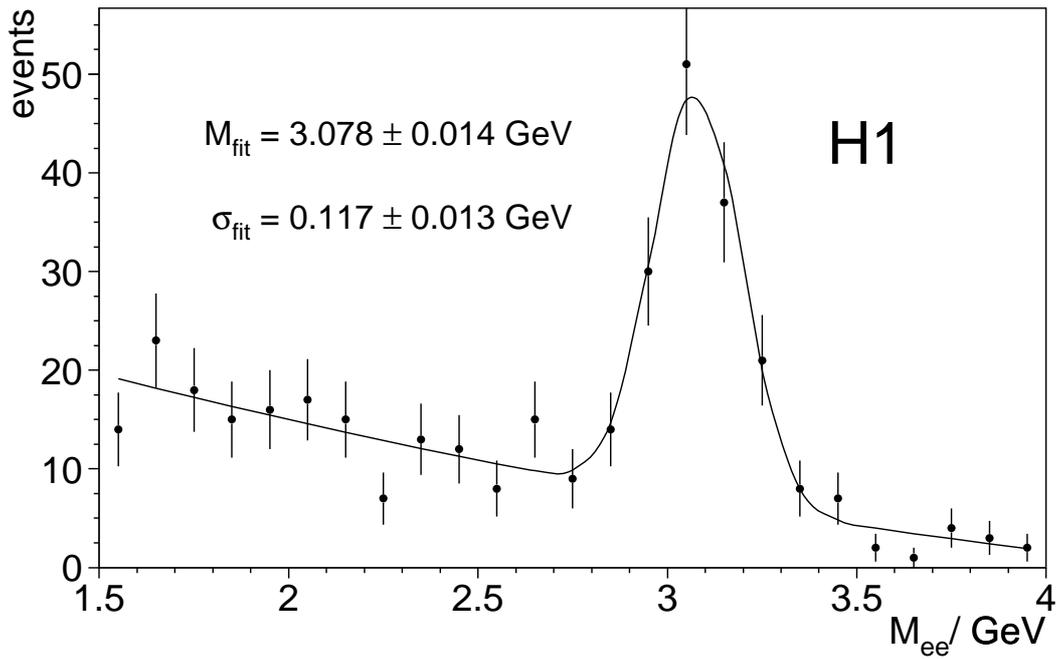,width=0.9\linewidth}
}
\caption{\label{fig:jpsi}Invariant mass distribution of the two 
electron candidate tracks for a special $J/\psi$ event selection.
The line indicates a fit to the data. $M_{fit}$ and $\sigma_{fit}$ 
correspond to the Gaussian mean and width of the peak. }
\end{figure}

The distribution of $M_{ee}$ for the {\MB} data is shown in
\Fig~\ref{fig:jpsi}.  A clear enhancement around the nominal $J/\psi$
mass is observed. The data are fitted with a sum of a Gaussian for the
signal peak and a second order polynomial to describe the background
shape.  The fit uses the binned maximum likelihood method. The
measured Gaussian peak position agrees with $M_{J/\psi}$ within
$1.3\,\sigma$. Based on this agreement the deviation from the nominal
energy scale is limited to be below $0.8\%$ at $68\%$ confidence level for
energies of about $6\,$GeV.

QED Compton events are used to check the calorimeter energy
scale in the intermediate energy region. For elastic events, the energy 
of the scattered electron is
related~\cite{qedc} to the polar angles of the scattered electron
\thetae\ and the photon $\theta_\gamma$ by
\begin{equation}
E^{\rm DA}_{\rm QEDC} = \frac{2 E_e \sin \theta_\gamma }%
{\sin \thetae + \sin \theta_\gamma  -
 \sin \left( \thetae + \theta_\gamma \right)}~.
\label{Eq:qedc}
\end{equation}
The comparison of the measured electron energy with 
$E^{\rm DA}_{\rm QEDC}$  tests
 the SpaCal energy scale linearity in the 
range $4-23$\,GeV.

For the QED Compton energy scale check, a bias free reconstruction of
the electron and photon angles is essential. Therefore in addition to
the basic QEDC event selection described in
\Sec\,\ref{sec:alignment}, both electron and photon SpaCal clusters
are required to be linked to BDC track segments.  This implies that
the photon  converted in the detector material in front of the
BDC. The electron cluster is identified by requiring a BST link.  The
photon cluster must have no associated signals, neither in the BST
nor in the CIP.

\begin{figure}
\centerline{%
\psfig{file=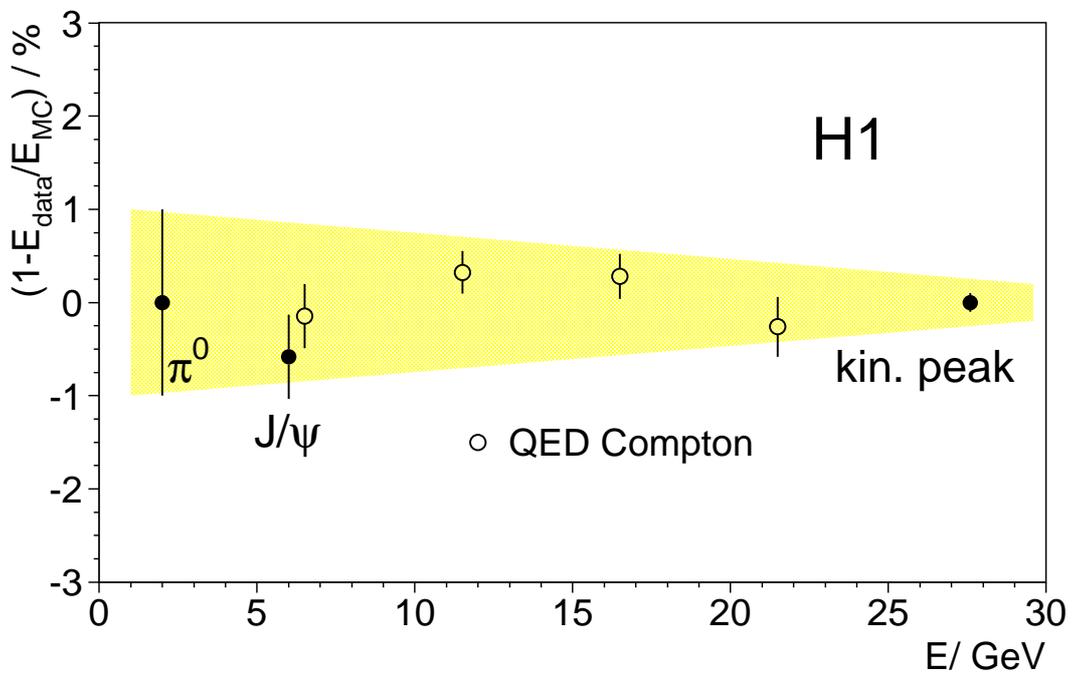,width=0.9\linewidth}
}
\caption{\label{fig:scalvse}Summary of SpaCal energy scale determination. The 
band indicates the uncertainty due to the scale difference between the
data and the simulation.}
\end{figure}

\begin{figure}
\centerline{%
\psfig{file=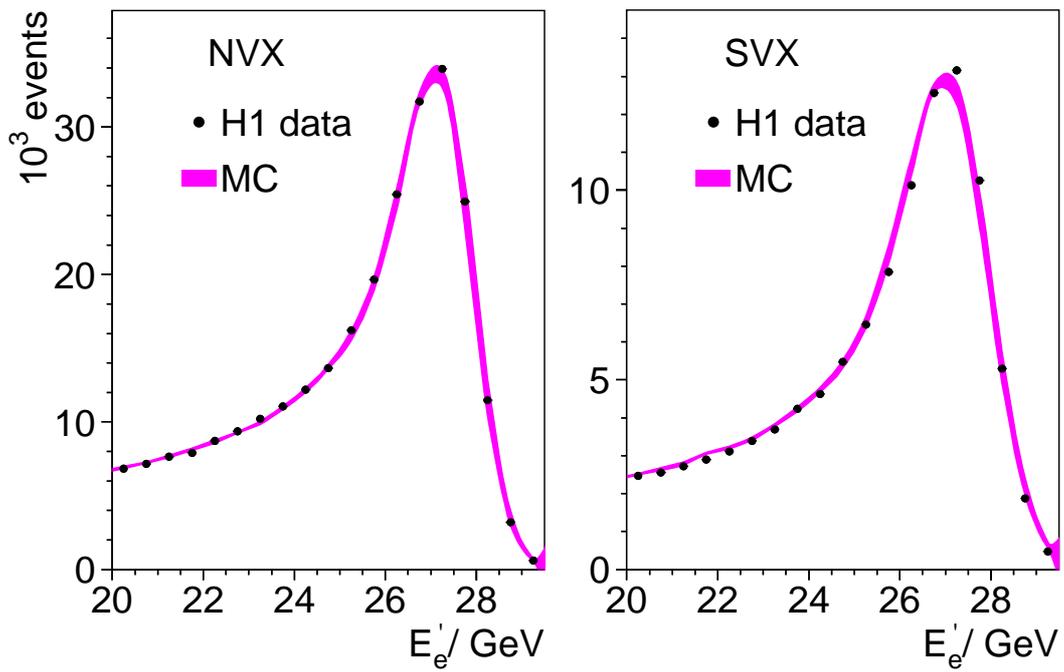,width=0.9\linewidth}
}
\caption{\label{fig:kinmbsvx} Distributions of the scattered electron energy \ee\ for the 
data and the MC simulation in the {\MB} (left) and the {\SVX} (right)  analyses.
The MC bands include the statistical uncertainty and the effect of a 
$\pm 0.2\%$ electromagnetic energy scale variation.}
\end{figure}

The results of  all calibration studies are summarised
in \Fig~\ref{fig:scalvse}.  
Both the $J/\psi$ and the various QEDC
energy scale determinations are inside the uncertainty band.  The
scattered electron energy distributions and the uncertainty bands
attached to the simulated \ee~ distributions in the kinematic peak
region are shown in \Fig~\ref{fig:kinmbsvx} for the {\MB} and the
{\SVX} analyses. The data are well described by the simulations.

\subsubsection{Calibration of the Hadronic Energy Scale \label{sec:hadcalib}}
The calibration of the calorimeters employed for the hadronic final
state energy measurement is based on kinematic constraints relating
the scattered electron to the hadronic final state. For the
calibration of the LAr calorimeter, conservation of the total
transverse momentum \pt\ is used. The SpaCal calibration makes use of
the conservation of \empz.

\paragraph{Calibration of the LAr calorimeter.}
The hadronic final state in the central and forward regions of the H1
detector is reconstructed using a combination of tracks and LAr 
energy deposits (see \Sec~\ref{sec:hfs}).
The LAr calibration coefficients are
determined for the eight calorimeter wheels, each subdivided into
eight octants, separately for the hadronic and electromagnetic
sections. There are thus $120$ calibration constants in
total\footnote{The most backward LAr wheel does not have a hadronic
  section.}, corresponding to the calorimeter segmentation in
rapidity, azimuthal angle, and depth.
The same calibration procedure is applied to the
data and the MC simulation.

To reduce the influence of the SpaCal on the  calibration of
the LAr, forward and central
 hadronic angles are selected: $13^\circ \le \thetah
\le 150^\circ$. 
The electron transverse momentum $P_{\perp}^e$ is determined
from the SpaCal energy and the $\theta_e$ measured by the BST.
 The photoproduction background is reduced to a negligible
level by requiring $\ee>20$\,GeV.

In the calibration procedure, a least squares minimisation of the following function is
performed
\begin{equation} 
 \label{eq:larcalib}
{\cal L}(\alpha_j ) 
= \sum 
 \left(P_{\perp}^e - \big|\mathbold{P}_{\perp}^{\rm Tr}+ \mathbold{P}_{\perp}^{\rm Sp} +
 \sum_j \alpha_j \mathbold{P}^j_\perp \big| \cdot \left|\cos(\phi_e-\phi_h)\right| \right)^2.
\end{equation}
Here the transverse momenta
$\mathbold{P}_{\perp}^{\rm Tr}$ and $\mathbold{P}_{\perp}^{\rm
  Sp}$ are the vector sums of the  contributions from the tracks and
the SpaCal, respectively, $\mathbold{P}^j_\perp$ are the vector sums
of the contributions from all cells of a calorimeter volume $j$, $\phi_h$ is
the azimuthal direction of the hadronic final state and $\alpha_j$ are
the calibration coefficients, which are free parameters.  The outer
summation is performed over all DIS events selected for the
calibration.

\begin{figure}
\centerline{%
\psfig{file=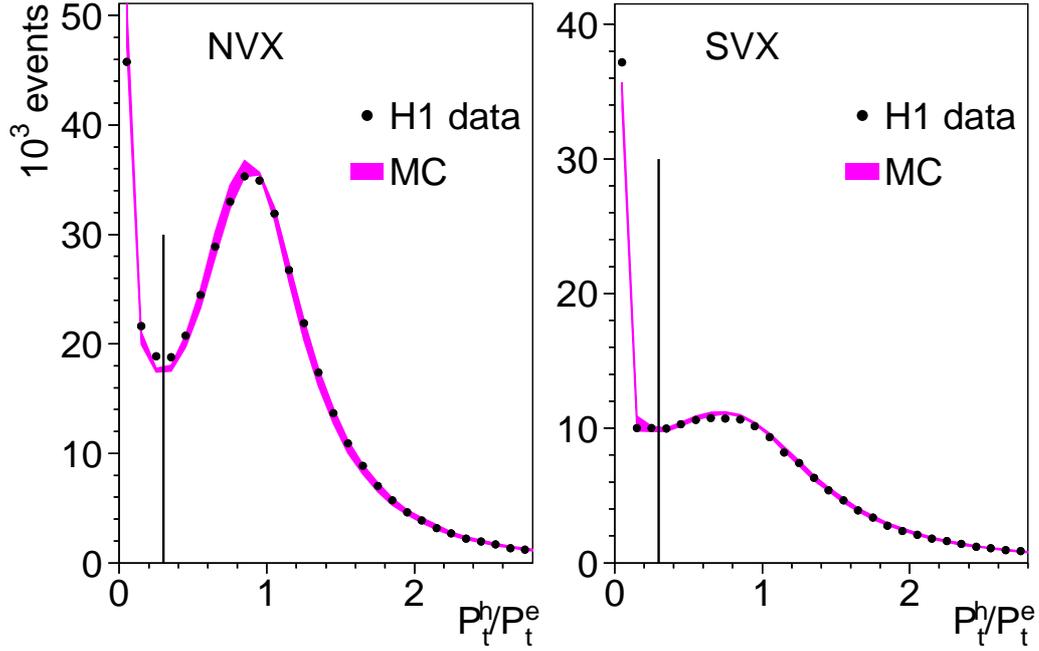,width=0.9\linewidth}
}
\caption{\label{fig:ptbal}Transverse momentum balance $P_{\perp}^h/P_{\perp}^e$ distribution for the
data and the MC simulation in the {\MB} and the {\SVX} analyses. 
The bands include the statistical uncertainty of
the simulation and the effect of the LAr hadronic scale uncertainty, see description in the
text. The vertical line indicates the analysis requirement $P_{\perp}^h/P_{\perp}^e > 0.3$.}
\end{figure}

The $\pt$ balance between the scattered electron and the calibrated
hadronic final state is studied as a function of various variables,
such as $P^e_\perp$, $\thetah$ and  $y_\Sigma$.
For central events,
where $y_{\Sigma} \geq 10^{-2}$, the simulation reproduces the
behaviour of the data within $2\%$ accuracy. 
At  lowest $y$, the hadronic final state is produced at small polar angles
and partially escapes the LAr acceptance.  In this
case, simulation and data agree within $10\%$. 
The systematic
uncertainty of the hadronic scale is therefore extrapolated linearly
in $\log y$,
from $10\%$ at $y_{\Sigma}=10^{-3}$ to $2\%$ at $y_{\Sigma} =
10^{-2}$.  
It is then set to $2\%$ for  $y_\Sigma \ge 0.01$.
\FFig~\ref{fig:ptbal} shows the overall $P_\perp$ balance
for the standard analysis selection. 
The vertical line at $P_\perp^h/P_\perp^e = 0.3$
indicates the analysis cut value. An increase in number of events
for $P_\perp^h/P_\perp^e < 0.3$ corresponds to very low $y<0.001$.
The  data
agree with the simulation within the hadronic energy scale uncertainty.

\paragraph{Hadronic Energy Calibration of the SpaCal.}
For large values of $y\gtrsim 0.4$, the contribution of the SpaCal to the total
\empz\ becomes larger than the combined contribution of the LAr calorimeter 
and tracks. Given the accurate knowledge of the SpaCal linearity 
after calibration (\Sec~\ref{sec:elmcalib}), the study of \empz\ as a function of \ee\
allows a check of both the linearity and the absolute scale of the
SpaCal hadronic measurement to be made.

\begin{figure}
\centerline{%
\psfig{file=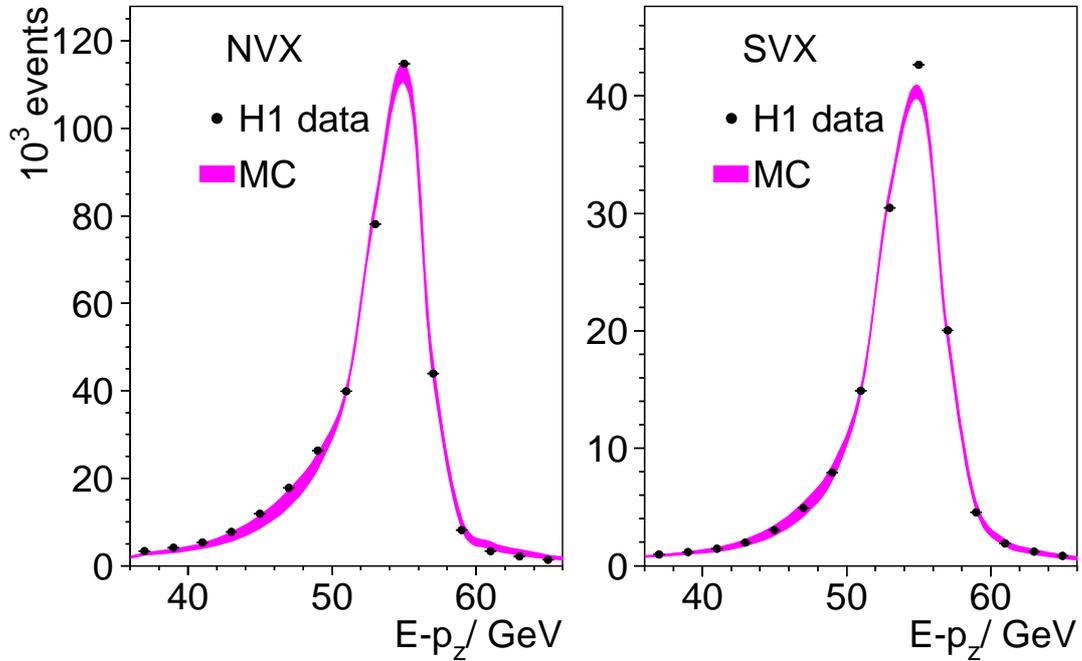,width=0.9\linewidth}
}
\caption{\label{fig:empz}\empz\  distribution for the data and the MC simulation
in the {\MB} and the {\SVX} analyses.
The bands include the statistical uncertainty of the simulation and the effect
of a $\pm 0.5$\,GeV variation of the SpaCal hadronic final state contribution.}
\end{figure}

The  \empz\  distribution is studied for $\ee>7$\,GeV in $\ee$ intervals of $1$\,GeV.
For each interval, a Kolmogorov-Smirnov test is performed to
estimate a possible shift in the \empz\ distribution between data and
simulation.  For the {\SVX} analysis, the data and the simulation agree
well within their statistical uncertainties, while for the {\MB} data sample
a global shift of $\sim 1$\,GeV is observed. This shift  is
applied in the {\MB} analysis to the simulated events.
An additional systematic uncertainty, $\Delta (\empz)_{\rm SpaCal} =
0.5$\,GeV, is considered for both {\SVX} and {\MB} analyses. 
\FFig~\ref{fig:empz} shows
the \empz\ distribution for the data and the simulation. The
uncertainty band includes a $\pm 0.5$\,GeV variation of the SpaCal
contribution to the total \empz.

\subsubsection{Calorimeter Noise Uncertainty}
\label{sec:larnoise}
For $y \lesssim 0.01$, even a small fake energy contribution in the
LAr can strongly affect the determination of $y_h$. Therefore, a
dedicated procedure is used to identify the LAr noise, as described in 
\Sec~\ref{sec:hfs}.  Samples of LAr electronic and beam
induced noise are recorded in special runs and added to the simulation.

\begin{figure}
\centerline{%
\psfig{file=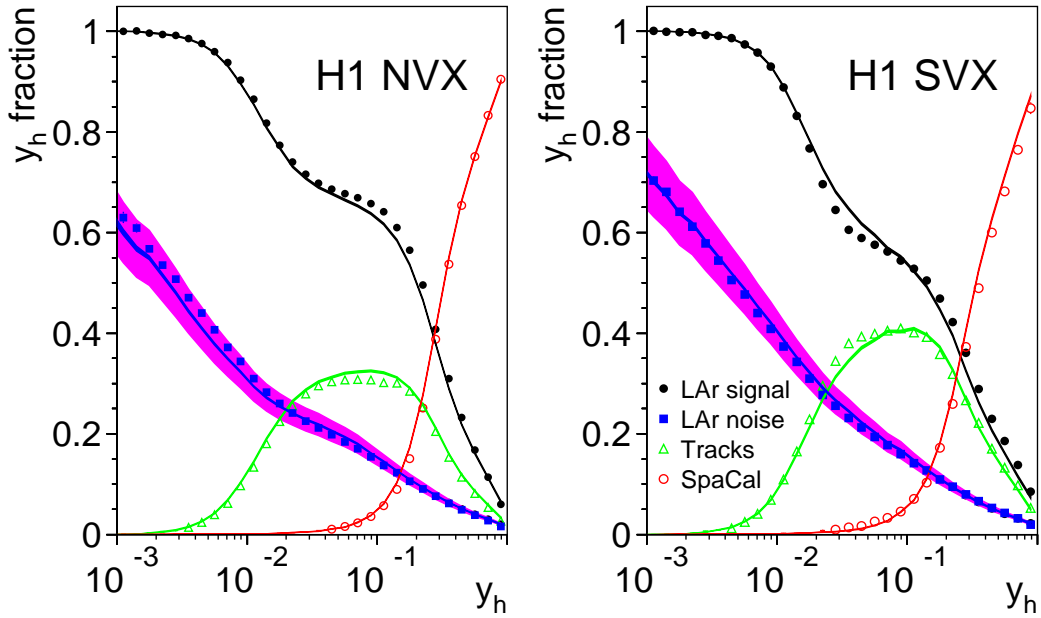,width=0.85\linewidth}
}
\caption{\label{fig:larnoise}
Relative contributions to the measured $y_h$
from the LAr (closed circles), tracks (triangles) and SpaCal (open circles) together 
with the subtracted LAr noise fractions (squares) in the {\MB} (left)
and {\SVX} (right) analyses. 
The distributions of simulated events are shown as curves. 
The shaded areas correspond to a $10\%$ systematic
uncertainty on the LAr noise description.}
\end{figure}

The uncertainty of the noise influence on the DIS cross section
measurement is determined  as a function of $y_h$ 
by studying the ratio $y_{h,\rm noise} /
(y_h + y_{h,\rm noise})$, where $y_{h,\rm noise}$ is defined as
$y_{h,\rm noise} = \sum_i (E_i - P_{z,i})/ 2E_e$ with the sum
extending over the identified noise cells only.
This comparison is shown in \Fig~\ref{fig:larnoise} for the
{\MB} and {\SVX} data samples together with contributions to $y_h$
from the tracks, LAr and SpaCal calorimeters.  
The noise fraction is described by the
simulation within $10\%$ accuracy which is taken as a systematic uncertainty.
Note that at high $y$ the noise fraction is small.
More details on the LAr noise uncertainty estimation can be found in~\cite{vargasth,behrendtth}.

\subsection{Background Subtraction}
\label{sec:bg}

\subsubsection{Methods}
\label{sec:bg:meth}
The dominant background source for this analysis arises from very low
$Q^2$ photoproduction events in which the scattered electron escapes
detection in the backward beam pipe and a particle from the hadronic
final state mimics the electron.  Other potential background sources
arise from non-$ep$ interactions.  They are studied using
non-colliding HERA bunches and are found to be negligible.

For a fraction of photoproduction events the scattered electron is
detected by the electron tagger of the luminosity system.  
These events are used to study the photoproduction background.
The acceptance of the electron tagger, which 
corresponds to the geometrical aperture of the detector as well as  
to the detection efficiency, is
determined using Bethe-Heitler $ep\to ep\gamma$ events \cite{Aid:1995bz}, in which the
scattered electron and the emitted photon are detected in the electron
and photon tagger, respectively, and is parametrised as a function of
$y$. The acceptance is large in the range $0.3 < y <0.6$.

The simulated photoproduction background (PHOJET) is normalised based on events
where the scattered electron is 
detected by the electron tagger and all
of the analysis selection
criteria\footnote{For the electron method, this selection excludes the
  \empz\ cut in order to increase the electron tagger acceptance.  In addition,
  to reduce the influence of overlapping DIS and Bethe-Heitler events, the absence of
  energy deposits in the photon tagger is required, and the total
  $(\empz)_{\rm tot} = \empz + 2E^e_{\rm tagger}$, where $E^e_{\rm tagger}$ is
the energy measured in the electron tagger, has to be less than
  $75$\,GeV.} are satisfied.  Two normalisation methods are used. In  the
first method the background is normalised globally and then subtracted
bin-by-bin
\begin{equation}
N_{\gamma p}^i = N^i_{\rm bg~MC} \cdot 
\frac{\textstyle N_{\rm tag}}
     {\textstyle N_{\rm bg~MC,~tag}},
\qquad
  N_{\rm DIS}^i = N_{\rm data}^i - N_{\gamma p}^i .
 \label{eq:gpglobal}
\end{equation}
Here, $N_{\rm DIS}^i$ ($N_{\gamma p}^i$) is the estimated number of 
DIS (photoproduction) events in 
the cross section measurement bin
$i$, $N_{\rm data}^i$ and $N^i_{\rm bg~MC}$ are the numbers of data and
PHOJET events in bin $i$, respectively, and $N_{\rm tag},N_{\rm
  bg~MC,~tag}$ are the total numbers of events detected using the electron
tagger in the data and the PHOJET simulation, respectively.

In the second method the background is normalised bin-wise using the
bin-averaged tagger acceptance $A_i$ and then subtracted in each bin
\begin{equation}
A_i = \frac{\textstyle N^i_{\rm bg~MC,~tag}}
     {\textstyle N^i_{\rm bg~MC}} ~, \qquad \label{eq:gplocal}
  N_{\rm DIS}^i = N_{\rm data}^i -
 \frac{\textstyle N^i_{\rm tag}}{\textstyle A_i}~,
\end{equation}
where $N^i_{\rm tag}$ and $N^i_{\rm bg~MC,~tag}$ are the numbers of events
detected by the electron tagger in bin $i$ in the data and the PHOJET
event sample, respectively. Both methods lead to a cancellation of
global selection uncertainties, while the second method
(\Eq~\ref{eq:gplocal}) also allows local uncertainties to cancel at
the expense of an increased statistical uncertainty.

\begin{figure}
\centerline{%
\epsfig{file=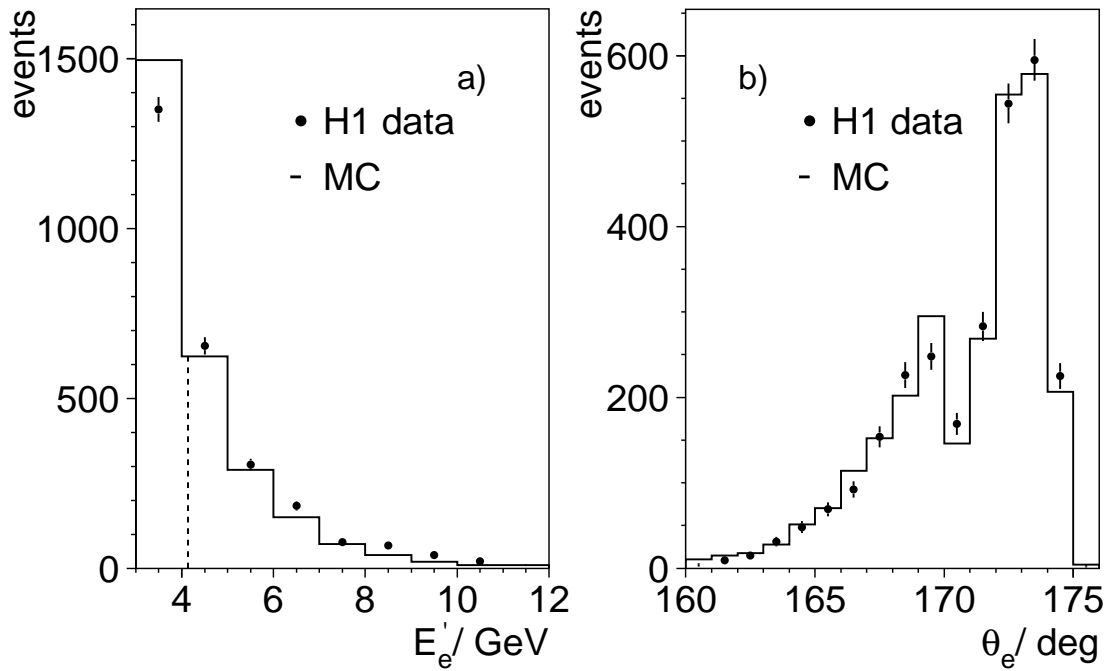,width=0.9\linewidth}
}
\caption{\label{fig:tagged}
Distribution of  {\ee} a) and {\thetae} b)
for photoproduction events detected in the electron tagger.
The plots are based on the \MB-S9 analysis. The dashed
line in a) corresponds to 
the minimum  \ee\  permitted  by the analysis cut
$y_e<0.85$.}
\end{figure}

For the \MB-S9 analysis, the global normalisation of the background 
 (\Eq~\ref{eq:gpglobal}) is used, 
since for this sample the $\ee$ and $\thetae$ distributions are well
reproduced by the simulation (\Fig~\ref{fig:tagged}). Furthermore
there is a direct control of the background normalisation as discussed
in the next section.  For the other analyses, a local bin-wise
normalisation is performed. As a cross check, both normalisation
methods are used for all samples, leading to cross section results
consistent within statistical uncertainties.

\subsubsection{Normalisation Uncertainty}

The photoproduction background normalisation  is 
checked for the {\MB-S9} analysis using electron
candidates associated with tracks of opposite charge to the lepton
beam charge, termed ``wrong charge'' tracks.  
Assuming
charge symmetry of the background tracks, the wrong charge track
sample gives an estimate of the remaining background in the
correct charge sample. 
The track charge can be 
measured for tracks which are reconstructed in the BST sector equipped
with $u$ strip detectors in addition to the $r$ detectors.  

In this method, any charge asymmetry
creates a bias.
In addition, the requirement of a $u$ strip track 
in the background study could modify the
normalisation  compared to the standard sample. 
The geometrical acceptance and efficiency $\epsilon$ of the $u$ strip track
reconstruction are first determined based on a high \ee\ sample in
which the background can be neglected.  The acceptance and efficiency
are well described by the simulation. The acceptance difference
between data and simulation is found to be $(2.0 \pm 1.3)\%$, while
the efficiency difference is determined to be $(0.2 \pm 0.5)\%$.

All events within the $u$ sector acceptance passing the \MB-S9
analysis cuts, $N_{\rm acc}$, are classified according to $N_{\rm
  acc}=N_{0}+N_{+}+N_{-}$, where $N_{0}$ denotes all events without a
linked $u$ track, $N_+$ is the number of all events with correct sign tracks 
(positive, as expected from the scattered positron) and
$N_-$ is the number of all events with wrong sign tracks. If $\kappa =
N_{+}^{\rm bg}/N_{-}^{\rm bg}$ is the background charge asymmetry ratio,
then the total number of background events in the $u$ sector geometrical
acceptance is
\begin{equation} \label{eq:phisec}
\begin{split}
 N_{\rm bg} & = N_{\rm acc} - N_{\rm sig} =
 N_{\rm acc} - \frac{N_+ - \kappa
 N_{-}}{\epsilon } = \\
            & = N_0 + N_- 
   \left( 1 +  \frac{\kappa}{\epsilon} \right)
   + N_+ \left(1 - \frac{1}{\epsilon}\right) ~.
\end{split}
\end{equation}
Here $N_{\rm acc} (N_{\rm sig})$ denotes the number of accepted (genuine DIS
signal) events.

The charge asymmetry of the background for the PHOJET
simulation is found to be $\kappa = 0.79\pm 0.06$. A dedicated
study of the origin of this asymmetry~\cite{h1alphas} showed that the
main effect is due to the difference between the proton and antiproton
interaction cross sections and the visible energy which they deposit
in the SpaCal. 
A larger value of $|E/p|$ is
expected for antiprotons
since they annihilate at the end of their paths. Indeed, for 
simulated events with $|E/p|>2$ the deviation
of $\kappa$ from unity is larger:
 $\kappa = 0.60 \pm 0.14$.  From the
data with $|E/p|>2$ a consistent value  $\kappa = 0.65\pm0.12$ is
measured\footnote{At low  energy,
 the contribution of DIS electrons
  with $|E/p|>2$ is negligible.}.  The charge asymmetry  
is also checked using events in which the scattered
electron is detected in the electron tagger. It is found to be $0.82
\pm 0.17$.  The PHOJET based asymmetry estimate is 
also consistent with the value
estimated in~\cite{h1alphas} using tagged events, $\kappa = 0.91\pm 0.04$.
In order to cover the findings on the charge asymmetry explained above, a value $\kappa =0.9 \pm 0.1$ is assumed for this analysis.

\begin{figure}
\centerline{%
\epsfig{file=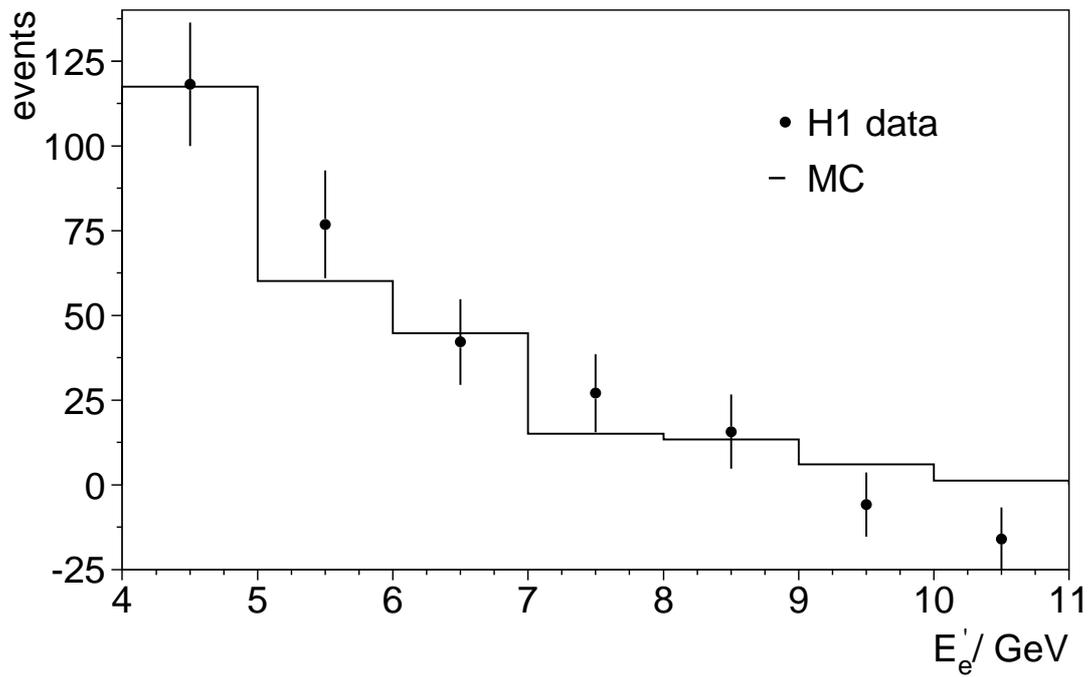,width=.9\linewidth}
}
\caption{\label{fig:phibg}
Distribution of \ee\ for background events, estimated using
wrong charge BST tracks (\Eq~\ref{eq:phisec})
for data and the PHOJET simulation. The simulated sample is
normalised using photoproduction events with the scattered 
electron detected in the electron tagger.}
\end{figure}

The ratio of the number of photoproduction events obtained using
\Eq~\ref{eq:phisec} to the estimated number of events based on 
the electron tagger, \Eq~\ref{eq:gpglobal},
 for the \ee\ range of the {\MB}-S9
analysis, is $r = 1.00 \pm 0.14_{\rm stat} \pm 0.05_{\rm asym}$. Here
the first error gives the statistical uncertainty and the second error
corresponds to the uncertainty in the background asymmetry
determination.  \FFig~\ref{fig:phibg} shows the distribution of \ee\
for the background events, estimated using $u$ sector tracks. 
The systematic uncertainty on the background normalisation is taken to
be $\pm 15\%$, based on the statistical uncertainty of the $u$ sector
sample and the uncertainty in the background charge asymmetry.

\subsection{Luminosity Determination}
\label{sec:lumi}
The luminosity measurement is based on Bethe-Heitler events detected
using the photon detector.
A precise luminosity measurement requires a good
understanding of the beam optics, of the photon detector acceptance
and its variation with changing beam conditions. The
uncertainties related to the acceptance are similar for the {\MB} and
the {\SVX} data.
         
The time structure of the $ep$ bunch crossings is characterised by the
main proton bunch accompanied by  satellite bunches.
Two such bunches are at $\pm
4.8$\,ns away from the nominal bunch and lead to $ep$ interactions at about $\pm 70$\,cm
from the mean vertex position.  The photon
detector is sensitive in  a time
window of about $12$\,ns for Bethe-Heitler events and thus does not
distinguish interactions at the nominal vertex position from satellite
bunch interactions. The luminosity measurement therefore requires the
fraction of satellite bunch interactions to be determined independently.
This is possible in H1 using TOF and PPU systems.

For the {\SVX} data, with the main bunch centred at $z=70$\,cm, 
the backward satellite is located at $z\sim 0\,$cm.
The backward satellite in this case gives a
larger contribution to the
 luminosity measurement than the forward satellite at $z\sim 140\,$cm.
The fraction of events in the backward 
satellite can be determined directly from the fraction of DIS events
with a reconstructed vertex 
around $z=0\,$cm and amounts to $2.7\%$.
A $3\%$ uncertainty is assigned to the luminosity measurement 
for the \mbox{\SVX} data, 
which covers the differences observed between
the methods of determining the satellite bunch fraction and
also includes uncertainties related to the photon detector. The same
procedure is performed to verify the 
contribution from the forward satellite at $+70\,$cm of
the {\MB} data sample. In this case the different methods are in
agreement within $0.7\%$, leading to a total luminosity uncertainty 
of $1.1\%$.

In the course of this analysis an extended reanalysis of the $1997$
data at $Q^{2} \le 12\, \mathrm{GeV^{2}}$, this sample termed B in~\cite{h1alphas}, was  performed, which reproduced the published
cross sections  in shape. These, however, are to be multiplied by a factor of
$1.034$ as the result  from an improved analysis of the
satellite bunch structure and the photon detector acceptance. 
This corresponds to a shift of two standard deviations
of the quoted luminosity measurement accuracy.

\subsection{Summary of Systematic Uncertainties}

\begin{table}
\centerline{%
\begin{tabular}{l c}
\hline\hline
\multicolumn{2}{c}{{\bfseries Correlated errors}} \\
{\bfseries Source} &  {\bfseries Uncertainty}\\
\hline
 \ee\ scale uncertainty & $0.2\%$ at $27.6$\,GeV to $1\%$ at $2$\,GeV linear in $E_e'$  \\
 \thetae \ uncertainty & $0.2\,$mrad (BST)  \\
                       & $0.5\,$mrad (BDC-Central vertex) \\  
 LAr scale uncertainty   & $10\%$ at $y=0.001$ to $2\%$ at $y=0.01$ linear in $\log y$\\
                         & $2\%$ for $y > 0.01$\\
 LAr noise contribution to \empz   & $10\%$ \\
 SpaCal hadronic scale & $0.5$\,GeV \\
 $\gamma p$ background normalisation & $15\%$ \\
Luminosity & $3\%$ (\SVX)  \\
           & $1.1\%$ (\MB) \\
\hline\hline
\multicolumn{2}{c}{{\bfseries Uncorrelated errors}} \\
{\bfseries Source} &  {\bfseries Uncertainty} \\
\hline
BST efficiency & $2\%$ (BST) \\ 
BDC efficiency  & $1.5\%$ (BDC-Central vertex)\\
Central Tracker vertex efficiency & $2-10\%$ (BDC-Central vertex)\\
Trigger efficiency & $0.9\%$ (\MB) \\
& $1.1\%$ (\MB-S9) \\
& $0.9\%$ (\SVX-BST) \\
& $0.7\%$ (\SVX-BDC) \\
Radiative corrections & $0.5\%$  \\
\hline\hline
\end{tabular}
}
\tablecaption{\label{tab:syssum}Summary of the systematic uncertainties. For the correlated
sources, the uncertainties are given in terms of the uncertainty
of the corresponding source. The effect on the
cross section measurement varies
 from bin to bin and is given in \Tab~\ref{tab:svxtable}-\ref{tab:mbtablesn}. For the
 uncorrelated sources, the uncertainties
are quoted in terms of the effect on the measured cross section directly and
the type of analysis is given in brackets.}
\end{table}

The systematic uncertainties are classified into two groups, bin-to-bin
correlated and uncorrelated systematic errors.  
For this analysis
the correlated sources are the electromagnetic and hadronic energy
scales, the electron scattering angle, the calorimeter
noise, the background subtraction and the normalisation uncertainty.  The
uncorrelated errors are related to various efficiencies and radiative
corrections.  A summary of the correlated and uncorrelated errors for
the present analysis is given in \Tab~\ref{tab:syssum}. 

The large overall contributions to the total error are due to the BST electron
track reconstruction efficiency and the Central Tracker vertex efficiency
uncertainty.
The correlated error sources affect the DIS cross section measurement
in a manner which depends on the kinematic domain. The most pronounced
variation arises with the inelasticity $y$. For high $y$, the uncertainty
is dominated by the photoproduction background (about $6\%$ for
$y=0.8$). For intermediate $y \sim 0.1$, the \ee\ scale uncertainty
becomes more prominent for the electron method (about $3\%$ cross
section uncertainty).  
 For $y < 0.01$, the dominant error source is the LAr noise
(up to $10\%$ cross section error). 

\subsection{Control Distributions}
\begin{figure}
\centerline{%
 \epsfig{file=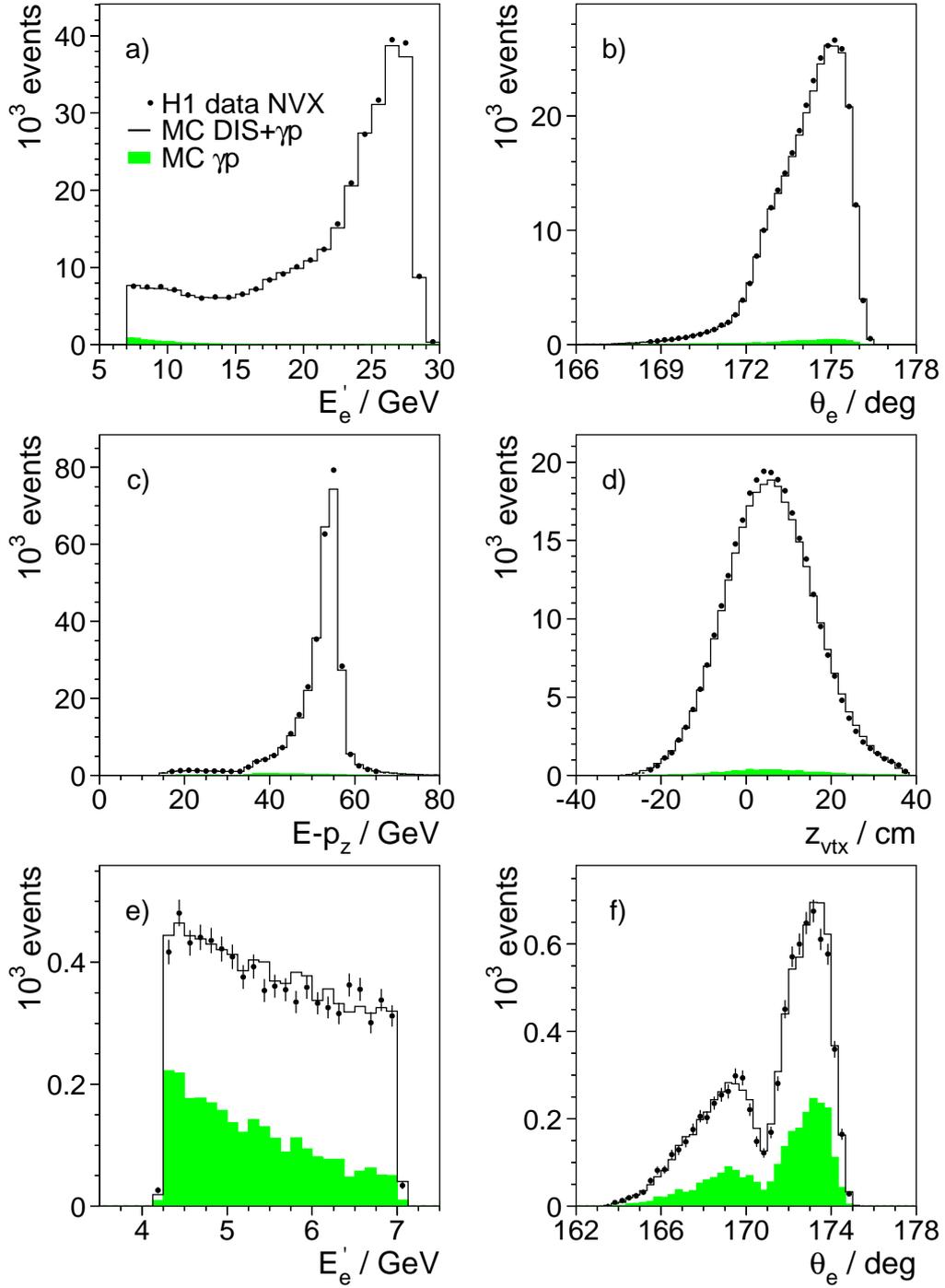,width=0.85\linewidth}
}
\caption{\label{fig:controlplotmbs0s9}
Distribution of events for the \MB-BST analysis: the
energy a) and the polar angle b) of the scattered positron;
 $\empz$ c) and the $z$ vertex position d).
Control distributions for the \MB-S9 analysis:
energy e) and polar angle f) of the scattered positron candidates.
The histograms represent the simulation of DIS and the photoproduction
background (shaded).}
\end{figure}

\begin{figure}
  \begin{minipage}[c]{0.5\textwidth}
    \epsfig{file=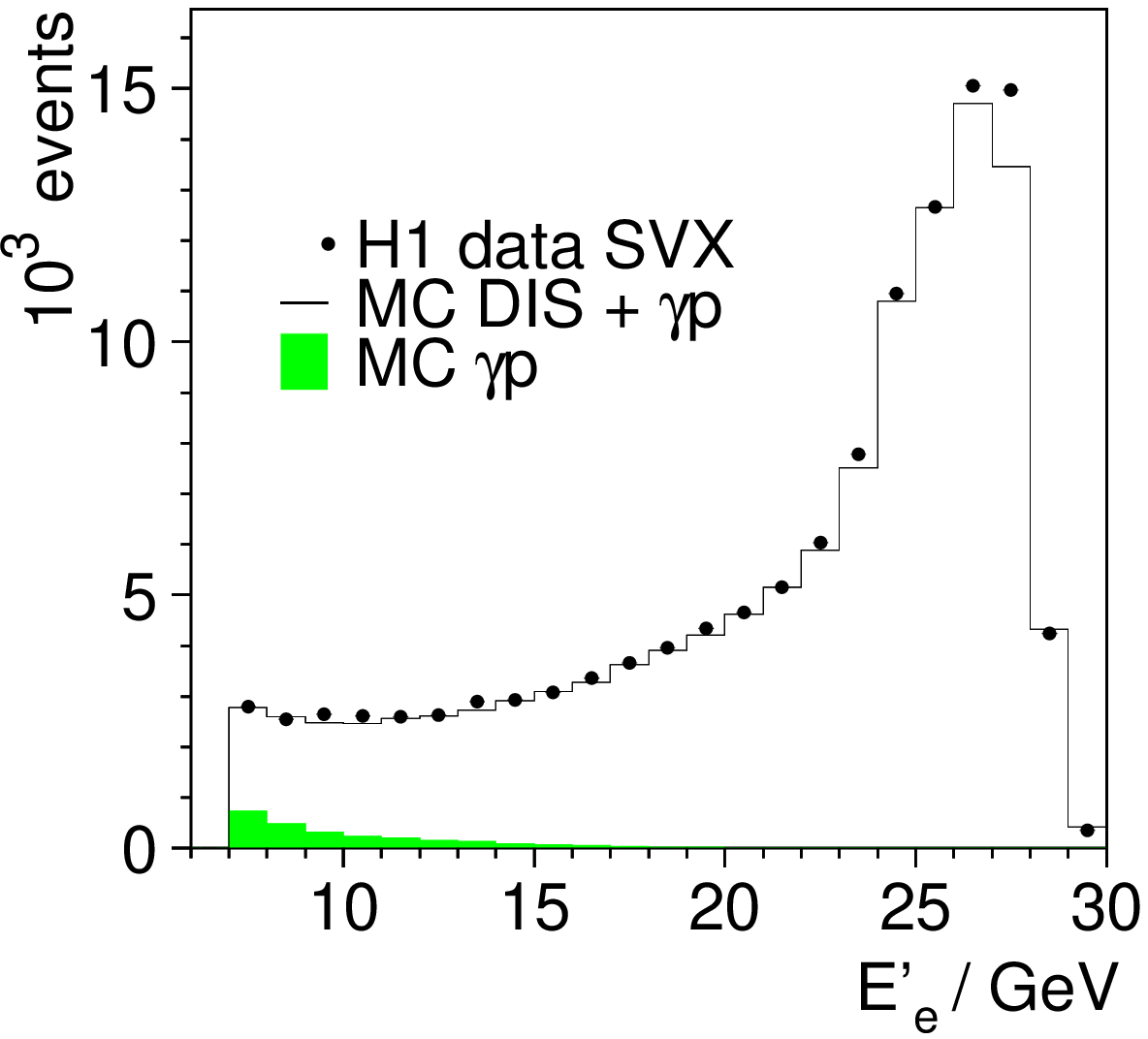, width=0.85\textwidth}
    \put(-52,54){a)}
  \end{minipage}
  \begin{minipage}[c]{0.5\textwidth}
    \epsfig{file=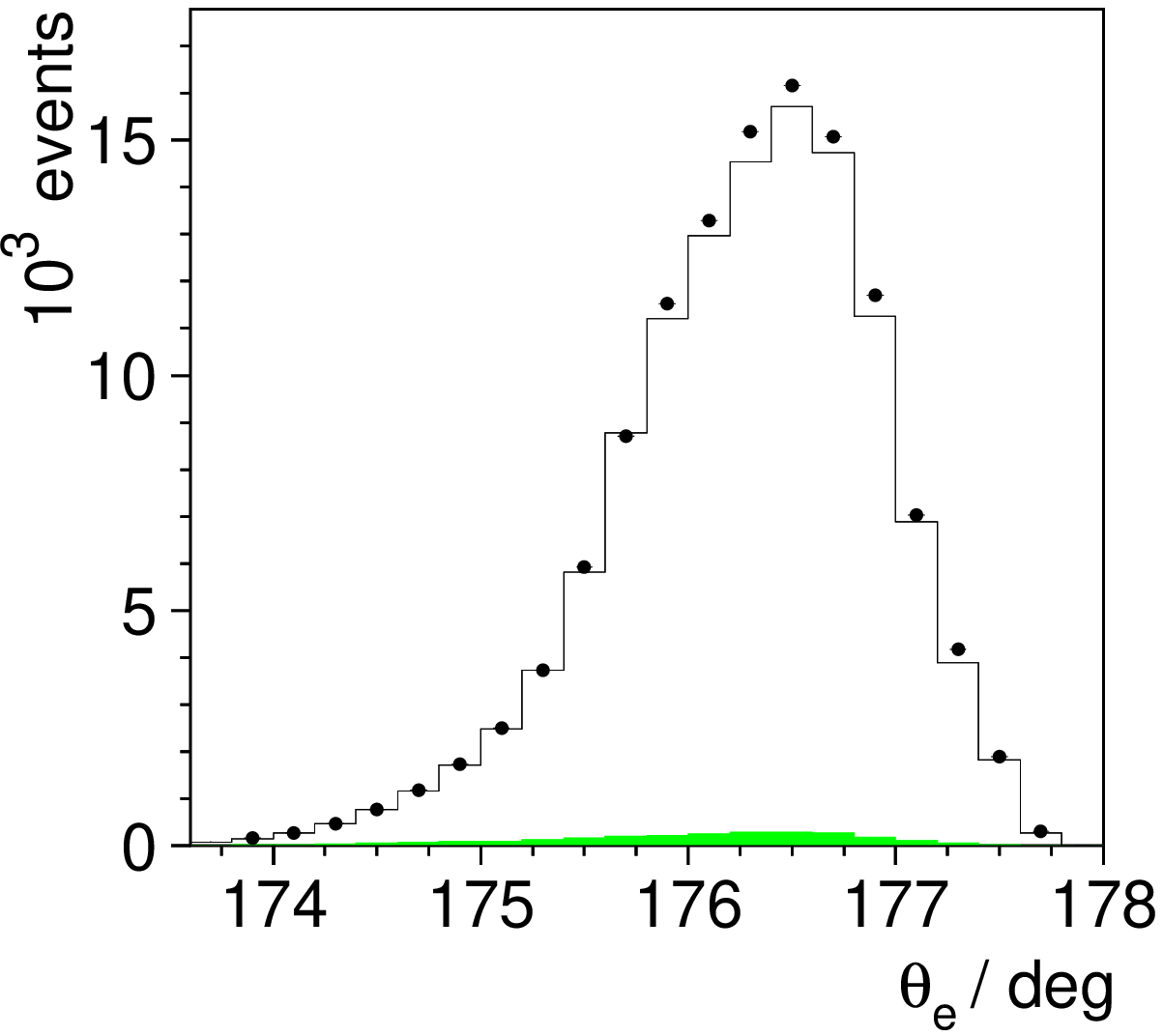, width=0.85\textwidth}
    \put(-52,54){b)}
  \end{minipage}
  \begin{minipage}[c]{0.5\textwidth}
    \epsfig{file=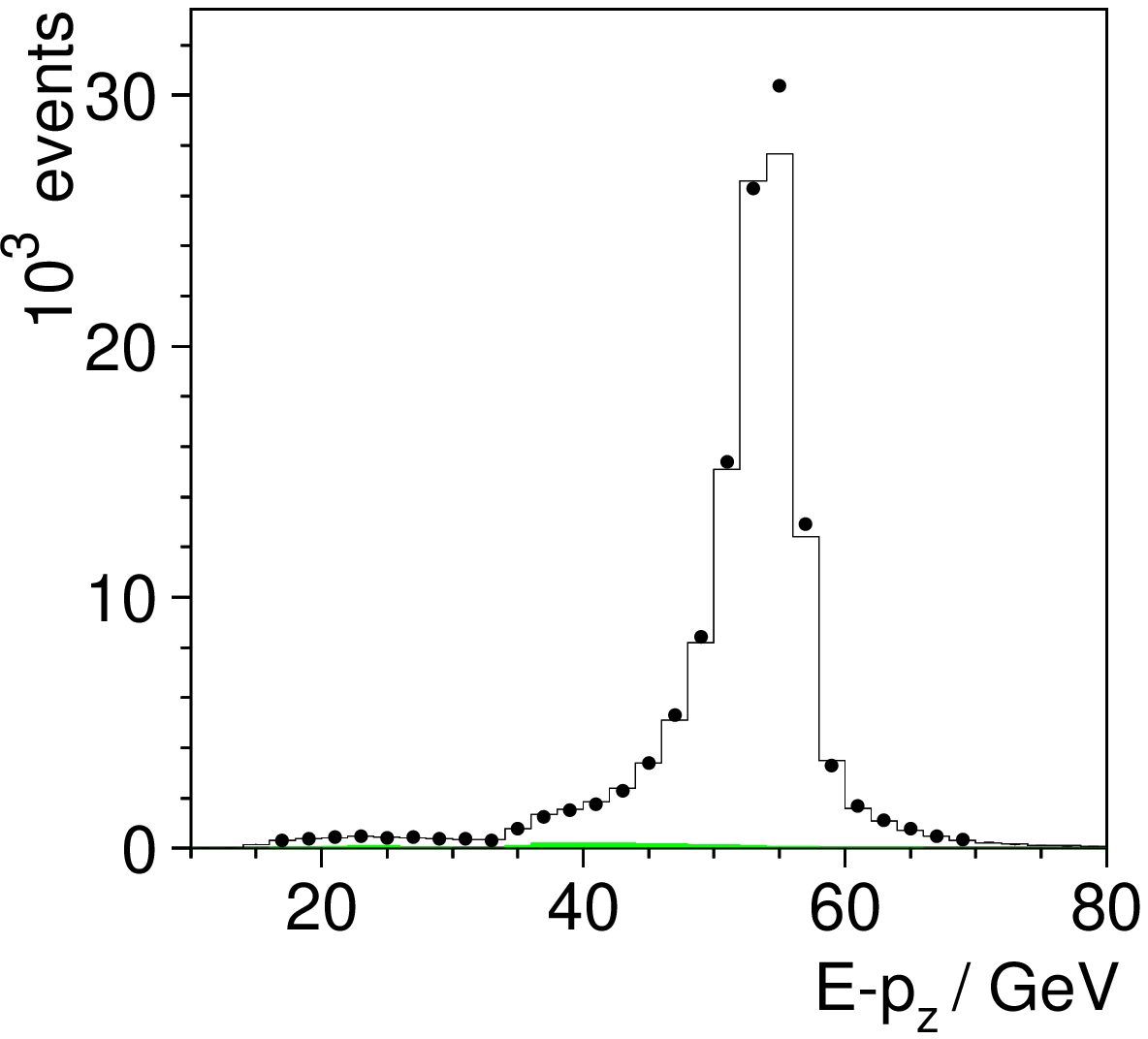, width=0.85\textwidth}
    \put(-52,54){c)}
  \end{minipage}
  \begin{minipage}[c]{0.5\textwidth}
    \epsfig{file=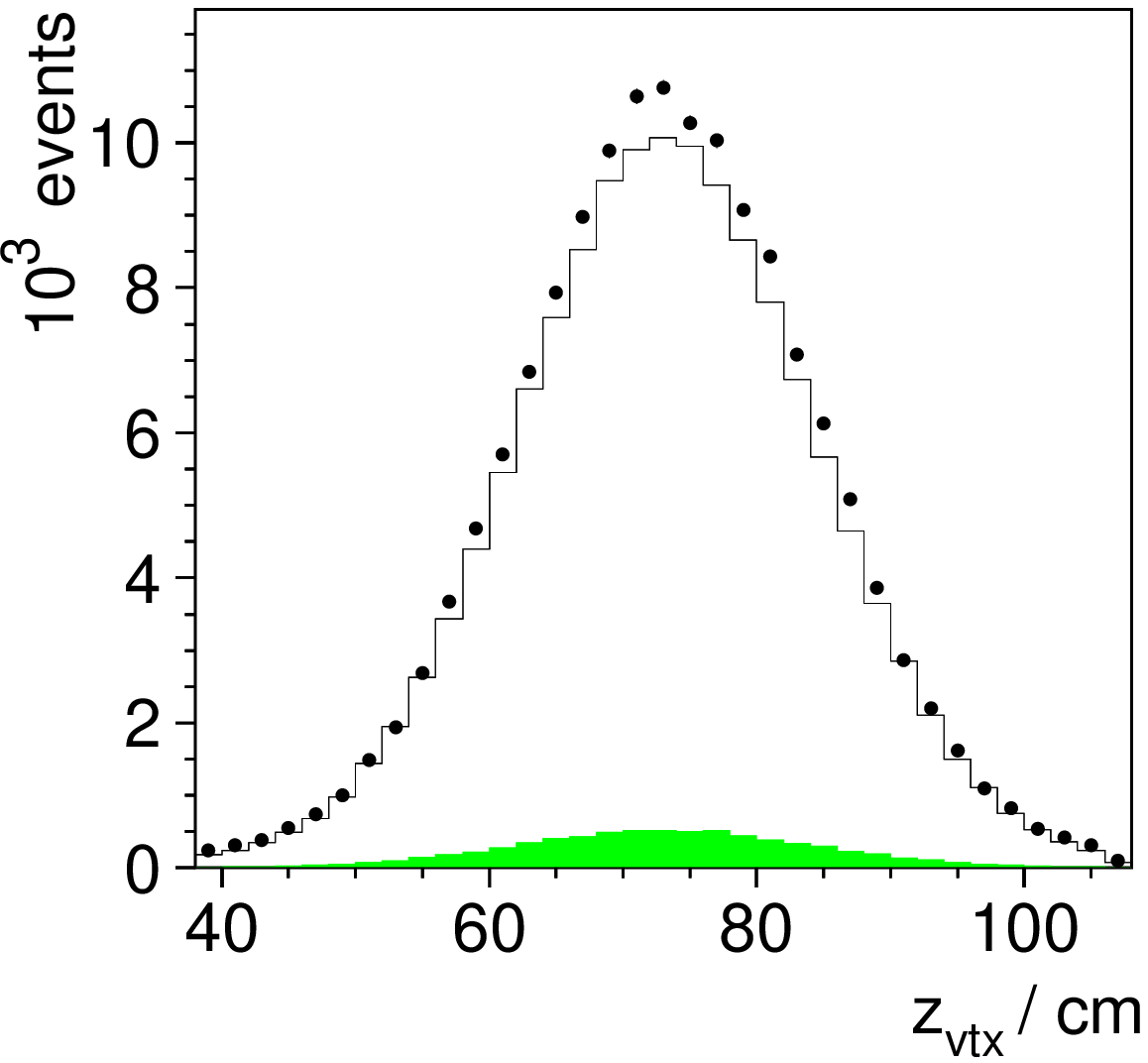, width=0.85\textwidth}
    \put(-52,54){d)}
  \end{minipage}
  \begin{minipage}[c]{0.5\textwidth}
    \epsfig{file=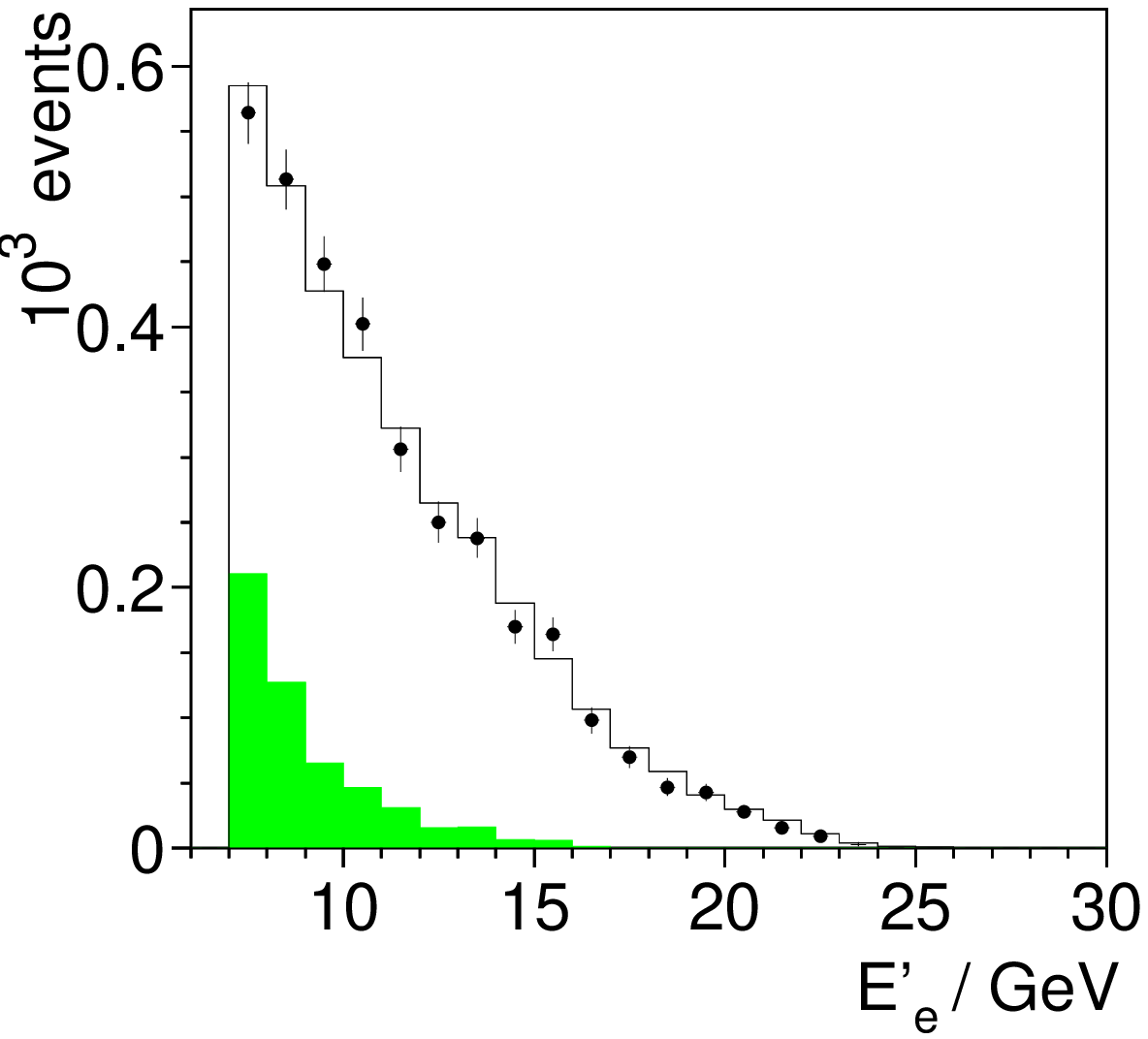, width=0.85\textwidth}
    \put(-10,54){e)}
  \end{minipage}
  \begin{minipage}[c]{0.5\textwidth}
    \epsfig{file=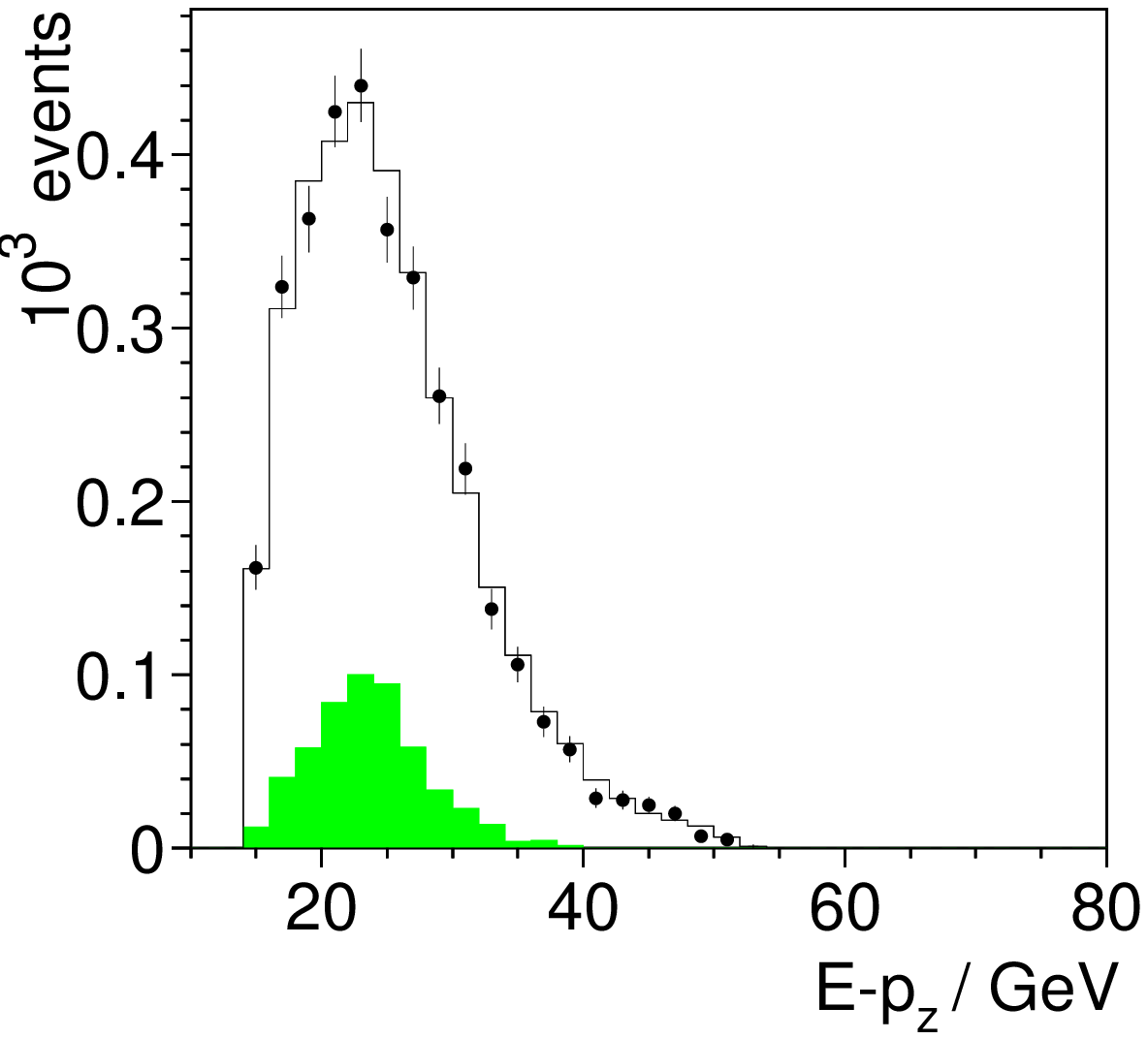, width=0.85\textwidth}
    \put(-10,54){f)}
  \end{minipage}
\caption{\label{fig:controlplotsvxbstbdc}
Distribution of events for the \SVX-BST
a-c,e,f) and \SVX-BDC d) analyses: the energy a) and the polar angle b) of the
scattered positron; $\empz$ c) and the $z$ vertex position d); the
energy e) and $\empz$ f) for the ISR bins.  The histograms represent
the simulation of DIS and the photoproduction background (shaded).}
\end{figure}

\begin{figure}[h]
\centerline{%
\epsfig{file=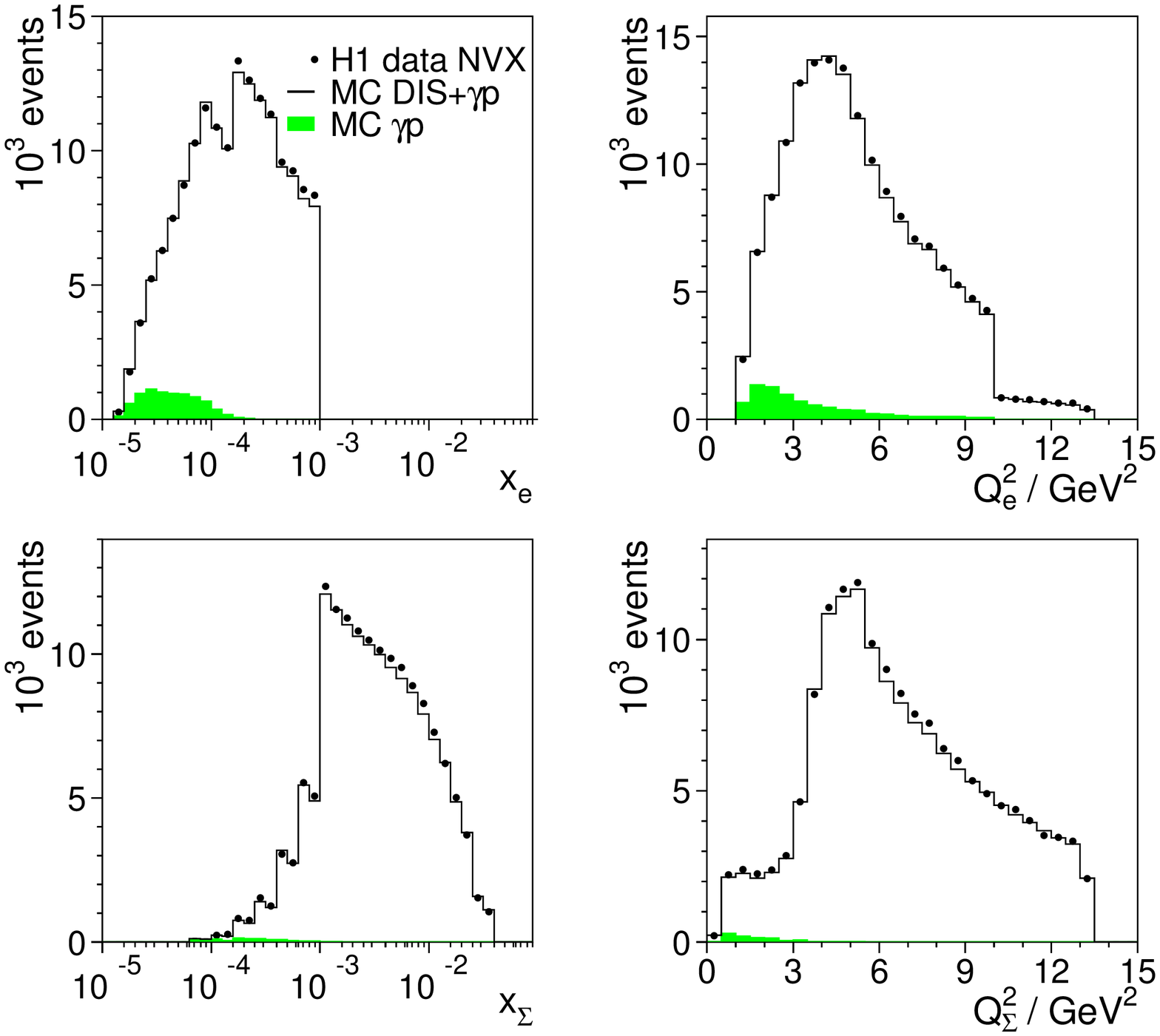,width=0.9\linewidth}
}
\caption{\label{fig:controlplotmb}
Distribution of Bjorken-$x$ and $Q^2$ using the electron (top) and 
sigma (bottom) reconstruction methods for the {\MB} data. 
The histograms represent
the simulation of DIS and the photoproduction background (shaded).}
\end{figure}

\begin{figure}[h]
  \begin{minipage}[c]{0.5\textwidth}
    \epsfig{file=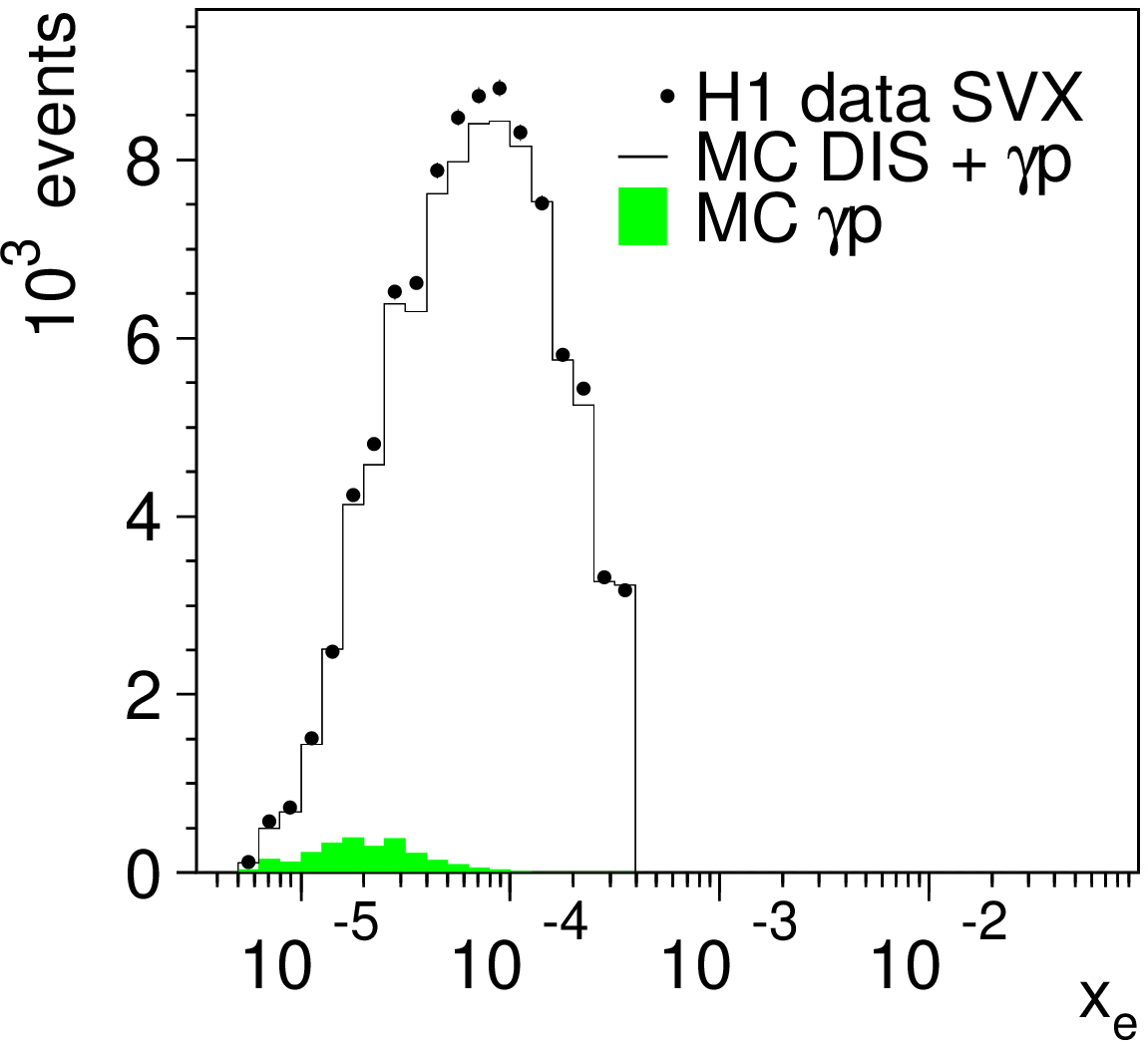, width=0.85\textwidth}
  \end{minipage}
  \begin{minipage}[c]{0.5\textwidth}
    \epsfig{file=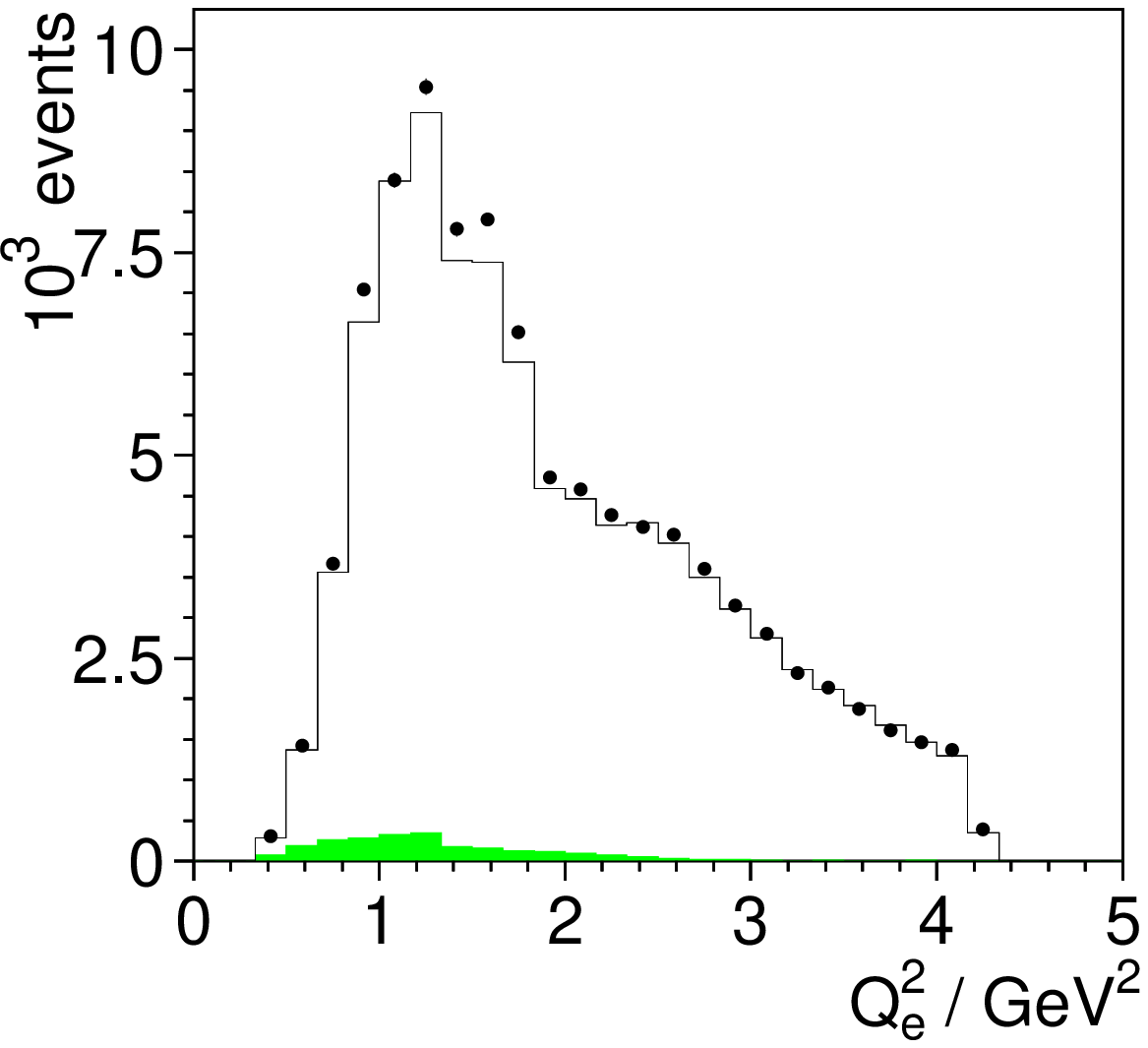, width=0.85\textwidth}
  \end{minipage}
  \begin{minipage}[c]{0.5\textwidth}
    \epsfig{file=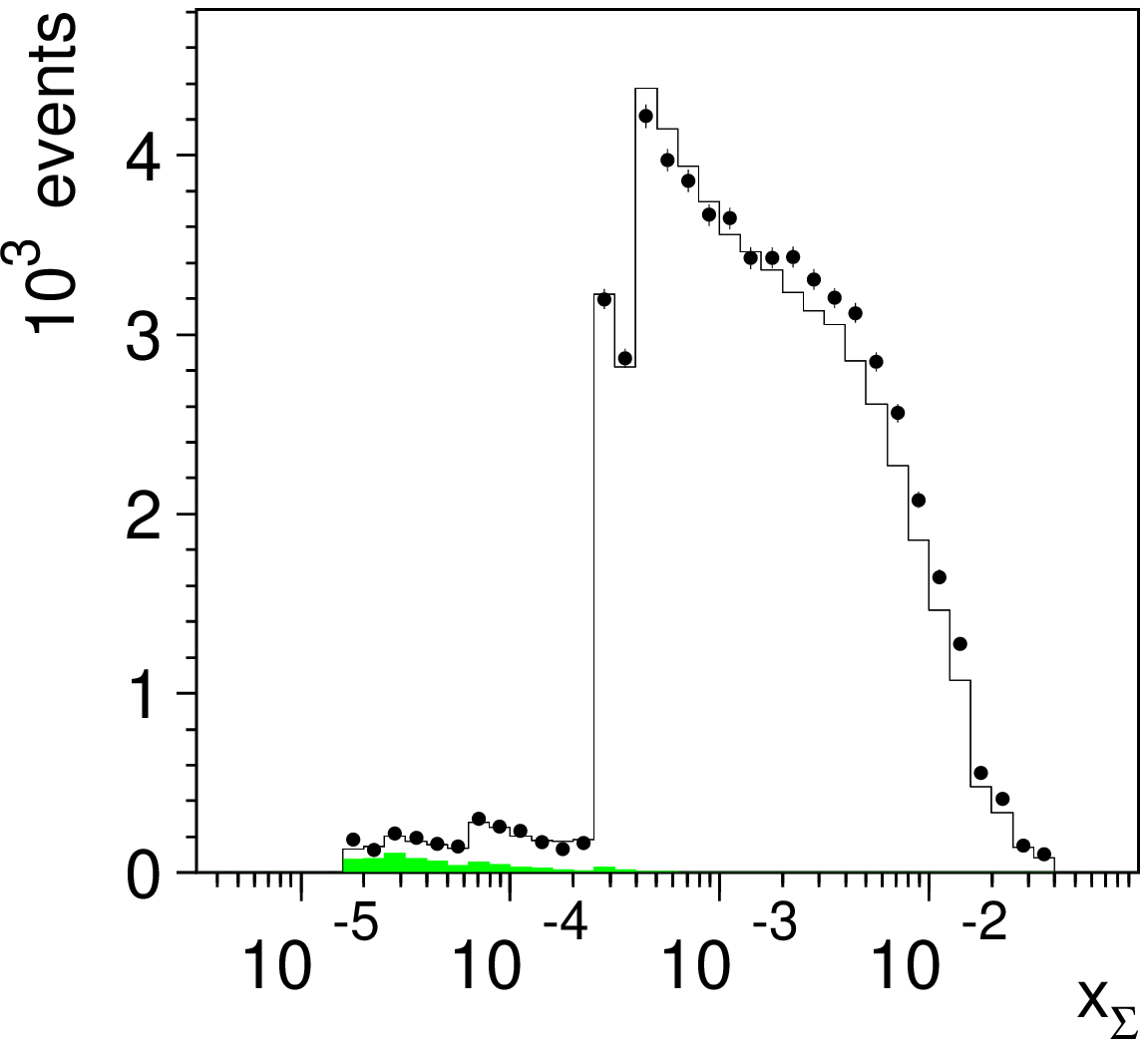, width=0.85\textwidth}
  \end{minipage}
  \begin{minipage}[c]{0.5\textwidth}
    \epsfig{file=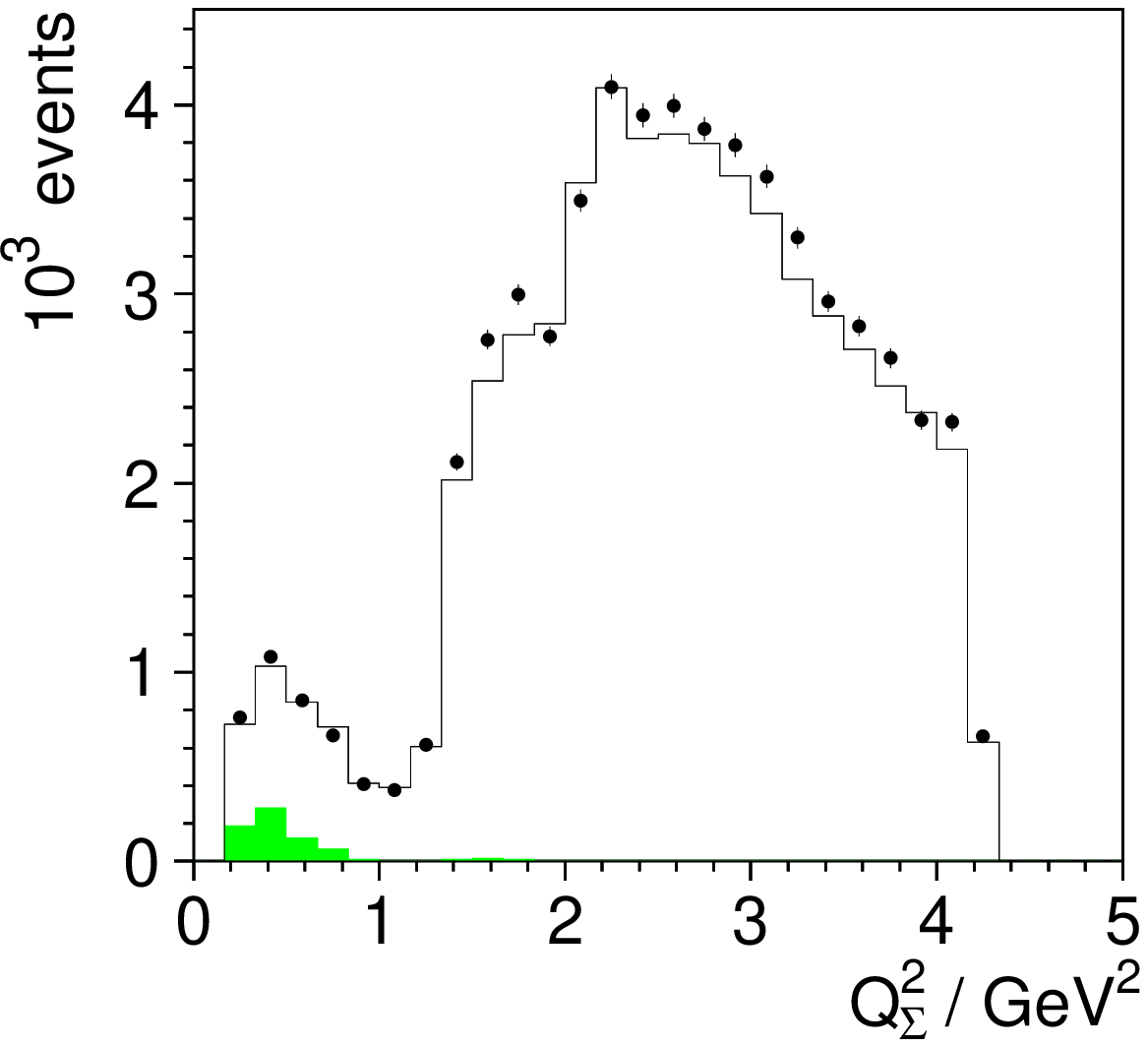, width=0.85\textwidth}
  \end{minipage}
\caption{\label{fig:controlplotsvx}
Distribution of Bjorken-$x$ and $Q^2$ using the electron (top) and
sigma (bottom) reconstruction methods for the {\SVX} data. 
The histograms represent
the simulation of DIS and the photoproduction background (shaded).
}
\end{figure}

Data and Monte Carlo simulation distributions of
important quantities for the events passing all selection criteria are
compared in  \Figs~\ref{fig:controlplotmbs0s9}~-~\ref{fig:controlplotsvx}. 
Only events corresponding to analysis bins passing the stability and purity
criteria are considered. The simulated distributions are normalised to
the measured luminosity. The DIS MC cross section prediction is reweighted to a
 parameterisation using the fractal model  introduced in
\Sec~\ref{subsec:fractalfit}.
 A rather good ({\MB}) to acceptable ({\SVX})
overall agreement is obtained in the description of
the data by the simulation.

\FFig~\ref{fig:controlplotmbs0s9}a-d) shows basic kinematic and vertex
distributions for the {\MB} analysis.
The background from photoproduction events  is
 very small. It is larger at lower scattered positron
energies $E_e'$ as can be seen in \Figs~\ref{fig:controlplotmbs0s9}
e) and f), which show the $E_e'$ and $\theta_e$ distributions for the
dedicated high $y$ analysis ({\MB}-S9).
In \Fig~\ref{fig:controlplotsvxbstbdc} basic kinematic distributions
for the \SVX-BST analysis a)-c), the \SVX-BDC analysis d) and the
\SVX-BST analysis considering events from ISR bins only  e), f) are shown.
The ISR distributions are very well reproduced by the simulation. 
The other SVX plots reveal a small normalisation
difference. \FFig~\ref{fig:controlplotmb}  shows the $x$
and $Q^2$ distributions for the two kinematic reconstruction methods, electron
and $\Sigma$, in the {\MB}  analysis. 
\FFig~\ref{fig:controlplotsvx} shows similar distributions for the {\SVX} 
analysis. Events are only
taken into account from bins which pass the stability and purity
criteria and are covered by the chosen method. 
For the SVX sample the data are less well described 
than for the {\MB} sample, but consistency is observed
within the total measurement uncertainty including a
$3\%$ normalisation error of the SVX data.

\subsection{Cross Checks\label{sec:xcheck}}

The stability of the cross section measurement is tested with a set of
dedicated cross checks which can be divided into three classes:
(i) checks for a given data set  and a given reconstruction
method, (ii) checks of the consistency between
the different reconstruction methods, and
(iii) checks of the  consistency between  the different data sets.

The consistency of the cross section measurement
for a given data set (e.g.\,{\MB}) and a given reconstruction
method (e.g.\,the electron method with $\theta_e$ measured by the
BST) is studied by splitting  the data into two approximately equal sub-samples
and comparing these sub-samples  to each other. The data are compared
as measured with the upper and the lower half of the SpaCal, for negative and
positive $z$-vertex positions, and dividing the sample into an early
and late data taking period.
These tests are sensitive to local effects like efficiency variation, energy
miscalibrations  and the stability of the luminosity
measurement. In such studies no significant
deficiencies in the data are observed.

\begin{figure}
\centerline{\epsfig{file=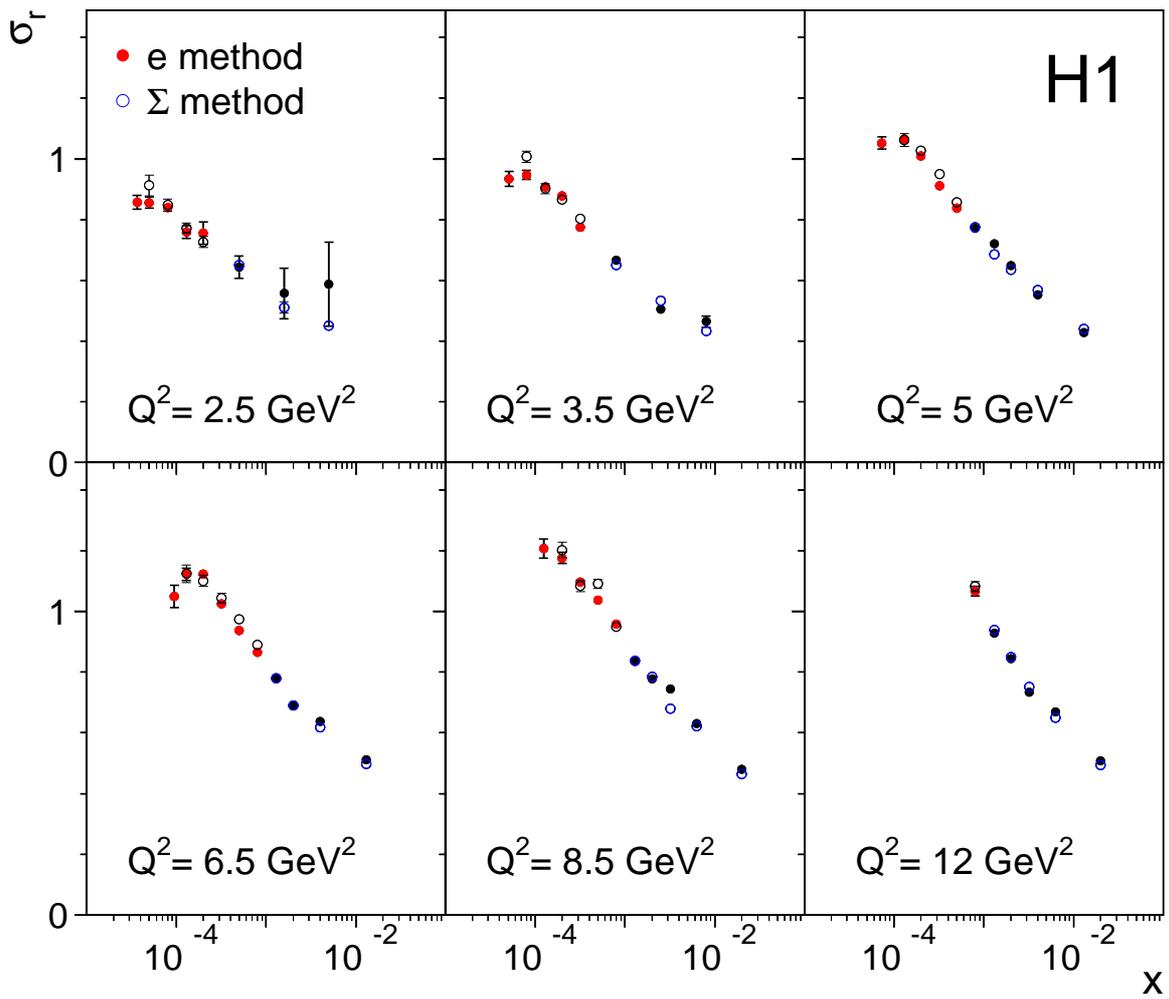,width=\linewidth}}
\caption{\label{fig:elsigmb}
Comparison of reduced cross sections as obtained  with the electron (closed circles) 
and $\Sigma$ (open circles) reconstruction methods, for the {\MB} data sample.
The errors represent statistical uncertainties only.}
\end{figure}

For the comparison of the  cross section
measurements for a given data set but using different reconstruction
methods, the test samples are strongly correlated.
The uncorrelated statistical uncertainty 
is estimated in this case by subdividing the
simulated events into a number of independent sub-samples of equal size. 
The measurement of the cross sections is repeated for each sub-sample
 and  the statistical uncertainty is calculated as the
luminosity rescaled RMS of the resulting variations of the cross
section measurements. Employing this technique, the cross section
measurements based on different triggers and different $\theta_e$
reconstruction methods (BDC and BST) are compared.
In most of the cases the measurements with each 
of the samples agree within the uncorrelated
statistical uncertainty. In a few cases the measurements agree
within the total  uncertainty only.
A particularly interesting test is the comparison of the cross section
measurement performed with the electron and sigma methods, since the
two methods have different sensitivities to  systematic error
sources.  The two methods can both be applied in many common bins where the
purity and stability of the measurement are high for both methods.
\FFig~\ref{fig:elsigmb} shows an example of this comparison, performed
for the \MB-BST data set.

The third class of cross checks compares the cross section
measurements performed with different data samples: {\SVX} is compared
to {\MB} and the new data are compared to the
previously published results.  This comparison is an integral part of
the cross section averaging procedure, as discussed subsequently.

\subsection{Cross Sections}
\begin{figure}
\begin{center}
\epsfig{file=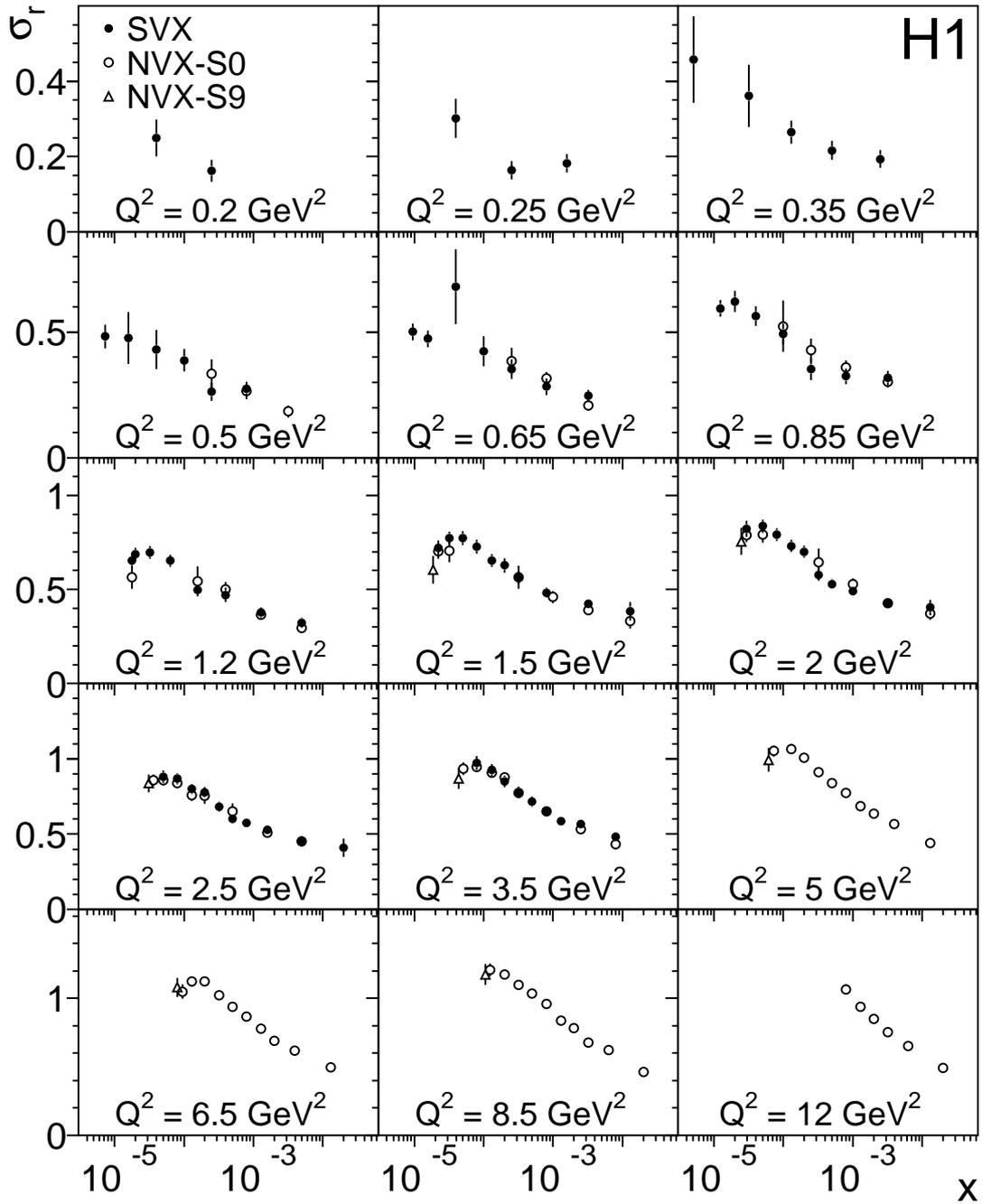,width=1.2\linewidth}
\end{center}
\caption{\label{fig:mbsvx0} Reduced inclusive $e^+p$ scattering cross
section as measured in the \MB-BST~(open circles), \MB-S9~(triangles) 
and {\SVX} (closed circles) analyses of the  $920$\,GeV data.
The errors represent the statistical and systematic uncertainties added in quadrature.}
\end{figure}

The cross section data measured from the {\SVX} and {\MB}
data samples are given in  \Tabs~\ref{tab:svxtable}-\ref{tab:mbtablesn}
and presented in \Fig~\ref{fig:mbsvx0}.
The uncertainty of the new data is typically $3-4$\% and larger
at the acceptance edges.
Lowest values of $Q^2$, down to $0.2$\,GeV$^2$, are reached with the
shifted vertex data.
The analysis of the {\SVX} data is mainly based on the BST
but complemented by an
independent analysis using the BDC at lower radii.
 For $Q^2$ between $0.5$ and $3.5$\,GeV$^2$,
the  {\MB} and {\SVX} cross section data overlap  
in their kinematic coverage and are observed to be in
agreement. The kinematic region of larger $Q^2 \ge 5$\,GeV$^2$
is covered by the nominal vertex data. 
The data at highest $y$, corresponding to smallest $x$,
are obtained  using the dedicated trigger S9
and can be seen in \Fig~\ref{fig:mbsvx0}
to be consistent with the behaviour of $\sigma_r$
towards small $x$.

\section{Combination of H1 Cross Section Measurements\label{sec:averaging}}
The new data cover a  kinematic region 
which overlaps with  data  sets taken at $820\,$GeV proton
beam energy in $1995$~\cite{Adloff:1997mf} and in 
$1997$ (sample B)~\cite{h1alphas}. 
The combination of all these data,
as described subsequently, provides a  single data set 
in the range
 $0.2 \leq Q^2 \leq 12$\,GeV$^2$ 
and 
$5 \cdot 10^{-6} \leq x \leq 0.02$.

\subsection{Procedure\label{subsec:aveprocedure}}
The combination of the data sets is based on the prescription introduced
in~\cite{glazov} which is applicable if the uncertainties of the measurements
do not depend on the central values. This procedure is described in 
\Sec~\ref{secsec:additive}. For the cross section measurements 
the estimated statistical and systematic uncertainties depend on the
central values.  
This leads to a modification  of the averaging procedure
as is described in \Sec~\ref{secsec:mult}. 

\def\nmbeelv{-0.32}
\def\nmbeele{0.56}
\def\nmbthev{1.03}
\def\nmbthee{0.63}
\def\nmbehadv{-0.23}
\def\nmbehade{0.62}
\def\nmbnoiv{-0.20}
\def\nmbnoie{0.62}
\def\nmbhspav{0.48}
\def\nmbhspae{0.34}
\def\nmbgpv{0.05}
\def\nmbgpe{0.56}
\def\nmblumiv{0.10}
\def\nmblumie{0.83}
\def\nsvxeelv{0.36}
\def\nsvxeele{0.67}
\def\nsvxthev{0.48}
\def\nsvxthee{0.78}
\def\nsvxehadv{-1.79}
\def\nsvxehade{0.71}
\def\nsvxnoiv{-1.13}
\def\nsvxnoie{0.46}
\def\nsvxhspav{-1.66}
\def\nsvxhspae{0.69}
\def\nsvxgpv{0.10}
\def\nsvxgpe{0.90}
\def\nsvxlumv{0.17}
\def\nsvxlume{0.48}
\def\ombeelv{1.19}
\def\ombeele{0.45}
\def\ombthev{-0.72}
\def\ombthee{0.73}
\def\ombehadv{0.06}
\def\ombehade{0.74}
\def\ombnoiv{-1.06}
\def\ombnoie{0.68}
\def\ombgpv{-0.10}
\def\ombgpe{0.92}
\def\omblumiv{0.66}
\def\omblumie{0.63}
\def\hnfeev{0.03}
\def\hnfeee{0.32}
\def\hnfthv{0.20}
\def\hnfthe{0.56}
\def\hnfehv{-0.09}
\def\hnfehe{0.88}
\def\hnfgpv{0.48}
\def\hnfgpe{0.76}
\def\hnfdiv{0.16}
\def\hnfdie{0.93}
\def\hnflumiv{-1.60}
\def\hnflumie{0.59}
\def\ochiave{86.2}
\def\odofave{125}
\def\mbeelv{-0.34}
\def\mbeele{0.83}
\def\mbthev{0.43}
\def\mbthee{0.76}
\def\mbehadv{0.65}
\def\mbehade{0.79}
\def\mbnoiv{0.24}
\def\mbnoie{0.79}
\def\mbhspav{0.98}
\def\mbhspae{0.44}
\def\mbgpv{0.10}
\def\mbgpe{0.59}
\def\mblumiv{0.16}
\def\mblumie{0.96}
\def\svxeelv{-0.26}
\def\svxeele{0.84}
\def\svxthev{0.19}
\def\svxthee{0.92}
\def\svxehadv{-0.84}
\def\svxehade{0.81}
\def\svxnoiv{-0.70}
\def\svxnoie{0.52}
\def\svxhspav{-0.67}
\def\svxhspae{0.81}
\def\svxgpv{0.20}
\def\svxgpe{0.90}
\def\svxlumv{-0.45}
\def\svxlume{0.60}
\def\chiave{19.5}
\def\dofave{ 39}
\subsubsection{Linear Averaging\label{secsec:additive}}
The averaging procedure is based on a $\chi^2$ minimisation.
For a single data set, 
 the $\chi^2$ function can be defined as
\begin{equation}
 \chi^2_{\rm exp}\left(\boldsymbol{m},\boldsymbol{a}\right) = 
 \sum_i
 \frac{\left[m^i 
- \sum_j \frac{\textstyle \partial {\mu^i}}{\textstyle \partial \alpha_j} 
\left(a_j-\alpha_j\right) -\mu^i \right]^2}
{\Delta^2_{i}} 
 + \sum_j \frac{ \left(a_j-\alpha_j\right)^2} {\Delta^2_{\alpha_j}} ~.
\label{eq:aveold}
\end{equation}
Here ${\mu^i}$ is the  measured central value  at a point $i$ 
with combined statistical and uncorrelated systematic uncertainty 
$\Delta_i = (\Delta^2_{i, {\rm stat}} + \Delta^2_{i,{\rm uncor}})^{1/2}$.
Further, $\alpha_j$ denotes the central value determined for
 a correlated systematic error  source of type $j$ with an uncertainty
$\Delta_{\alpha_j}$, while
$\partial {\mu^i}/\partial \alpha_j$ quantifies the sensitivity of the
measurement ${\mu^i}$ at the point $i$ to the systematic source $j$. 
The function $\chi^2_{\rm exp}$ depends on the set of
underlying physical quantities $m^i$ 
(denoted as the vector $\boldsymbol{m}$) and 
 the set of systematic uncertainties $a_j$ ($\boldsymbol{a}$).
For the reduced cross section  measurements one has ${\mu^i} = \sigma_r^i$,
$i$ denotes a $(x,Q^2)$ interval, and the summation over
$j$ extends over all correlated systematic sources.

Introducing the variables $b_j = (a_j-\alpha_j)/\Delta_{\alpha_j}$
and $\Gamma^i_j = (\partial {\mu^i}/\partial \alpha_j) \Delta_{\alpha_j}$,
\Eq~\ref{eq:aveold} can be written as
\begin{equation}
 \chi^2_{\rm exp}\left(\boldsymbol{m}, \boldsymbol{b}\right) = 
 \sum_i
 \frac{\left[m^i 
- \sum_j \Gamma^i_j  b_j - \mu^i \right]^2}
{\Delta^2_{i}} 
 + \sum_j b_j^2.
\label{eq:avebeta}
\end{equation}
If several analyses 
 provide a number of  measurements
at the same $(x,Q^2)$ values, they can be  
combined using the formula above, 
generalised for the case of multiple data sets. Then a total
$\chi^2$  function,  $\chi^2_{\rm tot}$, 
is built from the sum of the $\chi^2_{\rm exp}$ functions 
for each data set  according to
\begin{equation}
 \chi^2_{\rm tot}\left(\boldsymbol{m}, \boldsymbol{b}\right) = 
 \sum_e \sum_{i=1}^{N_M}
 \frac{\left[m^i-
 \sum_{j=1}^{N_S} \Gamma^i_{j,e}  b_{j} - \mu^i_e   \right]^2}
{\Delta^2_{i,e}} \DeltaM
 + \sum_{j=1}^{N_S} b_{j}^2,
\label{eq:avet}
\end{equation}
where the summation over $i$ ($j$) runs over  all $N_{M}$ measured points
(all $N_{S}$ systematic error sources) of all data sets considered.
The symbol $\DeltaM$  is equal to one
if data set
 $e$ contributes a measurement at the point $i$, otherwise
it is zero. 
Similarly, the symbol $ \Gamma^i_{j,e}$ equals to zero
if the measurement $i$ from the data set $e$ is insensitive 
to the systematic source $j$.
This definition of $\chi^2_{\rm tot}$ assumes
that the data sets $e$  are statistically uncorrelated.
The systematic error sources $b_j$, however, may be 
either uncorrelated (separate
sources) or  correlated across data sets 
(different data sets sharing a common source).

Since $\chi^2_{\rm tot}$ is  a quadratic form of 
$\boldsymbol{m}$ and $\boldsymbol{b}$, it may be rearranged
such that it takes a form similar to  \Eq~\ref{eq:aveold}
\begin{eqnarray}
\label{eq:offdiag}
\chi^2_{\rm tot}(\boldsymbol{m},\boldsymbol{a}) \equiv \chi^2_{\rm min}
&+&
 \sum_{i=1}^{N_M}
 \frac{\textstyle \left[m^i-
 \sum_j \frac{\textstyle \partial \mu^{i,{\rm ave}}}{\textstyle \partial \alpha_j} 
\left(a_j-\alpha_{j,{\rm ave}}\right) - \mu^{i,{\rm ave}} \right]^2}
{\textstyle \Delta^2_{i,{\rm ave}}}\\
 &+& \sum_{j=1}^{N_S} \sum_{k=1}^{N_S}
\Bigl(
a_j-\alpha_{j,{\rm ave}}
\Bigr)
\Bigl(
a_{k}-\alpha_{k,{\rm ave}}
\Bigr)
 \bigl(A_S'\bigr)_{jk}~. \nonumber
\end{eqnarray}
The data averaging procedure, described in detail in 
appendix~\ref{avemath}, determines the  average values $\mu^{i,{\rm ave}}$, 
the uncorrelated uncertainties $\Delta_{i,{\rm ave}}$,
the average systematic error source values $\alpha_{j, {\rm ave}}$,
the dependencies of $\mu^{i,ave}$ on $\alpha_j$,
$\partial\mu^{i,ave}/\partial\alpha_j$,
and the matrix $(A_S')_{jk}$.
The value of $\chi^2_{\rm min}$ corresponds to the minimum of \Eq~\ref{eq:avet}.
The ratio $\chi^2_{\rm min}/\dof$ is a measure 
of the consistency of the data sets.
The number of degrees of
freedom, $\dof$, is calculated as the difference between the total 
number of measurements and 
the number of the measured points $N_M$.
This procedure represents a
method to average data sets,
which allows correlations among the measurements
due to systematic uncertainties to be taken into account.

The matrix $(A'_S)_{jk}$ 
can be diagonalised and the $\chi^2$ function
takes a form similar to \Eq~\ref{eq:avebeta}
\begin{equation}
 \chi^2_{\rm tot}\left(\boldsymbol{m}, \boldsymbol{b'}\right) = 
\chi^2_{\rm min} + 
 \sum_{i=1}^{N_M}
 \frac{\left[m^i
- \sum_{j=1}^{N_S} \Gamma^{i,{\rm ave}}_j  b_j' -\mu^{i,{\rm ave}} \right]^2}
{\Delta^2_{i,{\rm ave}}} 
 + \sum_{j=1}^{N_S} (b_j')^2, \label{eq:final}
\end{equation}
 where $b'_j = \sum_k U_{jk} ( b_k - \beta_{k,{\rm ave}}
)D_{jj}$ and $\beta_{k,{\rm ave}} = \alpha_{k, {\rm ave}}/\Delta_{\alpha_k}$. The orthogonal 
matrix $U$  connecting the systematic sources before and
after averaging and the diagonal matrix $D$ are
given in appendix~\ref{avemath}.

\subsubsection{Implementation for the Cross Section Averaging\label{secsec:mult}}
The $\chi^2$ function of \Eq~\ref{eq:aveold} is  suitable for 
measurements in which the uncertainties are absolute, 
i.e. do not depend on the central value
of the measurement. 
However, for the H1 cross section data considered here,
the  correlated and uncorrelated systematic errors are to a good approximation 
proportional to the  central values (multiplicative errors),
whereas the statistical errors scale with the square roots of the expected
number of events. 
In this case the combination of the data sets using  
\Eq~\ref{eq:aveold}
leads to a small  bias to lower cross section values since the measurements
with lower central values have smaller
absolute uncertainties. To take this effect
into account, the $\chi^2$ definition is modified to
\begin{equation}
 \chi^2_{\rm exp}\left(\boldsymbol{m},\boldsymbol{b}\right) = 
 \sum_i
 \frac{\left[m^i
- \sum_j \gamma^i_j m^i b_j  - {\mu^i} \right]^2}
{ \textstyle \delta^2_{i,{\rm stat}}\mu^i\left(m^i -  \sum_j \gamma^i_j m^i b_j\right)+
\left(\delta_{i,{\rm uncor}}\,  m^i\right)^2}
 + \sum_j b^2_j.
\label{eq:ave}
\end{equation}
Here $\gamma^i_j = \Gamma^i_j/\mu^i$, 
$\delta_{i,{\rm stat}} = \Delta_{i, {\rm stat}}/\mu^i$ and 
$\delta_{i,{\rm uncor}} = \Delta_{i,{\rm uncor}}/\mu^i$ are relative
correlated systematic, statistical and uncorrelated systematic uncertainty,
respectively.
This $\chi^2$ definition is used for the 
averaging procedure and also for the phenomenological analysis of the data (see \Sec~\ref{sec:pheno}).
\EEq~\ref{eq:ave} is equivalent to the one used in previous H1 analyses~\cite{h1alphas} up to modifications in the denominator. 
In contrast to \Eq~\ref{eq:avebeta}, the $\chi^2$ function  
of \Eq~\ref{eq:ave} is not a simple
quadratic form with respect to $m^i$ and 
$ b_j$. The average is found in an iterative procedure: first \Eq~\ref{eq:avebeta}
is used to get an initial approximation for $\mu^{i,{\rm ave}}$ and $\beta_{j,{\rm ave}}$  which are used
to recalculate the errors as $\Gamma^i_j = \gamma^i_j \, \mu^{i,{\rm ave}}$ and
$\Delta^2_i = \delta^2_{i,{\rm stat}}\mu^i(\mu^{i,{\rm ave}} -  \sum_j \gamma^i_j \mu^{i,{\rm ave}} \beta_{j,{\rm ave}})+
(\delta_{i,{\rm uncor}}\,  \mu^{i,{\rm ave}})^2$.  Then the determination of $\mu^{i,{\rm ave}}$ is repeated.
Convergence is observed  after  two iterations.

For measurements with multiplicative errors the geometric mean instead of the
arithmetic mean can be used as an alternative, 
i.e. the average is performed for $\ln \sigma_r^i$. 
In this case the quadratic \Eq~\ref{eq:avebeta} can be used by replacing $\mu^i \to \ln \sigma_r^i$,
$\Delta_i \to (\delta_{i,{\rm stat}}^2+\delta^2_{i,{\rm uncor}})^{1/2} $ and $\Gamma^i_j \to \gamma^i_j$. This logarithmic
averaging procedure is used as a cross check.

For the {\MB} and {\SVX} analyses,
the measured cross section values $\sigma_r^i$,
the statistical and uncorrelated uncertainties $\delta_{i,{\rm stat}}$,
$\delta_{i,{\rm uncor}}$ and all correlated systematic  uncertainties $\gamma^i_j$ 
as well as the total error 
$\delta_{i,{\rm tot}}=[\delta_{i,{\rm stat}}^2+\delta_{i, {\rm 
uncor}}^2+\sum_j(\gamma^i_{j})^2]^{1/2}$
are given in \Tabs~\ref{tab:svxtable}-\ref{tab:mbtablesn}.
The average of the H1 data 
is reported in \Tabs~\ref{tab:all1}-\ref{tab:all3c}, where the 
average reduced cross sections $\sigma^{i,{\rm ave}}_r = \mu^{i,{\rm ave}}$, 
the statistical $\delta_{i,{\rm ave,stat}}$, uncorrelated
 $\delta_{i,{\rm ave,uncor}}$, correlated $\gamma^{i,{\rm ave}}_j$
and total 
$\delta_{i,{\rm ave,tot}}=[\delta_{i,{\rm ave,stat}}^2
+\delta_{i,{\rm ave,uncor}}^2+\sum_j (\gamma^{i,{\rm ave}}_j)^2]^{1/2}$ 
uncertainties are given.
The transformation matrix $U$ 
is given in \Tab~\ref{tab:delta}.  
The shifts of the central values of the 
systematic error sources, in units of the systematic errors 
$\alpha_{j,{\rm ave}}/\Delta_{\alpha j}$, are given in \Tab~\ref{tab:9799sys}.

\subsection{Compatibility of {\SVX} and {\MB} Data \label{sec:ave9900}}
The combination of the {\SVX} and {\MB} data depends upon assumptions on the 
correlation
between different data points, within a data set as well as across the 
data sets. For each data set, two types of systematic uncertainty 
are considered: fully correlated ones, which are treated as  $\alpha_j$
 in \Eq~\ref{eq:ave}, and fully uncorrelated ones, which are added
to the statistical uncertainties in quadrature and treated as 
$\delta_i$ in \Eq~\ref{eq:ave}.
Following the notation in \Tab~\ref{tab:syssum},
the six sources of correlated uncertainties are $E'_e$ scale, $\theta_e$,
LAr hadronic energy
scale, LAr noise, SpaCal hadronic scale and photoproduction background.
A further correlated uncertainty arises from the luminosity
measurements.
Concerning the relation between data sets, the 
systematic uncertainty of the luminosity measurement
is separated into a $0.5\%$ fully correlated 
theoretical uncertainty and an uncorrelated 
experimental part due to variations 
of beam and detector acceptance conditions.
The other systematic uncertainties are considered to be uncorrelated.

The  systematic uncertainties which are correlated
between data points can be assumed as either fully correlated,
uncorrelated or partially correlated between the {\MB} and the {\SVX} data. 
The reasons for correlations between data sets
are the similarity in the calibration procedure and the detector setup.
Uncorrelated effects arise
from  variations with time, differences between 
the kinematic ranges of the calibration samples,  the dead material, the detector illumination or the acceptance.
For each source the uncorrelated part is more important
and thus all sources are considered to be uncorrelated between
the {\MB} and the {\SVX} data.

To check the sensitivity of the averaged result to the correlation assumptions,
the average of the {\MB} and {\SVX} data, obtained by considering the six
systematic sources to be uncorrelated, is compared to $2^{6}-1$ other
possible assumptions
in which each source is either fully correlated or fully uncorrelated. 
Most of the cases lead to numerically small variations
for both the central values and the total errors of the average data. The only
significant variation is observed for the lowest $y$ points 
for $Q^2>2$~GeV$^2$, if the LAr noise is assumed to be correlated
between the {\MB} and {\SVX} data. Since the LAr noise, however,
is a time dependent uncorrelated source, no additional systematic uncertainty is 
assigned to the combined measurement.

The {\MB} and {\SVX} data sets are fully consistent, according to the 
averaging procedure, with 
$\chi^2_{\rm min}/\dof = \chiave/\dofave$. The shifts
 of the central values of the systematic uncertainties do not exceed one standard deviation.

\subsection{Global Combination of Low $\mathbold{Q^2}$ H1 Data \label{sec:ave979900} }

\begin{figure}
\begin{center}
\epsfig{file=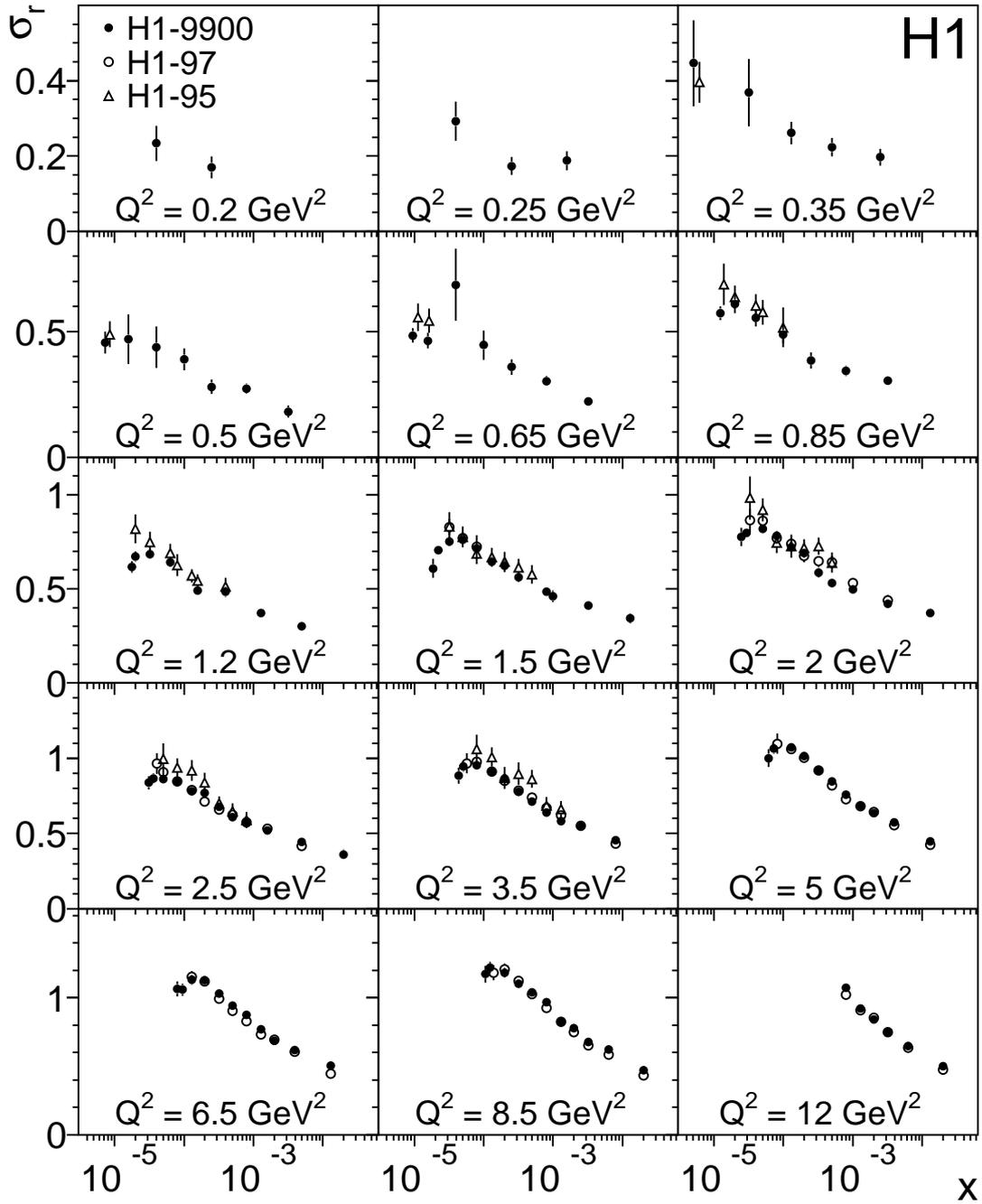,width=1.2\linewidth}
\end{center}
\caption{\label{fig:mbsvx0mb97} Reduced cross section $\sigma_{r}$.  Closed
circles: combined $1999$-$2000$  data taken at $E_p = 920$\,GeV;
Triangles: SVX data  
taken in $1995$\,\cite{Adloff:1997mf}; Open circles:
NVX data taken in $1997$\,\cite{h1alphas}.
The normalisation of the  $1997$ data has changed by $+3.4\%$, 
see \Sec~\ref{sec:lumi}. 
The 1995 and 1997 data were taken at $E_p =820$\,GeV
but are corrected here for comparison to 920\,GeV.
The errors represent the statistical and
systematic uncertainties added in quadrature.}
\end{figure}

The new data given in \Tabs~\ref{tab:svxtable}-\ref{tab:mbtablesn}
are combined with the previously published H1 data obtained
for a similar kinematic region.
The comparison of the present cross section data, obtained by averaging 
the \SVX\
and the {\MB} data,  with the published  cross section data,
is given in \Fig~\ref{fig:mbsvx0mb97}. 
The new data are in agreement with the published NVX97 data~\cite{h1alphas}
taking the $+3.4\%$ normalisation
shift of the published data (\Sec~\ref{sec:lumi})
into account. The data are also consistent with the SVX95 data~\cite{Adloff:1997mf}
within their rather large uncertainties.
 For the  combination of all data, the
systematic uncertainties are considered to  be uncorrelated
across the data sets.

The published H1 data~\cite{Adloff:1997mf,h1alphas} were taken with a  
proton beam energy of $E_p = 820$~GeV. Therefore a centre-of-mass energy 
(CME) correction, based on \Eq~\ref{sigred},
is applied when comparing to the previously published
cross section  according to
\begin{align}\label{eq:cmecorr}
  \sigma_{r}^{920}(x,Q^2) & = \sigma_{r}^{820}(x,Q^2) + 
  F^{\rm th}_L(x,Q^2) \left[  f(y_{820}) - f(y_{920}) \right]. 
\end{align}
Here  $\sigma_{r}^{920}(x,Q^2)$ is the 
reduced cross section  rescaled to  $E_p=920$~GeV;
$\sigma_{r}^{820}(x,Q^2)$
is the  measured reduced cross section
for $E_p=820$~GeV; $y_{820}$ and $y_{920}$ are the inelasticities for
the two proton beam energies  calculated as $y = Q^2/4 E_e E_p x$,
and  $F^{\rm th}_L(x,Q^2)$ is calculated using the fractal model for $F_2(x,Q^2)$ and
$R=0.5$.
This correction becomes large only at high $y$.
To avoid any sizeable effect of the energy dependence of 
$\sigma_{r}$ on the combination of the $820$ and $920$~GeV 
data, the combination of the points at the same $(x,Q^2)$ is restricted to a 
region of $y_{820} < 0.35$. At higher $y$ the measurements
are kept separately but they are affected by the combination procedure. 
The residual dependence on the $F_L$
assumption  for the average points is negligible.
For illustrative purposes, the CME correction is applied 
to all $820$~GeV data points 
in \Figs~\ref{fig:mbsvx0mb97}-\ref{fig:wplot}.

The H1 data sets are consistent with each other. If all
samples are averaged in a single step one obtains
$\chi^2_{\rm tot}/\dof=\ochiave/\odofave$.
Shifts of the central values of the 
systematic sources $\alpha_j/\Delta_{\alpha_j}$
are given in \Tab~\ref{tab:9799sys}.
The systematic shifts imposed by the averaging procedure
are mostly within one standard deviation. The most noticeable
effects are a downward shift of the normalisation of the
SVX95 data and a modification of the LAr hadronic energy scale
of the SVX data which corresponds to a small adjustment of
the SVX data at large $x$.

The combination of the H1 data
using the $\chi^2$ definition of \Eq~\ref{eq:ave}
has been compared to that using 
the $\chi^2$ definition of \Eq~\ref{eq:avebeta} and also
using the logarithmic averaging procedure.
For the bulk of the phase space, 
the definition of \Eq~\ref{eq:avebeta} would lead to a change of
typically $-0.7\%$, which increases
to $-2.0\%$ for the data at $Q^2 \le 0.5$~GeV$^2$.
For the logarithmic average
the difference compared to \Eq~\ref{eq:ave} is typically below $0.1\%$.
\begin{table}[tb]
\centerline{%
\begin{tabular}{l|rrrr}
\hline\hline
  {\bfseries Systematic Source} & \multicolumn{4}{c}{\bfseries Shift in $\sigma$}\\
                     & SVX95 & NVX97 &   {\MB}  &  {\SVX}  \\
\hline
  $E'_e$ scale          &$\hnfeev$ & $\ombeelv  $ & $\nmbeelv  $ &$ \nsvxeelv$ \\
  $\theta_e$            &$\hnfthv$ & $\ombthev$   & $\nmbthev$ &$ \nsvxthev$ \\
  LAr scale             &$\hnfehv$ & $\ombehadv$ & $\nmbehadv$ &$\nsvxehadv$ \\
  LAr noise             &---       & $\ombnoiv $  & $\nmbnoiv $ &$\nsvxnoiv $\\
  SpaCal hadronic scale &---&  ---        & $\nmbhspav$ &$\nsvxhspav$ \\
  $\gamma p$ background &$\hnfgpv$ & $\ombgpv$   & $\nmbgpv$   &$\nsvxgpv $\\
  Luminosity            &$\hnflumiv$& $\omblumiv$ & $\nmblumiv$ &$ \nsvxlumv $ \\
\hline\hline
\end{tabular}%
}
\tablecaption{\label{tab:9799sys}Shifts of the 
central values of the systematic 
uncertainties $\alpha_{j,{\rm ave}}/\delta_{\alpha_j}$ based on the 
average of the published $E_p=820$\,GeV  and
the new \MB-{\SVX} data.  For example, the quoted value
for the luminosity shift of the SVX95 sample, $\hnflumiv$, corresponds 
to a $\hnflumiv \times 3\% = 4.80\%$ downward shift of the SVX95 cross section
values.}
\end{table}

\subsection{Combined Cross Section Results}
\begin{figure}
\begin{center}
\epsfig{file=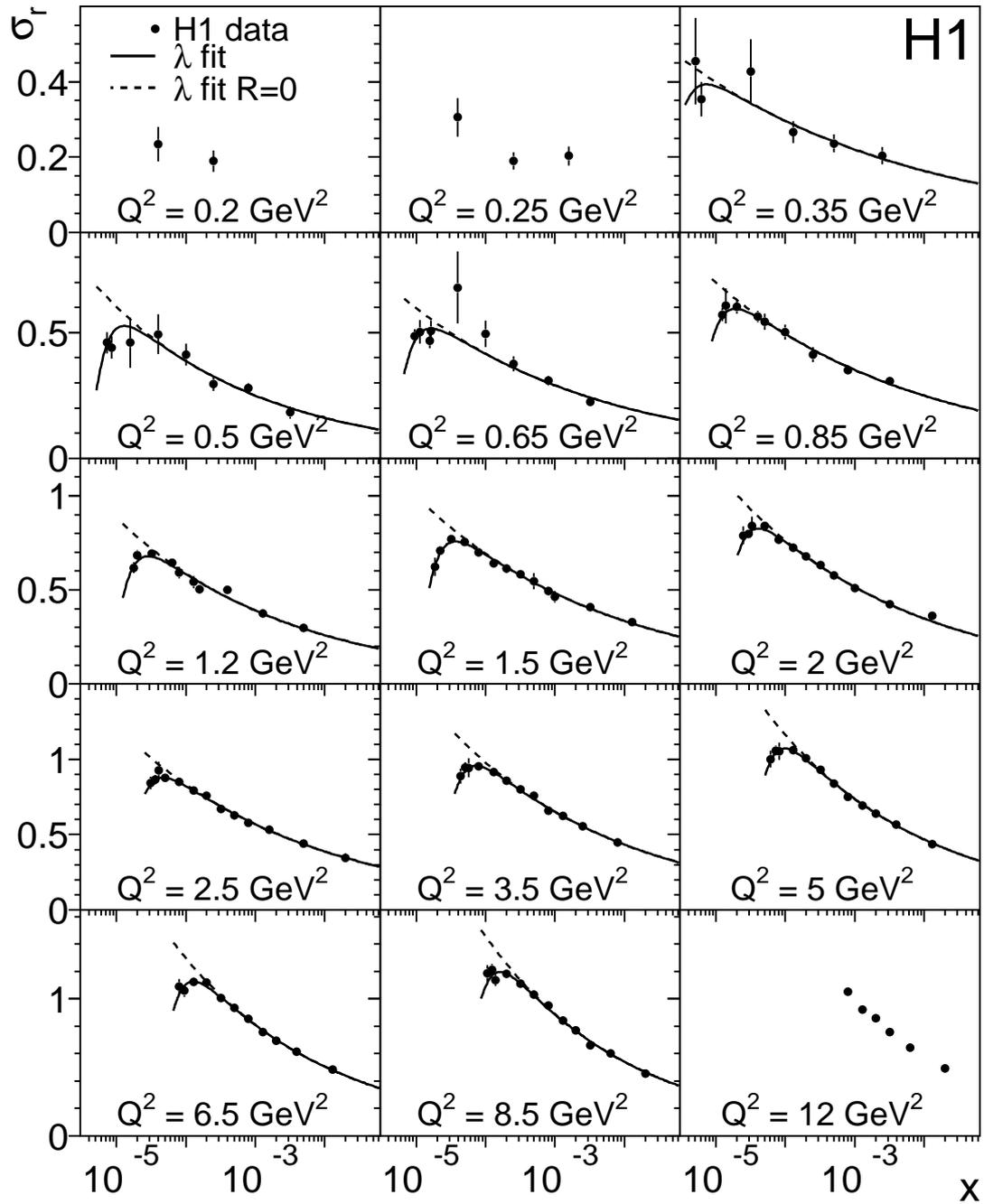,width=1.2\linewidth}
\end{center}
\caption{\label{fig:xsec-ave-xlam}
Reduced cross section $\sigma_{r}$, from the combined
low $Q^2$ H1 data, 
as a function of $x$ compared to the $\lambda$ fit result (solid line)
and to a $\lambda$ parameterisation with the same values of $c(Q^2)$ and
$\lambda(Q^2)$ but $R=0$  (dashed line). 
The errors represent the statistical and
systematic uncertainties added in quadrature.}
\end{figure}

\begin{figure}
\begin{center}
\epsfig{file=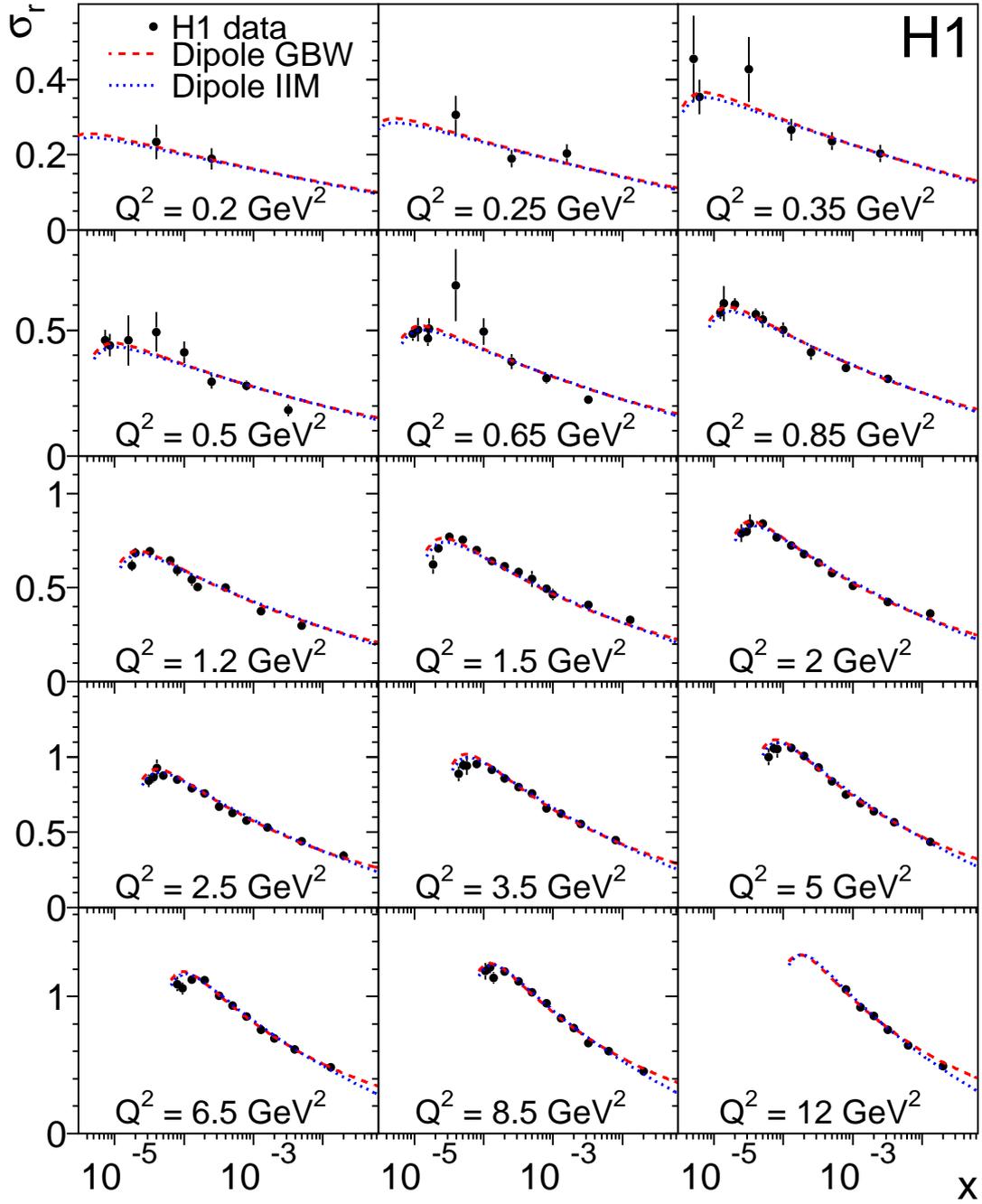,width=1.2\linewidth}
\end{center}
\caption{\label{fig:xsec-ave}
Reduced cross section $\sigma_{r}$, from the combined
low $Q^2$ H1 data, 
as a function of $x$ compared to the GBW and IIM models. 
The errors represent the statistical and
systematic uncertainties added in quadrature.}
\end{figure}

\begin{figure}
\begin{center}
\epsfig{file=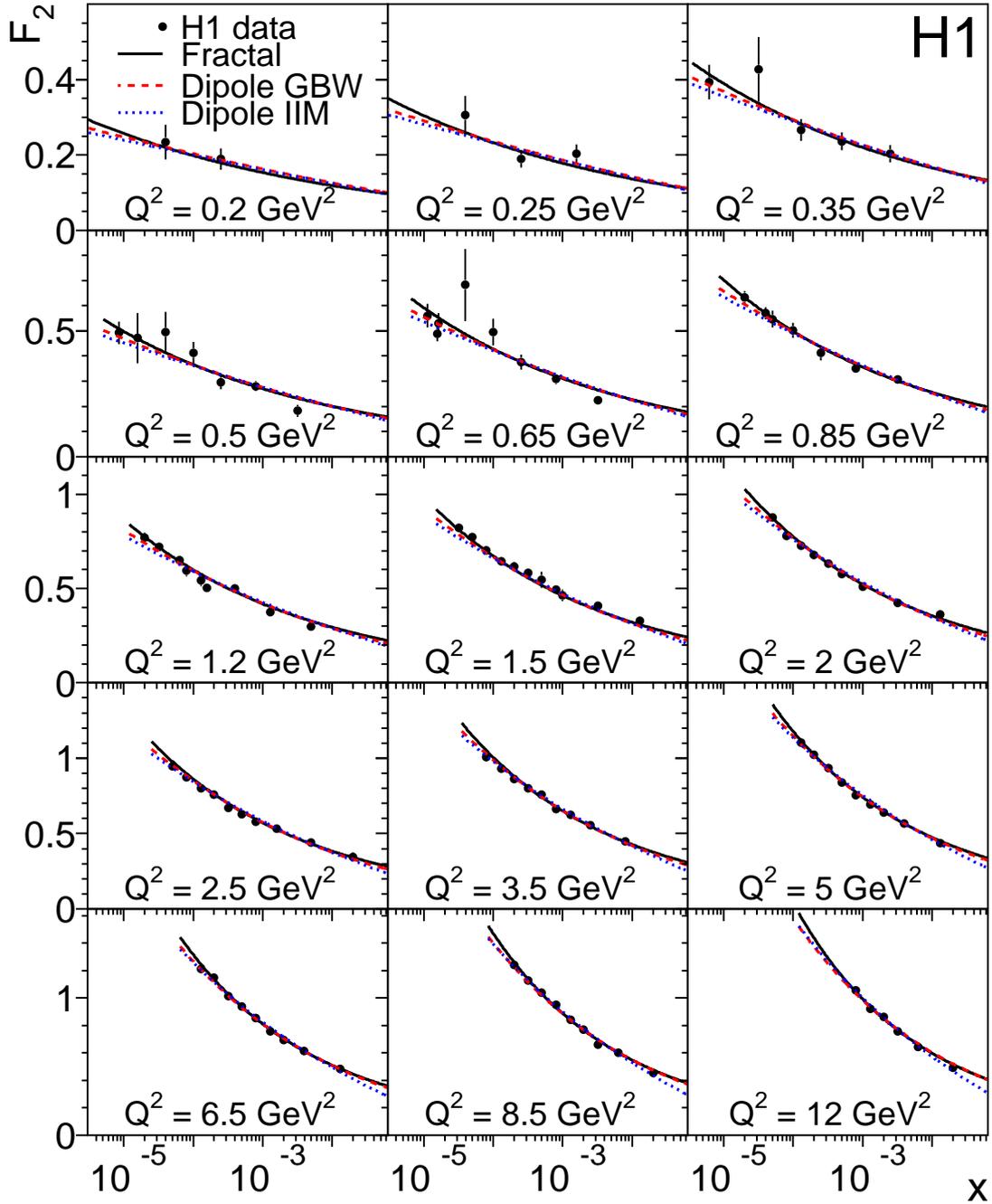,width=1.2\linewidth}
\end{center}
\caption{\label{fig:f2-ave}
Structure function $F_2$, from the combined
low $Q^2$ H1 data for $y<0.6$, 
as a function of $x$ compared to the fractal, the dipole GBW and the dipole IIM
 fit results. The errors represent the statistical and
systematic uncertainties added in quadrature.
}
\end{figure}

\begin{figure}
\centerline{%
\epsfig{file=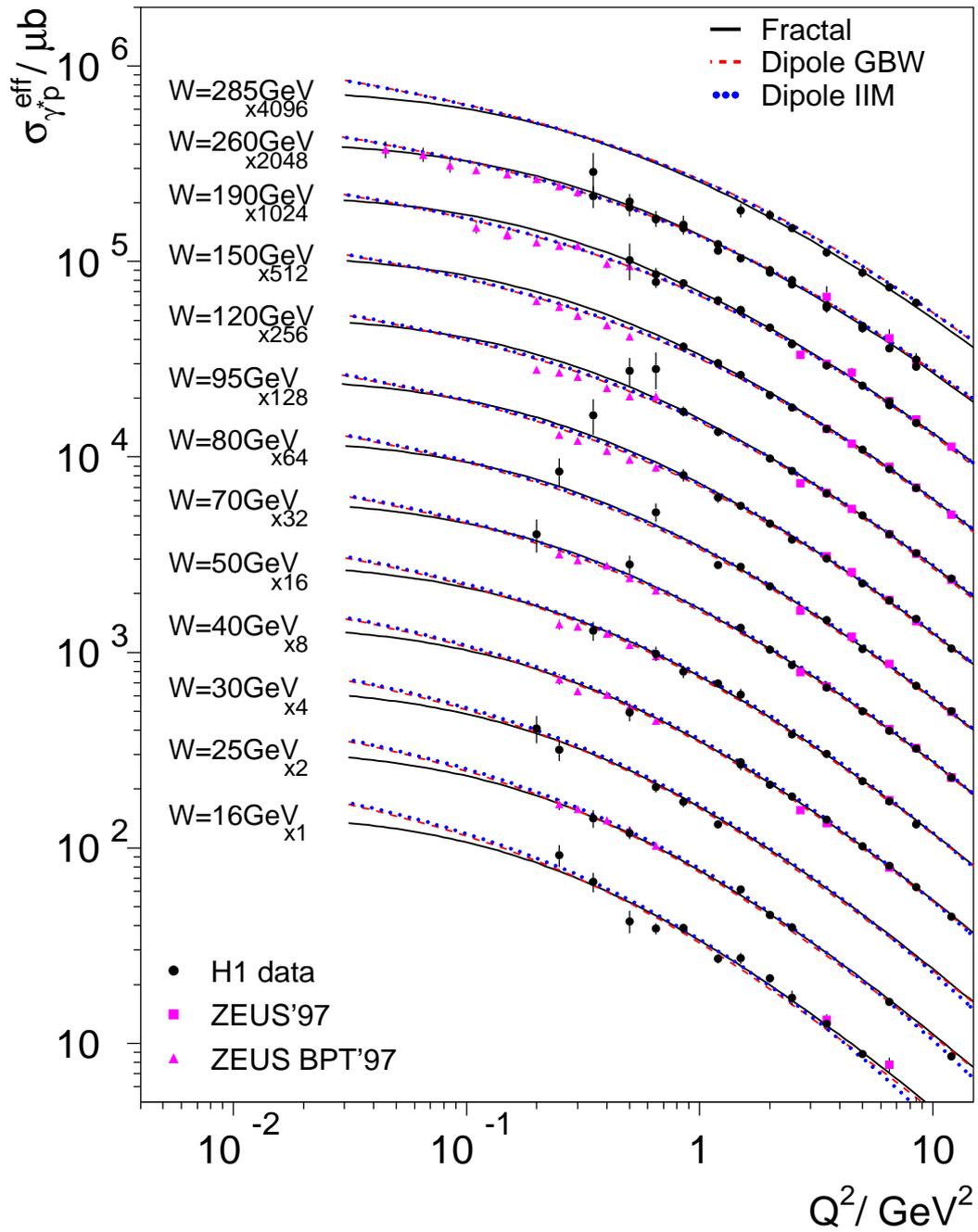,width=0.9\linewidth}
}
\caption{\label{fig:wplot}
Measurement of the virtual photon-proton cross section
$\sigma_{\gamma^* p}^{\rm eff}$ as a function of $Q^2$ at various values of $W$.
The cross sections for different $W$ values are multiplied with
the factors indicated in the figure. The errors represent the statistical and
systematic errors added in quadrature. The averaged H1 results are compared to
data obtained by the ZEUS experiment and to the fractal and dipole model  fit results.}
\end{figure}

The combined low $Q^2$ data and the resulting uncertainties are listed 
in \Tabs~\ref{tab:all1}-\ref{tab:all3c} and shown in \Figs~\ref{fig:mbsvx0mb97}-\ref{fig:wplot}. 
There are   $\frnpt$  data points. 
The total uncertainty 
in the central $Q^2,x$ region
of this measurement is about $2$\% but it becomes larger towards the edges of
the covered phase space. At high $y$, for example, the measurement
at a $Q^2$ value of a few GeV$^2$ has an uncertainty  of about $5$\%.

\FFigs~\ref{fig:xsec-ave-xlam} and \ref{fig:xsec-ave}
show the combined H1 reduced $ep$ cross section measurement
and different phenomenological descriptions as introduced
below. 
For all $Q^2$ bins, starting at large $x$ the reduced cross section 
first increases for $x\to 0$.
For $Q^2\ge 0.6$~GeV$^2$ there is a characteristic
turn over of the cross section observed
at the smallest $x$ values. This region,
for each $Q^2$, corresponds to highest inelasticity, 
$y = Q^2/(s x)$, and thus the turn over
at $y \approx 0.6$ can be attributed to the influence of the
longitudinal structure function $F_L$.

For $y<0.6$ the influence of the longitudinal structure function is
small and thus one can extract the structure function $F_2$ 
with only a small residual dependence on the assumption on $F_L$.
Using $R=0.5$,  $F_2$ is extracted and 
 shown in \Fig~\ref{fig:f2-ave}.  The structure function $F_2$
exhibits a steady increase as $x\to 0$ for all $Q^2$ bins.

\FFig~\ref{fig:wplot} shows the measurement of the virtual 
photon-proton effective cross section $\sigma_{\gamma^* p}^{\rm eff}$
as a function of $Q^2$ at various values of $W$. The H1 data are compared to
the data of ZEUS~\cite{Chekanov:2001qu,Breitweg:2000yn} and to different
models, as discussed below.
A good agreement between the data sets is observed. 
The H1 data extend the HERA measurements to higher
and lower $W$ and also cover the $Q^2\sim1$~GeV$^2$ region.

\section{Cross Section Analysis \label{sec:pheno}}
\subsection{Rise of $\boldsymbol{F_2}$  at Low $\boldsymbol{x}$ and Extraction of $\boldsymbol{R}$ \label{sec:lambda}}
\def\crlam{  7.9}
\def\mrlam{ 0.55}
\def\erlam{ 0.05}
The rise of the structure function $F_2$ towards low $x$ has 
previously been described
by a power law in $x$, $F_2 = c(Q^2) x^{-\lambda(Q^2)}$, where the exponent
$\lambda$ increases approximately logarithmically with $\ln Q^2$ for
$Q^2 \gtrsim 2$\,GeV$^2$~\cite{adloff-2001-520}. This simple parametrisation
has been shown to model the $ep$ data  well for $x < 0.01$.

This idea can be extended to fit the reduced cross section $\sigma_{r}$
in order simultaneously to extract the exponent $\lambda$ and 
to estimate the longitudinal structure function $F_L$. 
The measured $ep$ cross section is sensitive to the longitudinal structure
function $F_L$ only for large $y \gtrsim 0.5$, a region which corresponds to a limited $x$
range  for a given $Q^2$ value. 
Gluon dominance at low $x$ suggests
that the function $F_L$  may exhibit an $x$ dependence similar to  $F_2$.
In the subsequent studies  using this ansatz 
it is assumed  that $F_L$ is proportional to $F_2$ and that the coefficient
of proportionality depends only on $Q^2$.
For the extraction of $F_L$,
the ratio of $\sigma_L/\sigma_T=R$  is
used such that 
\begin{equation}
F_L(Q^2,x) = F_2(Q^2,x) \, \frac{R(Q^2)}{1+R(Q^2)}.  \label{eq:rdef}
\end{equation}
and
\begin{equation}
  \sigma_{r}(Q^2,x) 
= c(Q^2) x^{-\lambda (Q^2)}\left[1 - f(y) 
\frac{\textstyle R(Q^2)}{1+\textstyle R(Q^2)} \right].  \label{eq:lamfit}
\end{equation}

\begin{figure}
\begin{center}
\epsfig{file=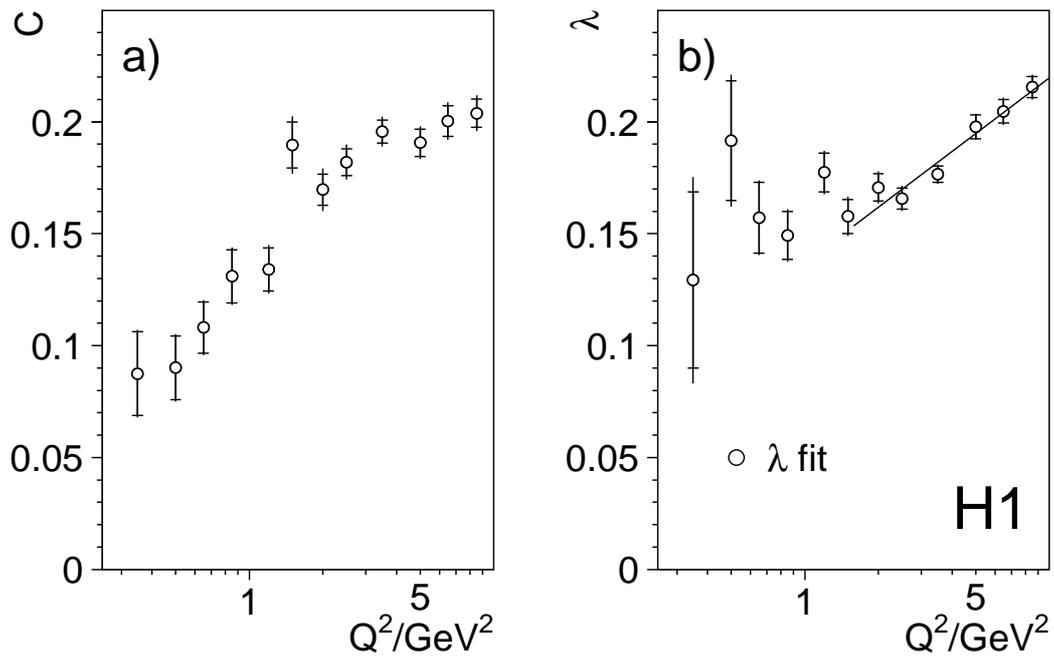,width=0.9\linewidth}
\end{center}
\caption{\label{fig:lambda} Coefficients $c$ and $\lambda$, as 
defined in \Eq~\ref{eq:lamfit}, determined from a fit to
the H1 data as a function of $Q^2$.
The inner  error bars represent uncorrelated systematic  
uncertainties. The outer error bars represent total
uncertainties. The line in b) shows a straight line fit for 
$Q^2\ge 2$~GeV$^2$. }
\end{figure}

\begin{figure}
\begin{center}
\epsfig{file=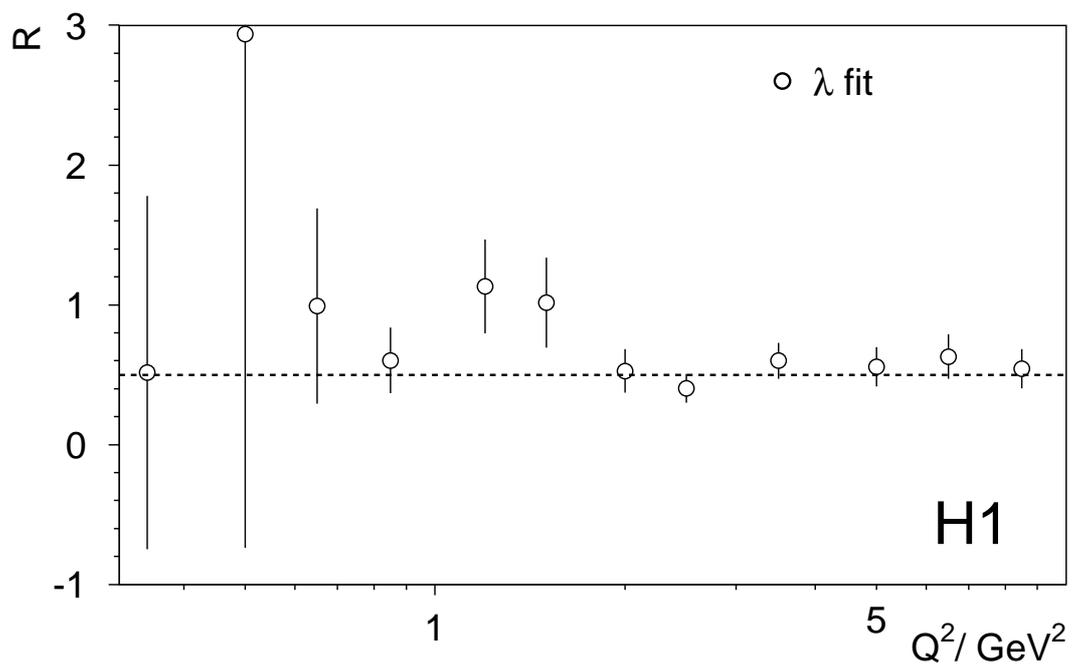,width=0.9\linewidth}
\end{center}
\caption{\label{fig:fl} 
Coefficient $R$ as a function  of $Q^2$ from a simple
parameterisation of the reduced cross section as defined in\,\Eq~\ref{eq:lamfit}.
The dashed line is drawn at $R=0.5$. The errors represent the total uncertainties.
}
\end{figure}

The combined $1995$-$2000$ H1 low $Q^2$ data are fitted following
\Eq~\ref{eq:lamfit} for each $Q^2$ bin. These fits 
describe the data very well, as is illustrated in \Fig~\ref{fig:xsec-ave-xlam}.
The results of these fits 
 are shown in \Figs~\ref{fig:lambda} and~\ref{fig:fl}. The fit results for 
$\lambda(Q^2)$ are given in \Tab~\ref{tab:lamfit}.
One can see in \Fig~\ref{fig:lambda}b) that the parameter $\lambda$ shows 
an approximately linear increase
as a function of $\ln Q^2$ for  $Q^2>2$~GeV$^2$ as has been
observed previously\,\cite{adloff-2001-520}.
For lower $Q^2$ the variation of $\lambda$
is diminished but relatively large uncertainties prevent definite conclusions.
The normalisation coefficient $c(Q^2)$ 
rises with increasing $Q^2$ for $Q^2 < 2$\,GeV$^2$ and is consistent with
a constant behaviour in the DIS region, as in\,\cite{adloff-2001-520}.

The values of the coefficient $R(Q^2)$
are consistent with no dependence on $Q^2$. The 
mean $R$ is $\mrlam \pm \erlam$ with\footnote{For the 
determination of the mean, $R$ values from different
$Q^2$ bins are assumed to be uncorrelated and total errors are used.}
$\chi^2/\dof=\crlam/(8-1)$. While the experimental error
is small there is a very strong model dependence, different parameterisations
for  $F_2$ leading to significant changes in $F_L$, see  \Sec~\ref{sec:modeldata}.   
The value of the average $R$ obtained in this model is
consistent within about one standard deviation with $R=0.5$ or 
 $\sigma_L = \frac{1}{2} \sigma_T$. This value of $R$ leads to an $F_L$ 
which is higher than the
first direct measurement of $F_L$ at low $x$ performed by the H1 
collaboration~\cite{h1fl}. The data in~\cite{h1fl} correspond however
to higher $Q^2$ values ($\ge 12$~GeV$^2$).

\begin{table}[tb]
\centerline{
\begin{tabular}{c|c|c|c|c|c}
\hline \hline
$Q^2$/GeV$^2$ & $\lambda$ & $\delta_{\rm \lambda, stat}$& $\delta_{\rm \lambda, uncor}$& $\delta_{\rm \lambda, cor}$& $\delta_{\rm \lambda, tot}$ \\
\hline
$  0.35 $ & $ 0.129 $ & $ 0.029 $ & $ 0.026 $ & $ 0.024 $ & $ 0.046 $ \\ 
$  0.50 $ & $ 0.192 $ & $ 0.022 $ & $ 0.016 $ & $ 0.012 $ & $ 0.030 $ \\ 
$  0.65 $ & $ 0.157 $ & $ 0.010 $ & $ 0.011 $ & $ 0.006 $ & $ 0.016 $ \\ 
$  0.85 $ & $ 0.149 $ & $ 0.009 $ & $ 0.009 $ & $ 0.005 $ & $ 0.014 $ \\ 
$  1.20 $ & $ 0.177 $ & $ 0.007 $ & $ 0.007 $ & $ 0.005 $ & $ 0.011 $ \\ 
$  1.50 $ & $ 0.158 $ & $ 0.004 $ & $ 0.006 $ & $ 0.004 $ & $ 0.008 $ \\ 
$  2.00 $ & $ 0.171 $ & $ 0.003 $ & $ 0.005 $ & $ 0.004 $ & $ 0.007 $ \\ 
$  2.50 $ & $ 0.166 $ & $ 0.002 $ & $ 0.005 $ & $ 0.003 $ & $ 0.006 $ \\ 
$  3.50 $ & $ 0.177 $ & $ 0.002 $ & $ 0.003 $ & $ 0.002 $ & $ 0.004 $ \\ 
$  5.00 $ & $ 0.198 $ & $ 0.003 $ & $ 0.004 $ & $ 0.002 $ & $ 0.005 $ \\ 
$  6.50 $ & $ 0.205 $ & $ 0.003 $ & $ 0.005 $ & $ 0.003 $ & $ 0.007 $ \\ 
$  8.50 $ & $ 0.216 $ & $ 0.003 $ & $ 0.005 $ & $ 0.003 $ & $ 0.007 $ \\ 
\hline \hline
\end{tabular}
}
\tablecaption{\label{tab:lamfit}Results of the fit (\Eq~\ref{eq:lamfit}) to the
combined H1 low $Q^2$ data on the exponent $\lambda$ with 
the statistical $\delta_{\rm \lambda, stat}$, uncorrelated systematic $\delta_{\rm \lambda, uncor}$, correlated systematic $\delta_{\rm \lambda, cor}$, and total uncertainties $\delta_{\rm \lambda, tot}$.
}
\end{table}

\subsection{Determination of $\boldsymbol{F_L}$ using the $\boldsymbol{y}$ Dependence of the Cross Section}
The turn-over of the measured DIS cross section for the highest $y$ values,
apparent in \Fig~\ref{fig:xsec-ave-xlam}, can be used for an extraction of the
longitudinal structure function $F_L$ using the so-called derivative
method~\cite{h1alphas}.
The derivative
of the reduced cross section with respect to $\ln y$ is
\begin{equation}
\left. \frac{{\rm d} \sigma_{r}}{{\rm d} \ln y}\right|_{Q^2=\text{const}} = - \frac{{\rm d} F_2}{{\rm d} \ln x}
- \frac{2y^2(2-y)}{(1+(1-y)^2)^2} F_L -  \frac{y^2}{1+(1-y)^2} \frac{{\rm d} F_L}{{\rm d} \ln x} ~.
\end{equation}
At high $y$ for a wide variety of models 
the term proportional to $ F_L$  becomes numerically larger 
than other contributions.
Therefore the extraction of  the derivative provides  means
for determining  $F_L$ at low $x$ and $Q^2$ with little phenomenological 
assumptions.

\begin{figure}
\centerline{%
\epsfig{file=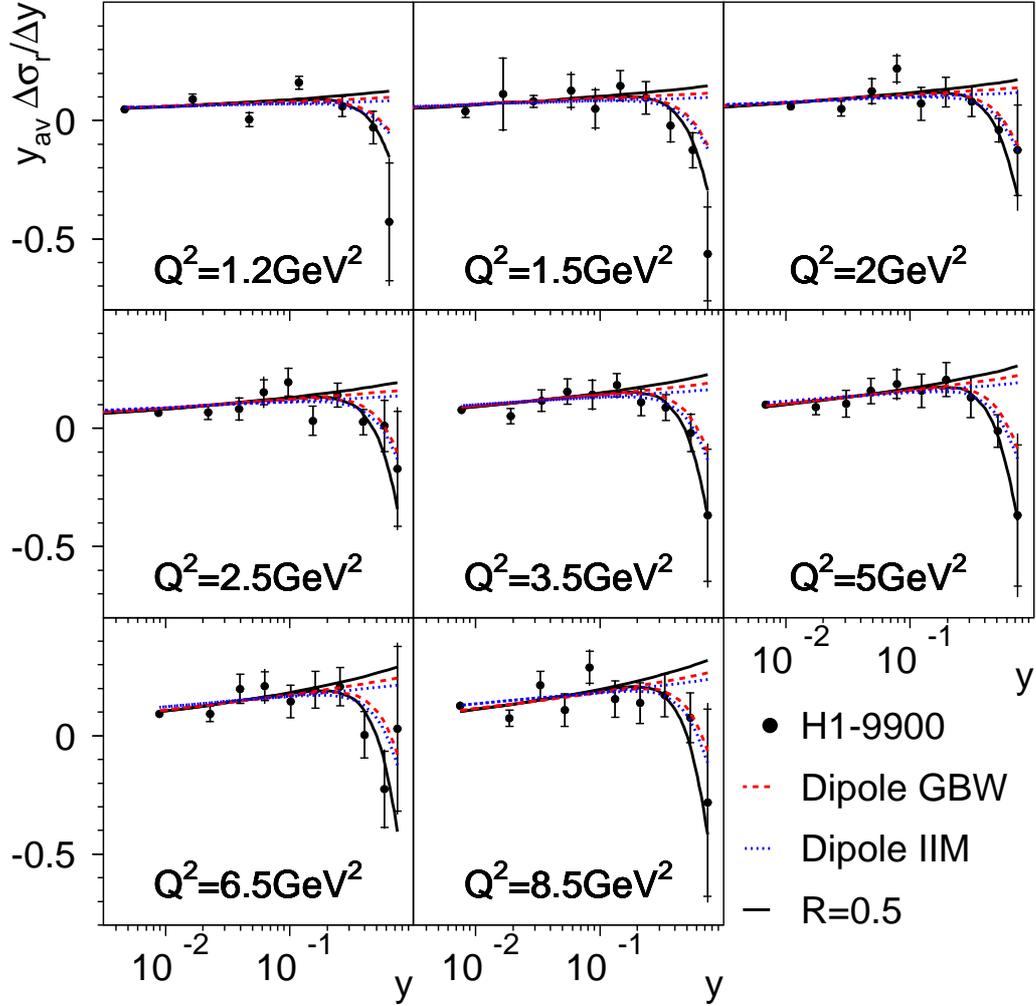,width=0.9\linewidth}
}
\caption{Derivative $y_{\rm av} \Delta \sigma_{r}/\Delta y$
for the combined $1999$-$2000$ H1 data compared to the predictions of the 
dipole models and the fractal model
for $F_2$ with an assumption $R=0.5$ to describe $F_L$, labeled $R=0.5$. 
The lines increasing as a function of $\ln y$ 
correspond to $F_L=0$ for these models. The lines turning
over at high $y$ correspond to the 
cross section predictions. 
The inner error bars represent statistical and uncorrelated
uncertainties added in quadrature, 
the outer error bars represent the total
uncertainties. 
\label{fig:dsdy}}
\end{figure}

Experimentally,  ${\rm d}\sigma_{r}/{\rm d}\ln y$ is approximated by
 $y_{\rm av} \Delta \sigma_{r}/ \Delta y$, which is
calculated for each pair of cross section measurements in neighbouring  bins.
Here $\Delta y$ is determined using bin centre values,
and $y_{\rm av}$ is  the logarithmic average value. Only the $E_p=920$\,GeV
data are used in this determination. 
The H1 data are illustrated in
\Fig~\ref{fig:dsdy} and  are compared
to the fractal and dipole models discussed below in \Secs~\ref{subsec:fractalfit} and \ref{sec:satur}.
Similar analysis using the
$E_p=820$\,GeV data was presented in\,\cite{h1alphas}.
The systematic uncertainties are evaluated by changing the cross sections
for each source of systematic uncertainty and repeating the calculation of
the cross section difference. For the model predictions, 
$\Delta \sigma_{r}/\Delta y$ is calculated in an analogous way and
using the same  binning as for the data.

For the extraction of the structure function $F_L$, the fractal fit,
introduced in \Sec~\ref{subsec:fractalfit},
 is used
to estimate the ${\rm d}F_2/{\rm d}\ln x$ contribution to $\Delta \sigma_{r}/\Delta y$, and also for the
bin centre correction.
To reduce the dependence on $F_2$, 
only $\Delta \sigma_{r}/\Delta y$ value corresponding to $y=0.735$
are used to determine $F_L$. 

\begin{figure}
\centerline{%
\epsfig{file=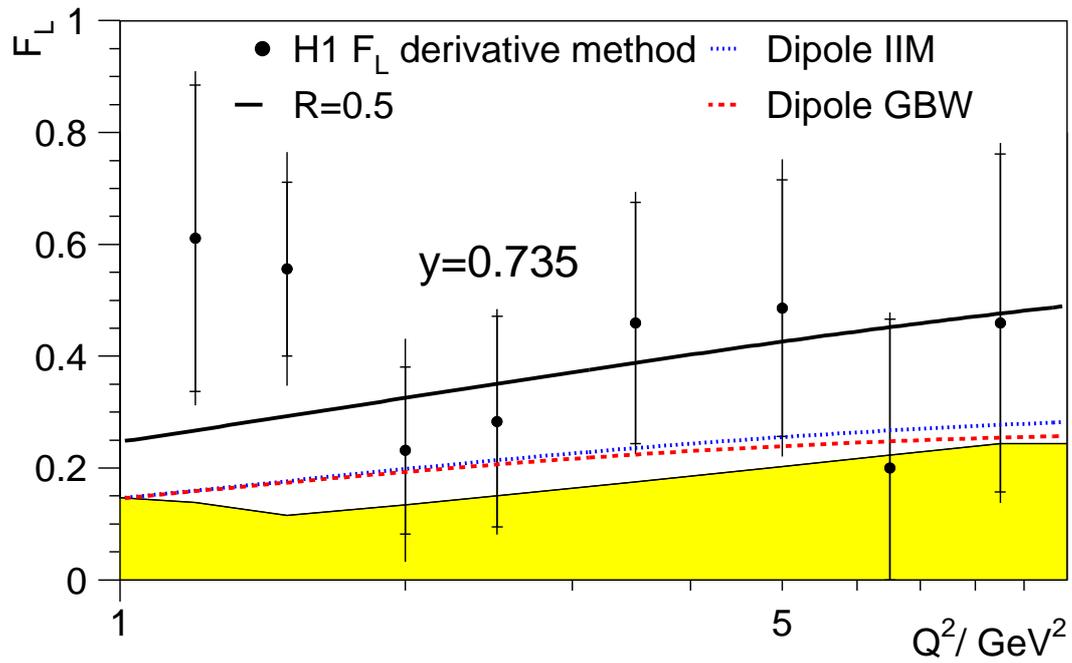,width=0.9\linewidth}
}
\caption{Structure function $F_L$ extracted using the
derivative method\label{fig:flder}. The solid line is drawn for $R=0.5$
assuming the fractal parameterisation for $F_2$.
The dashed (dotted) line corresponds to  
the dipole GBW (IIM)   model. 
The inner error bars represent statistical and uncorrelated
uncertainties added in quadrature, 
the outer error bars represent the total
uncertainties. 
The solid (yellow) band indicates the model uncertainty, see text.}
\end{figure}

The resulting longitudinal structure function  values are shown in \Fig~\ref{fig:flder}.
The derivative method is only weakly dependent on the model assumptions.
There are however large experimental
uncertainties, mostly due to statistics and
  the photoproduction
background at large $y$. 
The  $F_L$ data are consistent with  a constant $R=0.5$, 
as introduced above, and also with smaller values on $R$,
as obtained in the dipole models. 
The dependence of the measurement on the assumption made for $F_2$ is
estimated by a comparison with results obtained when assuming
$F_2$ to be  independent of $x$.
The difference between the extracted $F_L$ values is shown as  the band at the bottom of 
 \Fig~\ref{fig:flder}.

\section{Model Comparisons\label{sec:phenomod} \label{sec:modeldata}} 
In the following the combined data are analysed in the context
of the fractal model \cite{fractal} and two versions of the colour dipole 
model~\cite{Golec-Biernat:1998js,Iancu:2003ge},
which unlike pQCD may be applied to describe the transition region
from photoproduction to deep inelastic scattering. 
Fits are performed using  \Eq~\ref{eq:ave}.

\subsection{Fractal Fit\label{subsec:fractalfit}}
\def\frchi{155.3}
\def\frnpt{149}
\def\frnpa{  5}
\def\frdzv{ 0.75}
\def\frdze{ 0.03}
\def\frdov{ 0.052}
\def\frdoe{ 0.002}
\def\frdtv{ 1.08}
\def\frdte{ 0.00}
\def\frdev{-1.16}
\def\frdee{ 0.03}
\def\frqzv{ 0.093}
\def\frqze{ 0.010}
\def\frrrv{ 0.56}
\def\frrre{ 0.07}
\def\frdiagchi{128.0}
\def\frdiagnpt{149}
\def\frdiagnpa{  5}
\def\frdiagdzv{ 0.73}
\def\frdiagdze{ 0.04}
\def\frdiagdov{ 0.053}
\def\frdiagdoe{ 0.003}
\def\frdiagdtv{ 1.08}
\def\frdiagdte{ 0.00}
\def\frdiagdev{-1.17}
\def\frdiagdee{ 0.04}
\def\frdiagqzv{ 0.097}
\def\frdiagqze{ 0.016}
\def\frdiagrrv{ 0.64}
\def\frdiagrre{ 0.05}
\def\frsnchi{145.7}
\def\frsnnpt{142}
\def\frsnnpa{  5}
\def\frsndzv{ 0.77}
\def\frsndze{ 0.03}
\def\frsndov{ 0.052}
\def\frsndoe{ 0.002}
\def\frsndtv{ 1.08}
\def\frsndte{ 0.00}
\def\frsndev{-1.16}
\def\frsndee{ 0.02}
\def\frsnqzv{ 0.091}
\def\frsnqze{ 0.010}
\def\frsnrrv{ 0.65}
\def\frsnrre{ 0.09}
In the fractal ansatz~\cite{fractal} ,
the proton structure function $F_2$ is parameterised 
using five parameters $Q_0$ and $D_0$  to $D_3$ as
\begin{equation}
F_2(Q^2,x) = D_0 Q^2_0 \left(1+\frac{Q^2_0}{Q^2}\right)^{1-D_2}
\frac{x^{-D_2+1}}{1+D_3-D_1\ln x} \left(x^{-D_1\ln \left[1+\frac{Q^2}{Q^2_0} 
\right]} \left(1+\frac{Q^2}{Q^2_0} \right)^{D_3+1} -1 \right) ~.\label{eq:fract}
\end{equation}
The parameters of this model are determined with  a fit to the
cross section data, except for the parameter $D_2$, which governs the structure function
behaviour for the photoproduction regime and is fixed to $D_2 = 1.08$.
This parameterisation is used in the Monte Carlo reweighting procedure.
The fractal model~\cite{fractal} does not
provide predictions for  $F_L$.
The same prescription is
 followed as for the $\lambda$ parameterisation fit described in
\Sec~\ref{sec:lambda} taking  the $F_L$ contribution
to be proportional to $F_2$.

The values of $R$  are found to be consistent
with the $\lambda$ fit and with being independent of $Q^2$.
Thus for the fractal parameterisation of the 
reduced cross section, $R$ is taken to be a constant,
which results in the simple five parameter  representation
used in the present analysis.
The parameters of the fit   are given in \Tab~\ref{tab:fract}.
The fit describes the data well
with $\chi^2/\dof=\frchi/(\frnpt-\frnpa)$.
Similarly to the $\lambda$ fit, the value of $R=\frrrv \pm \frrre$ is
consistent within about one standard deviation with $R=0.5$. This agreement
with the $\lambda$ fit may be attributed to the structure function $F_2$
having a  power law-like $x$ dependence.
\begin{table}
\centerline{%
\begin{tabular}{c|c|c}
\hline\hline
  {\bfseries Parameter} & {\bfseries Value} & {\bfseries Uncertainty}\\
\hline
   $D_0$  (GeV$^{-2}$)  & $\frdzv$ & $\frdze$ \\
   $D_1$    & $\frdov$ & $\frdoe$ \\
   $D_3$    & $\frdev$ & $\frdee$ \\
   $Q^2_0$ (GeV$^2$)  & $\frqzv$ & $\frqze$ \\
   $R$      & $\frrrv$ & $\frrre$ \\
\hline\hline
\end{tabular}
}
\tablecaption{\label{tab:fract}Parameters of the  fractal fit 
and their total uncertainties. For the central fit $D_2$ is kept constant: $D_2=\frdtv$.
If the parameter $D_2$ is floated, the fit gives $D_2=1.061\pm 0.012$.}
\end{table}

\subsection{Dipole Model Fits\label{sec:saturation}}
\def\dpchi{183.1}
\def\dpnpt{149}
\def\dpnpa{  3}
\def\dpsat{ 0.29}
\def\dpszv{ 4.79}
\def\dpszvmb{ 24.5 } 
\def\dpsze{ 0.09}
\def\dpszemb{ 0.5 } 
\def\dpxlv{ 0.256}
\def\dpxle{ 0.003}
\def\dpxzv{ 0.60}
\def\dpxxe{ 0.03}
\def\dpxmv{ 0.00}
\def\dpxme{ 0.00}
\def\dprrv{ 0.00}
\def\dprre{ 0.00}
\def\drchi{170.9}
\def\drnpt{149}
\def\drnpa{  4}
\def\drsat{ 0.32}
\def\drszv{ 4.78}
\def\drsze{ 0.05}
\def\drxlv{ 0.261}
\def\drxle{ 0.003}
\def\drxzv{ 0.65}
\def\drxxe{ 0.01}
\def\drxmv{ 0.00}
\def\drxme{ 0.00}
\def\drrrv{ 0.54}
\def\drrre{ 0.15}
\def\dichi{178.2}
\def\dinpt{149}
\def\dinpa{  3}
\def\dpsat{ 0.22}
\def\diszv{ 0.605}
\def\disze{ 0.008}
\def\dixlv{ 0.260}
\def\dixle{ 0.003}
\def\dixzv{ 0.45}
\def\dixxe{ 0.03}
\def\dixmv{ 0.00}
\def\dixme{ 0.00}
\def\dirrv{ 0.00}
\def\dirre{ 0.00}
\def\dschi{175.9}
\def\dsnpt{149}
\def\dsnpa{  4}
\def\dpsat{ 0.18}
\def\dsszv{ 0.614}
\def\dssze{ 0.003}
\def\dsxlv{ 0.256}
\def\dsxle{ 0.002}
\def\dsxzv{ 0.37}
\def\dsxxe{ 0.01}
\def\dsxmv{ 0.00}
\def\dsxme{ 0.00}
\def\dsrrv{ 0.15}
\def\dsrre{ 0.14}
\label{sec:satur}
In the GBW  model~\cite{Golec-Biernat:1998js}  the dipole-proton
cross section $\hat{\sigma}$  (see \Sec~\ref{sec:models})
is given by
\begin{equation}
\hat{\sigma}(x,r) = \sigma_0
 \left\{1 - \exp \left[-r^2 / \left(4 r_0^2(x)\right) \right] \right\} ~,
\end{equation}
where $r$ corresponds to the transverse separation between the quark and 
the antiquark, and $r^2_0$ is an $x$ dependent scale parameter,
assumed to have the form
\begin{equation}
 r_0^2(x) \sim \left(x/x_0\right)^{\lambda} \label{eq:satrad}~.
\end{equation}
For small 
$r\ll r_0$,  $\hat{\sigma}$  is proportional to $r^2$ (colour transparency, 
$\hat{\sigma} \sim (r/2 r_0)^2$) while for $r\gg r_0$
the cross section approaches a constant value (saturation, 
$\hat{\sigma}\simeq \sigma_0$).
The boundary in the $(x,Q^2)$ plane which separates these regions
is described by the ``critical line'' at the $x$ dependent
saturation scale $Q^2_s(x) = 1/ r_0^2(x)$.
The model provides predictions for both $\sigma_T$ and $\sigma_L$
in terms of only three parameters, $\sigma_0$, $x_0$ and $\lambda$.

The fit to the reduced cross section with the dipole model of GBW (``GBW fit'')
yields a $\chi^2/\dof = \dpchi/(\dpnpt-\dpnpa)$, acceptable but worse
 than that for
the fractal model.
It has been suggested that improved models of $\hat{\sigma}$ 
lead to a  better description of the data and a variety of models has
been developed.
As an example, a fit using $\hat{\sigma}$
as proposed in the IIM model, with $N_0=0.7$ as defined in~\cite{Iancu:2003ge}, has been
performed. This fit
also has three free parameters
and gives  $\chi^2/\dof = \dichi/(\dinpt-\dinpa)$. The results of the two
dipole model fits are shown in \Figs~\ref{fig:xsec-ave}-\ref{fig:wplot} and 
\ref{fig:dsdy}-\ref{fig:flder}. The dipole model fit parameters are given in \Tabs~\ref{tab:dipol}
and \ref{tab:dipoliir}.
 
\begin{figure}
\centerline{%
\epsfig{file=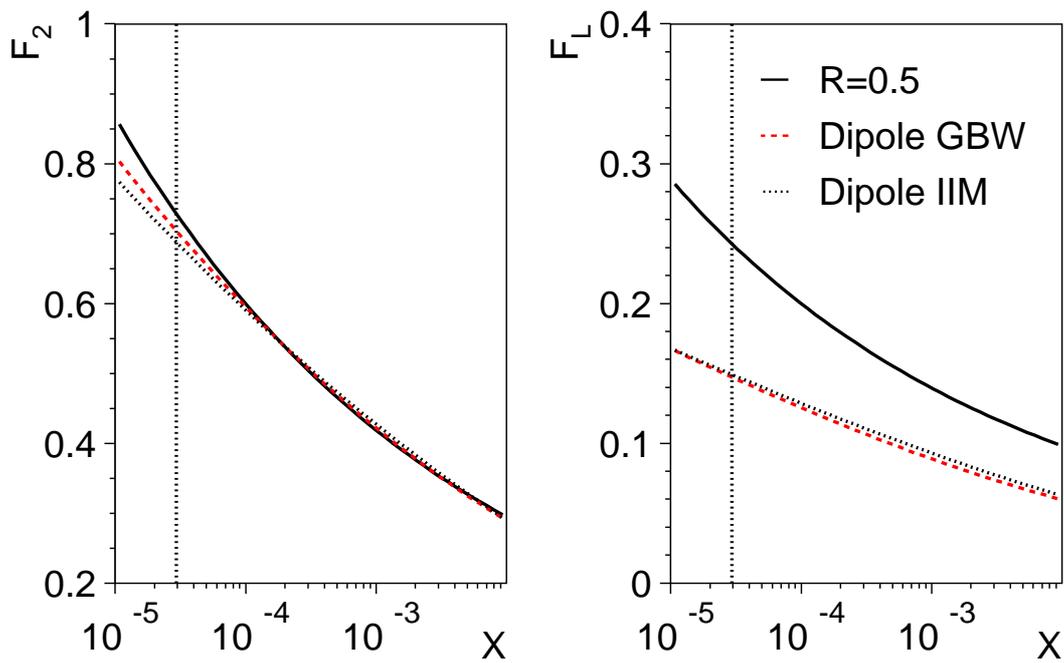,width=0.9\linewidth}
}
\caption{\label{fig:fractdipolf2fl}Comparison of the structure functions
$F_2$ (left) and $F_L$ (right) for $Q^2=1.2$~GeV$^2$ as a function of Bjorken 
$x$, for the fractal fit with $R=0.5$ (solid line), and the
predictions of the dipole models, GBW (dashed line) and
 IIM (dotted line), resulting from the fits to the H1 cross section data.
The vertical line indicates the value of $x=x_{s}$ for which the GBW dipole
model saturation radius is reached. }
\end{figure}

To trace the origin of the $\chi^2$ differences
between the models, predictions for
the structure functions $F_2$ and $F_L$ are compared individually.
As an example, \Fig~\ref{fig:fractdipolf2fl} shows the comparison between
the three models for the bin $Q^2=1.2$~GeV$^2$. The structure functions $F_2$ agree rather
well for the models considered for $x>x_{\rm s}=\dpsat \times 10^{-4}$,
where $x_{\rm s}$ corresponds to the saturation radius of the GBW dipole
model at the chosen $Q^2$ value.
 However, for $x<x_{\rm s}$ the dipole models show a softer $F_2$ dependence on $x$.
This holds in particular for the IIM dipole model. 
The main difference between the models is in the structure function $F_L$.
As shown in \Fig~\ref{fig:fractdipolf2fl},
the predictions of the dipole models are nearly half of the
result for $F_L$ obtained with the fractal model analysis.

\begin{table}[tb]
\centerline{%
\begin{tabular}{c|c|c}
\hline\hline
  {\bfseries Parameter} & {\bfseries Value} & {\bfseries Uncertainty} \\
\hline
   $\sigma_0$ (mb)  & \dpszvmb & \dpszemb \\
   $\lambda$    & \dpxlv & \dpxle \\
   $x_0$        & \dpxzv$\times 10^{-4}$ & \dpxxe$\times 10^{-4}$  \\
\hline\hline
\end{tabular}
}
\tablecaption{\label{tab:dipol}Parameters of the GBW dipole fit and their total uncertainties.}
\end{table}

\begin{table}[tb]
\centerline{%
\begin{tabular}{c|c|c}
\hline\hline
  {\bfseries Parameter} & {\bfseries Value} & {\bfseries Uncertainty} \\
\hline
   $R_{\rm IIM}$ (fm)   & \diszv & \disze \\
   $\lambda$    & \dixlv & \dixle \\
   $x_0$        & \dixzv $\times 10^{-4}$ & \dixxe$\times 10^{-4}$  \\
\hline\hline
\end{tabular}
}
\tablecaption{\label{tab:dipoliir}Parameters of the IIM  dipole fit
with $N_0 = 0.7$ and their total uncertainties.}
\end{table}

The  strict correlation  between $F_L$ and
$F_2$ predicted by the dipole models 
 could be broken  by higher twist
 effects~\cite{Bartels-2000-17}.
To  quantify the influence of the structure function $F_L$ 
another fit to the reduced cross section data
is performed, in which the $F_L$ prediction
of the  dipole model is scaled with an additional free parameter
$B_L$
\begin{equation}
  F_L(x,Q^2)  =  F^{\rm dipole}_L(x,Q^2)\,(1 + B_L) ~. \label{eq:dipolmod}
\end{equation}
With $B_L$ as a formal free parameter
the GBW fit returns $B_L$ deviating from $0$ by more than $3\,$ standard
deviations, $B_L=\drrrv \pm \drrre$.
The fit for the IIM model does not
yield a significant change for the $F_L$ prediction: $B_L=\dsrrv \pm \dsrre$.

To summarise, a steeper rise to smaller $x$  of the structure function $F_2$, 
together with a larger $R$ value as obtained in the 
fractal model fit with constant $R$, gives
the best description of the H1 data. However, a sufficiently 
softer rise of $F_2$ together with a smaller $F_L$, as predicted by the IIM model,
also describes the data well. For the GBW model, the rise of $F_2$ is 
rather steep such that the fit to the
data prefers a larger $F_L$, which is inconsistent
 with the prediction of the model.

\section{Summary\label{sec:summary}}
A new measurement is performed of the inclusive 
double differential cross section for neutral
current deep inelastic positron-proton scattering,
$e^+p \rightarrow e^+X$, in the region of small Bjorken\,$x$
and low absolute momentum transfers squared, $Q^2$.
The data were obtained with the H1 detector at
the $ep$ collider HERA  in two dedicated periods
of data taking at beam energies $E_e=27.5$\,GeV and
$E_p=920$\,GeV.  In the year $1999$, events were collected
with a dedicated trigger on low $Q^2$
DIS events at the  nominal interaction vertex position, 
corresponding to an integrated luminosity
of $2.1$\,pb$^{-1}$. In the year $2000$,  the interaction
vertex was shifted forward by $+ 70$\,cm 
in proton beam direction 
to access even smaller values of $Q^2$, and
data with an integrated luminosity of $0.505$\,pb$^{-1}$ was taken.

The measurement is performed in a wide range of 
inelasticity $y$, from $0.0015$ to $0.8$, and of Bjorken\,$x$,
from $5 \cdot 10^{-6}$ to $0.02$.
The data cover a $Q^2$ range from $0.2$ to $12$\,GeV$^2$,
with an overlap region of the nominal and the
shifted vertex data   
of $0.5 \leq Q^2 \leq 3.5$\,GeV$^2$, in 
which both measurements agree.
At low $Q^2$ the data analysed here
comprise the full statistics collected with the H1 experiment at $920$
 GeV.

The measurement obtained with the
$1999$ and the $2000$ data 
is combined with data collected
in the years $1995$ and $1997$, 
which were taken 
at $820$\,GeV proton beam energy in similar experimental conditions
and published previously.
This combination takes the correlation of
systematic uncertainties into account and
provides a new, single data set from
the H1 experiment, which supersedes all H1
data previously released in that kinematic region.  
The total uncertainty of the final reduced 
cross section measurement
is about $2$\% for a large part of the phase space.

The neutral current $ep$ cross section
at low $Q^2$  is governed by  two independent
proton structure functions, $F_2$ and $F_L$.
For $y < 0.6$, the influence of the longitudinal structure function
$F_L$ is small, and the data in this range are also presented as
a measurement of the proton structure function $F_2(x,Q^2)$.
For $y=0.735$,
using a method based on the derivative of the
cross section with respect to $\ln y$, 
the structure function $F_L(x,Q^2)$ is extracted
with minimum assumptions on the behaviour of
$F_2$.

In each $Q^2$ bin a simple parameterisation of the reduced cross section
in terms of a power law of $F_2(x,Q^2) \propto x^{-\lambda}$
and  $R=F_L/(F_2-F_L)$
describes the data  well.
The power $\lambda$ increases approximately logarithmically
with $Q^2$ at $Q^2 \gtrsim 2$\,GeV$^2$. The parameterisation is consistent
with a constant value of $R(x,Q^2) \simeq 0.5$, which implies
that  $F_L(x,Q^2) \simeq F_2(x,Q^2) /3$ 
under the assumption of a power law rise 
of $F_2$ towards low $x$. 

The transition region of DIS to photoproduction, $Q^2 \simeq 1$\,GeV$^2$,
cannot be analysed within perturbative QCD.
The data therefore are studied here within  phenomenological models. 
The structure function $F_2(x,Q^2)$ is analysed using 
a self similarity based ansatz within a fractal
model. The fractal $F_2$ parameterisation, combined
with a constant $R$, provides a  good description of
the measured cross section in the full range of phase space
covered by the data.

The Colour Dipole
Model predicts both structure functions $F_2$ and $F_L$
using a single characteristic dipole scattering cross section.
Two versions of the CDM, the
GBW model and the IIM model, are used in this analysis
and are found to generally describe the cross section
data well. The description of the data in the GBW model is 
observed to improve when the contribution of $F_L$ within 
this model is formally allowed to be enhanced.
The IIM model prediction on $F_L$ is similar to the 
GBW model. Owing to a softer rise of $F_2$  towards small $x$,
the IIM ansatz yet is able to describe the cross section
data better and no modification on the predicted $F_L$ is suggested
by the data.

For the region  $Q^2 \simeq 1$\,GeV$^2$,
in which the transition from photoproduction to
DIS takes place, the data as presented in this paper are
the most precise  result of the H1 Collaboration.

\section*{Acknowledgements}
\refstepcounter{pdfadd} \pdfbookmark[0]{Acknowledgements}{s:acknowledge}
%
We are grateful to the HERA machine group whose outstanding
efforts have made this experiment possible.
We thank the engineers and technicians for their work in constructing
and maintaining the H1 detector, our funding agencies for
financial support, the DESY technical staff for continual assistance
and the DESY directorate for support and for the hospitality
which they extend to the non DESY members of the collaboration.

\clearpage
\appendix
\section{Averaging Procedure \label{avemath}}
The $\chi^2$ function of  \Eq~\ref{eq:avet}
is to be minimised with respect to the sets $m^i$ and $b_j$.
This determines the averaged measurements
and uncertainties, $\mu^{i,{\rm ave}}$, $\Delta_{i,{\rm ave}}$, $\alpha_{j,{\rm ave}}$
and the matrix $A'_S$, used in
 \Eq~\ref{eq:offdiag}.

The minimum $\chi^2_{\rm min}$ in  \Eq~\ref{eq:avet} is found by solving a system of linear equations
obtained by requiring $\partial \chi^2/\partial m^i = 0$ 
and $\partial \chi^2/\partial b_j = 0$ which can be written in matrix form
\begin{equation}
  \left( \begin{array}{cc} A_M & A_{SM} \\ (A_{SM})^{\rm T} & A_S \end{array}  \right) \cdot \left( \begin{array}{c} M^{\rm ave}  \\ B^{\rm ave} \end{array} \right) 
 = \left( \begin{array}{c} C_M \\ C_S\end{array} \right). 
\end{equation}
Here the vector $M^{\rm ave}$ corresponds to all
measurements  and the  vector $B^{\rm ave}$  
corresponds to all  systematic error sources.
The matrix $A_M$ has a diagonal structure with $N_M$  diagonal elements
\begin{equation}
 A_M^{ii} = \sum_e \frac{\DeltaM}{\Delta^2_{i,e}} .
\end{equation} 
The other matrices have the  following elements
\begin{equation}
\begin{array}{rcl}
\textstyle A_{SM}^{ij} &=& -\sum_e \frac{\textstyle \Gamma^i_{j,e}}{\textstyle \Delta^2_{i,e}}~\DeltaM;\\
\textstyle A_{S}^{ij}  &=& \delta_{ij} + \sum_e \sum^{N_{M}}_k \frac{\textstyle \Gamma^k_{i,e} \Gamma^k_{j,e}}{\textstyle \Delta^2_{k,e}} \DeltaMK;\\
\textstyle C_M^i &=& \sum_e \frac{\textstyle \mu^i_e}{\textstyle \Delta^2_{i,e}}\DeltaM;\\
\textstyle C_S^{j} &=& -\sum_e \sum^{N_M}_k \frac{\textstyle \mu^k_e \Gamma^k_{j,e}}{\textstyle \Delta^2_{k,e}}\DeltaMK.
\end{array}
\end{equation}
Here $\delta_{ij}$ is the standard Kronecker symbol. Note that the matrix $A_{SM}$ has the dimension
$N_M \times N_S$ while the matrix $A_S$ is quadratic with $N_S \times N_S$ elements. 

\noindent
Using the method of the Schur complement, the solution is found as
\begin{equation}
\begin{array}{rcl}
\textstyle     A'_S &=& A_S - (A_{SM})^{\rm T}A_{M}^{-1}A_{SM} \\ 
\textstyle           B^{\rm ave}    &=& (A'_S)^{-1} \left( C_S - (A_{SM})^{\rm T}A_{M}^{-1} C_M \right)\\
\textstyle        M^{\rm ave} &=& A_M^{-1}\left[C_M - A_{SM} B^{\rm ave}  \right].
\end{array}
\end{equation}
Given the components of the vector $B^{\rm ave}$, 
$\beta_{j,{\rm ave}}=\alpha_{j,{\rm ave}}/\Delta_{\alpha_j}$,
the solution for $\mu^{i,{\rm ave}}$ can be written in explicit form
\begin{equation}\label{eq:avemean}
\mu^{i,{\rm ave}} = \frac{\textstyle \sum_e \left[ \left( \mu^i_e + \sum_j \Gamma^i_{j,e} \beta_{j,{\rm ave}}\right) \frac{\textstyle \DeltaM}{\textstyle \Delta^2_{i,e}}\right]}
{\sum_e \frac{\textstyle  \DeltaM }{\textstyle \Delta^2_{i,e}} }. 
\end{equation}
The uncorrelated uncertainty squared is determined by  the inverse of the elements of the diagonal matrix $A_M$
\begin{equation}\label{eq:avesig}
\Delta_{i,\rm ave}^2 = \frac{\textstyle 1}{\sum_e \frac{\textstyle\DeltaM }{\textstyle \Delta^2_{i,e}}}. 
\end{equation}
Similarly, the contributions from statistical and systematical uncertainties can be calculated
 \begin{equation}\label{eq:avedelta}
 \Delta_{i,\rm ave,\rm stat}^2 = \Delta_{i,\rm ave}^4 \ {\sum_e \frac{\textstyle\DeltaM}
 {\textstyle \Delta^4_{i,e}}} \Delta_{i,e,\rm stat}^2 \ , \quad
 \Delta_{i,\rm ave,\rm unc}^2 = \Delta_{i,\rm ave}^4 \ {\sum_e \frac{\textstyle
\DeltaM}
{\textstyle \Delta^4_{i,e}}} \Delta_{i,e,\rm unc}^2. 
 \end{equation}
\EEq~\ref{eq:avemean} and \Eq~\ref{eq:avesig} reproduce the standard formula 
for a statistically weighted average of several
uncorrelated measurements when all 
shifts of the systematic error sources are set to zero.

The non-diagonal nature of the matrix $A'_S$ expresses the fact
that 
the original sources of the systematic uncertainties 
are correlated with each other after averaging.
The matrix $A'_S$ can be decomposed to re-express \Eq~\ref{eq:avebeta} in terms
of diagonalised systematic error sources
\begin{equation}
  D D = U A'_S  U^{-1}~~~~~~  
\Gamma_{ave} = A_{SM} A^{-1}_M D^{-1} U^{-1}.
\end{equation}
Here $U$ is an orthogonal matrix composed of the eigenvectors of $A'_S$,  $D$\, is a 
diagonal matrix with corresponding square roots of eigenvalues as diagonal elements
and $\Gamma_{ave}$ represents the sensitivity of the
average result to these new sources. Its elements are the 
$\Gamma^{i,{\rm ave}}_{j}$ in 
\Eq~\ref{eq:final}.

\clearpage
\begin{flushleft}
\bibliography{desy08-171.bib}
\end{flushleft}
\clearpage
\begin{table}
\begin{scriptsize}
\begin{center}
\begin{tabular}{ccccrrrrrrrrr}
\hline\hline
$Q^2$ & $x$ & $y$ & $\sigma_{r}$ & 
$\delta_{\rm tot}$ &
$\delta_{\rm stat}$ & 
$\delta_{\rm uncor}$ & 
$\gamma_{E_e^{\prime}}$ & 
$\gamma_{\theta_e}$ & 
$\gamma_{E_{\rm had}}$ & 
$\gamma_{\rm noise}$ 
& $\gamma_{E^h_{\rm SpaCal}}$ 
& $\gamma_{\gamma p}$ \\    
 GeV$^2$ &  &  &   & \% &\% &\% &\% &\% &\% &\% &\% &\% \\
\hline
0.20  &$ 3.980 \times 10^{-5}$ & $ 4.948 \times 10^{-2}$ &  0.249   & 20.3 &  13.8 &  12.0 &  0.58 & -1.74 &  5.70 & -0.34 & -1.37 &  -6.44 \\ 
0.20  &$ 2.510 \times 10^{-4}$ & $ 7.845 \times 10^{-3}$ &  0.162   & 16.7 &  14.2 &  6.19 &  1.38 & -0.78 & -1.65 & -3.64 & -4.21 &  -1.68 \\ 
0.25  &$ 3.980 \times 10^{-5}$ & $ 6.184 \times 10^{-2}$ &  0.302   & 17.5 &  9.80 &  11.3 &  0.49 & -2.22 &  3.10 & -1.48 & -2.62 &  -7.69 \\ 
0.25  &$ 2.510 \times 10^{-4}$ & $ 9.806 \times 10^{-3}$ &  0.163   & 14.1 &  10.8 &  4.71 & -1.93 &  0.70 &  0.01 & -4.47 & -5.71 &  -1.42 \\ 
0.25  &$ 1.580 \times 10^{-3}$ & $ 1.558 \times 10^{-3}$ &  0.182   & 13.2 &  11.5 &  5.29 &  0.57 &  0.46 & -1.73 & -1.93 & -2.61 &  -0.30 \\ 
0.35  &$ 5.120 \times 10^{-6}$ & $ 6.726 \times 10^{-1}$ &  0.458   & 25.2 &  21.6 &  12.8 & -0.61 & -0.51 &  0.34 & -0.03 &  0.59 &  -2.45 \\ 
0.35  &$ 3.200 \times 10^{-5}$ & $ 1.077 \times 10^{-1}$ &  0.361   & 22.2 &  9.72 &  11.1 & -2.17 & -0.08 & -1.61 &  0.54 & -6.88 &  -14.78 \\ 
0.35  &$ 1.300 \times 10^{-4}$ & $ 2.651 \times 10^{-2}$ &  0.265   & 11.6 &  9.61 &  4.38 & -0.38 &  0.27 &  2.55 & -2.99 & -0.51 &  -2.46 \\ 
0.35  &$ 5.000 \times 10^{-4}$ & $ 6.892 \times 10^{-3}$ &  0.216   & 11.1 &  9.22 &  4.19 & -0.91 & -0.81 & -0.47 & -3.51 & -2.48 &  -0.53 \\ 
0.35  &$ 2.510 \times 10^{-3}$ & $ 1.373 \times 10^{-3}$ &  0.193   & 11.6 &  10.2 &  4.55 & -1.19 & -0.25 & -0.04 & -2.49 & -1.47 &  -0.08 \\ 
0.50  &$ 7.320 \times 10^{-6}$ & $ 6.726 \times 10^{-1}$ &  0.483   & 10.0 &  5.23 &  5.74 &  0.18 &  1.96 &  2.31 & -0.18 &  2.75 &  -4.84 \\ 
0.50  &$ 1.580 \times 10^{-5}$ & $ 3.116 \times 10^{-1}$ &  0.477   & 21.6 &  18.6 &  9.84 & -3.86 & -2.83 &  0.27 &  0.03 &  0.47 &  -0.19 \\ 
0.50  &$ 3.980 \times 10^{-5}$ & $ 1.237 \times 10^{-1}$ &  0.431   & 17.7 &  10.7 &  6.07 & -2.11 &  0.82 & -1.48 & -0.10 & -4.88 &  -11.49 \\ 
0.50  &$ 1.000 \times 10^{-4}$ & $ 4.923 \times 10^{-2}$ &  0.388   & 11.0 &  9.10 &  4.87 & -0.30 &  0.57 & -0.02 &  0.52 & -3.39 &  -1.75 \\ 
0.50  &$ 2.510 \times 10^{-4}$ & $ 1.961 \times 10^{-2}$ &  0.262   & 12.8 &  10.6 &  4.45 &  0.01 & -1.43 &  0.66 & -4.32 & -3.28 &  -1.27 \\ 
0.50  &$ 8.000 \times 10^{-4}$ & $ 6.154 \times 10^{-3}$ &  0.275   & 9.51 &  7.92 &  3.86 & -0.40 & -0.43 &  0.04 & -3.41 & -0.90 &  -0.22 \\ 
0.65  &$ 9.520 \times 10^{-6}$ & $ 6.726 \times 10^{-1}$ &  0.502   & 6.22 &  3.87 &  2.90 & -1.15 &  0.68 &  1.11 & -0.18 &  1.85 &  -2.98 \\ 
0.65  &$ 1.580 \times 10^{-5}$ & $ 4.050 \times 10^{-1}$ &  0.474   & 6.68 &  3.06 &  5.44 & -0.63 & -2.05 &  0.24 & -0.09 &  0.34 &  -0.92 \\ 
0.65  &$ 3.980 \times 10^{-5}$ & $ 1.608 \times 10^{-1}$ &  0.681   & 21.7 &  17.4 &  11.2 & -6.22 &  2.19 & -0.04 &  0.04 & -0.15 &   0.00 \\ 
0.65  &$ 1.000 \times 10^{-4}$ & $ 6.400 \times 10^{-2}$ &  0.424   & 13.2 &  5.59 &  5.85 & -1.89 & -2.22 & -1.17 &  0.52 & -9.22 &  -3.52 \\ 
0.65  &$ 2.510 \times 10^{-4}$ & $ 2.550 \times 10^{-2}$ &  0.353   & 10.6 &  8.94 &  4.04 & -0.78 & -1.00 & -0.48 &  0.49 & -3.39 &  -1.53 \\ 
0.65  &$ 8.000 \times 10^{-4}$ & $ 8.000 \times 10^{-3}$ &  0.283   & 10.5 &  7.57 &  3.61 & -1.74 &  0.53 &  1.06 & -5.94 & -0.75 &  -0.15 \\ 
0.65  &$ 3.200 \times 10^{-3}$ & $ 2.000 \times 10^{-3}$ &  0.246   & 10.1 &  8.92 &  4.05 & -1.83 &  0.36 & -0.68 &  1.63 & -0.36 &  -0.09 \\ 
0.85  &$ 1.244 \times 10^{-5}$ & $ 6.726 \times 10^{-1}$ &  0.594   & 5.01 &  2.48 &  2.52 & -1.16 & -0.22 &  1.23 & -0.17 &  1.77 &  -2.55 \\ 
0.85  &$ 2.000 \times 10^{-5}$ & $ 4.184 \times 10^{-1}$ &  0.623   & 6.24 &  1.94 &  5.36 & -0.98 & -2.27 &  0.16 & -0.03 &  0.34 &  -0.45 \\ 
0.85  &$ 3.980 \times 10^{-5}$ & $ 2.103 \times 10^{-1}$ &  0.564   & 6.24 &  2.03 &  5.37 & -0.98 & -2.23 &  0.01 & -0.08 &  0.00 &  -0.05 \\ 
0.85  &$ 1.000 \times 10^{-4}$ & $ 8.369 \times 10^{-2}$ &  0.493   & 7.75 &  4.98 &  5.78 & -0.38 & -0.81 &  0.16 & -0.77 &  0.63 &   0.00 \\ 
0.85  &$ 2.510 \times 10^{-4}$ & $ 3.334 \times 10^{-2}$ &  0.353   & 11.3 &  8.06 &  3.75 &  0.57 & -1.86 & -1.56 &  0.17 & -6.37 &  -1.05 \\ 
0.85  &$ 8.000 \times 10^{-4}$ & $ 1.046 \times 10^{-2}$ &  0.325   & 8.86 &  6.77 &  3.41 & -0.19 & -0.32 &  1.16 & -4.25 & -1.20 &  -0.26 \\ 
0.85  &$ 3.200 \times 10^{-3}$ & $ 2.615 \times 10^{-3}$ &  0.318   & 8.65 &  7.27 &  3.78 &  0.55 & -1.91 &  1.43 & -1.05 & -0.73 &  -0.04 \\ 
1.20  &$ 1.757 \times 10^{-5}$ & $ 6.726 \times 10^{-1}$ &  0.652   & 5.82 &  2.66 &  2.51 & -1.08 & -0.35 &  1.33 & -0.26 &  2.16 &  -3.57 \\ 
1.20  &$ 2.000 \times 10^{-5}$ & $ 5.907 \times 10^{-1}$ &  0.686   & 3.95 &  2.59 &  2.51 & -0.73 & -0.46 &  0.40 & -0.04 &  0.91 &  -0.93 \\ 
1.20  &$ 3.200 \times 10^{-5}$ & $ 3.692 \times 10^{-1}$ &  0.697   & 3.78 &  1.66 &  2.73 & -0.81 & -1.73 &  0.22 & -0.10 &  0.09 &  -0.59 \\ 
1.20  &$ 6.310 \times 10^{-5}$ & $ 1.872 \times 10^{-1}$ &  0.653   & 4.12 &  1.37 &  2.71 & -1.17 & -2.50 &  0.07 & -0.22 &  0.22 &  -0.02 \\ 
1.20  &$ 1.580 \times 10^{-4}$ & $ 7.478 \times 10^{-2}$ &  0.498   & 4.40 &  2.06 &  2.78 &  0.69 & -2.07 &  0.70 & -1.43 &  0.35 &   0.00 \\ 
1.20  &$ 3.980 \times 10^{-4}$ & $ 2.969 \times 10^{-2}$ &  0.471   & 7.52 &  5.21 &  3.16 & -2.04 &  0.11 & -1.35 & -0.02 & -3.65 &  -0.10 \\ 
1.20  &$ 1.300 \times 10^{-3}$ & $ 9.088 \times 10^{-3}$ &  0.378   & 6.85 &  5.08 &  3.09 & -2.10 & -0.39 &  1.38 & -2.00 & -1.05 &  -0.03 \\ 
1.20  &$ 5.000 \times 10^{-3}$ & $ 2.363 \times 10^{-3}$ &  0.322   & 8.23 &  6.55 &  3.54 & -1.54 & -1.25 &  2.10 &  1.84 & -0.71 &   0.00 \\ 
\hline\hline
\end{tabular}
\end{center}
\end{scriptsize}
\begin{small}
  \tablecaption{Reduced cross section $\sigma_{r}$, as measured with the
    {\SVX} data sample for  $0.2 \le Q^2\le 1.2$~GeV$^2$. The uncertainties are
    quoted in \% relative to $\sigma_r$.  $\delta_{\rm tot}$ is the total uncertainty
    determined as the quadratic sum of systematic and statistical
    uncertainties. $\delta_{\rm stat}$ is the statistical uncertainty.
    $\delta_{\rm uncor}$ represents the uncorrelated systematic
    uncertainty.  $\gamma_{E_e^{\prime}}$, $\gamma_{\theta_e}$,
    $\gamma_{E_{\rm had}}$, $\gamma_{\rm noise}$, $\gamma_{E^h_{\rm
        SpaCal}}$ and $\gamma_{\gamma p}$ are the bin-to-bin
    correlated systematic uncertainties in the cross section
    measurement due to uncertainties in the SpaCal electromagnetic
    energy scale, electron scattering angle, LAr calorimeter hadronic
    energy scale, LAr calorimeter noise, SpaCal hadronic energy scale
    and the photoproduction background, respectively. The luminosity uncertainty of
    $3$\% for the SVX data is not included in $\delta_{\rm tot}$. \label{tab:svxtable}}
\end{small} 
\end{table}

\begin{table}
\begin{scriptsize}
\begin{center}
\begin{tabular}{ccccrrrrrrrrr}
\hline\hline
$Q^2$ & $x$ & $y$ & $\sigma_{r}$ & 
$\delta_{\rm tot}$ &
$\delta_{\rm stat}$ & 
$\delta_{\rm uncor}$ & 
$\gamma_{E_e^{\prime}}$ & 
$\gamma_{\theta_e}$ & 
$\gamma_{E_{\rm had}}$ & 
$\gamma_{\rm noise}$ 
& $\gamma_{E^h_{\rm SpaCal}}$ 
& $\gamma_{\gamma p}$ \\    
 GeV$^2$ &  &  &   & \% &\% &\% &\% &\% &\% &\% &\% &\% \\
\hline
1.50  &$ 2.196 \times 10^{-5}$ & $ 6.726 \times 10^{-1}$ &  0.722   & 4.43 &  2.45 &  2.47 & -1.08 & -0.58 &  0.81 & -0.20 &  1.37 &  -1.86 \\ 
1.50  &$ 3.200 \times 10^{-5}$ & $ 4.615 \times 10^{-1}$ &  0.774   & 3.28 &  1.78 &  2.36 & -0.63 & -0.76 &  0.39 & -0.10 &  0.65 &  -0.68 \\ 
1.50  &$ 5.000 \times 10^{-5}$ & $ 2.954 \times 10^{-1}$ &  0.773   & 3.80 &  1.46 &  2.71 & -0.93 & -2.02 &  0.01 & -0.09 & -0.04 &  -0.13 \\ 
1.50  &$ 8.000 \times 10^{-5}$ & $ 1.846 \times 10^{-1}$ &  0.727   & 3.92 &  1.57 &  2.73 & -0.84 & -2.17 &  0.04 & -0.18 &  0.16 &  -0.02 \\ 
1.50  &$ 1.300 \times 10^{-4}$ & $ 1.136 \times 10^{-1}$ &  0.654   & 4.31 &  1.77 &  2.75 & -0.94 & -2.57 &  0.18 & -0.54 &  0.28 &  -0.01 \\ 
1.50  &$ 2.000 \times 10^{-4}$ & $ 7.384 \times 10^{-2}$ &  0.628   & 5.58 &  3.57 &  2.77 & -3.21 & -0.43 &  0.45 & -0.15 &  0.00 &   0.00 \\ 
1.50  &$ 3.200 \times 10^{-4}$ & $ 4.615 \times 10^{-2}$ &  0.564   & 4.78 &  1.90 &  2.76 & -0.45 & -2.40 & -1.13 &  0.11 & -2.08 &  -0.14 \\ 
1.50  &$ 8.000 \times 10^{-4}$ & $ 1.846 \times 10^{-2}$ &  0.483   & 4.32 &  2.38 &  2.47 & -1.61 & -0.99 &  0.33 & -1.23 & -1.30 &  -0.02 \\ 
1.50  &$ 3.200 \times 10^{-3}$ & $ 4.615 \times 10^{-3}$ &  0.424   & 5.02 &  2.69 &  2.56 & -1.41 & -0.56 &  2.17 & -2.06 & -0.31 &   0.00 \\ 
1.50  &$ 1.300 \times 10^{-2}$ & $ 1.136 \times 10^{-3}$ &  0.384   & 14.1 &  4.49 &  3.05 & -1.71 & -0.37 &  1.16 &  12.8 & -0.33 &   0.00 \\ 
2.00  &$ 2.928 \times 10^{-5}$ & $ 6.726 \times 10^{-1}$ &  0.822   & 4.28 &  2.19 &  2.39 & -0.61 & -0.93 &  1.07 & -0.18 &  1.51 &  -1.75 \\ 
2.00  &$ 5.000 \times 10^{-5}$ & $ 3.938 \times 10^{-1}$ &  0.837   & 3.10 &  1.62 &  2.33 & -0.84 & -0.76 &  0.27 & -0.22 &  0.20 &  -0.30 \\ 
2.00  &$ 8.000 \times 10^{-5}$ & $ 2.461 \times 10^{-1}$ &  0.791   & 3.03 &  1.63 &  2.34 & -0.94 & -0.32 &  0.29 & -0.15 &  0.00 &  -0.05 \\ 
2.00  &$ 1.300 \times 10^{-4}$ & $ 1.515 \times 10^{-1}$ &  0.731   & 3.28 &  1.81 &  2.36 & -1.33 & -0.07 &  0.37 & -0.12 &  0.00 &  -0.01 \\ 
2.00  &$ 2.000 \times 10^{-4}$ & $ 9.846 \times 10^{-2}$ &  0.700   & 3.58 &  1.97 &  2.39 & -1.73 & -0.28 &  0.46 & -0.11 &  0.00 &  -0.01 \\ 
2.00  &$ 3.200 \times 10^{-4}$ & $ 6.154 \times 10^{-2}$ &  0.578   & 4.39 &  2.14 &  2.40 & -0.73 & -0.65 & -0.78 & -0.16 & -2.71 &  -0.03 \\ 
2.00  &$ 5.000 \times 10^{-4}$ & $ 3.938 \times 10^{-2}$ &  0.528   & 3.95 &  2.41 &  2.43 & -1.46 & -0.07 & -0.61 & -0.28 & -1.16 &  -0.01 \\ 
2.00  &$ 1.000 \times 10^{-3}$ & $ 1.969 \times 10^{-2}$ &  0.490   & 3.79 &  1.86 &  2.36 & -1.38 & -0.11 &  0.67 & -1.62 & -0.61 &  -0.01 \\ 
2.00  &$ 3.200 \times 10^{-3}$ & $ 6.154 \times 10^{-3}$ &  0.424   & 4.65 &  1.63 &  2.34 & -1.35 & -0.07 &  2.41 & -2.41 & -0.25 &   0.00 \\ 
2.00  &$ 1.300 \times 10^{-2}$ & $ 1.515 \times 10^{-3}$ &  0.404   & 10.5 &  2.46 &  2.48 & -1.12 & -0.52 &  0.95 &  9.81 & -0.25 &   0.00 \\ 
2.50  &$ 5.000 \times 10^{-5}$ & $ 4.923 \times 10^{-1}$ &  0.881   & 3.68 &  2.28 &  2.40 & -0.96 & -0.47 &  0.65 & -0.18 &  0.68 &  -0.75 \\ 
2.50  &$ 8.000 \times 10^{-5}$ & $ 3.077 \times 10^{-1}$ &  0.869   & 3.08 &  1.66 &  2.34 & -0.80 & -0.74 &  0.30 &  0.03 &  0.04 &  -0.14 \\ 
2.50  &$ 1.300 \times 10^{-4}$ & $ 1.893 \times 10^{-1}$ &  0.800   & 3.04 &  1.63 &  2.34 & -0.85 & -0.53 &  0.33 & -0.03 &  0.00 &  -0.01 \\ 
2.50  &$ 2.000 \times 10^{-4}$ & $ 1.231 \times 10^{-1}$ &  0.777   & 3.25 &  1.63 &  2.34 & -1.44 & -0.47 &  0.36 & -0.16 &  0.00 &  -0.01 \\ 
2.50  &$ 3.200 \times 10^{-4}$ & $ 7.692 \times 10^{-2}$ &  0.683   & 4.03 &  1.71 &  2.35 & -2.69 & -0.52 &  0.48 & -0.15 &  0.00 &  -0.01 \\ 
2.50  &$ 5.000 \times 10^{-4}$ & $ 4.923 \times 10^{-2}$ &  0.601   & 3.45 &  1.90 &  2.36 &  0.15 & -0.85 & -0.10 & -0.81 & -1.14 &  -0.01 \\ 
2.50  &$ 8.000 \times 10^{-4}$ & $ 3.077 \times 10^{-2}$ &  0.574   & 3.31 &  1.96 &  2.38 & -0.23 & -0.41 &  0.35 & -0.95 & -0.49 &   0.00 \\ 
2.50  &$ 1.580 \times 10^{-3}$ & $ 1.558 \times 10^{-2}$ &  0.527   & 3.99 &  1.44 &  2.32 & -0.20 & -0.42 &  1.17 & -2.60 & -0.27 &   0.00 \\ 
2.50  &$ 5.000 \times 10^{-3}$ & $ 4.923 \times 10^{-3}$ &  0.448   & 4.10 &  1.29 &  2.31 & -0.19 & -0.58 &  2.66 & -1.53 & -0.21 &   0.00 \\ 
2.50  &$ 2.000 \times 10^{-2}$ & $ 1.231 \times 10^{-3}$ &  0.409   & 16.8 &  2.30 &  2.44 & -0.14 & -0.63 &  0.73 &  16.4 & -0.18 &   0.00 \\ 
3.50  &$ 8.000 \times 10^{-5}$ & $ 4.307 \times 10^{-1}$ &  0.971   & 3.75 &  2.35 &  2.42 & -1.09 & -1.15 &  0.32 & -0.14 &  0.22 &  -0.25 \\ 
3.50  &$ 1.300 \times 10^{-4}$ & $ 2.651 \times 10^{-1}$ &  0.925   & 3.21 &  1.81 &  2.36 & -0.50 & -1.04 &  0.34 & -0.05 &  0.00 &  -0.05 \\ 
3.50  &$ 2.000 \times 10^{-4}$ & $ 1.723 \times 10^{-1}$ &  0.852   & 3.20 &  1.78 &  2.35 & -1.02 & -0.64 &  0.35 & -0.08 &  0.00 &  -0.02 \\ 
3.50  &$ 3.200 \times 10^{-4}$ & $ 1.077 \times 10^{-1}$ &  0.779   & 3.44 &  1.80 &  2.36 & -1.53 & -0.71 &  0.40 & -0.14 &  0.00 &  -0.01 \\ 
3.50  &$ 5.000 \times 10^{-4}$ & $ 6.892 \times 10^{-2}$ &  0.716   & 3.49 &  1.96 &  2.38 &  0.39 & -0.73 &  0.02 & -0.88 & -1.11 &  -0.01 \\ 
3.50  &$ 8.000 \times 10^{-4}$ & $ 4.307 \times 10^{-2}$ &  0.651   & 3.59 &  2.02 &  2.38 &  0.36 & -0.66 &  0.45 & -1.45 & -0.56 &   0.00 \\ 
3.50  &$ 1.300 \times 10^{-3}$ & $ 2.651 \times 10^{-2}$ &  0.588   & 3.65 &  2.09 &  2.39 &  0.36 & -0.88 &  0.37 & -1.46 & -0.30 &   0.00 \\ 
3.50  &$ 2.510 \times 10^{-3}$ & $ 1.373 \times 10^{-2}$ &  0.566   & 4.57 &  1.48 &  2.33 &  0.23 & -0.63 &  1.87 & -3.04 & -0.22 &   0.00 \\ 
3.50  &$ 8.000 \times 10^{-3}$ & $ 4.307 \times 10^{-3}$ &  0.481   & 3.76 &  1.38 &  2.32 &  0.30 & -0.73 &  2.48 &  0.20 & -0.17 &   0.00 \\ 
\hline\hline
\end{tabular}
\end{center}
\end{scriptsize}
\begin{small}
  \tablecaption{Reduced cross section $\sigma_{r}$, as measured with the
    {\SVX} data sample for  $1.5 \le Q^2 \le 3.5$~GeV$^2$. The uncertainties are
    quoted in \% relative to $\sigma_r$.  $\delta_{\rm tot}$ is the total uncertainty
    determined as the quadratic sum of systematic and statistical
    uncertainties. $\delta_{\rm stat}$ is the statistical uncertainty.
    $\delta_{\rm uncor}$ represents the uncorrelated systematic
    uncertainty.  $\gamma_{E_e^{\prime}}$, $\gamma_{\theta_e}$,
    $\gamma_{E_{\rm had}}$, $\gamma_{\rm noise}$, $\gamma_{E^h_{\rm
        SpaCal}}$ and $\gamma_{\gamma p}$ are the bin-to-bin
    correlated systematic uncertainties in the cross section
    measurement due to uncertainties in the SpaCal electromagnetic
    energy scale, electron scattering angle, LAr calorimeter hadronic
    energy scale, LAr calorimeter noise, SpaCal hadronic energy scale
    and the photoproduction background, respectively. The luminosity uncertainty of
    $3$\% for the SVX data is not included in $\delta_{\rm tot}$. \label{tab:svxtableb}}
\end{small} 
\end{table}

\begin{table}
\begin{scriptsize}
\begin{center}
\begin{tabular}{ccccrrrrrrrrr}
\hline\hline
$Q^2$ & $x$ & $y$ & $\sigma_{r}$ & 
$\delta_{\rm tot}$ &
$\delta_{\rm stat}$ & 
$\delta_{\rm uncor}$ & 
$\gamma_{E_e^{\prime}}$ & 
$\gamma_{\theta_e}$ & 
$\gamma_{E_{\rm had}}$ & 
$\gamma_{\rm noise}$ 
& $\gamma_{E^h_{\rm SpaCal}}$ 
& $\gamma_{\gamma p}$ \\    
 GeV$^2$ &  &  &   & \% &\% &\% &\% &\% &\% &\% &\% &\% \\
\hline
0.50  &$ 2.510 \times 10^{-4}$ & $ 1.968 \times 10^{-2}$ &  0.334   & 19.6 &  14.4 &  10.0 &  1.91 & -1.41 &  1.76 &  0.43 & -6.60 &  -4.90 \\ 
0.50  &$ 8.000 \times 10^{-4}$ & $ 6.176 \times 10^{-3}$ &  0.266   & 11.7 &  9.14 &  6.46 & -0.46 & -0.75 & -1.13 & -2.77 & -0.84 &  -1.10 \\ 
0.50  &$ 3.200 \times 10^{-3}$ & $ 1.544 \times 10^{-3}$ &  0.184   & 13.5 &  11.3 &  6.39 & -0.76 &  0.65 & -0.94 &  0.61 & -3.15 &  -0.59 \\ 
0.65  &$ 2.510 \times 10^{-4}$ & $ 2.559 \times 10^{-2}$ &  0.385   & 14.2 &  10.8 &  6.38 & -0.77 &  0.64 &  0.25 & -2.71 & -5.02 &  -3.39 \\ 
0.65  &$ 8.000 \times 10^{-4}$ & $ 8.029 \times 10^{-3}$ &  0.315   & 8.75 &  6.62 &  4.88 &  0.06 &  0.57 & -1.69 & -1.78 & -1.52 &  -0.46 \\ 
0.65  &$ 3.200 \times 10^{-3}$ & $ 2.007 \times 10^{-3}$ &  0.209   & 9.47 &  7.68 &  4.59 & -0.03 &  0.68 & -0.76 & -2.43 & -1.64 &  -0.14 \\ 
0.85  &$ 1.000 \times 10^{-4}$ & $ 8.399 \times 10^{-2}$ &  0.523   & 20.5 &  14.2 &  5.30 & -1.12 & -0.62 & -0.92 & -1.13 & -4.02 &  -13.04 \\ 
0.85  &$ 2.510 \times 10^{-4}$ & $ 3.346 \times 10^{-2}$ &  0.428   & 11.9 &  9.45 &  4.93 & -0.63 &  0.71 &  0.07 & -1.49 & -3.32 &  -3.71 \\ 
0.85  &$ 8.000 \times 10^{-4}$ & $ 1.050 \times 10^{-2}$ &  0.359   & 8.15 &  6.42 &  4.13 & -0.51 &  0.22 & -0.98 & -1.97 & -0.54 &  -1.63 \\ 
0.85  &$ 3.200 \times 10^{-3}$ & $ 2.625 \times 10^{-3}$ &  0.302   & 7.22 &  5.82 &  3.98 & -0.36 &  0.27 & -0.90 & -0.10 & -1.17 &  -0.11 \\ 
1.20  &$ 1.757 \times 10^{-5}$ & $ 6.750 \times 10^{-1}$ &  0.563   & 10.1 &  6.54 &  3.79 & -1.95 &  1.82 & -0.15 & -0.16 &  1.00 &  -6.00 \\ 
1.20  &$ 1.580 \times 10^{-4}$ & $ 7.505 \times 10^{-2}$ &  0.542   & 15.8 &  10.2 &  4.06 & -0.24 &  0.26 & -0.90 & -0.93 & -7.76 &  -8.15 \\ 
1.20  &$ 3.980 \times 10^{-4}$ & $ 2.979 \times 10^{-2}$ &  0.501   & 8.02 &  6.08 &  4.14 &  0.00 &  0.38 &  0.18 & -0.94 & -2.70 &  -1.37 \\ 
1.20  &$ 1.300 \times 10^{-3}$ & $ 9.121 \times 10^{-3}$ &  0.364   & 7.27 &  4.98 &  4.45 &  0.07 & -0.11 & -1.48 & -2.36 & -0.50 &  -0.50 \\ 
1.20  &$ 5.000 \times 10^{-3}$ & $ 2.372 \times 10^{-3}$ &  0.295   & 7.64 &  6.03 &  3.75 &  1.51 & -0.06 & -1.63 &  0.84 & -1.51 &  -0.25 \\ 
1.50  &$ 2.196 \times 10^{-5}$ & $ 6.750 \times 10^{-1}$ &  0.703   & 5.78 &  3.08 &  2.53 & -0.95 &  0.49 & -0.10 & -0.11 &  0.90 &  -3.94 \\ 
1.50  &$ 3.200 \times 10^{-5}$ & $ 4.632 \times 10^{-1}$ &  0.706   & 8.41 &  6.46 &  4.21 & -0.66 &  2.58 & -0.07 & -0.09 &  0.67 &  -1.95 \\ 
1.50  &$ 3.200 \times 10^{-4}$ & $ 4.632 \times 10^{-2}$ &  0.565   & 10.7 &  7.50 &  3.18 & -1.08 &  0.09 & -0.14 & -0.68 & -5.64 &  -3.71 \\ 
1.50  &$ 1.000 \times 10^{-3}$ & $ 1.482 \times 10^{-2}$ &  0.459   & 6.94 &  5.24 &  3.74 & -1.41 &  0.73 & -0.93 & -1.45 & -1.04 &  -0.44 \\ 
1.50  &$ 3.200 \times 10^{-3}$ & $ 4.632 \times 10^{-3}$ &  0.390   & 6.13 &  4.43 &  3.29 & -0.53 &  0.04 & -2.05 & -1.51 & -0.63 &   0.00 \\ 
1.50  &$ 1.300 \times 10^{-2}$ & $ 1.140 \times 10^{-3}$ &  0.331   & 11.5 &  6.93 &  4.32 &  1.21 & -0.49 & -1.26 &  7.65 & -1.98 &   0.00 \\ 
2.00  &$ 2.928 \times 10^{-5}$ & $ 6.750 \times 10^{-1}$ &  0.788   & 4.45 &  2.00 &  2.28 & -1.08 &  0.19 & -0.13 & -0.11 &  1.10 &  -2.85 \\ 
2.00  &$ 5.000 \times 10^{-5}$ & $ 3.953 \times 10^{-1}$ &  0.792   & 5.31 &  4.25 &  2.58 & -0.91 &  1.42 & -0.16 & -0.08 &  0.29 &  -0.73 \\ 
2.00  &$ 3.200 \times 10^{-4}$ & $ 6.176 \times 10^{-2}$ &  0.645   & 12.2 &  3.48 &  2.70 & -2.66 &  1.50 &  0.91 &  0.92 & -10.8 &  -1.58 \\ 
2.00  &$ 1.000 \times 10^{-3}$ & $ 1.976 \times 10^{-2}$ &  0.527   & 5.93 &  4.55 &  3.36 &  0.77 & -0.05 & -0.20 & -0.41 & -1.43 &  -0.52 \\ 
2.00  &$ 3.200 \times 10^{-3}$ & $ 6.176 \times 10^{-3}$ &  0.426   & 5.80 &  3.93 &  3.06 & -0.63 &  0.51 & -2.41 & -1.00 & -1.16 &  -0.05 \\ 
2.00  &$ 1.300 \times 10^{-2}$ & $ 1.520 \times 10^{-3}$ &  0.372   & 9.24 &  5.78 &  3.79 & -0.68 &  0.63 & -0.80 &  5.80 & -1.55 &   0.00 \\ 
2.50  &$ 3.660 \times 10^{-5}$ & $ 6.750 \times 10^{-1}$ &  0.857   & 4.42 &  2.29 &  2.29 & -0.70 & -0.32 & -0.21 & -0.12 &  0.96 &  -2.73 \\ 
2.50  &$ 5.000 \times 10^{-5}$ & $ 4.941 \times 10^{-1}$ &  0.856   & 3.39 &  1.99 &  2.26 & -1.01 &  0.00 & -0.19 & -0.10 &  0.52 &  -1.05 \\ 
2.50  &$ 8.000 \times 10^{-5}$ & $ 3.088 \times 10^{-1}$ &  0.839   & 3.01 &  1.63 &  2.29 & -0.76 &  0.66 & -0.27 & -0.05 &  0.12 &  -0.23 \\ 
2.50  &$ 1.300 \times 10^{-4}$ & $ 1.900 \times 10^{-1}$ &  0.759   & 4.67 &  2.73 &  2.62 & -1.39 &  2.32 & -0.31 &  0.00 &  0.00 &  -0.05 \\ 
2.50  &$ 2.000 \times 10^{-4}$ & $ 1.235 \times 10^{-1}$ &  0.756   & 7.06 &  4.84 &  3.65 & -1.19 &  3.38 & -0.38 &  0.00 &  0.00 &  -0.10 \\ 
2.50  &$ 5.000 \times 10^{-4}$ & $ 4.941 \times 10^{-2}$ &  0.651   & 8.65 &  1.99 &  2.36 & -2.61 &  1.76 &  0.86 &  1.25 & -7.27 &  -0.55 \\ 
2.50  &$ 1.580 \times 10^{-3}$ & $ 1.564 \times 10^{-2}$ &  0.511   & 5.86 &  3.52 &  2.92 & -2.34 &  2.37 & -0.16 &  0.61 & -1.40 &  -0.10 \\ 
2.50  &$ 5.000 \times 10^{-3}$ & $ 4.941 \times 10^{-3}$ &  0.451   & 5.91 &  3.27 &  2.82 & -1.43 &  2.21 & -2.87 & -0.75 & -0.79 &   0.00 \\ 
3.50  &$ 5.124 \times 10^{-5}$ & $ 6.750 \times 10^{-1}$ &  0.935   & 4.27 &  2.17 &  2.25 & -0.87 & -0.14 & -0.16 & -0.11 &  1.03 &  -2.57 \\ 
3.50  &$ 8.000 \times 10^{-5}$ & $ 4.323 \times 10^{-1}$ &  0.947   & 2.89 &  1.49 &  2.20 & -0.85 & -0.20 & -0.20 & -0.09 &  0.38 &  -0.57 \\ 
3.50  &$ 1.300 \times 10^{-4}$ & $ 2.660 \times 10^{-1}$ &  0.908   & 2.63 &  1.21 &  2.21 & -0.67 &  0.00 & -0.35 & -0.01 &  0.00 &  -0.07 \\ 
3.50  &$ 2.000 \times 10^{-4}$ & $ 1.729 \times 10^{-1}$ &  0.879   & 2.83 &  1.42 &  2.26 & -0.83 &  0.35 & -0.30 &  0.00 &  0.00 &  -0.03 \\ 
3.50  &$ 3.200 \times 10^{-4}$ & $ 1.081 \times 10^{-1}$ &  0.775   & 3.60 &  1.75 &  2.32 & -1.70 &  1.21 & -0.45 &  0.00 &  0.00 &   0.00 \\ 
3.50  &$ 8.000 \times 10^{-4}$ & $ 4.323 \times 10^{-2}$ &  0.651   & 4.34 &  1.08 &  2.20 & -1.44 &  0.67 &  0.30 &  0.38 & -3.17 &  -0.12 \\ 
3.50  &$ 2.510 \times 10^{-3}$ & $ 1.378 \times 10^{-2}$ &  0.533   & 3.64 &  1.66 &  2.31 & -1.54 &  1.14 & -0.80 & -0.19 & -0.89 &  -0.01 \\ 
3.50  &$ 8.000 \times 10^{-3}$ & $ 4.323 \times 10^{-3}$ &  0.433   & 4.44 &  1.68 &  2.31 & -1.56 &  1.20 & -2.71 &  0.46 & -0.36 &  -0.01 \\ 
\hline\hline
\end{tabular}
\end{center}
\end{scriptsize}
\begin{small}
  \tablecaption{Reduced cross section $\sigma_{r}$, as measured with the
    {\MB}-BST data sample for  $0.5 \le Q^2 \le 3.5$~GeV$^2$. The uncertainties are
    quoted in \% relative to $\sigma_r$.  $\delta_{\rm tot}$ is the total uncertainty
    determined as the quadratic sum of systematic and statistical
    uncertainties. $\delta_{\rm stat}$ is the statistical uncertainty.
    $\delta_{\rm uncor}$ represents the uncorrelated systematic
    uncertainty.  $\gamma_{E_e^{\prime}}$, $\gamma_{\theta_e}$,
    $\gamma_{E_{\rm had}}$, $\gamma_{\rm noise}$, $\gamma_{E^h_{\rm
        SpaCal}}$ and $\gamma_{\gamma p}$ are the bin-to-bin
    correlated systematic uncertainties in the cross section
    measurement due to uncertainties in the SpaCal electromagnetic
    energy scale, electron scattering angle, LAr calorimeter hadronic
    energy scale, LAr calorimeter noise, SpaCal hadronic energy scale
    and the photoproduction background, respectively. The luminosity uncertainty of
    $1.1$\% for the NVX data is not included in $\delta_{\rm tot}$. \label{tab:mbtablea}}
\end{small} 
\end{table}

\begin{table}
\begin{scriptsize}
\begin{center}
\begin{tabular}{ccccrrrrrrrrr}
\hline\hline
$Q^2$ & $x$ & $y$ & $\sigma_{r}$ & 
$\delta_{\rm tot}$ &
$\delta_{\rm stat}$ & 
$\delta_{\rm uncor}$ & 
$\gamma_{E_e^{\prime}}$ & 
$\gamma_{\theta_e}$ & 
$\gamma_{E_{\rm had}}$ & 
$\gamma_{\rm noise}$ 
& $\gamma_{E^h_{\rm SpaCal}}$ 
& $\gamma_{\gamma p}$ \\    
 GeV$^2$ &  &  &   & \% &\% &\% &\% &\% &\% &\% &\% &\% \\
\hline
5.00  &$ 7.320 \times 10^{-5}$ & $ 6.750 \times 10^{-1}$ &  1.052   & 3.26 &  1.60 &  2.21 & -0.75 & -0.31 & -0.22 & -0.12 &  0.84 &  -1.33 \\ 
5.00  &$ 1.300 \times 10^{-4}$ & $ 3.801 \times 10^{-1}$ &  1.066   & 2.72 &  1.33 &  2.20 & -0.79 & -0.32 & -0.26 & -0.07 &  0.09 &  -0.14 \\ 
5.00  &$ 2.000 \times 10^{-4}$ & $ 2.470 \times 10^{-1}$ &  1.009   & 2.62 &  1.13 &  2.20 & -0.75 & -0.22 & -0.40 &  0.00 &  0.00 &  -0.03 \\ 
5.00  &$ 3.200 \times 10^{-4}$ & $ 1.544 \times 10^{-1}$ &  0.911   & 2.79 &  1.20 &  2.21 & -1.15 & -0.17 & -0.32 &  0.00 &  0.00 &  -0.01 \\ 
5.00  &$ 5.000 \times 10^{-4}$ & $ 9.881 \times 10^{-2}$ &  0.838   & 3.11 &  1.27 &  2.22 & -1.72 & -0.04 & -0.40 &  0.00 &  0.00 &   0.00 \\ 
5.00  &$ 8.000 \times 10^{-4}$ & $ 6.176 \times 10^{-2}$ &  0.775   & 3.50 &  1.29 &  2.23 & -0.27 & -0.09 &  0.17 & -0.40 & -2.32 &  -0.02 \\ 
5.00  &$ 1.300 \times 10^{-3}$ & $ 3.801 \times 10^{-2}$ &  0.686   & 2.91 &  1.39 &  2.24 & -0.46 &  0.07 & -0.18 & -0.53 & -0.99 &  -0.04 \\ 
5.00  &$ 2.000 \times 10^{-3}$ & $ 2.470 \times 10^{-2}$ &  0.636   & 2.84 &  1.45 &  2.26 & -0.69 &  0.26 & -0.24 & -0.06 & -0.53 &  -0.01 \\ 
5.00  &$ 3.980 \times 10^{-3}$ & $ 1.241 \times 10^{-2}$ &  0.569   & 3.18 &  1.08 &  2.20 & -0.50 & -0.04 & -1.73 & -0.86 & -0.33 &  -0.01 \\ 
5.00  &$ 1.300 \times 10^{-2}$ & $ 3.801 \times 10^{-3}$ &  0.440   & 3.90 &  1.13 &  2.20 & -0.43 &  0.05 & -2.50 &  1.62 & -0.26 &   0.00 \\ 
6.50  &$ 9.515 \times 10^{-5}$ & $ 6.750 \times 10^{-1}$ &  1.050   & 4.71 &  2.96 &  2.31 & -0.65 & -0.19 & -0.20 & -0.14 &  0.82 &  -2.63 \\ 
6.50  &$ 1.300 \times 10^{-4}$ & $ 4.941 \times 10^{-1}$ &  1.122   & 2.98 &  1.67 &  2.22 & -0.80 & -0.45 & -0.23 & -0.11 &  0.37 &  -0.31 \\ 
6.50  &$ 2.000 \times 10^{-4}$ & $ 3.211 \times 10^{-1}$ &  1.122   & 2.70 &  1.25 &  2.20 & -0.84 & -0.24 & -0.31 & -0.03 &  0.02 &  -0.06 \\ 
6.50  &$ 3.200 \times 10^{-4}$ & $ 2.007 \times 10^{-1}$ &  1.024   & 2.70 &  1.19 &  2.20 & -0.85 & -0.40 & -0.39 &  0.00 &  0.00 &  -0.01 \\ 
6.50  &$ 5.000 \times 10^{-4}$ & $ 1.285 \times 10^{-1}$ &  0.937   & 2.79 &  1.22 &  2.21 & -1.08 & -0.41 & -0.31 &  0.00 &  0.00 &  -0.01 \\ 
6.50  &$ 8.000 \times 10^{-4}$ & $ 8.029 \times 10^{-2}$ &  0.865   & 3.50 &  1.25 &  2.22 & -2.31 & -0.51 & -0.39 &  0.00 &  0.00 &   0.00 \\ 
6.50  &$ 1.300 \times 10^{-3}$ & $ 4.941 \times 10^{-2}$ &  0.780   & 3.06 &  1.32 &  2.23 &  0.19 & -0.22 & -0.41 & -0.96 & -1.20 &  -0.03 \\ 
6.50  &$ 2.000 \times 10^{-3}$ & $ 3.211 \times 10^{-2}$ &  0.691   & 2.80 &  1.36 &  2.24 &  0.48 & -0.55 & -0.03 & -0.33 & -0.59 &   0.00 \\ 
6.50  &$ 3.980 \times 10^{-3}$ & $ 1.614 \times 10^{-2}$ &  0.618   & 2.79 &  1.00 &  2.19 &  0.07 & -0.22 & -1.09 & -0.82 & -0.28 &   0.00 \\ 
6.50  &$ 1.300 \times 10^{-2}$ & $ 4.941 \times 10^{-3}$ &  0.497   & 3.52 &  0.98 &  2.18 &  0.03 & -0.24 & -2.39 &  0.89 & -0.24 &   0.00 \\ 
8.50  &$ 1.244 \times 10^{-4}$ & $ 6.750 \times 10^{-1}$ &  1.207   & 3.60 &  2.26 &  2.28 & -0.65 & -0.33 & -0.21 & -0.12 &  0.88 &  -1.15 \\ 
8.50  &$ 2.000 \times 10^{-4}$ & $ 4.200 \times 10^{-1}$ &  1.176   & 2.87 &  1.52 &  2.22 & -0.90 & -0.27 & -0.22 & -0.09 &  0.20 &  -0.14 \\ 
8.50  &$ 3.200 \times 10^{-4}$ & $ 2.625 \times 10^{-1}$ &  1.097   & 2.76 &  1.29 &  2.21 & -0.94 & -0.31 & -0.30 &  0.00 &  0.00 &  -0.01 \\ 
8.50  &$ 5.000 \times 10^{-4}$ & $ 1.680 \times 10^{-1}$ &  1.036   & 2.71 &  1.30 &  2.22 & -0.71 & -0.39 & -0.28 &  0.00 &  0.00 &   0.00 \\ 
8.50  &$ 8.000 \times 10^{-4}$ & $ 1.050 \times 10^{-1}$ &  0.959   & 3.05 &  1.32 &  2.23 & -1.53 & -0.30 & -0.38 &  0.00 &  0.00 &   0.00 \\ 
8.50  &$ 1.300 \times 10^{-3}$ & $ 6.461 \times 10^{-2}$ &  0.837   & 3.13 &  1.41 &  2.24 &  0.43 & -0.55 & -0.26 & -0.97 & -1.14 &   0.00 \\ 
8.50  &$ 2.000 \times 10^{-3}$ & $ 4.200 \times 10^{-2}$ &  0.784   & 2.89 &  1.43 &  2.25 &  0.47 & -0.47 & -0.20 & -0.53 & -0.68 &   0.00 \\ 
8.50  &$ 3.200 \times 10^{-3}$ & $ 2.625 \times 10^{-2}$ &  0.679   & 2.91 &  1.49 &  2.26 &  0.37 & -0.44 & -0.45 & -0.66 & -0.42 &   0.00 \\ 
8.50  &$ 6.310 \times 10^{-3}$ & $ 1.331 \times 10^{-2}$ &  0.621   & 3.09 &  1.08 &  2.20 &  0.16 & -0.34 & -1.59 & -0.89 & -0.29 &   0.00 \\ 
8.50  &$ 2.000 \times 10^{-2}$ & $ 4.200 \times 10^{-3}$ &  0.464   & 3.99 &  1.20 &  2.21 &  0.41 & -0.50 & -1.77 &  2.45 & -0.27 &   0.00 \\ 
12.00  &$ 8.000 \times 10^{-4}$ & $ 1.482 \times 10^{-1}$ &  1.067   & 3.05 &  1.45 &  2.25 & -1.40 & -0.34 & -0.24 &  0.00 &  0.00 &   0.00 \\ 
12.00  &$ 1.300 \times 10^{-3}$ & $ 9.121 \times 10^{-2}$ &  0.938   & 3.31 &  1.54 &  2.26 &  0.48 & -0.36 & -0.29 & -0.80 & -1.54 &  -0.01 \\ 
12.00  &$ 2.000 \times 10^{-3}$ & $ 5.929 \times 10^{-2}$ &  0.850   & 3.00 &  1.58 &  2.27 &  0.47 & -0.42 & -0.08 & -0.67 & -0.67 &   0.00 \\ 
12.00  &$ 3.200 \times 10^{-3}$ & $ 3.706 \times 10^{-2}$ &  0.752   & 2.98 &  1.63 &  2.29 &  0.44 & -0.39 & -0.40 & -0.55 & -0.40 &   0.00 \\ 
12.00  &$ 6.310 \times 10^{-3}$ & $ 1.879 \times 10^{-2}$ &  0.650   & 2.89 &  1.21 &  2.22 &  0.46 & -0.47 & -0.91 & -0.77 & -0.31 &   0.00 \\ 
12.00  &$ 2.000 \times 10^{-2}$ & $ 5.929 \times 10^{-3}$ &  0.494   & 3.45 &  1.25 &  2.22 &  0.54 & -0.56 & -1.75 &  1.32 & -0.26 &   0.00 \\ 
\hline\hline
\end{tabular}
\end{center}
\end{scriptsize}
\begin{small}
  \tablecaption{Reduced cross section $\sigma_{r}$, as measured with the
    {\MB}-BST data sample for  $ 5 \le Q^2 \le 12$~GeV$^2$. The uncertainties are
    quoted in \% relative to $\sigma_r$.  $\delta_{\rm tot}$ is the total uncertainty
    determined as the quadratic sum of systematic and statistical
    uncertainties. $\delta_{\rm stat}$ is the statistical uncertainty.
    $\delta_{\rm uncor}$ represents the uncorrelated systematic
    uncertainty.  $\gamma_{E_e^{\prime}}$, $\gamma_{\theta_e}$,
    $\gamma_{E_{\rm had}}$, $\gamma_{\rm noise}$, $\gamma_{E^h_{\rm
        SpaCal}}$ and $\gamma_{\gamma p}$ are the bin-to-bin
    correlated systematic uncertainties in the cross section
    measurement due to uncertainties in the SpaCal electromagnetic
    energy scale, electron scattering angle, LAr calorimeter hadronic
    energy scale, LAr calorimeter noise, SpaCal hadronic energy scale
    and the photoproduction background, respectively. The luminosity uncertainty of
    $1.1$\% for the NVX data is not included in $\delta_{\rm tot}$. \label{tab:mbtableb}}
\end{small} 
\end{table}

\begin{table}
\begin{scriptsize}
\begin{center}
\begin{tabular}{ccccrrrrrrrrr}
\hline\hline
$Q^2$ & $x$ & $y$ & $\sigma_{r}$ & 
$\delta_{\rm tot}$ &
$\delta_{\rm stat}$ & 
$\delta_{\rm uncor}$ & 
$\gamma_{E_e^{\prime}}$ & 
$\gamma_{\theta_e}$ & 
$\gamma_{E_{\rm had}}$ & 
$\gamma_{\rm noise}$ 
& $\gamma_{E^h_{\rm SpaCal}}$ 
& $\gamma_{\gamma p}$ \\    
 GeV$^2$ &  &  &   & \% &\% &\% &\% &\% &\% &\% &\% &\% \\
\hline
1.50  &$ 1.853 \times 10^{-5}$ & $ 8.000 \times 10^{-1}$ &  0.605   & 12.0 &  3.18 &  3.48 &  0.49 &  0.81 & -0.02 &  0.20 &  1.19 &  -10.97 \\ 
2.00  &$ 2.470 \times 10^{-5}$ & $ 8.000 \times 10^{-1}$ &  0.756   & 9.23 &  2.35 &  2.70 & -1.47 &  1.36 & -0.03 &  0.12 &  2.15 &  -7.98 \\ 
2.50  &$ 3.088 \times 10^{-5}$ & $ 8.000 \times 10^{-1}$ &  0.837   & 7.11 &  2.46 &  2.67 & -1.17 & -0.44 & -0.05 &  0.21 &  0.12 &  -5.98 \\ 
3.50  &$ 4.323 \times 10^{-5}$ & $ 8.000 \times 10^{-1}$ &  0.871   & 7.99 &  3.10 &  2.83 & -0.86 & -0.56 &  0.62 & -0.04 &  2.32 &  -6.28 \\ 
5.00  &$ 6.176 \times 10^{-5}$ & $ 8.000 \times 10^{-1}$ &  0.993   & 7.70 &  3.12 &  2.78 & -1.70 & -0.72 & -0.72 &  0.14 &  0.50 &  -6.13 \\ 
6.50  &$ 8.029 \times 10^{-5}$ & $ 8.000 \times 10^{-1}$ &  1.080   & 6.42 &  3.11 &  2.64 & -0.55 &  0.91 &  0.00 & -0.46 & -1.62 &  -4.54 \\ 
8.50  &$ 1.050 \times 10^{-4}$ & $ 8.000 \times 10^{-1}$ &  1.174   & 6.22 &  3.73 &  2.80 & -0.28 &  1.06 & -0.89 &  0.00 & -1.12 &  -3.71 \\ 
\hline\hline
\end{tabular}
\end{center}
\end{scriptsize}
\begin{small}
  \tablecaption{Reduced cross section $\sigma_{r}$, as measured with the
    {\MB}-S9 data sample. The uncertainties are
    quoted in \% relative to $\sigma_r$.  $\delta_{\rm tot}$ is the total uncertainty
    determined as the quadratic sum of systematic and statistical
    uncertainties. $\delta_{\rm stat}$ is the statistical uncertainty.
    $\delta_{\rm uncor}$ represents the uncorrelated systematic
    uncertainty.  $\gamma_{E_e^{\prime}}$, $\gamma_{\theta_e}$,
    $\gamma_{E_{\rm had}}$, $\gamma_{\rm noise}$, $\gamma_{E^h_{\rm
        SpaCal}}$ and $\gamma_{\gamma p}$ are the bin-to-bin
    correlated systematic uncertainties in the cross section
    measurement due to uncertainties in the SpaCal electromagnetic
    energy scale, electron scattering angle, LAr calorimeter hadronic
    energy scale, LAr calorimeter noise, SpaCal hadronic energy scale
    and the photoproduction background, respectively. The luminosity uncertainty of
    $1.1$\% for the NVX data is not included in $\delta_{\rm tot}$. \label{tab:mbtablesn}}
\end{small} 
\end{table}

\begin{table}
\begin{tiny}
\begin{center}
\begin{tabular}{l|cccccccccc}
\hline
\hline
$\#$ &  $Q^2$  & $x$ & $y$ & $F^{\rm th}_L$ & $\sigma^{\rm ave}_{\rm r}$ & $F_2$ & $\delta_{\rm ave,stat}$ & $\delta_{\rm ave,uncor}$ & $\delta_{\rm ave,tot}  $ & CME\\
     & GeV$^2$ &   &       &                &                  &     & \%  &    \%              &   \% & GeV \\
\hline
    1  &   0.2  &$ 0.398 \times 10^{-4}$ &  0.050  & 0.08  & 0.230  & 0.230  & 14.3 &  12.0 & 19.98 &    319 \\
    2  &   0.2  &$ 0.251 \times 10^{-3}$ &  0.008  & 0.06  & 0.190  & 0.190  & 13.1 &  6.18 & 15.03 &    319 \\
    3  &  0.25  &$ 0.398 \times 10^{-4}$ &  0.062  & 0.09  & 0.300  & 0.300  & 9.84 &  11.3 & 16.82 &    319 \\
    4  &  0.25  &$ 0.251 \times 10^{-3}$ &  0.010  & 0.07  & 0.191  & 0.191  &10.00 &  4.70 & 12.05 &    319 \\
    5  &  0.25  &$ 0.158 \times 10^{-2}$ &  0.002  & 0.06  & 0.203  & 0.203  & 10.8 &  5.29 & 12.37 &    301 \\
    6  &  0.35  &$ 0.512 \times 10^{-5}$ &  0.675  & ---  &  0.450  & ---  &  21.7 &  12.8 & 25.34 &    319 \\
    7  &  0.35  &$ 0.610 \times 10^{-5}$ &  0.634  & ---  &  0.357  & ---  &  5.74 &  11.0 & 13.50 &    301 \\
    8  &  0.35  &$ 0.320 \times 10^{-4}$ &  0.108  & 0.12  & 0.410  & 0.411  & 9.12 &  11.1 & 20.36 &    319 \\
    9  &  0.35  &$ 0.130 \times 10^{-3}$ &  0.027  & 0.10  & 0.264  & 0.264  & 9.62 &  4.38 & 10.99 &    319 \\
   10  &  0.35  &$ 0.500 \times 10^{-3}$ &  0.007  & 0.08  & 0.237  & 0.237  & 8.81 &  4.19 & 10.08 &    319 \\
   11  &  0.35  &$ 0.251 \times 10^{-2}$ &  0.001  & 0.07  & 0.204  & 0.204  & 9.93 &  4.55 & 11.08 &    319 \\
   12  &   0.5  &$ 0.732 \times 10^{-5}$ &  0.675  & ---  &  0.449  & ---  &  5.42 &  5.74 &  9.44 &    319 \\
   13  &   0.5  &$ 0.860 \times 10^{-5}$ &  0.642  & ---  &  0.442  & ---  &  3.75 &  9.17 & 10.69 &    301 \\
   14  &   0.5  &$ 0.158 \times 10^{-4}$ &  0.313  & 0.16  & 0.461  & 0.472  & 19.0 &  9.84 & 21.61 &    319 \\
   15  &   0.5  &$ 0.398 \times 10^{-4}$ &  0.124  & 0.15  & 0.478  & 0.480  & 10.1 &  6.07 & 16.25 &    319 \\
   16  &   0.5  &$ 0.100 \times 10^{-3}$ &  0.049  & 0.13  & 0.411  & 0.411  & 8.85 &  4.87 & 10.57 &    319 \\
   17  &   0.5  &$ 0.251 \times 10^{-3}$ &  0.020  & 0.11  & 0.296  & 0.296  & 8.37 &  4.20 &  9.74 &    319 \\
   18  &   0.5  &$ 0.800 \times 10^{-3}$ &  0.006  & 0.10  & 0.280  & 0.280  & 5.92 &  3.44 &  7.07 &    319 \\
   19  &   0.5  &$ 0.320 \times 10^{-2}$ &  0.002  & 0.08  & 0.183  & 0.183  & 11.4 &  6.39 & 13.12 &    301 \\
   20  &  0.65  &$ 0.952 \times 10^{-5}$ &  0.675  & ---  &  0.479  & ---  &  3.96 &  2.90 &  5.85 &    319 \\
   21  &  0.65  &$ 0.112 \times 10^{-4}$ &  0.641  & ---  &  0.504  & ---  &  3.74 &  8.21 &  9.89 &    301 \\
   22  &  0.65  &$ 0.158 \times 10^{-4}$ &  0.407  & 0.20  & 0.466  & 0.490  & 3.09 &  5.44 &  6.51 &    319 \\
   23  &  0.65  &$ 0.164 \times 10^{-4}$ &  0.438  & 0.19  & 0.510  & 0.538  & 3.02 &  7.28 &  8.33 &    301 \\
   24  &  0.65  &$ 0.398 \times 10^{-4}$ &  0.161  & 0.17  & 0.678  & 0.681  & 17.5 &  11.2 & 21.16 &    319 \\
   25  &  0.65  &$ 0.100 \times 10^{-3}$ &  0.064  & 0.15  & 0.500  & 0.500  & 5.14 &  5.84 & 10.70 &    319 \\
   26  &  0.65  &$ 0.251 \times 10^{-3}$ &  0.026  & 0.13  & 0.376  & 0.376  & 6.79 &  3.46 &  7.98 &    319 \\
   27  &  0.65  &$ 0.800 \times 10^{-3}$ &  0.008  & 0.11  & 0.308  & 0.308  & 4.94 &  3.02 &  6.17 &    319 \\
   28  &  0.65  &$ 0.320 \times 10^{-2}$ &  0.002  & 0.09  & 0.225  & 0.225  & 5.81 &  3.15 &  6.76 &    319 \\
   29  &  0.85  &$ 0.124 \times 10^{-4}$ &  0.675  & ---  &  0.565  & ---  &  2.54 &  2.52 &  4.50 &    319 \\
   30  &  0.85  &$ 0.138 \times 10^{-4}$ &  0.675  & ---  &  0.614  & ---  &  5.20 &  9.45 & 12.17 &    301 \\
   31  &  0.85  &$ 0.200 \times 10^{-4}$ &  0.420  & 0.22  & 0.612  & 0.641  & 1.96 &  5.36 &  5.99 &    319 \\
   32  &  0.85  &$ 0.200 \times 10^{-4}$ &  0.469  & 0.22  & 0.596  & 0.634  & 2.65 &  4.98 &  6.27 &    301 \\
   33  &  0.85  &$ 0.398 \times 10^{-4}$ &  0.211  & 0.20  & 0.567  & 0.573  & 1.65 &  3.39 &  4.13 &    319 \\
   34  &  0.85  &$ 0.500 \times 10^{-4}$ &  0.168  & 0.20  & 0.546  & 0.549  & 2.92 &  4.52 &  5.97 &    319 \\
   35  &  0.85  &$ 0.100 \times 10^{-3}$ &  0.084  & 0.18  & 0.499  & 0.500  & 2.78 &  3.59 &  5.98 &    319 \\
   36  &  0.85  &$ 0.251 \times 10^{-3}$ &  0.033  & 0.15  & 0.414  & 0.414  & 5.88 &  2.98 &  7.31 &    319 \\
   37  &  0.85  &$ 0.800 \times 10^{-3}$ &  0.010  & 0.13  & 0.350  & 0.350  & 4.61 &  2.66 &  5.60 &    319 \\
   38  &  0.85  &$ 0.320 \times 10^{-2}$ &  0.003  & 0.11  & 0.307  & 0.307  & 4.56 &  2.81 &  5.49 &    301 \\
   39  &   1.2  &$ 0.176 \times 10^{-4}$ &  0.675  & ---  &  0.608  & ---  &  2.54 &  2.14 &  4.65 &    319 \\
   40  &   1.2  &$ 0.200 \times 10^{-4}$ &  0.593  & ---  &  0.671  & ---  &  2.62 &  2.51 &  3.94 &    319 \\
   41  &   1.2  &$ 0.200 \times 10^{-4}$ &  0.663  & ---  &  0.741  & ---  &  3.60 &  8.36 &  9.98 &    301 \\
   42  &   1.2  &$ 0.320 \times 10^{-4}$ &  0.371  & 0.26  & 0.689  & 0.714  & 1.67 &  2.73 &  3.55 &    319 \\
   43  &   1.2  &$ 0.320 \times 10^{-4}$ &  0.414  & 0.26  & 0.705  & 0.738  & 2.68 &  4.55 &  5.83 &    301 \\
   44  &   1.2  &$ 0.631 \times 10^{-4}$ &  0.188  & 0.23  & 0.647  & 0.652  & 1.18 &  2.25 &  3.09 &    319 \\
   45  &   1.2  &$ 0.800 \times 10^{-4}$ &  0.148  & 0.22  & 0.594  & 0.597  & 2.18 &  4.02 &  5.24 &    319 \\
   46  &   1.2  &$ 0.130 \times 10^{-3}$ &  0.091  & 0.21  & 0.543  & 0.544  & 2.43 &  4.97 &  5.78 &    319 \\
   47  &   1.2  &$ 0.158 \times 10^{-3}$ &  0.075  & 0.20  & 0.503  & 0.504  & 1.67 &  2.30 &  3.24 &    319 \\
   48  &   1.2  &$ 0.398 \times 10^{-3}$ &  0.030  & 0.17  & 0.502  & 0.502  & 2.88 &  2.67 &  4.26 &    319 \\
   49  &   1.2  &$ 0.130 \times 10^{-2}$ &  0.009  & 0.14  & 0.374  & 0.374  & 3.58 &  2.62 &  4.74 &    319 \\
   50  &   1.2  &$ 0.500 \times 10^{-2}$ &  0.002  & 0.12  & 0.298  & 0.298  & 4.51 &  2.60 &  5.47 &    319 \\
   51  &   1.5  &$ 0.185 \times 10^{-4}$ &  0.800  & ---  &  0.610  & ---  &  3.17 &  3.48 &  7.93 &    319 \\
   52  &   1.5  &$ 0.220 \times 10^{-4}$ &  0.675  & ---  &  0.702  & ---  &  1.94 &  1.78 &  3.31 &    319 \\
   53  &   1.5  &$ 0.320 \times 10^{-4}$ &  0.463  & 0.29  & 0.756  & 0.804  & 1.77 &  2.12 &  3.08 &    319 \\
   54  &   1.5  &$ 0.320 \times 10^{-4}$ &  0.518  & 0.29  & 0.801  & 0.864  & 1.20 &  3.20 &  4.47 &    301 \\
   55  &   1.5  &$ 0.500 \times 10^{-4}$ &  0.296  & 0.27  & 0.759  & 0.775  & 1.06 &  1.97 &  2.62 &    319 \\
   56  &   1.5  &$ 0.800 \times 10^{-4}$ &  0.185  & 0.25  & 0.699  & 0.705  & 1.26 &  2.15 &  2.95 &    319 \\
   57  &   1.5  &$ 0.130 \times 10^{-3}$ &  0.114  & 0.23  & 0.643  & 0.644  & 1.49 &  2.42 &  3.32 &    319 \\
   58  &   1.5  &$ 0.200 \times 10^{-3}$ &  0.074  & 0.22  & 0.615  & 0.616  & 2.40 &  2.59 &  3.97 &    319 \\
   59  &   1.5  &$ 0.320 \times 10^{-3}$ &  0.046  & 0.20  & 0.584  & 0.584  & 1.60 &  2.18 &  3.30 &    319 \\
   60  &   1.5  &$ 0.500 \times 10^{-3}$ &  0.030  & 0.19  & 0.548  & 0.548  & 2.51 &  7.05 &  7.74 &    319 \\
   61  &   1.5  &$ 0.800 \times 10^{-3}$ &  0.019  & 0.17  & 0.495  & 0.495  & 2.35 &  2.47 &  3.80 &    319 \\
   62  &   1.5  &$ 0.100 \times 10^{-2}$ &  0.015  & 0.17  & 0.463  & 0.463  & 5.22 &  3.74 &  6.61 &    319 \\
   63  &   1.5  &$ 0.320 \times 10^{-2}$ &  0.005  & 0.14  & 0.409  & 0.409  & 2.32 &  2.03 &  3.51 &    301 \\
   64  &   1.5  &$ 0.130 \times 10^{-1}$ &  0.001  & 0.11  & 0.327  & 0.327  & 3.99 &  2.49 &  7.00 &    319 \\
\hline
\hline
\end{tabular}
\end{center}
\end{tiny}
\tablecaption{ Combined H1 reduced cross section $\sigma^{\rm ave}_{\rm r}$  for  
$0.2 \le Q^2\le 1.5$~GeV$^2$. 
The uncertainties are quoted in \% relative to $\sigma^{\rm ave}_{\rm r}$.
$F^{\rm th}_L$ represents the structure function $F_L$ used for the CME
 correction (\Eq~\ref{eq:cmecorr}) and to calculate the structure function $F_2$.
$\delta_{\rm ave,stat}$ ($\delta_{\rm ave,uncor}$) represents the statistical (uncorrelated systematic)  uncertainty. 
$\delta_{\rm ave,tot}$ 
is the total uncertainty calculated as a sum of uncorrelated uncertainty and all correlated sources in quadrature.
A global normalisation uncertainty of $0.5\%$ is not included in $\delta_{\rm ave,tot}$.
CME stands for the centre-of-mass energy of the measurement.
\label{tab:all1}}
\end{table}

\begin{sidewaystable}
\begin{tiny}
\begin{tabular}{c|rrrrrrrrrrrrrrrrrrrrrrrrrr}
\hline
\hline
$\#$  & $\gamma^{\rm ave}_1$ &$\gamma^{\rm ave}_2$ &$\gamma^{\rm ave}_3$ &$\gamma^{\rm ave}_4$ &$\gamma^{\rm ave}_5$ &
   $\gamma^{\rm ave}_6$ &$\gamma^{\rm ave}_7$ &$\gamma^{\rm ave}_8$ &$\gamma^{\rm ave}_9$ &$\gamma^{\rm ave}_{10}$ &
   $\gamma^{\rm ave}_{11}$ &$\gamma^{\rm ave}_{12}$ &$\gamma^{\rm ave}_{13}$ &$\gamma^{\rm ave}_{14}$ &$\gamma^{\rm ave}_{15}$ &
   $\gamma^{\rm ave}_{16}$ &$\gamma^{\rm ave}_{17}$ &$\gamma^{\rm ave}_{18}$ &$\gamma^{\rm ave}_{19}$ &$\gamma^{\rm ave}_{20}$  & 
   $\gamma^{\rm ave}_{21}$ &$\gamma^{\rm ave}_{22}$ &$\gamma^{\rm ave}_{23}$ &$\gamma^{\rm ave}_{24}$ &$\gamma^{\rm ave}_{25}$ &
   $\gamma^{\rm ave}_{26}$ \\
  & \permil &\permil &\permil &\permil &\permil &\permil &\permil &\permil &\permil &\permil &\permil &\permil &\permil &\permil &\permil &\permil &\permil &\permil &\permil &\permil &\permil &\permil &\permil &\permil &\permil &\permil \\
\hline
    1 &   -8.47 &    4.05 &   13.89 &  -47.47 &   28.59 &  -29.31 &    8.41 &    9.60 &   11.37 &    0.70 &    2.45 &   10.13 &    4.13 &    4.53 &   -7.00 &   16.08 &    5.55 &   -2.12 &    5.17 &   -4.79 &   -1.39 &    0.57 &    3.25 &   -0.76 &   -1.67 &    6.96\\
    2 &   -8.44 &    0.03 &    2.05 &  -16.70 &    4.06 &    3.02 &    8.85 &    4.87 &   -8.92 &    4.13 &    7.94 &    1.45 &   17.70 &    1.42 &  -13.72 &  -13.88 &   -0.90 &    1.39 &   -7.65 &    2.47 &   -9.60 &   -0.56 &    1.05 &   -1.06 &   -1.12 &    5.38\\
    3 &   -8.46 &    5.18 &   15.49 &  -60.99 &   27.90 &  -21.88 &   11.74 &    5.01 &    3.26 &    2.87 &    4.43 &    8.68 &    9.12 &    2.88 &  -10.26 &    4.44 &    1.89 &   -3.94 &    2.73 &   -2.79 &   -3.82 &    0.68 &    2.74 &   -0.82 &   -1.69 &    7.03\\
    4 &   -8.45 &   -1.19 &    1.31 &  -12.83 &   11.91 &   -7.49 &   20.51 &    8.35 &   -7.96 &   -2.77 &   22.68 &  -11.42 &   11.97 &   -7.76 &   -6.84 &  -12.41 &    0.96 &    2.99 &   -8.42 &    3.25 &  -10.50 &   -1.33 &    2.29 &   -1.41 &   -1.41 &    6.91\\
    5 &   -8.46 &   -1.54 &   -0.42 &   -4.36 &   -2.37 &    8.37 &    2.08 &    9.28 &   -6.33 &    2.73 &    6.68 &   -1.39 &   10.58 &   -1.36 &   -7.56 &  -11.91 &    0.62 &    1.23 &   -5.17 &    1.79 &   -4.26 &   -0.14 &    0.76 &   -0.69 &   -0.87 &    4.36\\
    6 &   -8.45 &    1.47 &    4.81 &  -19.08 &    3.08 &    1.98 &   -1.34 &    3.89 &    0.63 &    4.12 &   -1.29 &    2.26 &   -4.14 &   -1.06 &   -1.34 &   -0.65 &    1.81 &   -2.91 &    2.88 &   -1.95 &    1.69 &    0.89 &    1.76 &   -0.02 &   -1.03 &    4.82\\
    7 &   -8.66 &   19.34 &    9.32 &    4.74 &    8.68 &    7.41 &  -10.14 &   -4.02 &   -0.07 &    0.70 &   26.17 &   10.09 &  -11.30 &   25.55 &    5.77 &   -5.90 &    1.03 &   -1.40 &   -1.20 &    1.14 &    0.98 &    6.38 &    7.17 &  -15.10 &    2.23 &   -1.74\\
    8 &   -8.42 &    7.01 &   28.05 & -123.68 &   41.80 &  -15.78 &   17.04 &   18.13 &   -7.89 &   -1.98 &   22.54 &   -2.54 &   22.13 &  -12.02 &   -7.29 &  -21.45 &   -2.31 &  -11.93 &    4.21 &    0.70 &    4.74 &    2.76 &    0.80 &   -0.93 &   -1.77 &    6.19\\
    9 &   -8.46 &    0.28 &    4.59 &  -17.69 &    9.30 &   -6.18 &    5.69 &   10.17 &    0.15 &    1.78 &    6.25 &    0.08 &   -4.02 &   -1.68 &   -5.16 &    7.06 &    1.47 &    0.40 &    0.05 &   -2.06 &   -7.68 &   -1.15 &    3.32 &   -0.83 &   -1.23 &    6.19\\
   10 &   -8.46 &    0.26 &    0.04 &   -5.57 &    2.86 &    0.54 &   11.44 &   -1.48 &   -6.32 &    3.04 &    8.08 &   -3.79 &    4.03 &   -2.11 &   -6.41 &   -8.73 &    0.29 &    1.19 &   -5.28 &    1.42 &   -8.00 &   -0.65 &    2.12 &   -0.71 &   -1.22 &    6.19\\
   11 &   -8.47 &   -0.34 &   -0.56 &   -0.84 &    0.85 &    1.46 &    7.59 &    1.31 &   -3.73 &    2.29 &    6.72 &   -4.19 &   -0.09 &   -2.91 &   -3.31 &   -5.60 &    1.41 &    1.01 &   -3.33 &    0.59 &   -5.04 &   -0.41 &    2.10 &   -0.52 &   -1.12 &    5.73\\
   12 &   -8.44 &    0.29 &    9.73 &  -35.80 &    6.26 &    3.70 &  -10.82 &   23.30 &    2.89 &    2.91 &   -0.79 &    4.32 &  -13.10 &   -3.38 &   -0.91 &   13.08 &    0.79 &   -4.09 &    7.84 &   -5.04 &    0.10 &   -0.19 &    2.90 &   -0.32 &   -0.84 &    4.16\\
   13 &   -8.66 &   21.09 &    3.43 &    2.40 &    0.05 &    1.81 &   -4.02 &   -3.78 &    0.51 &    0.33 &   20.31 &    9.38 &   -7.66 &   15.14 &    2.38 &   -7.80 &    4.12 &   -1.96 &   -0.15 &   -6.68 &    1.36 &   -7.56 &    1.17 &   -8.39 &    1.49 &   -1.27\\
   14 &   -8.48 &    3.46 &    0.47 &   -0.95 &    2.62 &   -4.76 &   11.15 &  -19.63 &    1.55 &    3.96 &   -0.89 &   -4.72 &  -11.40 &   -3.11 &    5.69 &   -5.85 &    3.61 &   -2.83 &    1.84 &   -0.76 &    4.21 &    1.70 &    2.20 &    0.56 &   -1.39 &    6.60\\
   15 &   -8.43 &    4.57 &   21.52 &  -96.02 &   29.98 &   -7.24 &   11.43 &   18.94 &   -7.85 &   -1.25 &   19.35 &   -4.28 &   13.28 &  -11.98 &   -4.93 &  -16.82 &   -2.15 &   -9.41 &    3.50 &    0.34 &    2.78 &    1.81 &    1.20 &   -0.81 &   -1.49 &    5.64\\
   16 &   -8.47 &   -0.86 &    3.21 &  -14.11 &    7.55 &   -4.49 &    3.90 &   13.42 &    1.00 &   -1.32 &    9.85 &   -1.87 &   13.02 &   -2.64 &   -3.47 &   -8.49 &    4.41 &    0.24 &   -1.94 &    0.79 &    2.78 &    1.05 &    0.52 &   -0.60 &   -1.05 &    4.50\\
   17 &   -8.47 &    0.90 &    1.81 &  -10.87 &    6.19 &   -4.16 &   10.70 &   -1.34 &   -3.16 &    2.24 &    4.90 &   -0.34 &    7.40 &    1.23 &   -4.87 &   -5.96 &   -0.04 &    7.70 &   -5.29 &   -1.76 &   -8.17 &   -0.62 &    5.40 &    0.89 &    0.46 &    4.72\\
   18 &   -8.46 &   -0.12 &   -0.41 &   -1.73 &    0.12 &    2.30 &    6.67 &    0.72 &   -6.06 &    6.10 &    4.41 &   -0.39 &   -2.26 &   -0.93 &   -6.12 &   -0.98 &   -0.77 &    4.02 &   -2.33 &   -0.09 &   -3.58 &   -0.06 &    0.94 &    0.06 &    0.31 &    3.43\\
   19 &   -8.48 &   -0.13 &    0.14 &    1.36 &    1.68 &   -3.36 &    1.87 &    2.48 &    2.66 &    2.81 &    0.02 &    0.58 &   -0.46 &    0.58 &   -1.66 &   -4.41 &    1.11 &    0.75 &    2.51 &   -0.03 &   -0.55 &   -0.28 &    5.49 &    3.61 &    4.55 &   -0.73\\
   20 &   -8.45 &    0.84 &    5.86 &  -22.11 &    3.12 &    3.62 &   -4.22 &   10.18 &    1.17 &    3.16 &   -0.05 &    0.34 &  -11.86 &   -4.00 &    1.27 &    4.83 &    1.37 &   -3.47 &    5.19 &   -3.16 &    1.36 &    0.38 &    2.47 &   -0.04 &   -0.96 &    4.73\\
   21 &   -8.65 &   20.32 &    3.93 &    2.78 &    2.24 &    3.05 &   -5.19 &   -3.74 &    0.38 &    0.55 &   20.73 &    9.06 &   -8.13 &   16.77 &    2.97 &   -7.18 &    3.44 &   -1.80 &   -0.34 &   -4.92 &    1.25 &   -4.46 &    2.44 &   -9.73 &    1.63 &   -1.36\\
   22 &   -8.47 &    2.20 &    1.96 &   -7.04 &    1.11 &   -0.54 &    2.14 &   -6.90 &    1.55 &    5.53 &   -4.53 &    3.52 &   -2.39 &    2.13 &   -1.85 &   -2.62 &    2.79 &   -1.88 &    1.19 &   -1.35 &    1.67 &    1.22 &    1.57 &    0.16 &   -1.13 &    5.29\\
   23 &   -8.60 &   17.09 &   -8.57 &   -1.73 &   -3.98 &   -1.64 &    0.24 &   -1.41 &    1.38 &    2.84 &    5.70 &    2.20 &   -3.48 &    4.08 &   -0.37 &   -3.64 &    3.89 &   -0.68 &    0.22 &   -2.30 &    0.72 &    0.31 &    4.07 &  -11.34 &    1.74 &   -1.56\\
   24 &   -8.47 &   -1.45 &   -0.74 &    1.62 &    0.81 &    2.74 &    8.57 &    5.51 &   -1.63 &   -4.27 &   16.47 &  -20.04 &  -17.65 &  -17.26 &   13.57 &   -7.04 &    3.48 &   -1.72 &    2.01 &    0.29 &    5.31 &    0.52 &    2.48 &    0.08 &   -1.08 &    5.62\\
   25 &   -8.47 &    2.45 &    6.38 &  -31.57 &   23.73 &  -22.45 &   25.69 &   -2.50 &   -1.25 &   -4.12 &   20.69 &   -6.10 &   33.65 &   -3.25 &   -6.44 &  -26.34 &    5.80 &   -0.20 &   -7.40 &    4.79 &    4.15 &    2.58 &   -0.51 &   -0.84 &   -1.72 &    6.54\\
   26 &   -8.47 &    0.69 &    1.95 &  -10.69 &    5.12 &   -2.52 &    8.31 &    1.20 &   -2.17 &    2.61 &    5.27 &   -1.29 &    6.70 &    0.25 &   -4.87 &   -7.55 &    3.59 &    4.91 &   -3.17 &   -0.93 &    3.90 &    1.35 &    3.44 &    1.90 &    1.61 &    2.75\\
   27 &   -8.47 &   -0.53 &   -0.73 &    0.23 &    1.70 &    0.53 &    9.59 &    2.56 &   -4.82 &    2.94 &    6.83 &   -4.26 &   -7.21 &   -2.28 &   -7.14 &   -0.15 &   -1.88 &    3.65 &   -1.02 &    0.16 &   -6.45 &   -0.93 &    2.24 &    0.34 &    0.86 &    3.38\\
   28 &   -8.47 &   -0.35 &   -0.56 &    0.36 &   -0.63 &    2.74 &    4.23 &    1.95 &   -1.82 &    0.93 &    3.68 &   -2.31 &   -2.55 &   -1.35 &   -4.59 &   -2.15 &    0.31 &    1.73 &    1.17 &   -0.82 &    5.09 &    1.38 &    0.84 &    1.12 &    1.28 &    1.51\\
   29 &   -8.45 &    1.53 &    5.16 &  -18.70 &    3.31 &    1.29 &   -2.22 &    4.47 &    1.99 &    3.96 &   -1.96 &    1.47 &  -11.17 &   -2.20 &    0.92 &    4.49 &    1.87 &   -3.25 &    4.75 &   -3.07 &    1.44 &    0.59 &    2.45 &    0.05 &   -1.04 &    5.06\\
   30 &   -8.70 &   23.84 &   10.36 &    5.53 &    6.16 &    5.84 &   -9.16 &   -5.70 &    0.01 &   -1.55 &   30.75 &   15.38 &  -12.02 &   23.47 &    4.49 &   -9.88 &    3.79 &   -2.20 &   -0.78 &   -6.07 &    1.57 &   -6.03 &    2.46 &  -10.56 &    1.81 &   -1.45\\
   31 &   -8.47 &    2.26 &    1.04 &   -3.27 &    0.34 &   -0.80 &    3.24 &   -9.51 &    1.66 &    5.48 &   -4.58 &    2.61 &   -3.13 &    1.92 &   -0.93 &   -3.44 &    3.08 &   -1.71 &    0.93 &   -1.14 &    2.10 &    1.32 &    1.56 &    0.24 &   -1.15 &    5.42\\
   32 &   -8.61 &   18.05 &   -7.99 &   -1.01 &   -1.26 &   -0.16 &   -0.97 &   -1.35 &    0.94 &    2.93 &    7.86 &    1.83 &   -3.62 &    7.39 &    0.73 &   -3.78 &    3.22 &   -1.06 &    0.16 &   -2.86 &    0.79 &   -1.16 &    3.13 &   -9.74 &    1.53 &   -1.36\\
   33 &   -8.48 &    4.42 &   -1.99 &   -0.46 &   -0.97 &   -0.59 &    2.40 &   -5.61 &    1.56 &    3.93 &   -1.05 &    1.99 &   -2.33 &    0.01 &   -1.63 &   -3.62 &    4.04 &   -0.48 &    0.25 &   -0.76 &    1.14 &    2.56 &    3.85 &   -7.64 &    0.64 &    1.30\\
   34 &   -8.47 &   -0.30 &    2.10 &    0.44 &   -3.54 &   -1.65 &    3.26 &   -4.98 &    2.41 &    1.35 &   -0.36 &    6.63 &   -3.37 &   -9.21 &   -5.03 &   -4.90 &    7.60 &    0.55 &    0.67 &   -3.16 &    0.67 &    0.65 &    5.53 &  -15.11 &    2.29 &   -2.13\\
   35 &   -8.47 &    0.31 &    4.11 &   -5.19 &   -6.12 &    1.27 &    1.30 &    5.71 &   -0.09 &   12.30 &   -5.40 &  -16.02 &   -0.58 &   17.29 &    6.25 &    5.63 &   -3.93 &    3.08 &   -2.00 &    9.65 &   -0.45 &   16.99 &    7.93 &   -6.56 &    1.04 &    1.00\\
   36 &   -8.47 &    0.70 &    1.64 &  -10.06 &    6.23 &   -4.71 &    9.66 &    0.29 &   -1.57 &    1.96 &    5.60 &    0.03 &   16.26 &    2.56 &   -7.16 &  -13.29 &    4.58 &    5.94 &   -6.66 &    1.56 &    2.20 &    1.15 &    1.81 &    1.22 &    1.08 &    2.73\\
   37 &   -8.47 &   -0.07 &    0.12 &   -2.35 &    1.87 &   -0.19 &    7.08 &    1.88 &   -3.63 &    4.38 &    4.05 &   -0.74 &   -3.24 &    0.11 &   -6.71 &    0.99 &    0.65 &    4.88 &   -3.20 &    0.43 &   -4.44 &   -0.60 &    1.12 &    0.15 &    0.54 &    3.08\\
   38 &   -8.47 &    0.36 &    0.10 &    1.05 &    1.75 &   -3.55 &    2.87 &   -0.56 &    2.01 &    2.88 &   -0.51 &    2.68 &    0.72 &    2.00 &   -3.79 &   -0.34 &    1.29 &    0.86 &    0.55 &    0.04 &   -0.93 &    0.21 &    1.36 &    0.75 &    0.95 &    2.14\\
   39 &   -8.45 &    2.26 &    6.77 &  -24.90 &    4.39 &    1.50 &   -2.50 &    3.92 &    2.41 &    4.89 &   -3.89 &    1.71 &  -12.77 &   -0.87 &    0.71 &    4.36 &    4.00 &   -0.73 &    1.74 &   -1.45 &    1.19 &   -0.02 &    1.97 &    0.46 &   -0.33 &    3.85\\
   40 &   -8.46 &    0.65 &    1.86 &   -6.42 &   -0.75 &    3.25 &   -1.89 &    2.34 &    1.30 &    4.00 &   -2.04 &    1.28 &   -5.72 &   -0.73 &   -0.22 &   -0.07 &    2.54 &   -1.65 &    2.11 &   -1.76 &    1.69 &    0.75 &    1.74 &    0.05 &   -0.96 &    4.72\\
   41 &   -8.65 &   19.02 &    3.66 &    3.59 &    5.94 &    4.61 &   -6.00 &   -4.30 &    0.16 &    0.33 &   20.90 &    9.48 &   -8.11 &   16.48 &    2.88 &   -7.48 &    3.63 &   -1.84 &   -0.29 &   -5.55 &    1.28 &   -5.67 &    1.86 &   -8.97 &    1.54 &   -1.29\\
   42 &   -8.47 &    1.68 &    1.25 &   -4.31 &    0.72 &   -0.48 &    2.54 &   -5.52 &    1.46 &    4.70 &   -2.90 &    2.12 &   -2.03 &    1.26 &   -1.32 &   -3.13 &    3.06 &   -1.39 &    0.71 &   -1.06 &    1.71 &    1.13 &    1.53 &    0.11 &   -1.11 &    5.25\\
   43 &   -8.58 &   13.51 &   -6.83 &   -1.18 &   -1.40 &   -0.20 &   -0.45 &   -1.46 &    1.28 &    3.32 &    4.69 &    1.21 &   -3.48 &    4.63 &   -0.05 &   -2.66 &    3.21 &   -0.40 &    0.09 &   -0.24 &    0.55 &    3.91 &    5.50 &  -12.79 &    1.88 &   -1.66\\
   44 &   -8.47 &    2.22 &    0.22 &    0.06 &   -1.15 &   -1.11 &    4.10 &   -9.48 &    1.56 &    4.51 &   -3.05 &    2.71 &   -3.06 &   -0.46 &   -1.70 &   -4.29 &    4.16 &   -0.91 &    0.41 &   -1.56 &    1.40 &    0.86 &    2.51 &   -3.84 &   -0.21 &    3.32\\
   45 &   -8.47 &   -1.72 &    1.85 &    0.72 &    1.37 &    1.19 &    0.80 &   -3.89 &    1.92 &    2.81 &   -0.58 &    3.26 &   -3.55 &   -3.91 &   -2.92 &   -2.39 &    5.00 &    0.80 &    0.22 &    1.74 &    0.32 &    8.73 &    8.37 &  -17.40 &    2.47 &   -2.21\\
   46 &   -8.47 &   -0.17 &    1.27 &   -0.73 &   -1.94 &   -0.14 &    4.34 &    0.00 &    0.50 &    6.35 &   -2.71 &   -8.02 &    2.23 &    3.73 &    0.48 &   -0.47 &    1.76 &   -1.37 &    0.86 &   -3.27 &    0.54 &   -3.55 &    0.52 &   -4.62 &    0.76 &   -0.79\\
   47 &   -8.47 &    0.91 &    0.45 &   -0.64 &   -0.55 &    0.22 &    2.09 &   -3.37 &    0.86 &    6.94 &   -5.11 &    2.27 &   -0.47 &    4.82 &   -3.90 &   -0.38 &    2.49 &    1.82 &   -0.81 &   -1.70 &   -1.85 &    0.01 &    2.70 &   -1.26 &    0.15 &    3.21\\
   48 &   -8.47 &   -0.39 &   -0.73 &   -1.57 &    3.31 &    1.37 &    5.23 &    1.37 &   -1.41 &    1.38 &    4.91 &   -6.20 &    4.31 &   -0.08 &   -0.36 &   -6.73 &    1.58 &    1.83 &   -1.94 &    0.69 &    1.77 &    1.75 &    2.49 &   -1.38 &    0.83 &    1.73\\
   49 &   -8.47 &   -0.04 &   -0.37 &    0.82 &    2.94 &   -3.09 &    8.60 &    0.29 &   -1.85 &    2.71 &    5.70 &   -2.84 &   -4.57 &   -1.95 &   -3.82 &    0.16 &    0.37 &    3.44 &   -0.22 &   -0.06 &   -0.56 &    0.34 &    0.56 &    0.06 &    0.21 &    3.39\\
   50 &   -8.48 &    0.32 &    0.61 &    2.45 &    3.93 &   -7.98 &    2.06 &    1.24 &    8.00 &   -1.04 &   -0.84 &    0.21 &   -0.57 &    0.75 &    1.25 &   -3.12 &    0.95 &    2.57 &    4.41 &   -0.82 &    2.32 &    0.86 &    1.35 &    0.67 &    0.48 &    2.35\\
   51 &   -8.45 &    3.77 &    9.23 &  -29.52 &    3.30 &    2.81 &   -5.20 &    2.53 &   11.83 &    5.11 &  -16.00 &   -3.97 &  -18.62 &    7.34 &   10.88 &   -6.95 &   16.43 &   35.69 &  -19.63 &    5.70 &    0.80 &   -3.19 &   -0.64 &    1.91 &    1.75 &   -1.58\\
   52 &   -8.46 &    1.44 &    3.59 &  -12.55 &    1.39 &    1.58 &   -1.29 &    1.02 &    2.21 &    4.21 &   -3.19 &    0.65 &   -8.34 &   -0.10 &    1.71 &    1.08 &    4.70 &    3.06 &   -1.91 &    0.49 &    1.05 &   -0.20 &    0.58 &    0.42 &    0.05 &    2.56\\
   53 &   -8.46 &    0.80 &    1.49 &   -4.84 &   -0.54 &    2.39 &   -0.68 &    0.56 &    2.13 &    2.79 &   -2.49 &    0.96 &   -4.78 &    0.54 &   -1.40 &   -0.62 &    3.37 &   -0.89 &    0.46 &   -0.71 &    1.37 &    0.50 &    1.37 &    0.17 &   -0.67 &    4.09\\
   54 &   -8.65 &   20.32 &   12.14 &    4.13 &    1.88 &    6.94 &    0.93 &   -2.78 &   -0.14 &    3.49 &    3.77 &   -1.86 &    2.24 &   -1.59 &   -2.25 &    0.39 &    6.11 &    1.73 &    4.71 &   -0.14 &   -1.24 &    0.27 &    1.25 &   -0.11 &   -2.26 &   -1.57\\
   55 &   -8.47 &    4.02 &    0.50 &   -0.36 &   -0.62 &    0.76 &    2.62 &   -5.38 &   -0.07 &    3.58 &   -0.77 &    0.76 &   -0.68 &    0.72 &   -1.28 &   -2.75 &    4.33 &   -0.12 &    1.68 &   -0.61 &    0.73 &    0.90 &    1.79 &   -1.70 &   -1.00 &    2.11\\
   56 &   -8.47 &    0.86 &    0.50 &   -0.07 &   -1.58 &    0.22 &    3.02 &   -7.42 &    2.14 &    4.55 &   -2.01 &    2.48 &   -1.88 &   -0.90 &   -2.34 &   -3.74 &    4.70 &   -0.50 &    0.81 &   -1.75 &    0.88 &    0.83 &    2.14 &   -3.35 &   -0.43 &    2.58\\
   57 &   -8.47 &    1.76 &   -0.40 &    0.48 &    1.50 &    0.05 &    4.06 &   -9.57 &    0.86 &    5.55 &   -3.82 &    1.09 &   -1.97 &    2.02 &   -1.48 &   -3.47 &    3.18 &   -1.28 &    0.34 &   -2.79 &    0.70 &   -2.24 &    0.56 &   -0.14 &   -0.77 &    4.10\\
   58 &   -8.47 &    0.33 &    0.67 &    0.60 &   -1.50 &   -1.74 &    6.77 &   -2.79 &    0.96 &    2.27 &    1.91 &   -5.61 &   -4.44 &   -5.07 &    2.10 &   -3.96 &    4.39 &   -1.39 &    1.25 &   -4.56 &    2.18 &   -5.80 &   -0.34 &   -0.16 &   -0.53 &    3.19\\
   59 &   -8.47 &    1.15 &    0.52 &   -2.31 &    0.17 &   -1.24 &    6.11 &   -6.12 &   -0.41 &    4.87 &   -0.93 &    0.56 &    6.18 &    1.02 &   -3.14 &   -9.45 &    3.90 &    0.73 &   -1.76 &   -1.99 &    2.02 &   -2.19 &    0.74 &    0.40 &    0.18 &    3.06\\
   60 &   -8.47 &   -0.02 &   -0.71 &    0.26 &    1.13 &    0.47 &    4.47 &   -1.74 &    0.49 &    4.78 &   -2.20 &   -4.58 &    2.27 &   -0.83 &   -1.30 &   -2.88 &    4.07 &   -1.58 &    1.25 &   -7.91 &    0.86 &  -11.42 &   -2.45 &   -1.80 &    0.50 &   -0.63\\
   61 &   -8.47 &    0.52 &   -0.26 &    0.19 &    2.39 &   -2.63 &    8.11 &   -3.04 &   -0.48 &    2.09 &    4.58 &   -3.30 &   -0.35 &   -1.97 &   -1.10 &   -5.22 &    3.09 &    0.36 &   -1.91 &    0.19 &   -1.19 &    0.42 &    1.85 &   -0.24 &   -1.19 &    5.81\\
   62 &   -8.47 &   -0.04 &   -0.31 &    0.03 &    0.90 &   -0.21 &    5.50 &    0.66 &   -3.43 &    5.31 &    3.31 &    0.36 &   -3.80 &    0.06 &   -7.25 &    1.77 &    1.88 &    0.65 &   -0.95 &    1.72 &    2.21 &    0.26 &    0.44 &    1.87 &    3.16 &   -0.49\\
   63 &   -8.47 &   -0.03 &   -0.22 &    2.39 &    3.48 &   -5.32 &    6.97 &    1.21 &    0.52 &    3.37 &    4.01 &   -1.22 &   -5.44 &   -1.03 &   -3.25 &    2.64 &    1.71 &    1.73 &    0.51 &   -0.30 &   -2.30 &   -0.11 &    1.42 &    0.01 &   -0.02 &    4.16\\
   64 &   -8.50 &    0.05 &    2.81 &    4.38 &    3.81 &  -14.06 &  -15.68 &    9.14 &   25.44 &   -5.63 &   -8.31 &    3.97 &    9.46 &    0.23 &   18.29 &   -8.77 &   11.73 &   -4.45 &   10.74 &   -0.87 &   22.17 &    3.93 &   -0.06 &    1.60 &    0.11 &    0.71\\
\hline
\hline
\end{tabular}
\end{tiny}
\tablecaption{
Correlated systematic uncertainties given in per mill
relative to the cross section
measurement reported in \Tab~\ref{tab:all1}. The coefficients
 $\gamma^{\rm ave}_1,\gamma^{\rm ave}_{26}$ represent diagonalised correlated systematic uncertainties (see \Sec~\ref{subsec:aveprocedure}).
\label{tab:all1c}}
\end{sidewaystable}

\begin{table}
\begin{tiny}
\begin{center}
\begin{tabular}{l|cccccccccc}
\hline
\hline
$\#$ &  $Q^2$  & $x$ & $y$ & $F^{\rm th}_L$ & $\sigma^{\rm ave}_{\rm r}$ & $F_2$ & $\delta_{\rm ave,stat}$ & $\delta_{\rm ave,uncor}$ & $\delta_{\rm ave,tot}  $ & CME\\
     & GeV$^2$ &   &       &                &                  &     & \%  &    \%              &   \% & GeV \\
\hline
   65  &   2.0  &$ 0.247 \times 10^{-4}$ &  0.800  & ---  &  0.775  & ---  &  2.32 &  2.70 &  6.03 &    319 \\
   66  &   2.0  &$ 0.293 \times 10^{-4}$ &  0.675  & ---  &  0.792  & ---  &  1.49 &  1.65 &  2.86 &    319 \\
   67  &   2.0  &$ 0.327 \times 10^{-4}$ &  0.675  & ---  &  0.839  & ---  &  1.82 &  5.21 &  6.34 &    301 \\
   68  &   2.0  &$ 0.500 \times 10^{-4}$ &  0.395  & 0.32  & 0.825  & 0.861  & 1.61 &  1.86 &  2.71 &    319 \\
   69  &   2.0  &$ 0.500 \times 10^{-4}$ &  0.442  & 0.32  & 0.856  & 0.903  & 0.92 &  2.45 &  3.00 &    301 \\
   70  &   2.0  &$ 0.800 \times 10^{-4}$ &  0.247  & 0.29  & 0.768  & 0.780  & 0.91 &  1.64 &  2.19 &    319 \\
   71  &   2.0  &$ 0.130 \times 10^{-3}$ &  0.152  & 0.27  & 0.726  & 0.730  & 1.05 &  1.69 &  2.32 &    319 \\
   72  &   2.0  &$ 0.200 \times 10^{-3}$ &  0.099  & 0.25  & 0.679  & 0.680  & 1.09 &  1.78 &  2.51 &    319 \\
   73  &   2.0  &$ 0.320 \times 10^{-3}$ &  0.062  & 0.23  & 0.634  & 0.635  & 1.15 &  1.55 &  2.49 &    319 \\
   74  &   2.0  &$ 0.500 \times 10^{-3}$ &  0.040  & 0.21  & 0.578  & 0.578  & 1.33 &  1.94 &  2.83 &    319 \\
   75  &   2.0  &$ 0.100 \times 10^{-2}$ &  0.020  & 0.19  & 0.510  & 0.510  & 1.15 &  1.69 &  2.42 &    319 \\
   76  &   2.0  &$ 0.320 \times 10^{-2}$ &  0.006  & 0.15  & 0.424  & 0.424  & 1.26 &  1.78 &  2.77 &    319 \\
   77  &   2.0  &$ 0.130 \times 10^{-1}$ &  0.002  & 0.12  & 0.361  & 0.361  & 2.40 &  2.11 &  5.34 &    301 \\
   78  &   2.5  &$ 0.309 \times 10^{-4}$ &  0.800  & ---  &  0.835  & ---  &  2.46 &  2.67 &  5.06 &    319 \\
   79  &   2.5  &$ 0.366 \times 10^{-4}$ &  0.675  & ---  &  0.860  & ---  &  2.29 &  2.29 &  3.74 &    319 \\
   80  &   2.5  &$ 0.409 \times 10^{-4}$ &  0.675  & ---  &  0.920  & ---  &  1.56 &  6.21 &  6.98 &    301 \\
   81  &   2.5  &$ 0.500 \times 10^{-4}$ &  0.494  & 0.35  & 0.861  & 0.930  & 1.51 &  1.65 &  2.51 &    319 \\
   82  &   2.5  &$ 0.500 \times 10^{-4}$ &  0.552  & 0.35  & 0.895  & 0.984  & 1.20 &  2.09 &  3.13 &    301 \\
   83  &   2.5  &$ 0.800 \times 10^{-4}$ &  0.309  & 0.32  & 0.856  & 0.877  & 0.69 &  1.17 &  1.72 &    319 \\
   84  &   2.5  &$ 0.130 \times 10^{-3}$ &  0.190  & 0.30  & 0.795  & 0.801  & 0.73 &  1.14 &  1.73 &    319 \\
   85  &   2.5  &$ 0.200 \times 10^{-3}$ &  0.124  & 0.27  & 0.758  & 0.760  & 0.92 &  1.53 &  2.09 &    319 \\
   86  &   2.5  &$ 0.320 \times 10^{-3}$ &  0.077  & 0.25  & 0.671  & 0.672  & 0.92 &  1.68 &  2.28 &    319 \\
   87  &   2.5  &$ 0.500 \times 10^{-3}$ &  0.049  & 0.23  & 0.630  & 0.631  & 0.90 &  1.39 &  2.09 &    319 \\
   88  &   2.5  &$ 0.800 \times 10^{-3}$ &  0.031  & 0.21  & 0.578  & 0.578  & 1.02 &  1.77 &  2.30 &    319 \\
   89  &   2.5  &$ 0.158 \times 10^{-2}$ &  0.016  & 0.19  & 0.534  & 0.534  & 0.87 &  1.54 &  2.13 &    319 \\
   90  &   2.5  &$ 0.500 \times 10^{-2}$ &  0.005  & 0.16  & 0.439  & 0.439  & 1.01 &  1.69 &  2.59 &    319 \\
   91  &   2.5  &$ 0.200 \times 10^{-1}$ &  0.001  & 0.12  & 0.342  & 0.342  & 2.52 &  2.45 &  8.69 &    319 \\
   92  &   3.5  &$ 0.432 \times 10^{-4}$ &  0.800  & ---  &  0.877  & ---  &  3.09 &  2.83 &  5.75 &    319 \\
   93  &   3.5  &$ 0.512 \times 10^{-4}$ &  0.675  & ---  &  0.940  & ---  &  2.16 &  2.25 &  3.61 &    319 \\
   94  &   3.5  &$ 0.573 \times 10^{-4}$ &  0.675  & ---  &  0.931  & ---  &  2.00 &  6.18 &  6.94 &    301 \\
   95  &   3.5  &$ 0.800 \times 10^{-4}$ &  0.432  & 0.38  & 0.954  & 1.007  & 1.29 &  1.64 &  2.33 &    319 \\
   96  &   3.5  &$ 0.800 \times 10^{-4}$ &  0.483  & 0.38  & 0.950  & 1.020  & 1.00 &  1.75 &  2.67 &    301 \\
   97  &   3.5  &$ 0.130 \times 10^{-3}$ &  0.266  & 0.35  & 0.918  & 0.934  & 0.66 &  1.06 &  1.60 &    319 \\
   98  &   3.5  &$ 0.200 \times 10^{-3}$ &  0.173  & 0.32  & 0.859  & 0.865  & 0.69 &  1.07 &  1.64 &    319 \\
   99  &   3.5  &$ 0.320 \times 10^{-3}$ &  0.108  & 0.29  & 0.800  & 0.802  & 0.74 &  1.12 &  1.70 &    319 \\
  100  &   3.5  &$ 0.500 \times 10^{-3}$ &  0.069  & 0.27  & 0.759  & 0.760  & 0.83 &  1.31 &  1.91 &    319 \\
  101  &   3.5  &$ 0.800 \times 10^{-3}$ &  0.043  & 0.25  & 0.661  & 0.662  & 0.69 &  1.15 &  1.71 &    319 \\
  102  &   3.5  &$ 0.130 \times 10^{-2}$ &  0.027  & 0.22  & 0.626  & 0.626  & 0.89 &  1.36 &  1.98 &    319 \\
  103  &   3.5  &$ 0.251 \times 10^{-2}$ &  0.014  & 0.20  & 0.556  & 0.556  & 0.64 &  1.11 &  1.69 &    319 \\
  104  &   3.5  &$ 0.800 \times 10^{-2}$ &  0.004  & 0.16  & 0.448  & 0.448  & 0.84 &  1.48 &  2.32 &    319 \\
  105  &   5.0  &$ 0.618 \times 10^{-4}$ &  0.800  & ---  &  0.990  & ---  &  3.13 &  2.78 &  5.61 &    319 \\
  106  &   5.0  &$ 0.732 \times 10^{-4}$ &  0.675  & ---  &  1.056  & ---  &  1.60 &  2.21 &  3.02 &    319 \\
  107  &   5.0  &$ 0.818 \times 10^{-4}$ &  0.675  & ---  &  1.047  & ---  &  2.08 &  4.85 &  6.07 &    301 \\
  108  &   5.0  &$ 0.130 \times 10^{-3}$ &  0.380  & 0.41  & 1.066  & 1.108  & 1.33 &  2.20 &  2.76 &    319 \\
  109  &   5.0  &$ 0.130 \times 10^{-3}$ &  0.425  & 0.41  & 1.053  & 1.108  & 1.02 &  1.68 &  2.28 &    301 \\
  110  &   5.0  &$ 0.200 \times 10^{-3}$ &  0.247  & 0.37  & 1.011  & 1.025  & 0.74 &  1.19 &  1.75 &    319 \\
  111  &   5.0  &$ 0.320 \times 10^{-3}$ &  0.154  & 0.34  & 0.931  & 0.936  & 0.80 &  1.28 &  1.81 &    319 \\
  112  &   5.0  &$ 0.500 \times 10^{-3}$ &  0.099  & 0.31  & 0.839  & 0.841  & 0.80 &  1.28 &  1.83 &    319 \\
  113  &   5.0  &$ 0.800 \times 10^{-3}$ &  0.062  & 0.28  & 0.753  & 0.754  & 0.82 &  1.29 &  1.84 &    319 \\
  114  &   5.0  &$ 0.130 \times 10^{-2}$ &  0.038  & 0.25  & 0.696  & 0.696  & 0.85 &  1.31 &  1.93 &    319 \\
  115  &   5.0  &$ 0.200 \times 10^{-2}$ &  0.025  & 0.23  & 0.639  & 0.639  & 0.88 &  1.31 &  1.89 &    319 \\
  116  &   5.0  &$ 0.398 \times 10^{-2}$ &  0.012  & 0.20  & 0.569  & 0.569  & 0.67 &  1.22 &  1.81 &    319 \\
  117  &   5.0  &$ 0.130 \times 10^{-1}$ &  0.004  & 0.16  & 0.438  & 0.438  & 0.80 &  1.82 &  2.60 &    319 \\
\hline
\hline
\end{tabular}
\end{center}
\end{tiny}
\tablecaption{ Combined H1 reduced cross section $\sigma^{\rm ave}_{\rm r}$  for  $2\le Q^2\le 5$~GeV$^2$. 
The uncertainties are quoted in \% relative to $\sigma^{\rm ave}_{\rm r}$.
$F^{\rm th}_L$ represents the structure function $F_L$ used for the CME
 correction (\Eq~\ref{eq:cmecorr}) and to calculate the structure function $F_2$.
$\delta_{\rm ave,stat}$ ($\delta_{\rm ave,uncor}$) represents the statistical (uncorrelated systematic)  uncertainty. 
$\delta_{\rm ave,tot}$ 
is the total uncertainty calculated as a sum of uncorrelated uncertainty and all correlated sources in quadrature.
A global normalisation uncertainty of $0.5\%$ is not included in $\delta_{\rm ave,tot}$.
CME stands for the centre-of-mass energy of the measurement.
\label{tab:all2}}
\end{table}

\begin{sidewaystable}
\begin{tiny}
\begin{tabular}{c|rrrrrrrrrrrrrrrrrrrrrrrrrr}
\hline
\hline
$\#$  & $\gamma^{\rm ave}_1$ &$\gamma^{\rm ave}_2$ &$\gamma^{\rm ave}_3$ &$\gamma^{\rm ave}_4$ &$\gamma^{\rm ave}_5$ &
   $\gamma^{\rm ave}_6$ &$\gamma^{\rm ave}_7$ &$\gamma^{\rm ave}_8$ &$\gamma^{\rm ave}_9$ &$\gamma^{\rm ave}_{10}$ &
   $\gamma^{\rm ave}_{11}$ &$\gamma^{\rm ave}_{12}$ &$\gamma^{\rm ave}_{13}$ &$\gamma^{\rm ave}_{14}$ &$\gamma^{\rm ave}_{15}$ &
   $\gamma^{\rm ave}_{16}$ &$\gamma^{\rm ave}_{17}$ &$\gamma^{\rm ave}_{18}$ &$\gamma^{\rm ave}_{19}$ &$\gamma^{\rm ave}_{20}$  & 
   $\gamma^{\rm ave}_{21}$ &$\gamma^{\rm ave}_{22}$ &$\gamma^{\rm ave}_{23}$ &$\gamma^{\rm ave}_{24}$ &$\gamma^{\rm ave}_{25}$ &
   $\gamma^{\rm ave}_{26}$ \\
  & \permil &\permil &\permil &\permil &\permil &\permil &\permil &\permil &\permil &\permil &\permil &\permil &\permil &\permil &\permil &\permil &\permil &\permil &\permil &\permil &\permil &\permil &\permil &\permil &\permil &\permil \\
\hline
   65 &  -8.45 &   2.93 &   7.08 & -22.67 &   2.14 &   2.65 &  -4.10 &   1.48 &   7.64 &   6.42 & -10.74 &  -2.44 & -15.86 &   4.69 &   6.62 &  -1.76 &  17.36 &  20.72 & -18.15 &   8.18 &   0.83 &  -3.09 &  -2.60 &   1.13 &   1.51 &  -1.28\\
   66 &  -8.46 &   1.33 &   3.06 & -10.30 &   1.08 &   0.83 &  -1.52 &   0.18 &   2.05 &   4.65 &  -3.35 &   1.92 &  -6.78 &   0.40 &   1.43 &   2.12 &   4.52 &   2.63 &  -2.21 &   0.96 &   0.72 &  -0.23 &  -0.17 &   0.24 &   0.12 &   2.09\\
   67 &  -8.65 &  16.65 &  17.02 &   5.22 &   2.96 &   3.20 &  -6.82 &  -4.37 &  -1.78 &   2.03 &  10.40 &  -1.38 &  -1.14 &   5.29 &  -1.14 &  -2.75 &   6.23 &   0.05 &   2.70 &  -4.25 &  -1.60 &  -1.02 &  -0.99 &   0.12 &  -1.15 &  -1.24\\
   68 &  -8.47 &   0.51 &   0.65 &  -1.95 &  -0.57 &   1.64 &   0.76 &  -0.14 &   1.56 &   2.16 &  -0.82 &   0.04 &  -3.32 &  -0.02 &  -1.30 &  -1.36 &   3.29 &  -0.85 &  -0.36 &   0.05 &   1.04 &   0.38 &   1.06 &   0.21 &  -0.38 &   3.60\\
   69 &  -8.55 &   7.89 &   3.89 &   1.58 &   0.45 &   0.84 &  -0.63 &  -1.59 &   1.39 &   3.68 &   2.72 &  -1.41 &  -0.21 &  -1.15 &  -2.14 &  -1.17 &   3.86 &   0.10 &   1.32 &  -1.87 &  -1.05 &   1.15 &   0.79 &  -0.67 &  -1.62 &  -1.44\\
   70 &  -8.47 &   0.29 &   0.06 &   0.20 &  -0.62 &   1.17 &   0.34 &  -0.02 &   1.94 &   3.44 &   1.87 &  -1.11 &  -1.63 &  -0.91 &  -1.12 &  -1.82 &   3.91 &  -0.22 &   0.77 &  -1.59 &  -0.14 &   1.76 &   1.19 &  -0.93 &  -1.17 &   1.18\\
   71 &  -8.48 &  -0.25 &   0.08 &   0.44 &  -0.88 &   0.85 &   0.73 &   0.05 &   2.18 &   2.91 &   2.36 &  -0.77 &  -2.56 &  -2.39 &  -0.97 &  -2.05 &   4.09 &  -0.10 &   0.73 &  -1.40 &   0.21 &   1.81 &   1.80 &  -1.99 &  -0.98 &   1.24\\
   72 &  -8.47 &   0.16 &  -0.51 &   0.96 &   1.08 &   2.02 &   5.83 &   1.96 &   2.61 &   2.94 &   1.53 &  -1.69 &  -3.07 &  -0.72 &   1.12 &  -2.69 &   4.57 &  -1.37 &   0.22 &  -3.61 &  -0.41 &   1.60 &  -0.30 &   0.52 &  -1.16 &   1.44\\
   73 &  -8.48 &  -0.37 &  -0.63 &   0.31 &   1.69 &  -0.33 &   5.39 &   1.71 &   0.26 &   2.97 &   3.01 &  -1.17 &   4.00 &  -0.09 &  -1.66 &  -6.90 &   4.42 &  -0.49 &  -1.35 &  -3.44 &  -0.12 &   0.98 &   3.73 &   2.24 &   1.41 &   0.82\\
   74 &  -8.47 &  -0.47 &  -0.32 &  -0.37 &  -1.67 &   2.06 &   5.90 &   2.25 &  -3.24 &   0.19 &   2.24 &  -2.87 &  -0.09 &  -0.49 &   0.66 &  -5.59 &   7.25 &  -0.63 &   1.17 &  -3.24 &  -1.27 &   2.32 &  -1.16 &   0.51 &  -0.81 &   1.66\\
   75 &  -8.47 &  -0.29 &  -0.14 &   0.98 &   1.20 &  -1.16 &   4.49 &   2.88 &  -0.96 &   0.63 &   1.95 &  -1.65 &  -3.14 &  -0.88 &  -0.38 &  -2.68 &   2.98 &  -0.12 &  -1.16 &  -2.96 &  -2.42 &   2.35 &   0.26 &   0.50 &  -0.61 &   2.04\\
   76 &  -8.47 &  -0.33 &   0.09 &   2.89 &   3.92 &  -6.03 &   5.98 &   4.75 &   0.86 &   1.99 &   2.92 &  -1.63 &  -6.35 &  -1.01 &  -2.15 &   1.35 &   2.32 &   0.15 &  -1.26 &  -2.56 &  -3.68 &   1.33 &   0.86 &   0.25 &  -0.08 &   3.28\\
   77 &  -8.48 &   0.07 &   2.19 &   3.47 &   2.59 & -10.31 & -12.26 &   7.77 &  19.82 &  -3.09 &  -6.44 &   4.04 &   7.82 &   1.01 &  12.15 &  -5.37 &  12.20 &  -4.28 &   7.70 &  -0.94 &  21.83 &   3.98 &  -0.54 &   1.37 &  -0.07 &   1.41\\
   78 &  -8.46 &   1.84 &   4.77 & -15.40 &   1.73 &   0.03 &  -3.07 &   1.02 &   2.68 &  10.09 &  -6.20 &   0.94 &  -8.09 &   1.38 &   9.33 &  -2.03 &  10.32 &  17.91 & -11.40 &   3.73 &   0.30 &  -1.69 &   0.22 &   1.89 &   2.49 &  -0.99\\
   79 &  -8.46 &   0.95 &   2.36 &  -7.68 &   0.57 &   0.27 &  -1.35 &  -0.25 &   1.32 &   4.95 &  -2.34 &   1.21 &  -4.60 &  -0.10 &   4.38 &   0.60 &   4.61 &   8.20 &  -5.85 &   3.24 &   0.20 &  -0.77 &  -2.25 &   0.22 &   0.96 &  -0.36\\
   80 &  -8.64 &  17.04 &  12.77 &   4.46 &   2.83 &   2.62 &  -3.61 &  -2.45 &  -0.89 &   2.96 &   1.25 &  -5.00 &   1.58 &  -4.23 &  -3.60 &  -0.61 &   7.65 &   0.83 &   3.12 &  -3.02 &  -3.04 &   4.20 &  -0.90 &   1.61 &  -1.91 &  -1.54\\
   81 &  -8.46 &   0.48 &   1.22 &  -3.92 &   0.06 &   0.76 &  -0.26 &   0.52 &   0.84 &   3.33 &  -0.36 &   0.94 &  -3.77 &  -0.95 &   0.89 &   0.99 &   2.95 &   0.39 &  -0.97 &   0.77 &   0.69 &   0.06 &   0.01 &   0.21 &   0.23 &   2.09\\
   82 &  -8.59 &   9.90 &   7.63 &   2.40 &   0.93 &   3.54 &  -1.07 &  -2.87 &   5.04 &   5.59 &   4.95 &  -1.84 &   0.45 &  -2.31 &  -3.08 &  -0.81 &   4.43 &   0.41 &   1.82 &  -2.15 &  -1.60 &   1.44 &   0.32 &   1.12 &  -2.45 &  -1.66\\
   83 &  -8.46 &   1.55 &   1.07 &   0.01 &  -0.14 &   0.78 &   0.21 &  -0.91 &   2.11 &   2.91 &   1.61 &  -0.53 &  -1.10 &  -0.54 &  -1.58 &  -0.79 &   3.08 &  -0.38 &   0.43 &  -0.47 &  -0.19 &   1.07 &   0.56 &   0.28 &  -0.99 &   0.23\\
   84 &  -8.47 &  -0.12 &   0.15 &   0.11 &  -0.34 &   0.21 &   0.21 &  -0.49 &   1.38 &   2.00 &   1.57 &  -0.86 &  -1.82 &  -0.22 &  -2.04 &  -1.37 &   4.35 &  -0.87 &   0.42 &  -0.98 &  -0.58 &   2.08 &   0.71 &  -0.10 &  -0.99 &   0.06\\
   85 &  -8.47 &   0.22 &   0.25 &   0.70 &   0.06 &   0.52 &   2.87 &   0.28 &   1.95 &   1.95 &   0.76 &  -0.14 &  -2.89 &  -1.40 &  -1.25 &  -1.72 &   2.67 &  -0.90 &  -0.17 &  -0.19 &   0.78 &   0.55 &   1.90 &  -1.12 &  -1.00 &   1.13\\
   86 &  -8.47 &   0.08 &  -0.51 &   0.77 &   1.19 &   0.58 &   4.35 &  -1.09 &   1.71 &   2.94 &   2.50 &  -2.13 &  -3.54 &  -1.85 &   1.27 &  -1.74 &   2.13 &  -0.52 &   0.71 &  -0.37 &   1.17 &   0.11 &   1.77 &  -0.05 &  -1.71 &   1.65\\
   87 &  -8.47 &  -0.19 &  -0.38 &   0.57 &  -0.28 &   0.05 &   2.86 &   0.85 &  -0.35 &   2.72 &   0.87 &   0.35 &   2.83 &   0.70 &  -2.13 &  -3.04 &   3.35 &  -1.30 &  -0.27 &  -1.59 &  -0.38 &  -0.84 &   5.00 &   2.56 &   1.34 &   0.57\\
   88 &  -8.47 &  -0.25 &  -0.28 &   0.51 &  -0.51 &   0.91 &   3.21 &   1.90 &   0.29 &   2.93 &   0.93 &   0.63 &  -0.17 &  -0.01 &  -1.66 &  -1.53 &   2.45 &   0.04 &  -0.07 &  -0.97 &  -0.78 &  -0.27 &   1.07 &   0.25 &  -1.62 &   1.44\\
   89 &  -8.47 &  -0.17 &  -0.31 &   0.79 &   0.53 &   0.32 &   4.24 &   2.12 &   0.33 &   2.35 &   1.30 &   0.74 &  -2.17 &   0.31 &  -2.83 &   0.26 &   2.68 &  -0.80 &  -1.18 &  -0.20 &  -2.92 &   0.03 &   1.96 &   0.63 &  -0.88 &   1.66\\
   90 &  -8.48 &  -0.14 &   0.31 &   2.92 &   3.92 &  -6.34 &   4.30 &   4.82 &   4.73 &   3.28 &   0.90 &   1.01 &  -4.45 &   1.01 &  -4.64 &   2.25 &   2.39 &  -0.56 &  -1.40 &  -1.27 &  -2.03 &   0.64 &   1.12 &   0.38 &  -0.23 &   2.73\\
   91 &  -8.51 &  -0.05 &   3.62 &   5.10 &   2.44 & -14.55 & -24.91 &  14.68 &  31.99 &  -5.24 & -13.43 &  11.07 &  19.11 &   5.27 &  16.83 &  -5.88 &  20.59 &  -5.17 &  13.72 &  -3.40 &  48.22 &   9.22 &  -4.88 &   1.94 &  -0.51 &  -0.65\\
   92 &  -8.45 &   2.27 &   5.42 & -18.65 &   0.50 &   2.91 &  -4.76 &  -1.17 &   2.95 &   7.20 &  -7.19 &   0.58 &  -9.04 &   1.19 &  11.18 &   0.99 &  11.06 &  18.82 & -14.80 &   5.01 &   0.50 &  -1.84 &  -3.66 &  -0.12 &   0.20 &  -0.68\\
   93 &  -8.46 &   0.93 &   2.28 &  -7.45 &   0.45 &   0.50 &  -1.29 &  -0.36 &   1.29 &   4.59 &  -2.14 &   1.08 &  -4.62 &  -0.09 &   3.83 &   0.97 &   5.05 &   7.01 &  -6.05 &   3.52 &   0.23 &  -0.83 &  -2.28 &   0.22 &   0.97 &  -0.36\\
   94 &  -8.62 &  14.45 &  11.11 &   3.86 &   2.27 &   0.47 &  -5.96 &  -3.19 &  -2.03 &   2.80 &   1.24 &  -4.45 &   1.66 &  -4.05 &  -3.88 &   0.08 &   5.06 &   1.25 &   2.79 &  -1.03 &  -1.71 &   2.13 &   0.34 &   1.27 &  -2.15 &  -1.33\\
   95 &  -8.46 &   0.51 &   0.54 &  -1.60 &  -0.16 &   0.11 &   0.72 &  -1.61 &   0.40 &   3.28 &  -0.06 &   1.21 &  -2.02 &  -0.87 &   0.90 &   0.10 &   2.20 &   0.12 &  -0.90 &   0.94 &   0.65 &   0.23 &  -0.26 &   0.20 &   0.31 &   1.87\\
   96 &  -8.55 &   8.15 &   6.46 &   2.97 &   2.19 &  -1.33 &  -1.15 &   0.34 &  -6.57 &  -0.64 &  -0.53 &  -1.98 &  -0.43 &  -0.30 &  -1.02 &  -1.50 &   5.67 &   0.01 &   1.92 &  -2.44 &  -1.93 &   2.64 &  -0.58 &   0.68 &  -1.32 &  -1.14\\
   97 &  -8.46 &   0.45 &   0.45 &   0.25 &  -0.21 &  -0.25 &   0.26 &  -0.89 &   1.48 &   3.17 &   1.40 &   0.21 &  -0.75 &  -0.57 &  -1.26 &  -0.85 &   2.42 &  -0.34 &   0.36 &  -0.74 &  -0.34 &   0.85 &   0.38 &   0.41 &  -1.01 &   0.02\\
   98 &  -8.47 &  -0.38 &  -0.20 &   0.03 &  -0.35 &  -0.01 &   0.35 &  -0.85 &   1.88 &   2.98 &   1.95 &  -0.16 &  -1.34 &  -0.72 &  -1.28 &  -0.90 &   2.75 &  -0.51 &   0.31 &  -0.49 &  -0.32 &   1.42 &   0.76 &  -0.01 &  -0.98 &   0.01\\
   99 &  -8.47 &   0.04 &   0.10 &   0.80 &   0.60 &  -0.57 &   2.40 &   0.24 &   1.00 &   2.66 &   0.64 &   1.12 &  -2.14 &  -1.13 &  -1.21 &  -0.37 &   0.66 &  -1.14 &  -0.75 &   1.72 &   1.15 &  -0.39 &   2.18 &  -0.46 &  -1.08 &   0.26\\
  100 &  -8.47 &  -0.18 &  -0.33 &   0.41 &   0.06 &   0.24 &   1.86 &   0.51 &   1.55 &   4.08 &   1.34 &   1.52 &   1.16 &  -0.32 &  -2.45 &  -1.01 &  -0.25 &   0.37 &  -0.48 &   0.96 &   0.46 &  -1.75 &   2.10 &   0.32 &  -2.32 &   0.35\\
  101 &  -8.47 &  -0.29 &  -0.52 &   0.32 &  -0.40 &   0.43 &   1.67 &  -0.08 &  -2.43 &   1.47 &   1.20 &   0.91 &   1.19 &   0.55 &  -1.52 &  -0.51 &   2.95 &  -0.05 &   1.24 &   0.03 &  -0.74 &  -0.26 &   2.94 &   1.35 &  -0.03 &   0.12\\
  102 &  -8.47 &  -0.03 &  -0.43 &   0.67 &   0.14 &   1.08 &   3.36 &   0.94 &   1.64 &   4.21 &   0.60 &   2.20 &  -0.18 &   0.05 &  -2.05 &  -0.41 &  -0.26 &   0.14 &  -0.52 &   0.70 &  -0.20 &  -1.09 &   2.18 &   0.31 &  -2.32 &   0.42\\
  103 &  -8.47 &  -0.18 &  -0.37 &   0.72 &   0.48 &  -0.45 &   3.39 &   0.87 &  -0.43 &   2.59 &   1.47 &   1.43 &  -1.28 &   0.33 &  -2.77 &   1.08 &   1.78 &   0.08 &   0.34 &   1.03 &  -1.51 &  -0.62 &   1.96 &   0.68 &  -1.20 &   0.44\\
  104 &  -8.48 &  -0.12 &   0.68 &   2.99 &   3.46 &  -7.43 &   0.68 &   5.19 &   4.35 &   3.65 &  -0.92 &   2.05 &  -3.06 &   1.30 &  -3.22 &   0.14 &   3.08 &  -1.61 &   0.05 &   0.10 &  -0.13 &   0.73 &   0.11 &   0.59 &   0.44 &   1.53\\
  105 &  -8.46 &   1.80 &   4.92 & -14.97 &   2.59 &  -1.94 &  -2.92 &   2.30 &   1.72 &  14.68 &  -5.70 &   1.97 &  -9.50 &   0.82 &   9.43 &  -2.10 &  11.20 &  17.94 & -11.17 &   5.22 &   0.25 &  -1.82 &  -1.08 &   1.70 &   2.67 &  -0.99\\
  106 &  -8.46 &   0.50 &   1.24 &  -4.03 &   0.10 &   0.05 &  -0.67 &  -0.67 &   0.33 &   3.67 &  -0.41 &   1.63 &  -2.41 &  -0.96 &   2.76 &   1.49 &   2.55 &   3.52 &  -3.39 &   2.73 &   0.06 &  -0.40 &  -2.26 &   0.00 &   0.83 &  -0.17\\
  107 &  -8.66 &  19.06 &  14.15 &   5.00 &   3.27 &   3.75 &  -3.04 &  -2.46 &   0.80 &   3.95 &   1.40 &  -5.04 &   1.88 &  -5.00 &  -3.91 &  -0.38 &   7.23 &   0.85 &   2.97 &  -2.86 &  -2.95 &   4.10 &  -0.70 &   1.67 &  -2.10 &  -1.62\\
  108 &  -8.46 &   0.04 &   0.14 &  -0.41 &  -0.13 &  -0.54 &   0.28 &  -0.74 &  -0.67 &   2.96 &   1.41 &   2.03 &  -0.01 &  -1.63 &   1.27 &   1.54 &   0.48 &  -0.21 &  -0.63 &   1.62 &  -0.07 &  -0.01 &  -1.03 &   0.39 &   1.29 &  -0.09\\
  109 &  -8.51 &   2.68 &   2.06 &   0.99 &   0.37 &  -0.45 &  -0.48 &  -0.85 &   0.05 &   2.70 &   1.99 &  -0.99 &  -0.54 &  -1.07 &  -1.83 &  -1.01 &   3.54 &   0.06 &   1.14 &  -1.47 &  -1.32 &   1.98 &   0.36 &   1.11 &  -2.17 &  -1.38\\
  110 &  -8.46 &  -0.04 &   0.00 &   0.10 &  -0.23 &  -0.48 &   0.02 &  -0.94 &   1.92 &   3.66 &   2.58 &   0.32 &  -0.60 &  -1.15 &  -1.10 &  -0.31 &   1.76 &  -0.26 &   0.36 &  -0.29 &  -0.74 &   1.08 &   0.20 &   0.90 &  -1.19 &  -1.02\\
  111 &  -8.46 &  -0.05 &  -0.01 &   0.61 &   0.35 &  -1.54 &   0.83 &   0.12 &  -1.36 &   2.69 &   0.93 &   1.92 &  -0.43 &  -1.04 &  -0.39 &   0.65 &  -0.41 &  -0.39 &  -0.19 &   2.04 &   0.67 &  -1.07 &   1.36 &   0.55 &  -1.14 &  -0.75\\
  112 &  -8.47 &  -0.17 &  -0.13 &   0.42 &   0.10 &  -1.25 &   0.56 &  -0.32 &  -0.92 &   3.65 &   1.67 &   1.70 &  -0.43 &  -1.17 &  -0.75 &   0.83 &   0.54 &  -0.74 &  -0.12 &   2.12 &   0.51 &  -0.91 &   1.31 &   0.70 &  -1.10 &  -0.84\\
  113 &  -8.47 &  -0.20 &  -0.40 &   0.66 &  -0.05 &  -0.45 &   1.66 &  -0.58 &  -0.76 &   2.03 &   1.70 &   1.40 &   0.58 &  -0.79 &  -1.28 &   0.20 &  -1.27 &   0.84 &   0.77 &   0.67 &   0.92 &  -0.84 &   3.05 &   1.21 &  -0.81 &  -0.89\\
  114 &  -8.47 &  -0.10 &  -0.57 &   0.25 &  -0.22 &   1.30 &   3.40 &  -0.59 &   0.98 &   3.05 &   2.16 &   2.04 &   0.21 &  -0.91 &  -1.49 &   0.95 &  -1.13 &   0.84 &   0.89 &   2.09 &   1.16 &  -1.42 &   2.54 &   0.88 &  -1.46 &  -0.95\\
  115 &  -8.47 &  -0.30 &  -0.47 &  -0.23 &  -0.72 &   0.43 &   1.45 &  -1.12 &  -2.73 &   0.58 &   1.64 &   0.76 &   0.18 &   0.15 &  -0.83 &   0.34 &   2.86 &   0.57 &   2.39 &   1.24 &  -0.30 &  -0.08 &   1.12 &   0.90 &  -1.02 &  -0.89\\
  116 &  -8.47 &  -0.07 &  -0.26 &   1.33 &   1.09 &  -1.28 &   3.77 &   1.55 &   0.41 &   4.00 &   1.27 &   2.10 &  -1.76 &  -0.70 &  -2.08 &  -0.30 &  -1.38 &   0.41 &   0.59 &   1.64 &   0.85 &  -0.61 &   1.06 &   0.65 &  -1.22 &  -0.90\\
  117 &  -8.48 &  -0.12 &   0.86 &   2.97 &   3.00 &  -7.66 &  -1.74 &   4.35 &   5.83 &   4.88 &  -1.40 &   1.72 &  -3.17 &  -0.97 &   0.24 &  -4.68 &  -1.40 &  -1.78 &   1.56 &   2.23 &  -1.66 &  -0.25 &  -0.32 &   0.67 &   0.69 &  -0.50\\
\hline
\hline
\end{tabular}
\end{tiny}
\tablecaption{
Correlated systematic uncertainties given in per mill
relative to the cross section
measurement reported in \Tab~\ref{tab:all2}. The coefficients
 $\gamma^{\rm ave}_1,\gamma^{\rm ave}_{26}$ represent diagonalised correlated systematic uncertainties (see \Sec~\ref{subsec:aveprocedure}).
\label{tab:all2c}}
\end{sidewaystable}

\begin{table}
\begin{tiny}
\begin{center}
\begin{tabular}{l|cccccccccc}
\hline
\hline
$\#$ &  $Q^2$  & $x$ & $y$ & $F^{\rm th}_L$ & $\sigma^{\rm ave}_{\rm r}$ & $F_2$ & $\delta_{\rm ave,stat}$ & $\delta_{\rm ave,uncor}$ & $\delta_{\rm ave,tot}  $ & CME\\
     & GeV$^2$ &   &       &                &                  &     & \%  &    \%              &   \% & GeV \\
\hline
  118  &   6.5  &$ 0.803 \times 10^{-4}$ &  0.800  & ---  &  1.083  & ---  &  3.11 &  2.64 &  4.85 &    319 \\
  119  &   6.5  &$ 0.951 \times 10^{-4}$ &  0.675  & ---  &  1.053  & ---  &  2.95 &  2.31 &  4.16 &    319 \\
  120  &   6.5  &$ 0.130 \times 10^{-3}$ &  0.494  & 0.45  & 1.123  & 1.211  & 1.67 &  2.22 &  2.97 &    319 \\
  121  &   6.5  &$ 0.130 \times 10^{-3}$ &  0.552  & 0.45  & 1.124  & 1.239  & 1.53 &  1.73 &  3.03 &    301 \\
  122  &   6.5  &$ 0.200 \times 10^{-3}$ &  0.321  & 0.41  & 1.123  & 1.152  & 1.25 &  2.20 &  2.72 &    319 \\
  123  &   6.5  &$ 0.200 \times 10^{-3}$ &  0.359  & 0.41  & 1.117  & 1.155  & 1.09 &  1.62 &  2.24 &    301 \\
  124  &   6.5  &$ 0.320 \times 10^{-3}$ &  0.201  & 0.37  & 1.006  & 1.015  & 0.84 &  1.16 &  1.75 &    319 \\
  125  &   6.5  &$ 0.500 \times 10^{-3}$ &  0.128  & 0.34  & 0.936  & 0.939  & 0.86 &  1.26 &  1.86 &    319 \\
  126  &   6.5  &$ 0.800 \times 10^{-3}$ &  0.080  & 0.31  & 0.854  & 0.855  & 0.87 &  1.26 &  1.91 &    319 \\
  127  &   6.5  &$ 0.130 \times 10^{-2}$ &  0.049  & 0.28  & 0.758  & 0.758  & 0.90 &  1.28 &  1.87 &    319 \\
  128  &   6.5  &$ 0.200 \times 10^{-2}$ &  0.032  & 0.26  & 0.694  & 0.694  & 0.92 &  1.29 &  1.89 &    319 \\
  129  &   6.5  &$ 0.398 \times 10^{-2}$ &  0.016  & 0.22  & 0.616  & 0.617  & 0.69 &  1.19 &  1.76 &    319 \\
  130  &   6.5  &$ 0.130 \times 10^{-1}$ &  0.005  & 0.18  & 0.482  & 0.482  & 0.73 &  1.80 &  2.44 &    319 \\
  131  &   8.5  &$ 0.105 \times 10^{-3}$ &  0.800  & ---  &  1.178  & ---  &  3.72 &  2.80 &  5.20 &    319 \\
  132  &   8.5  &$ 0.124 \times 10^{-3}$ &  0.675  & ---  &  1.211  & ---  &  2.26 &  2.28 &  3.44 &    319 \\
  133  &   8.5  &$ 0.139 \times 10^{-3}$ &  0.675  & ---  &  1.136  & ---  &  2.07 &  1.82 &  4.54 &    301 \\
  134  &   8.5  &$ 0.200 \times 10^{-3}$ &  0.420  & 0.46  & 1.178  & 1.239  & 1.52 &  2.22 &  2.88 &    319 \\
  135  &   8.5  &$ 0.200 \times 10^{-3}$ &  0.469  & 0.46  & 1.182  & 1.261  & 1.38 &  1.64 &  2.59 &    301 \\
  136  &   8.5  &$ 0.320 \times 10^{-3}$ &  0.262  & 0.41  & 1.112  & 1.131  & 0.91 &  1.25 &  1.86 &    319 \\
  137  &   8.5  &$ 0.500 \times 10^{-3}$ &  0.168  & 0.37  & 1.033  & 1.039  & 0.95 &  1.18 &  1.81 &    319 \\
  138  &   8.5  &$ 0.800 \times 10^{-3}$ &  0.105  & 0.34  & 0.950  & 0.953  & 0.95 &  1.28 &  1.90 &    319 \\
  139  &   8.5  &$ 0.130 \times 10^{-2}$ &  0.065  & 0.30  & 0.842  & 0.842  & 0.99 &  1.30 &  1.94 &    319 \\
  140  &   8.5  &$ 0.200 \times 10^{-2}$ &  0.042  & 0.28  & 0.773  & 0.773  & 1.00 &  1.30 &  1.93 &    319 \\
  141  &   8.5  &$ 0.320 \times 10^{-2}$ &  0.026  & 0.25  & 0.663  & 0.663  & 1.04 &  1.32 &  1.99 &    319 \\
  142  &   8.5  &$ 0.631 \times 10^{-2}$ &  0.013  & 0.22  & 0.604  & 0.604  & 0.79 &  1.24 &  1.83 &    319 \\
  143  &   8.5  &$ 0.200 \times 10^{-1}$ &  0.004  & 0.17  & 0.456  & 0.456  & 0.88 &  1.82 &  2.67 &    319 \\
  144  &  12.0  &$ 0.800 \times 10^{-3}$ &  0.148  & 0.38  & 1.053  & 1.058  & 1.07 &  1.30 &  1.99 &    319 \\
  145  &  12.0  &$ 0.130 \times 10^{-2}$ &  0.091  & 0.34  & 0.923  & 0.924  & 1.10 &  1.30 &  1.97 &    319 \\
  146  &  12.0  &$ 0.200 \times 10^{-2}$ &  0.059  & 0.31  & 0.861  & 0.861  & 1.11 &  1.33 &  2.00 &    319 \\
  147  &  12.0  &$ 0.320 \times 10^{-2}$ &  0.037  & 0.28  & 0.757  & 0.757  & 1.14 &  1.34 &  2.02 &    319 \\
  148  &  12.0  &$ 0.631 \times 10^{-2}$ &  0.019  & 0.24  & 0.646  & 0.646  & 0.88 &  1.24 &  1.86 &    319 \\
  149  &  12.0  &$ 0.200 \times 10^{-1}$ &  0.006  & 0.19  & 0.490  & 0.490  & 0.93 &  1.83 &  2.51 &    319 \\
\hline
\hline
\end{tabular}
\end{center}
\end{tiny}
\tablecaption{ Combined H1 reduced cross section $\sigma^{\rm ave}_{\rm r}$  for  $6.5\le Q^2\le 12$~GeV$^2$. 
The uncertainties are quoted in \% relative to $\sigma^{\rm ave}_{\rm r}$.
$F^{\rm th}_L$ represents the structure function $F_L$ used for the CME
 correction (\Eq~\ref{eq:cmecorr}) and to calculate the structure function $F_2$.
$\delta_{\rm ave,stat}$ ($\delta_{\rm ave,uncor}$) represents the statistical (uncorrelated systematic)  uncertainty. 
$\delta_{\rm ave,tot}$ 
is the total uncertainty calculated as a sum of uncorrelated uncertainty and all correlated sources in quadrature.
A global normalisation uncertainty of $0.5\%$ is not included in $\delta_{\rm ave,tot}$.
CME stands for the centre-of-mass energy of the measurement.
\label{tab:all3}}
\end{table}

\begin{sidewaystable}
\begin{tiny}
\begin{tabular}{c|rrrrrrrrrrrrrrrrrrrrrrrrrr}
\hline
\hline
 $\#$& $\gamma^{\rm ave}_1$ &$\gamma^{\rm ave}_2$ &$\gamma^{\rm ave}_3$ &$\gamma^{\rm ave}_4$ &$\gamma^{\rm ave}_5$ &
   $\gamma^{\rm ave}_6$ &$\gamma^{\rm ave}_7$ &$\gamma^{\rm ave}_8$ &$\gamma^{\rm ave}_9$ &$\gamma^{\rm ave}_{10}$ &
   $\gamma^{\rm ave}_{11}$ &$\gamma^{\rm ave}_{12}$ &$\gamma^{\rm ave}_{13}$ &$\gamma^{\rm ave}_{14}$ &$\gamma^{\rm ave}_{15}$ &
   $\gamma^{\rm ave}_{16}$ &$\gamma^{\rm ave}_{17}$ &$\gamma^{\rm ave}_{18}$ &$\gamma^{\rm ave}_{19}$ &$\gamma^{\rm ave}_{20}$  & 
   $\gamma^{\rm ave}_{21}$ &$\gamma^{\rm ave}_{22}$ &$\gamma^{\rm ave}_{23}$ &$\gamma^{\rm ave}_{24}$ &$\gamma^{\rm ave}_{25}$ &
   $\gamma^{\rm ave}_{26}$ \\
  & \permil &\permil &\permil &\permil &\permil &\permil &\permil &\permil &\permil &\permil &\permil &\permil &\permil &\permil &\permil &\permil &\permil &\permil &\permil &\permil &\permil &\permil &\permil &\permil &\permil &\permil \\
\hline
  118 &  -8.46 &   1.37 &   3.17 & -11.27 &   1.39 &   1.70 &   1.22 &   0.86 &   2.94 &   3.14 &  -4.40 &  -1.73 &  -7.26 &   3.19 &   0.65 &  -2.69 &   7.29 &  14.18 &  -7.90 &   1.22 &   1.47 &  -1.01 &   3.45 &   3.17 &   3.62 &  -1.11\\
  119 &  -8.46 &   0.93 &   2.26 &  -7.40 &   0.54 &   0.40 &  -1.09 &  -0.24 &   1.49 &   4.29 &  -2.28 &   0.96 &  -4.56 &   0.08 &   3.74 &   0.48 &   4.45 &   7.88 &  -5.60 &   3.07 &   0.26 &  -0.74 &  -1.96 &   0.33 &   1.05 &  -0.37\\
  120 &  -8.46 &   0.11 &   0.30 &  -1.03 &  -0.17 &  -0.41 &   0.00 &  -0.89 &  -0.86 &   3.43 &   1.27 &   2.20 &  -0.25 &  -1.76 &   1.91 &   1.85 &   0.70 &   0.43 &  -1.18 &   1.85 &  -0.04 &  -0.04 &  -1.59 &   0.15 &   1.05 &  -0.06\\
  121 &  -8.59 &  10.47 &   7.68 &   2.70 &   1.52 &   2.48 &  -1.22 &  -1.95 &   1.85 &   4.01 &   2.36 &  -2.94 &   0.75 &  -3.10 &  -2.97 &  -0.64 &   5.55 &   0.49 &   2.18 &  -2.22 &  -2.23 &   3.14 &  -0.14 &   1.46 &  -2.26 &  -1.59\\
  122 &  -8.46 &   0.01 &   0.08 &  -0.12 &  -0.06 &  -0.69 &   0.35 &  -0.60 &  -0.53 &   2.91 &   1.46 &   1.98 &  -0.03 &  -1.59 &   0.96 &   1.40 &   0.56 &  -0.69 &  -0.43 &   1.71 &  -0.11 &  -0.03 &  -0.89 &   0.47 &   1.39 &  -0.11\\
  123 &  -8.50 &   1.34 &   1.04 &   0.58 &   0.03 &  -0.48 &  -0.36 &  -1.00 &   0.99 &   3.18 &   2.42 &  -0.64 &  -0.63 &  -0.98 &  -1.87 &  -0.91 &   2.88 &   0.06 &   0.93 &  -1.14 &  -1.05 &   1.60 &   0.65 &   1.07 &  -2.30 &  -1.39\\
  124 &  -8.46 &  -0.45 &  -0.22 &   0.18 &  -0.13 &  -1.32 &  -0.10 &  -0.54 &  -0.01 &   2.96 &   2.05 &   0.49 &  -0.69 &  -0.83 &  -0.57 &  -0.43 &   2.06 &  -0.36 &   0.38 &  -0.49 &  -0.84 &   1.24 &  -0.05 &   0.83 &  -0.97 &  -0.93\\
  125 &  -8.47 &   0.42 &   0.34 &   0.90 &   0.62 &  -1.60 &   0.89 &   0.28 &  -0.78 &   3.33 &   0.85 &   2.24 &  -0.29 &  -1.48 &  -0.21 &   0.78 &  -1.42 &  -0.30 &  -0.55 &   2.22 &   0.93 &  -1.36 &   1.57 &   0.48 &  -1.24 &  -0.72\\
  126 &  -8.47 &  -0.16 &  -0.12 &   0.55 &   0.12 &  -1.64 &   0.40 &  -0.31 &  -2.32 &   4.68 &   1.84 &   2.28 &  -0.07 &  -1.54 &  -0.07 &   1.30 &   0.68 &  -0.91 &  -0.21 &   2.51 &   0.64 &  -1.14 &   1.36 &   0.68 &  -0.96 &  -0.79\\
  127 &  -8.47 &  -0.15 &  -0.39 &   0.67 &   0.16 &  -0.34 &   2.57 &  -0.18 &  -1.07 &   1.93 &   1.56 &   1.46 &  -0.25 &  -0.56 &  -1.76 &   0.28 &  -1.50 &   1.45 &   1.07 &   0.88 &   0.99 &  -0.63 &   1.90 &   0.81 &  -1.09 &  -0.82\\
  128 &  -8.47 &  -0.20 &  -0.55 &  -0.10 &  -0.69 &   0.95 &   2.17 &  -1.19 &  -2.02 &   0.46 &   1.55 &   1.40 &   0.82 &  -0.21 &  -0.21 &   0.48 &   0.35 &   1.92 &   2.57 &   1.03 &   0.19 &  -0.46 &   1.43 &   0.64 &  -1.41 &  -0.79\\
  129 &  -8.47 &  -0.03 &  -0.42 &   0.92 &   0.64 &   0.12 &   4.65 &   1.14 &  -0.15 &   2.55 &   1.11 &   2.21 &  -1.06 &  -0.39 &  -1.29 &  -0.07 &  -0.97 &   0.94 &   1.13 &   1.29 &   0.67 &  -0.46 &   1.06 &   0.59 &  -1.42 &  -0.89\\
  130 &  -8.48 &  -0.19 &   0.52 &   2.59 &   2.43 &  -6.37 &  -0.78 &   3.29 &   4.07 &   4.47 &  -0.50 &   1.58 &  -2.72 &  -0.82 &  -0.58 &  -3.79 &  -2.07 &   0.11 &   2.60 &   1.96 &  -1.09 &  -0.17 &  -0.54 &   0.50 &   0.44 &  -0.44\\
  131 &  -8.47 &   1.14 &   2.93 &  -8.35 &   2.27 &  -0.86 &   0.79 &   2.55 &   5.45 &   2.64 &  -4.75 &  -1.59 &  -7.97 &   3.23 &  -0.57 &  -4.39 &   5.86 &  11.44 &  -4.64 &   2.62 &   0.47 &  -1.14 &   1.96 &   2.49 &   3.18 &  -0.93\\
  132 &  -8.46 &   0.46 &   1.11 &  -3.62 &   0.02 &   0.09 &  -0.64 &  -0.79 &   0.37 &   3.16 &  -0.27 &   1.63 &  -2.11 &  -1.03 &   2.63 &   1.55 &   2.05 &   3.14 &  -2.99 &   2.60 &   0.02 &  -0.33 &  -2.38 &  -0.12 &   0.69 &  -0.12\\
  133 &  -8.70 &  24.85 &  18.62 &   7.45 &   5.55 &   2.79 &  -3.21 &  -0.87 &  -2.55 &   2.64 &  -1.56 &  -4.85 &   2.24 &  -6.11 &  -3.80 &   0.10 &   5.30 &   0.87 &   2.28 &  -1.70 &  -2.06 &   2.95 &  -0.09 &   1.41 &  -2.01 &  -1.37\\
  134 &  -8.46 &   0.06 &   0.15 &  -0.56 &  -0.20 &  -0.36 &   0.25 &  -0.85 &  -0.81 &   3.03 &   1.48 &   2.02 &  -0.12 &  -1.67 &   1.20 &   1.85 &   0.85 &  -0.55 &  -0.99 &   1.84 &  -0.02 &  -0.05 &  -1.19 &   0.34 &   1.25 &  -0.09\\
  135 &  -8.54 &   6.02 &   4.54 &   1.91 &   1.12 &   0.06 &  -0.36 &  -0.51 &  -2.38 &   1.40 &   1.01 &  -1.83 &  -0.24 &  -1.17 &  -1.65 &  -1.24 &   5.60 &   0.13 &   1.95 &  -2.34 &  -2.10 &   3.03 &  -0.41 &   1.22 &  -1.85 &  -1.37\\
  136 &  -8.46 &  -0.08 &   0.01 &   0.06 &  -0.27 &  -0.79 &  -0.14 &  -0.85 &   0.14 &   2.98 &   2.32 &   0.20 &  -0.56 &  -0.89 &  -0.58 &  -0.36 &   2.91 &  -0.49 &   0.52 &  -0.78 &  -1.14 &   1.65 &  -0.36 &   0.92 &  -0.76 &  -0.94\\
  137 &  -8.47 &  -0.47 &  -0.26 &   0.05 &  -0.28 &  -0.98 &  -0.11 &  -0.77 &   0.43 &   2.90 &   2.22 &   0.42 &  -0.56 &  -0.87 &  -0.56 &  -0.38 &   2.06 &  -0.29 &   0.42 &  -0.59 &  -0.89 &   1.31 &  -0.08 &   0.84 &  -0.97 &  -0.93\\
  138 &  -8.46 &  -0.09 &  -0.04 &   0.61 &   0.28 &  -1.64 &   0.72 &   0.02 &  -2.19 &   3.17 &   1.14 &   2.06 &  -0.30 &  -1.12 &  -0.16 &   0.86 &   0.21 &  -0.56 &  -0.04 &   2.19 &   0.58 &  -1.02 &   1.23 &   0.59 &  -0.96 &  -0.73\\
  139 &  -8.47 &  -0.16 &  -0.30 &   0.73 &   0.15 &  -0.91 &   1.43 &  -0.57 &  -0.57 &   2.31 &   1.71 &   1.45 &   0.02 &  -1.04 &  -1.69 &   0.45 &  -2.85 &   1.74 &   0.66 &   0.83 &   1.28 &  -0.87 &   2.05 &   0.69 &  -1.16 &  -0.74\\
  140 &  -8.47 &  -0.10 &  -0.33 &   0.45 &  -0.04 &  -0.10 &   1.94 &  -0.55 &  -0.27 &   1.66 &   1.39 &   1.69 &   0.22 &  -0.81 &  -0.93 &   0.46 &  -2.07 &   1.61 &   1.11 &   1.30 &   0.95 &  -1.02 &   1.87 &   0.56 &  -1.41 &  -0.75\\
  141 &  -8.47 &  -0.19 &  -0.48 &   0.09 &  -0.50 &   0.27 &   2.12 &  -0.96 &  -3.53 &   0.28 &   1.36 &   1.30 &   0.36 &   0.03 &  -0.81 &   0.56 &   0.41 &   2.14 &   2.91 &   1.43 &   0.40 &  -0.58 &   1.13 &   0.54 &  -1.24 &  -0.71\\
  142 &  -8.47 &  -0.12 &  -0.25 &   1.19 &   0.84 &  -1.51 &   2.91 &   1.00 &  -0.60 &   3.00 &   1.19 &   1.75 &  -1.40 &  -0.52 &  -1.95 &  -0.36 &  -1.55 &   1.29 &   1.41 &   1.41 &   0.74 &  -0.50 &   0.81 &   0.52 &  -1.18 &  -0.78\\
  143 &  -8.48 &  -0.09 &   1.02 &   2.43 &   2.38 &  -7.17 &  -3.89 &   3.29 &   7.94 &   2.93 &  -2.21 &   1.91 &  -1.14 &  -1.44 &   3.37 &  -5.15 &  -2.86 &  -1.06 &   1.68 &   1.42 &  -2.33 &  -0.45 &   0.29 &   0.43 &   0.06 &  -0.41\\
  144 &  -8.46 &  -0.12 &   0.11 &   0.51 &   0.13 &  -2.48 &  -0.89 &  -0.51 &  -3.53 &   2.40 &   0.95 &   1.63 &  -0.08 &  -1.08 &  -0.17 &   1.00 &   0.17 &  -0.34 &   0.07 &   2.24 &   0.63 &  -1.15 &   1.23 &   0.50 &  -0.84 &  -0.62\\
  145 &  -8.47 &  -0.16 &  -0.21 &   0.95 &   0.40 &  -1.56 &   1.84 &   0.28 &  -2.89 &   0.39 &   0.73 &   1.01 &  -0.30 &  -0.15 &  -1.20 &  -0.50 &  -1.04 &   1.44 &   1.17 &  -0.44 &   0.46 &   0.25 &   1.48 &   0.89 &  -0.56 &  -0.71\\
  146 &  -8.47 &  -0.04 &  -0.23 &   0.55 &   0.12 &  -0.47 &   2.03 &  -0.24 &  -1.47 &   0.84 &   0.92 &   1.60 &   0.03 &  -0.60 &  -0.86 &   0.41 &  -1.98 &   1.51 &   0.82 &   0.89 &   0.94 &  -0.74 &   1.61 &   0.53 &  -1.19 &  -0.68\\
  147 &  -8.47 &  -0.16 &  -0.29 &   0.28 &  -0.23 &  -0.51 &   1.12 &  -0.89 &  -1.95 &   0.84 &   1.27 &   1.24 &   0.15 &  -0.43 &  -1.08 &   0.47 &  -1.05 &   1.87 &   1.91 &   1.49 &   0.71 &  -0.89 &   1.38 &   0.47 &  -1.28 &  -0.67\\
  148 &  -8.47 &  -0.01 &  -0.31 &   0.96 &   0.65 &  -0.42 &   3.82 &   0.95 &  -0.55 &   2.02 &   0.85 &   2.12 &  -0.83 &  -0.48 &  -1.05 &  -0.10 &  -1.80 &   1.38 &   1.15 &   1.14 &   0.77 &  -0.54 &   1.05 &   0.47 &  -1.35 &  -0.77\\
  149 &  -8.48 &  -0.07 &   0.66 &   2.34 &   2.22 &  -5.85 &  -1.40 &   3.01 &   5.28 &   2.94 &  -1.32 &   1.93 &  -1.56 &  -1.16 &   1.53 &  -3.92 &  -3.41 &   0.24 &   1.78 &   1.08 &  -1.30 &  -0.21 &  -0.08 &   0.33 &   0.03 &  -0.38\\
\hline
\hline
\end{tabular}
\end{tiny}
\tablecaption{
Correlated systematic uncertainties given in per mill
relative to the cross section
measurement reported in \Tab~\ref{tab:all3}. The coefficients
 $\gamma^{\rm ave}_1,\gamma^{\rm ave}_{26}$ represent diagonalised correlated systematic uncertainties (see \Sec~\ref{subsec:aveprocedure}).
\label{tab:all3c}}
\end{sidewaystable}

\begin{sidewaystable}
\begin{tiny}
\begin{tabular}{c|rrrrrrr|rrrrrrr|rrrrrr|rrrrrr}
 \hline\hline
    & \multicolumn{7}{c}{\MB} 
    & \multicolumn{7}{c|}{\SVX} 
    & \multicolumn{6}{c}{\MB-97} 
    & \multicolumn{6}{c}{\SVX-95} 
  \\
  & $b_{\ee}$ &$b_{\theta_e}$ &$b_{E_{\rm had}}$ &
  $b_{\rm noise}$ &$b_{E^h_{\rm SpaCal}}$ &
   $b_{\gp}$ &$b_{\cal L}$ 
  & $b_{\ee}$ &$b_{\theta_e}$ &$b_{E_{\rm had}}$ &
  $b_{\rm noise}$ &$b_{E^h_{\rm SpaCal}}$ &
   $b_{\gp}$ &$b_{\cal L}$ 
  & $b_{\ee}$ & $b_{\theta_e}$ &$b_{E_{\rm had}}$ &
  $b_{\rm noise}$  &
   $b_{\gp}$ &$b_{\cal L}$ 
  & $b_{\ee}$ &$b_{\theta_e}$ &$b_{E_{\rm had}}$ &
   $b_{\gp}$ &   $b_{\rm dif}$  &
  $b_{\cal L}$ \\
 \hline
 $\delta_{ 1}$ &    0.0 &    0.0 &    0.0 &    0.0 &    0.0 &    0.0 &  -77.0 &    0.0 &    0.0 &    0.0 &    0.0 &    0.0 &    0.0 &  -28.2 &    0.1 &    0.1 &   -0.1 &    0.1 &    0.6 &  -49.8 &    0.0 &    0.0 &    0.0 &    0.1 &   -0.4 &  -28.2 \\ 
 $\delta_{ 2}$ &    0.5 &    1.3 &   -1.1 &    0.5 &    1.3 &   -3.2 &    0.9 &   -2.6 &  -10.8 &   -0.5 &    0.0 &    1.9 &   -5.7 &   -2.3 &    3.0 &    2.4 &   -3.8 &    1.8 &  -69.0 &   -1.5 &    1.9 &    1.8 &    0.2 &   -0.6 &   70.8 &    0.0 \\ 
 $\delta_{ 3}$ &    0.6 &    1.7 &    0.7 &    4.3 &    2.7 &   -8.3 &    1.1 &    0.7 &   -1.2 &    2.4 &    2.2 &    0.4 &  -20.6 &   -0.3 &    3.2 &    4.8 &    0.8 &   -5.7 &  -52.5 &    1.2 &   -2.6 &   -8.5 &  -26.1 &  -54.5 &  -53.7 &   -5.2 \\ 
 $\delta_{ 4}$ &   -1.1 &   -4.5 &   14.6 &    1.2 &   -7.6 &   27.0 &   -5.4 &   -0.7 &    2.7 &   10.6 &    2.5 &    2.2 &   87.9 &    2.6 &    6.3 &    6.8 &   -8.3 &   -4.0 &  -21.3 &    6.8 &   -0.9 &   -0.7 &   13.2 &   -9.4 &  -12.2 &   -0.3 \\ 
 $\delta_{ 5}$ &   -0.3 &    0.6 &   14.5 &    1.9 &   -1.9 &   -3.9 &   -4.9 &   -9.1 &   -8.4 &   26.1 &    1.8 &  -21.6 &  -24.4 &  -10.7 &    7.3 &    8.2 &  -12.3 &   -5.2 &  -17.7 &    5.4 &   -0.8 &   20.1 &   76.7 &   10.3 &  -22.0 &   14.5 \\ 
 $\delta_{ 6}$ &    4.4 &   10.2 &  -29.0 &  -13.9 &    4.0 &   -1.2 &    6.2 &   10.7 &   22.1 &  -48.5 &   -9.6 &   28.9 &    5.2 &   17.2 &  -11.6 &  -25.1 &  -15.2 &   36.0 &  -15.4 &  -24.4 &   -0.8 &   11.5 &   32.6 &   -2.9 &  -13.4 &    9.1 \\ 
 $\delta_{ 7}$ &   -0.2 &    6.7 &    8.0 &  -27.6 &   -8.2 &    4.6 &    2.3 &  -27.8 &  -21.3 &   10.7 &  -16.7 &  -26.1 &   -0.6 &  -11.0 &    1.5 &   -3.2 &  -58.2 &   47.7 &    2.0 &   -2.2 &    4.2 &   -7.8 &  -26.1 &    7.1 &   -5.5 &    9.2 \\ 
 $\delta_{ 8}$ &   -1.1 &    3.0 &   25.2 &    6.8 &   -3.7 &   -4.2 &  -13.8 &   27.2 &   72.9 &   12.5 &    8.5 &   -8.7 &  -11.1 &   24.8 &    9.9 &   17.2 &  -34.5 &    3.3 &   -2.0 &    7.2 &    3.8 &   10.3 &  -12.1 &    4.4 &    7.7 &    0.2 \\ 
 $\delta_{ 9}$ &   22.9 &   24.9 &    9.6 &   35.1 &    5.5 &   -7.7 &   14.7 &    3.3 &   -4.9 &   27.5 &   21.8 &   -0.8 &    5.1 &    3.8 &   -8.5 &  -65.6 &  -23.1 &  -18.2 &    1.0 &  -25.1 &   -2.0 &   -4.6 &   -4.0 &    1.3 &    1.2 &    0.0 \\ 
 $\delta_{10}$ &  -45.6 &  -44.5 &   36.1 &  -10.8 &   -5.8 &  -14.4 &  -20.2 &   10.0 &  -16.1 &  -11.4 &   -6.7 &   14.4 &   -2.4 &   14.2 &   -1.8 &  -38.8 &   -4.8 &  -13.0 &   -4.4 &    6.3 &    3.9 &   12.2 &   -9.1 &    5.1 &   -3.7 &   29.7 \\ 
 $\delta_{11}$ &  -15.0 &  -12.7 &   -1.5 &  -22.9 &   -5.5 &   18.8 &    5.7 &  -26.6 &   33.1 &    7.4 &  -10.9 &  -35.9 &   -3.1 &   -2.4 &  -16.0 &  -32.6 &   17.0 &   -1.0 &   14.4 &   -0.9 &   -3.0 &   -2.6 &   16.1 &  -51.4 &   22.6 &  -12.0 \\ 
 $\delta_{12}$ &  -12.5 &  -25.6 &    3.8 &    5.1 &    0.2 &    4.3 &   11.3 &   43.0 &  -23.5 &    5.9 &    8.2 &   11.7 &   -8.1 &    6.6 &   14.3 &    1.6 &  -25.5 &   18.9 &   13.6 &    4.7 &  -15.5 &  -37.2 &   24.6 &  -21.2 &    9.9 &  -45.3 \\ 
 $\delta_{13}$ &    0.0 &  -25.9 &  -29.4 &   13.0 &  -12.3 &   23.3 &    4.1 &   42.7 &   -7.1 &   -8.2 &   17.5 &  -61.3 &    1.0 &    0.3 &   -1.1 &   -7.4 &   13.8 &   12.5 &  -12.8 &  -13.5 &    9.8 &    8.5 &  -11.1 &   17.5 &   -9.0 &   12.2 \\ 
 $\delta_{14}$ &   11.8 &   22.2 &    4.3 &   -1.7 &   -2.0 &   -9.5 &  -10.3 &   33.6 &  -27.6 &    8.1 &    3.8 &    2.2 &    6.2 &    0.4 &   -4.4 &   14.1 &   -2.1 &   12.5 &   25.3 &   -0.5 &    8.6 &   46.0 &   -0.9 &  -52.3 &   19.0 &   28.8 \\ 
 $\delta_{15}$ &   -4.7 &  -51.0 &  -25.6 &   49.9 &    7.0 &  -17.9 &    7.0 &  -37.7 &    9.3 &    3.3 &   18.7 &   11.7 &    6.2 &  -10.8 &    2.1 &   14.4 &  -18.4 &   16.1 &    9.1 &   -8.5 &    4.3 &   19.8 &   -1.4 &  -14.6 &    5.2 &    6.7 \\ 
 $\delta_{16}$ &  -21.6 &   -7.1 &  -31.2 &  -29.5 &   15.8 &   12.2 &   13.0 &   16.2 &   11.5 &   64.6 &   -5.7 &   32.4 &   -1.8 &  -13.8 &    7.7 &   -8.1 &   11.5 &   10.9 &   -6.9 &  -12.3 &    1.1 &   19.7 &   -9.9 &   11.9 &   -5.6 &    0.2 \\ 
 $\delta_{17}$ &  -42.7 &   26.2 &   -7.8 &   14.5 &    9.2 &  -29.3 &  -16.2 &   -1.6 &   -1.8 &   18.8 &   19.5 &  -10.3 &   10.8 &   20.0 &  -50.2 &    9.1 &   15.6 &   30.1 &   -4.1 &   10.1 &   -4.1 &  -25.9 &    5.8 &    4.5 &   -1.2 &    6.5 \\ 
 $\delta_{18}$ &   43.2 &  -28.0 &   10.2 &  -27.1 &   -8.3 &  -68.0 &    6.8 &    8.2 &    6.9 &   12.4 &   -6.8 &  -12.4 &   19.6 &   -0.2 &   -0.6 &   -4.8 &   21.3 &   15.3 &   -4.1 &   -4.7 &   -5.2 &   -3.5 &    2.1 &    5.2 &   -2.0 &  -10.0 \\ 
 $\delta_{19}$ &   28.2 &  -13.7 &   44.6 &   10.9 &  -19.1 &   38.3 &    0.0 &   -7.1 &    1.8 &    8.9 &   18.4 &   25.5 &  -17.5 &    0.1 &  -20.1 &   -1.5 &   32.6 &   45.8 &   -4.4 &   -3.4 &    2.1 &   -2.2 &   -1.4 &    3.6 &   -2.2 &    5.8 \\ 
 $\delta_{20}$ &  -28.0 &   20.4 &   30.4 &   17.2 &   31.3 &   -8.7 &   17.2 &   -4.5 &    0.4 &  -14.8 &   -3.8 &  -17.4 &    7.3 &   -8.0 &   39.4 &   -8.6 &   24.0 &   30.5 &   -1.4 &   -7.6 &  -23.6 &   32.9 &   -6.4 &    5.8 &   -3.0 &  -25.5 \\ 
 $\delta_{21}$ &   -4.1 &    5.8 &   -6.0 &  -40.2 &   -3.3 &   -3.4 &   -4.3 &  -18.8 &   -3.0 &   -8.1 &   81.0 &    1.7 &   -0.8 &   12.4 &   32.2 &   -2.2 &    1.6 &   -1.3 &    3.4 &   -2.9 &    1.2 &   -2.8 &    0.8 &   -3.1 &    1.3 &    4.6 \\ 
 $\delta_{22}$ &    6.1 &   -9.3 &   -1.9 &  -14.5 &   -4.7 &    7.9 &    2.4 &   -2.4 &   -8.7 &   -4.6 &   15.6 &   -1.6 &   -2.2 &    6.3 &  -46.5 &    5.7 &  -17.8 &  -15.2 &   -5.2 &   10.3 &  -56.2 &   44.2 &   -9.9 &   10.4 &   -4.0 &  -31.1 \\ 
 $\delta_{23}$ &   -4.0 &   13.9 &  -26.0 &   18.5 &  -66.4 &   -6.0 &  -19.3 &   -7.5 &    0.1 &   11.0 &  -13.2 &    6.9 &   -2.9 &   13.7 &   33.0 &  -16.4 &   10.0 &   11.6 &    0.3 &   19.9 &  -36.7 &    3.1 &   -2.1 &    0.0 &    1.3 &    3.9 \\ 
 $\delta_{24}$ &  -13.7 &   10.0 &   -0.1 &    1.6 &  -35.7 &   -9.8 &    9.1 &   -3.2 &   -4.9 &   -3.2 &    5.1 &    5.4 &    0.7 &   -1.8 &  -10.9 &   -8.0 &   -1.7 &    1.1 &   -3.8 &   16.0 &   64.7 &   28.9 &    0.5 &    5.9 &   -6.3 &  -51.0 \\ 
 $\delta_{25}$ &  -28.4 &    7.5 &   18.0 &   -3.2 &  -45.0 &   -9.8 &   39.7 &    3.5 &    4.8 &   -3.4 &    1.5 &    2.7 &    1.8 &  -14.6 &   -8.8 &   27.6 &   -1.4 &  -15.2 &    2.1 &  -59.0 &  -10.5 &   -7.2 &    0.5 &   -1.9 &    1.6 &   10.7 \\ 
 $\delta_{26}$ &    5.8 &   -7.2 &   -1.2 &   -0.3 &    7.5 &    8.5 &   -2.0 &  -19.7 &  -20.0 &   16.0 &  -18.6 &   -7.1 &   -3.6 &   80.1 &    8.8 &   14.4 &    6.1 &   -5.0 &    1.5 &  -35.3 &    9.5 &    8.4 &    0.3 &    0.2 &   -0.9 &  -12.3 \\ 
 \hline\hline
\end{tabular}
\end{tiny}
\tablecaption{\label{tab:delta}Orthogonal transition matrix $U_{jk}$ from
 the original ($b_{\ee}$, etc)
to the diagonalised ($b^{\rm ave}_1,b^{\rm ave}_{26}$) 
systematic sources for the 
averaged H1 data. The matrix ellements are given in \%. }
\end{sidewaystable}
\clearpage




\clearpage



\clearpage


\end{document}